\documentclass[a4paper,11pt]{article}
\pdfoutput=1 
\usepackage{jheppub} 
\usepackage[T1]{fontenc} 
\usepackage{graphicx}
\usepackage{hyperref}
\usepackage{amsmath}
\usepackage{multirow}
\usepackage{slashed}
\usepackage{bm}
\usepackage{xcolor}

\title{\boldmath A Comprehensive Analysis of $B_s \to D_s^{**}\ell\nu_\ell$ Decays Within and Beyond the Standard Model}

\author{Karthik Jain,}
\author{Tarun Kumar,}
\author[1]{Barilang Mawlong \note{Corresponding author.}}
\author{and Shantanu Sahoo}

\affiliation{School of Physics,  University of Hyderabad, Hyderabad-500046,  India}

\emailAdd{19phph08@uohyd.ac.in}
\emailAdd{21phph18@uohyd.ac.in}
\emailAdd{barilang@uohyd.ac.in}
\emailAdd{shantanusahoo.ra@uohyd.ac.in}

\abstract{ We examine the exclusive semileptonic decays $B_s \to D_s^{**} \ell \nu_\ell$, with $D_s^{**} =$ $\bigl\{D_{s0}^*,D_{s1}^*,D_{s1},D_{s2}^*\bigr\}$, within the Standard Model and beyond, using form factors evaluated in the Heavy Quark Effective Theory, including corrections up to $ \mathcal{O}(\alpha_s, \Lambda/{m_Q})$. A data-driven approach is employed to extract Heavy Quark Effective Theory parameters, and the resulting synthetic data are used to parameterize the form factors via the $z$-expansion. With the resulting form factor information across the full kinematic region, we compute various observables derived from the two-fold angular decay distribution, and predict precise lepton flavor universality ratios: $R_{D_{s0}^*}= 0.158(20)$, $R_{D_{s1}^*}= 0.045(5)$, $R_{D_{s1}}= 0.073(4)$, $R_{D_{s2}^*} = 0.066(9)$. We also analyse potential new physics effects using the Weak Effective Theory and the Standard Model Effective Field Theory, performing a global analysis considering both real and complex Wilson coefficients. Furthermore, we investigate new physics contributions arising from the general Two Higgs Doublet Model. We evaluate the sensitivity of decay observables to new physics, highlighting their potential to probe deviations from the Standard Model in future measurements. Notably, the scalar and tensor new physics operators induce large sensitivity, with some observables deviating by more than $2 \sigma$ from Standard Model predictions. }


\begin{document} 
\maketitle
\flushbottom

\section{Introduction}\label{Introduction}

One of the most compelling developments in flavor physics is the measurement of the lepton flavor universality (LFU) ratios, $R(D^{(*)})$ $\equiv$ $\mathcal{B}(\bar{B} \to D^{(*)}\tau^-\bar{\nu}_\tau)/\mathcal{B}(\bar{B} \to D^{(*)}\ell^-\bar{\nu}_\ell)$, with $\ell = e,\mu$, which currently exhibit a combined deviation of about $3.8\sigma$ from the Standard Model (SM) expectations~\cite{HFLAV}. These anomalies in $b \to c\ell \nu_\ell$ transitions, particularly in channels involving the $\tau$ lepton, provide hints of possible new physics (NP). This further motivates for an in-depth analysis of the charged-current sector by studying additional decay channels that follow the same quark-level transition. Among them, the $B_{s}$ meson and its phenomenology have attracted interest from both the phenomenological and experimental communities over the past decades \cite{Albrecht:2024oyn,HeavyFlavorAveragingGroupHFLAV:2024ctg,Bernlochner:2016bci,Gubernari:2023rfu,Harrison:2023dzh}. 

In this work, we focus on the $B_s \to D_s^{**}\ell \nu_\ell$ decays, where the $D_{s0}^*$ is a scalar particle, the $D_{s1}^*$ and $D_{s1}$ are axial-vector particles, and the $D_{s2}^*$ is a tensor particle. The $D_{sJ}^{(*)}$ states have been observed to have very narrow decay widths \cite{BaBar:2006eep,BaBar:2011vbs,BESIII:2018fpo,LHCb:2014ott,LHCb:2011rmd,BaBar:2006gme}, which is advantageous since broad resonances are difficult to distinguish from continuum backgrounds in experimental analyses \cite{Becirevic:2012te}. Regarding collider production capabilities, the Belle experiment produced $(6.53 \pm 0.66) \times 10^6$ $B_s\bar{B_s}$ pairs during its run which concluded in 2010. The Belle II experiment, expected to run until 2031, is projected to collect $0.5 \times 10^9$ $B_s\bar{B_s}$ pairs, nearly 50 times more than Belle. Similarly, the LHCb experiment is predicted to produce $4600 \times 10^9$ $B_s$ mesons by the end of Runs 5 and 6, about 50 times more than the $84 \times 10^9$ $B_s$ mesons produced in Run 2 \cite{Belle:2015ftp,Bernlochner:2021vlv}.
Given the enormous increase in data on $B_s$ mesons, the $B_s \to D_s^{**} \ell \nu_\ell$ (with $\ell=\tau, \mu, e$) decays are anticipated to be measured in the near future. Furthermore, the $B_s \to D_{s1}^{*}\ell  \nu_\ell$ and $B_s \to D_{s2}^{*}\ell  \nu_\ell$ modes constitute important backgrounds for the $R(D^{(*)})$ measurements. Since the $D_s^{**}$ states predominantly decay into $D^{(*)}K$, they can mimic $R(D^{(*)})$ signals at colliders. Currently, there is a $3\%$ correction to the measurement of $R(D^*)$ at LHCb due to feed-down from $B_s \to D_s^{**}\ell \nu_\ell$, with a relative uncertainty of about $50\%$. With increased production of $B_s$ mesons, these decays can introduce considerable uncertainties in the $R(D^{(*)})$ measurements. Therefore, to improve the precision of the $R(D^{(*)})$ measurements, it is crucial to determine the $B_s \to D_s^{**}\ell \nu_\ell$ branching fractions precisely~\cite{Bernlochner:2021vlv}. Additionally, the structure of the $D_{s0}^*$ and $D_{s1}^*$ states is uncertain, as there is ambiguity in their interpretation, whether they are simple $q\bar{q}$ states or $DK$ molecular states~\cite{Huang:2004et}. Precise analysis of the $B_s \to D_s^{**}\ell\nu_\ell$ decays may shed further light on the nature of these states. For the aforementioned reasons, the focus on the semileptonic $B_s \to D_s^{**}\ell\nu_\ell$ decays is crucial.
 
To carry out a phenomenological analysis of the $B_s \to D_s^{**}\ell\nu_\ell$ transitions, it is essential to determine the non-perturbative inputs, that is the form factors in the full kinematic region. In the absence of Lattice QCD (LQCD) results for the $B_s \to D_s^{**}$ transition form factors, our study relies on the available form factor calculations within the Heavy Quark Effective Theory (HQET) framework. Following~\cite{Bernlochner:2016bci,Bernlochner:2017jxt,Bernlochner:2017jka}, we utilize HQET form factors that incorporate upto $\mathcal{O}( \alpha_s, \Lambda/m_{Q})$ corrections to the leading order. The numerical values of the form factor parameters are determined through a data-driven analysis. Currently, there are no measurements for the branching fractions of $B_s \to D_s^{**}\ell\nu_\ell$. However, the available measurements of $B \to D^{**}\ell\nu_\ell$ can be utilized as done in~\cite{Bernlochner:2016bci,Bernlochner:2017jxt}. New measurements from Belle~\cite{Belle:2022yzd} related to the $B \to D^{**}\ell\nu_\ell$ cascade decays are available, and thus a data-driven analysis of the form factors is performed in this work based on these updated measurements. Using the synthetic data generated with the HQET parameters, we also perform a $\chi^2$ analysis to parameterize the form factors with the $z$- parametrization. A model-independent study was carried out in ref.~\cite{Gubernari:2023rfu} within the Light Cone Sum Rule (LCSR) framework. The focus was exclusively on the $B_s \to D_{s0}^*\ell\nu_\ell$ channels, providing the transition form factor using the LCSR method, which is reliable at $q^2 = 0$. In our work, we perform a combined analysis of the $B_s \to D_{s0}^*\ell\nu_\ell$ mode with LCSR and HQET to estimate the form factor shape in the entire physical $q^2$ region.

In the second part of the analysis, we probe the NP sensitivity of the observables of $B_s \to D_s^{**}\ell\nu_\ell$, where NP is assumed to be present only in the decays involving the $\tau$ lepton. The analysis considers benchmark points of the NP couplings determined from current data of $b \to c\tau\nu_\tau$ modes. The $B$ physics processes occur at the bottom-quark mass scale $\mu_b$ $\sim 4.2$ GeV and are described by an effective Lagrangian obtained by integrating out the heavy SM and NP degrees of freedom. This description is called the low-energy Effective Field Theory (LEFT), also known as the Weak Effective Theory (WET) \cite{Buchalla:1995vs}. Since this Effective Field Theory (EFT) is defined below the electroweak (EW) scale, it is invariant only under $SU(3)_C \otimes U(1)_{em}$. If NP effects of operators beyond the EW scale are to be analysed, the Standard Model Effective Field Theory (SMEFT) provides an interesting model-independent tool for doing that. The SMEFT is an extension of the SM where higher-dimensional operators are added to the SM Lagrangian~\cite{Grzadkowski:2010es}. These operators are constructed such that they have the same field content as the SM and are also invariant under the full SM $SU(3)_C \otimes SU(2)_L \otimes U(1)_Y$  gauge group. They are suppressed by powers of the high-energy scale $\Lambda$, representing the energy scale at which NP beyond the SM becomes relevant. The SM Lagrangian is of dimension-4 and the first higher-dimensional operators arise at dimension-5. However, these are not relevant to our current work as they only contribute towards neutrino mass generation. Thus, the lowest dimension operators relevant to semileptonic (leptonic) $b \to c\ell\nu_\ell$ transitions, which appear at dimension-6, are considered. We do not include higher-dimensional operators beyond dimension-6 as their effects are suppressed by the scale $\Lambda$. In this work, we perform a global analysis to extract the SMEFT Wilson coefficients (WCs) of dimension-6 operators, and we explore the NP sensitivity for the considered decay modes.

Further, we examine the extension of the SM in a model-dependent scenario by considering the Two Higgs Doublet Model (2HDM)~\cite{Branco:2011iw}. This extension is realized by introducing an additional $SU(2)_L$ doublet into the Yukawa Lagrangian of the SM. In literature, the 2HDM has been applied for various motivations, such as providing new sources of Charge-Parity (CP) violation, which in turn could explain the baryon asymmetry observed in the universe \cite{Fromme:2006cm,Cline:2011mm,Basler:2021kgq}. Similarly, the $\mu-\tau$ flavor violations in the 2HDM may also account for the muon $(g-2)$ anomaly \cite{Omura:2015nja,Omura:2015xcg,Tobe:2016qhz}. The 2HDM is further attractive because the new Higgs particles are within the direct search limits of the experimental colliders. In our work, the interest in this model stems from the fact that it also offers rich phenomenology in the flavor sector \cite{Iguro:2018qzf,Iguro:2023jju}. Various types of 2HDMs are defined based on the couplings between the Higgs doublets and the fermions. For example, in the Type I 2HDM, all the fermions only couple to one of the Higgs doublets, whereas in Type II 2HDM, the up quarks couple to one of the Higgs doublets and the down quarks and leptons couple to the other doublet. However, these models alone cannot account for the observed flavor anomalies \cite{BaBar:2012obs}. On the other hand, the Type III 2HDM with generic flavor structures, or the General 2HDM (G2HDM), allows both Higgs doublets to couple to all the fermions. It has been shown that the G2HDM can simultaneously explain $R(D)$ and $R(D^*)$ anomalies \cite{Kumar:2022rcf}. Consequently, in this work, we test the NP sensitivity of the $B_s \to D_s^{**}\ell\nu_\ell$ modes within the G2HDM framework.

The outline of this article is as follows. In section~\ref{sec:theor}, we present the theoretical framework for the $B_s \to D_s^{**}\ell\nu_\ell$ decays, including the hadronic matrix elements and the observables relevant to these decays. In section~\ref{sec:phen}, we introduce the HQET framework and present our fit results for the HQET form factor parameters. The model-independent description of form factors using the z-parametrization is given in section~\ref{sec:form_factor}, and the SM predictions for various observables are detailed in section~\ref{sec:SM_pred}. The NP sensitivity test of the observables is presented in section~\ref{sec:NP}, and the conclusions of this work are given in section~\ref{sec:concl}.

\section{Theoretical Framework}\label{sec:theor}
The effective Lagrangian including all possible four-fermion operators for the $b \to c \ell \nu_\ell$ transitions, describing these processes in the Weak Effective Theory, is given by \cite{Dassinger:2008as, Tanaka:2012nw},
\begin{eqnarray}\label{eq:eff_ham}
\mathcal{L}_{\text{eff}} = -\frac{4 G_F}{\sqrt{2}}  V_{cb} \left[(1 + C_{V_1}) \mathcal{O}_{V_1} + C_{V_2} \mathcal{O}_{V_2} + C_{S_1} \mathcal{O}_{S_1} + C_{S_2} \mathcal{O}_{S_2} + C_T \mathcal{O}_T\right]\, , \label{Leff}
\end{eqnarray} where $G_F$ is the Fermi constant, $V_{cb}$ is the CKM matrix element and the Wilson coefficients $C_i'$s are zero in the Standard Model. Assuming the neutrinos to be left-handed, the WET operators are defined as follows
\begin{eqnarray}
\mathcal{O}_{V_1} &=& \left(\bar{c}_L \gamma^\mu b_L\right) \left(\bar{\ell}_L \gamma_\mu \nu_{\ell L}\right)\, , \hspace{1cm}
\mathcal{O}_{V_2} = \left(\bar{c}_R \gamma^\mu b_R\right) \left(\bar{\ell}_L \gamma_\mu \nu_{\ell L}\right)\, , \nonumber \\
\mathcal{O}_{S_1} &=& \left(\bar{c}_L b_R\right) \left(\bar{\ell}_R \nu_{\ell L}\right)\, , \hspace{2cm}
\mathcal{O}_{S_2} = \left(\bar{c}_R b_L\right) \left(\bar{\ell}_R \nu_{\ell L}\right)\, , \nonumber \\
\mathcal{O}_{T} &=& \left(\bar{c}_R \sigma^{\mu\nu} b_L\right) \left(\bar{\ell}_R \sigma_{\mu\nu} \nu_{\ell L}\right)\, . \label{eq:wet_ops}
\end{eqnarray}
The WCs $C_i(\mu)$ are defined at $\mu=m_b$ and they contain information about the heavy degrees of freedom which are integrated out. With this effective Lagrangian, we can define some low-energy flavor observables that can be measured in the near future. In the following sections, we will briefly define and discuss these observables.

\subsection{Hadronic Matrix Elements and Form Factors}
In general, the heavy-to-light transition hadronic matrix elements of the currents appearing in the Lagrangian are parametrized in terms of non-perturbative form factors. For the $B_s \to D_{s0}^* \ell \nu_\ell$ transitions, the hadronic matrix elements are given by \cite{Gubernari:2023rfu,Cheng:2023knr}
\begin{eqnarray}\label{eq:hadmat_sca}
	\langle D_{s0}^*(k) \vert \bar{c} \gamma_\mu \gamma_5 b \vert B_s(p) \rangle &=& i \left[(p+k)_\mu - \frac{m_{B_s}^2 - m_{D_{s0}^*}^2}{q^2}q_\mu\right] f_+(q^2) + i q_\mu\frac{m_{B_s}^2 - m_{D_{s0}^*}^2}{q^2} f_0(q^2)\, , \nonumber\\
	\langle D_{s0}^*(k) \vert \bar{c} \sigma_{\mu\nu} \gamma_5 q^\mu b \vert B_s(p) \rangle &=& -\frac{f_T(q^2)}{m_{B_s} + m_{D_{s0}^*}} \left[ q^2(p + k)_\nu - (m_{B_s}^2 - m_{D_{s0}^*}^2)q_\nu \right] \, ,
\end{eqnarray}
where $p_\mu$ and $k_\mu$ are the four momenta of the parent hadron and daughter hadron, respectively, and $q_\mu = p_\mu - k_\mu$ is the momentum transferred to the lepton pair. Here, $f_+(q^2)$ and $f_0(q^2)$ are the form factors describing the hadronic transitions in the SM, and $f_T(q^2)$ is the tensor form factor arising from the tensor current beyond the SM. The vector current transition is zero as it cannot induce the required parity flip for the $0^- \to 0^+$ transition. To avoid the divergence at $q^2 = 0$, a kinematic constraint $f_+(0) = f_{0}(0)$ is imposed.


For the $B_s \to D_{s1}^{(*)} \ell \nu_\ell$ decays, the hadronic transition matrix elements are given by \cite{Gubernari:2022hrq,Di:2025hdu}
\begin{eqnarray}\label{eq:hadmat_axvec}
\langle D_{s1}^{(*)}(k,\epsilon) \vert \bar{c} \gamma_\mu \gamma_5 b \vert B_s(p) \rangle &=& \varepsilon_{\mu\nu\alpha\beta}\epsilon^{*\nu}k^\alpha(p-k)^\beta \frac{2 A(q^2)}{m_{B_s} + m_{D_{s1}^{(*)}}}\, , \nonumber\\
\langle D_{s1}^{(*)}(k,\epsilon) \vert \bar{c} \gamma_\mu b \vert B_s(p) \rangle &=& -i \epsilon_\mu^* (m_{B_s} + m_{D_{s1}^{(*)}}) V_1(q^2) + i (p+k)_\mu (\epsilon^*\cdot q) \frac{V_2 (q^2)}{m_{B_s} + m_{D_{s1}^{(*)}}} \nonumber\\ &&+ i q_\mu (\epsilon^*\cdot q)\frac{2 m_{D_{s1}^{(*)}}}{q^2} (V_3 (q^2) - V_0(q^2))\, , \nonumber \\
\langle D_{s1}^{(*)}(k,\epsilon) \vert \bar{c} \sigma_{\mu\nu} \gamma_5 q^\nu b \vert B_s(p) \rangle &=& -2 T_1(q^2) \varepsilon_{\mu\nu\alpha\beta} \epsilon^{*\nu} p^\alpha k^\beta \, , \nonumber \\
\langle D_{s1}^{(*)}(k,\epsilon) \vert \bar{c} \sigma_{\mu\nu} q^\nu b \vert B_s(p) \rangle &=& -i T_2(q^2) \left[(m_{B_s}^2 - m^2_{D_{s1}^{(*)}}) \epsilon^{*}_{\mu} - (\epsilon^*\cdot q)(p+k)_\mu \right] \nonumber \\ && -i T_3(q^2)(\epsilon^*\cdot q)\left[q_\mu - \frac{q^2}{m_{B_s}^2 - m^2_{D_{s1}^{(*)}}}(p+k)_\mu\right]\, ,
\end{eqnarray}
where $\epsilon_\mu$ is the polarization vector of the daughter hadron. We follow the convention $\varepsilon_{0123} = +1$ in this work. The functions $V_1(q^2)$, $V_2(q^2)$, $V_3(q^2)$, $V_0 (q^2)$, $A(q^2)$ are the form factors which describe SM transitions, and $T_1(q^2)$, $T_2(q^2)$, $T_3(q^2)$ are the tensor form factors. Although the $B_s \to D_{s1}^{(*)}$ transition is a $0^- \to 1^+$ transition, the vector current transition here does not become zero as the polarization vector of the daughter hadron allows for the construction of necessary Lorentz structures to produce non-zero matrix elements. The form factor $V_3(q^2)$ is not independent and is related to the $V_1(q^2)$ and $V_2(q^2)$ form factors as
\begin{eqnarray}\label{eq:kin_const_vec}
2 m_{D_{s1}^{(*)}} V_3(q^2) = (m_{B_s} + m_{D_{s1}^{(*)}}) V_1(q^2) - (m_{B_s} - m_{D_{s1}^{(*)}}) V_2(q^2)\, ,
\end{eqnarray}
with the  kinematic constraint $V_0(0) = V_3(0)$ imposed to avoid singularities at $q^2 = 0$.

For the $B_s \to D_{s2}^* \ell \nu_\ell$ transitions, the hadronic matrix elements are given by  \cite{Mandal:2019vwq}
\begin{eqnarray}\label{eq:hadmat_tens}
\langle D_{s2}^*(k,\epsilon) \vert \bar{c} \gamma_\mu b \vert B_s(p) \rangle &=& \varepsilon_{\mu\nu\alpha\beta}\epsilon^{*\nu}k^\alpha p^\beta \frac{2i V(q^2)}{m_{B_s} + m_{D_{s2}^*}} \, ,\nonumber\\
\langle D_{s2}^*(k,\epsilon) \vert \bar{c} \gamma_\mu \gamma_5 b \vert B_s(p) \rangle &=& 2 m_{D_{s2}^*} A_0(q^2) \frac{\epsilon^*.q}{q^2}q_\mu + (m_{B_s} + m_{D_{s2}^*}) A_1(q^2) \left[\epsilon^*_\mu - \frac{\epsilon^*\cdot q}{q^2}q_\mu\right] \nonumber \\ && -A_2(q^2) \frac{\epsilon^*\cdot q}{(m_{B_s} + m_{D_{s2}^*}) A_1(q^2)} \left[(p+k)_\mu - \frac{m^2_{B_s} - m^2_{D_{s2}^*}}{q^2}q_\mu \right]\, ,\nonumber\\
\langle D_{s2}^{(*)}(k,\epsilon) \vert \bar{c} \sigma_{\mu\nu} q^\nu b \vert B_s(p) \rangle &=& \varepsilon_{\mu \nu \rho \sigma} \epsilon_T^{*\nu} k^\rho p^\sigma 2 T_1(q^2)\, , \nonumber \\
\langle D_{s2}^{(*)}(k,\epsilon) \vert \bar{c} \sigma_{\mu\nu} \gamma_5 q^\nu b \vert B_s(p) \rangle &=& \left[(m_{B_s}^2 - m_{D_{s2}^*}^2) \epsilon_T^{*\mu} - (\epsilon_T^*\cdot q)(p + k)_\mu\right] T_2 (q^2) \nonumber \\ && - (\epsilon_T^*\cdot q) \left[q_\mu - \frac{q^2}{m_{B_s}^2 - m_{D_{s2}^*}^2}(p + k)_\mu\right] T_3(q^2)\, ,
\end{eqnarray} 
where $\epsilon_T^{\mu} = \epsilon^{\mu\nu}q_\nu/m_{B_s}$ and $\epsilon^{\mu\nu}$ is the spin-2 polarization tensor of the daughter hadron which is symmetric and traceless, and is constructed from the spin-1 polarization vector. In the above equations, $V(q^2)$, $A_0(q^2)$, $A_1(q^2)$, $A_2(q^2)$ are the form factors which appear in the SM, and $T_1(q^2)$, $T_2(q^2)$, $T_3(q^2)$ are the tensor form factors. At $q^2=0$, the form factors $A_1(q^2)$, $A_2(q^2)$ and $A_0(q^2)$ satisfy the kinematic relation 
\begin{eqnarray}\label{eq:kin_const_ten}
2 m_{D_{s2}^*} A_0(0) = (m_{B_s} + m_{D_{s2}^*}) A_1(0) - (m_{B_s} - m_{D_{s2}^*}) A_2(0)\, .
\end{eqnarray}
For the scalar and pseudoscalar transitions, the corresponding hadronic matrix elements can be derived using the equations of motion provided in \cite{Sakaki:2013bfa}. These matrix elements, and hence the associated form factors, are expressed in terms of the vector and axial-vector form factors. 

\subsection{Decay Distribution and Related Observables}\label{sec:obs}

The differential decay rate for the semileptonic decay $B_s \to M \ell \nu_\ell$ can be constructed from the currents appearing in the Lagrangian by convoluting the transition probability $|\mathcal{M}|^2$ with the three-body phase space $d\phi_3$ as \cite{Tanaka:2012nw}
\begin{eqnarray}\label{eq:convolution_int}
d\Gamma^{\lambda_\ell}_{\lambda_M} = \frac{1}{2 m_{B_s}} \sum_{\lambda_\ell,\lambda_M} \vert \mathcal{M}^{\lambda_\ell,\lambda_M} (q^2,\cos\theta_\ell) \vert^2 d\Phi_3\, ,
\end{eqnarray}
where $\theta_\ell$ is the angle between the charged lepton ($\ell$) and the opposite direction of motion of the daughter meson ($M$) in the rest frame of the $W^*$ boson, $\lambda_\ell$ is the helicity of $\ell$ and $\lambda_M$ is the helicity of $M$. The three-body phase space is calculated as
\begin{eqnarray}
d\Phi_3 = \frac{\sqrt{Q_+Q_-}}{256 \pi^3 m_{B_s}^2} \left(1 - \frac{m_\ell^2}{q^2}\right) dq^2 d\cos\theta_\ell \, ,
\end{eqnarray}
with $Q_\pm = (m_{B_s} \pm m_{D_s^{**}})^2 - q^2$.  In the SM, $\mathcal{M}^{\lambda_\ell,\lambda_M}$ is given as
\begin{eqnarray}
\mathcal{M}_{\text{SM}}^{\lambda_\ell\lambda_M} = \frac{G_F}{\sqrt{2}} V_{cb} \sum_\lambda \eta_\lambda H^{\lambda_M}_{V_1,\lambda} L^{\lambda_\ell}_{\lambda} \, \,,
\end{eqnarray}
 where $H$ and $L$ are the hadronic and leptonic helicity amplitudes, and $\eta_\lambda$ is the metric factor that is related to the metric tensor $g_{\mu \nu}$. The remaining decay amplitudes associated with new physics contributions are given by
\begin{eqnarray}
	\mathcal{M}_{V_i}^{\lambda_\ell,\lambda_M} &=& \frac{G_F}{\sqrt{2}} V_{cb} C_{V_i} \sum_\lambda \eta_\lambda H^{\lambda_M}_{V_1,\lambda} L^{\lambda_\ell}_{\lambda} \hspace{1cm} (i = 1,2)\, , \nonumber\\
	\mathcal{M}_{S_i}^{\lambda_\ell,\lambda_M} &=& -\frac{G_F}{\sqrt{2}} V_{cb} C_{S_i} H^{\lambda_M}_{S_1,\lambda} L^{\lambda_\ell}_{\lambda} \hspace{1cm} (i = 1,2)\, , \nonumber\\
	\mathcal{M}_{T}^{\lambda_\ell,\lambda_M} &=& -\frac{G_F}{\sqrt{2}} V_{cb} C_T \sum_{\lambda,\lambda^\prime} \eta_\lambda \eta_{\lambda^\prime} H^{\lambda_M}_{\lambda \lambda^\prime} L^{\lambda_\ell}_{\lambda\lambda^\prime} \, \, ,
\end{eqnarray}
where $\lambda$ and $\lambda^\prime$ are the helicities of the virtual vector boson. The hadronic and leptonic helicity amplitudes are obtained by contracting the respective matrix elements of the relevant operators with the polarization vector of the $W^*$ boson. Further details on these amplitudes can be found in~\cite{Tanaka:2012nw}. The hadronic helicity amplitudes are related to the underlying form factors and the explicit expressions for them are provided in appendix~\ref{app:ang_coeff}. Using the convolution integral defined in eq.~\ref{eq:convolution_int}, one can derive the total decay distribution. The two-fold differential decay rate then has the general form
\begin{eqnarray}\label{eq:ang_decay_dist}
\frac{d^2\Gamma}{dq^2 d\cos\theta_\ell} = a(q^2) + b(q^2)\cos\theta_\ell + c(q^2)\cos^2\theta_\ell \, \, .
\end{eqnarray}
The coefficients $a(q^2)$, $b(q^2)$ and $c(q^2)$ for each decay mode are given in appendix~\ref{app:ang_coeff}. The angular decay distribution of eq.~\ref{eq:ang_decay_dist} can be used to construct additional low-energy observables. The decay rate is obtained by integrating over $\cos\theta_\ell$, and within the SM it is given by
\begin{eqnarray}\label{eq:sm_decay_rate}
\frac{d\Gamma}{dq^2} &=& \frac{G_F^2 V_{cb}^2 q^2}{192 \pi^3 m_{B_s}^3} \sqrt{\lambda(m_{B_s}^2,m_{D_s^{**}}^2,q^2)} \left(1 - \frac{m_\ell^2}{q^2}\right)^2  \nonumber \\
&& \times  \Bigg[\biggl(1 + \frac{m_\ell^2}{2q^2}\biggr)\left(|H_0|^2 + |H_+|^2 + |H_-|^2\right)+ \frac{3}{2}\frac{m_\ell^2}{q^2}|H_t|^2\Bigg] \nonumber\\
&=& 2 \left[a(q^2) + \frac{c(q^2)}{3}\right] \, ,
\end{eqnarray}
where the K{\"a}llen function $\lambda(m_{B_s}^2,m_{D_s^{**}}^2,q^2) = Q_+ Q_-$ .
In the SM, there are four allowed hadronic helicity amplitudes, $H_\pm$, $H_0$, $H_t$, corresponding to the four helicity states of the virtual $W^*$ boson. These amplitudes satisfy the relation $\lambda_1 = \lambda_2 - \lambda$, where $\lambda_1$, $\lambda_2$ and $\lambda$ denote the helicities of the parent hadron, the daughter hadron and the $W^*$ boson, respectively. For light leptons, the helicity amplitude $H_t$ is suppressed in the SM due to the factor $m_\ell^2/q^2$ associated with the amplitude.

Apart from the decay rate, other useful observables can also be constructed from the two-fold angular decay distribution defined in eq.~\ref{eq:ang_decay_dist}. In the decay rate distribution, the angular information of the process is lost. The angular information can be recovered by defining the forward-backward asymmetry on the lepton side which is given as
\begin{eqnarray}
\mathcal{A}^{D_s^{**}}_{fb,\ell}(q^2) &=& \frac{\int_{0}^{1}\frac{d^2\Gamma}{dq^2d\cos\theta_\ell}d\cos\theta_\ell - \int_{-1}^{0}\frac{d^2\Gamma}{dq^2d\cos\theta_\ell}d\cos\theta_\ell}{\int_{-1}^{1}\frac{d^2\Gamma}{dq^2d\cos\theta_\ell}d\cos\theta_\ell} \nonumber\\
&=& \left(\frac{d\Gamma}{dq^2}\right)^{-1} \times  b(q^2) \, \, ,
\end{eqnarray}
the convexity parameter as
\begin{eqnarray}
C_{D^{**}_s}^\ell(q^2) = \frac{1}{d\Gamma/dq^2}\frac{d^2}{d(\cos\theta_\ell)^2} \left(\frac{d^2\Gamma}{dq^2d\cos\theta_\ell}\right) =\left(\frac{d\Gamma}{dq^2}\right)^{-1} \times 2 c(q^2) \, \, .
\end{eqnarray}
We also define the lepton polarization asymmetry as
\begin{eqnarray}
P^{D^{**}_s}_\ell(q^2) = \frac{\frac{d\Gamma^{\lambda_\ell = 1/2}(q^2)}{dq^2} - \frac{d\Gamma^{\lambda_\ell = -1/2}(q^2)}{dq^2}}{\frac{d\Gamma^{\lambda_\ell = 1/2}(q^2)}{dq^2} + \frac{d\Gamma^{\lambda_\ell = -1/2}(q^2)}{dq^2}} \, \, ,
\end{eqnarray}
the LFU ratio as
\begin{eqnarray}
R_{D_s^{**}}(q^2) = \frac{d\Gamma^\tau(q^2)}{dq^2} \bigg/ \frac{d\Gamma^\mu(q^2)}{dq^2} \, \, ,
\end{eqnarray}
and the longitudinal polarization of the daughter hadron is given as
\begin{eqnarray}
F_L^{D^{**}_s}(q^2) = \frac{\frac{d\Gamma^{\lambda_{D_s^{**}} = 0}(q^2)}{dq^2}}{\frac{d\Gamma^{\lambda_{D_s^{**}} = 0}(q^2)}{dq^2} + \frac{d\Gamma^{\lambda_{D_s^{**}} = +1}(q^2)}{dq^2} + \frac{d\Gamma^{\lambda_{D_s^{**}} = -1}(q^2)}{dq^2}} \, \, .
\end{eqnarray}
These observables are linearly independent and provide complementary information about the decay of interest. They are related to the hadronic helicity amplitudes, which in turn depend solely on the shape of the form factors. Therefore, precise determination of the form factor shapes is essential for predicting these observables. These observables will serve as important tests of the SM once they are measured in future experiments. Any deviations from the predicted value could provide hints of potential NP beyond the SM. 

\section{Form Factor Shape in HQET Framework}~\label{sec:phen}
In this section, we estimate the form factors relevant for the $B_s \to D_s^{**}\ell\nu_\ell$ decay modes. All observables are expressed in terms of helicity amplitudes, which depend on the transition form factors. Since LQCD inputs for the $B_s \to D_s^{**} \ell \nu_\ell $ form factors are not yet available, a model-independent analysis relies on the Heavy Quark Effective Theory framework. At present, corrections up to $\mathcal{O}(\alpha_s, \Lambda/m_{b, c})$ have been calculated for these decay modes, and we incorporate them into the relevant form factors to ensure an up-to-date analysis. We briefly discuss the HQET framework and the extraction of associated HQET parameters. We then present the resulting shapes of the form factors across the full kinematic region.   
\subsection{Overview of HQET Framework}
The HQET is a very reliable framework for studying the long-distance dynamics of hadrons containing a heavy quark ($Q$) $-$ a beauty ($b$) or charm ($c$) quark. As these quarks are heavy, the velocity of the hadron can be approximated to be equal to that of the heavy quark. The momentum of the heavy quark is expressed as~\cite{Neubert:1993mb,DiRisi:2023npw}
\begin{eqnarray}
p^\mu_Q = m_Q v^\mu + k^\mu \, ,
\end{eqnarray}
where $v^\mu$ is the four-velocity of the hadron and $k^\mu$ is the residual momentum. This allows decomposing the heavy quark spinor $Q(x)$ into two component spinors $h_v(x)$ and $H_v(x)$. The QCD Lagrangian for the heavy quark takes the form
\begin{eqnarray}
\mathcal{L}_{\text{QCD}} = \bar{h}_v i v \cdot D h_v + \bar{H}_v (-i v \cdot D - 2 m_Q) H_v + \bar{h}_v i \vec{\slashed{D}_\perp} H_v + \bar{H}_v i \vec{\slashed{D}_\perp} h_v\, ,
\end{eqnarray}
where $D^\mu$ is the gauge covariant derivative of QCD and $D^\mu_\perp=D^\mu-(v\cdot D) v^\mu$ denotes the transverse derivative orthogonal to the heavy quark velocity. This QCD Lagrangian is equivalent to having a massless fermion with field $h_v$ and a heavy fermion field $H_v$ carrying mass $2 m_Q$. After integrating out the heavy field, one obtains
\begin{eqnarray}
\mathcal{L}_{\text{HQET}} = \bar{h}_v i v \cdot D_s h_v + \mathcal{O}(1/m_Q)\, ,
\end{eqnarray}
where the subscript $s$ denotes the soft gluon fields after integrating out the hard gluons. This is the leading order Lagrangian in the HQET. By including the first order power corrections to the heavy quark mass, the Lagrangian modifies as
\begin{eqnarray}\label{eq:hqet_lag}
\mathcal{L}_{\text{HQET}} = \bar{h}_v i v \cdot D_s h_v + \frac{1}{2 m_Q} [\bar{h}_v (i D_s)^2 h_v + C_{mag}(\mu) \frac{g}{2} \bar{h}_v \sigma_{\mu\nu} G^{\mu\nu}_s h_v]\, ,
\end{eqnarray}
where the first term in the parentheses in the $1/m_Q$ correction is the kinetic energy operator, whereas the second term is the chromomagnetic operator, and the corresponding interaction strength is represented by $g$. The computation of transition matrix elements, and consequently the form factors, is essential for describing the $B_s \to D_s^{**}$ decays that proceed via the underlying $b\to c$ transition. Within the HQET framework, the SM hadronic matrix elements for the $B_s \to D_s^{**}$ transition mediated by vector and axial-vector currents have been explicitly calculated, as detailed in \cite{Bernlochner:2016bci,Bernlochner:2017jxt}. The relevant (axial)-vector hadronic transition matrix elements are expressed as
\begin{align}
\langle D_{s0}^*(v^\prime)|\bar{c}\gamma_\mu\gamma_5 b|B_s(v)\rangle &= \sqrt{m_{D^*_{s0}}m_{B_s}}[g_+(w)(v_\mu + v^\prime_{\mu}) + g_-(w)(v_\mu-v^\prime_{\mu})]\, ,\nonumber\\
\langle D_{s1}^*(v^\prime)|\bar{c}\gamma_\mu b|B_s(v)\rangle &= \sqrt{m_{D_{s1}^*}m_{B_s}}[g_{V_1}(w)\epsilon^{*}_\mu + (g_{V_2}(w)v_\mu + g_{V_3}(w)v^{\prime}_\mu)(\epsilon^*\cdot v)] \, ,\nonumber \\
\langle D_{s1}^*(v^\prime)|\bar{c}\gamma_\mu\gamma_5 b|B_s(v)\rangle &= i\sqrt{m_{D_{s1}^*}m_{B_s}}g_A(w)\varepsilon_{\mu\alpha\beta\gamma}\epsilon^{*\alpha} v^\beta v^{\prime\gamma}\, ,\nonumber\\
\langle D_{s1}(v^\prime)|\bar{c}\gamma_\mu b|B_s(v)\rangle &= \sqrt{m_{D_{s1}}m_{B_s}}[f_{V_1}(w)\epsilon^{*}_\mu + (f_{V_2}(w)v_\mu + f_{V_3}(w)v^{\prime}_\mu)(\epsilon^*\cdot v)]\, , \nonumber \\
\langle D_{s1}(v^\prime)|\bar{c}\gamma_\mu\gamma_5 b|B_s(v)\rangle &= i\sqrt{m_{D_{s1}}m_{B_s}}f_A(w)\varepsilon_{\mu\alpha\beta\gamma}\epsilon^{*\alpha} v^\beta v^{\prime\gamma}\, ,\nonumber\\
\langle D_{s2}^*(v^\prime)|\bar{c}\gamma_\mu b|B_s(v)\rangle &= i\sqrt{m_{D_{s2}^*}m_{B_s}}k_V(w)\varepsilon_{\mu\alpha\beta\gamma}\epsilon^{*\alpha\sigma}v_\sigma v^\beta v^{\prime\gamma}\, , \nonumber \\ 
\langle D_{s2}^*(v^\prime)|\bar{c}\gamma_\mu\gamma_5 b|B_s(v)\rangle &= \sqrt{m_{D_{s2}^*}m_{B_s}}[k_{A_1}(w)\epsilon^{*}_{\mu\alpha}v^\alpha + (k_{A_2}(w)v_\mu + k_{A_3}(w)v^{\prime}_\mu)\epsilon^*_{\alpha\beta}v^\alpha v^\beta]\, .
\end{align}
In the above equations, $v_\mu$ and $v^\prime_\mu$ are the four-velocities of the parent particle and daughter particle, respectively. The form factors denoted by $g_i$, $f_i$, and $k_i$ are functions of the dimensionless quantity $w$, which is related to the momentum transfer $q^2$ as
\begin{eqnarray}
w = v.v^\prime = \frac{m_{B_s}^2 + m_{D_s^{**}}^2 - q^2}{2 m_{B_s} m_{D_s^{**}}}\, .
\end{eqnarray}
In the HQET framework, all transition form factors can be expressed in terms of a single universal Isgur-Wise (IW) function that depends on the spin-parity of the light degrees of freedom, $s^{\pi_\ell}_\ell$. For decays to the $s^{\pi_\ell}_\ell = 1/2^+$ states ($D_{s0}^*$ and $D_{s1}^*$), the decay dynamics are governed by the IW function given by
\begin{eqnarray} 
\zeta(w) = \zeta(1) + (w-1)\zeta^\prime(1)\, .
\end{eqnarray}
The decays to the $s^{\pi_\ell}_\ell = 3/2^+$ states ($D_{s1}$ and $D_{s2}^*$) are described by the IW function given by
\begin{eqnarray}
\tau(w) = \tau(1) + (w-1)\tau^\prime(1)\, .
\end{eqnarray}
In the HQET framework, the contributions of $\zeta(w)$ and $\tau(w)$ vanish at $w=1$ \cite{Bernlochner:2016bci}. Therefore, at zero recoil, the form factor contributions arise entirely from the $\mathcal{O}(\Lambda/m_{c,b})$ correction and are expressed in terms of subleading IW functions. At order $\mathcal{O}(\Lambda/m_{b,c})$, corrections arise from both the heavy quark current operator and the HQET Lagrangian in eq.~\ref{eq:hqet_lag}, which contains matrix elements of the chromomagnetic and kinetic energy operators. The corrections to the heavy quark current arise from local $1/m_Q$ insertions into the current and are parametrized by the subleading IW functions $\zeta_1$ and $\tau_{1,2}$, whereas the corrections to the chromomagnetic operator originate from non-local $1/m_Q$ insertions into the Lagrangian, that violate spin symmetry, and are described by the functions $\chi^{c,b}_{1,2}$ and $\eta^{c,b}_{1,2,3}$. The spin symmetry conserving kinetic energy operator also receives corrections from non-local $1/m_Q$ insertions in the Lagrangian, which are parametrized by the functions $\chi_{ke}^{c,b}$ and $\eta_{ke}^{c,b}$. Additionally, short-distance QCD corrections appear through the $\mathcal{O}(\alpha_s)$ terms that renormalize the WCs of the effective local operators, during the matching of the HQET Lagrangian to the QCD Lagrangian. Importantly, the chromomagnetic and kinetic energy corrections do not complicate any individual form factor. Instead, they always enter as fixed linear combinations across all form factors \cite{Bernlochner:2017jxt}. Therefore, they can be absorbed into a redefinition of the leading IW functions. This absorption shifts the effective normalization and slope of the leading IW functions. Then the leading order IW functions $\zeta$ and $\tau$ are redefined by the replacement 
\begin{eqnarray}
\zeta + \epsilon_c \chi_{ke}^c + \epsilon_b[\chi_{ke}^b + 6 \chi_1^b - 2(w+1)\chi_2^b] &\to& \zeta\, , \nonumber\\
\tau + \epsilon_c \eta^c_{ke} + \epsilon_b[\eta_{ke}^b + 6 \eta_1^b - 2(w-1)\eta_2^b + \eta_3^b] &\to& \tau\, .
\end{eqnarray} 
As a result, the form factors for the $s^{\pi_\ell}_\ell = 1/2^+$ states are defined using 5 independent parameters $\{\zeta(1),\zeta^\prime(1),\zeta_1,\chi_1,\chi_2\}$, and those for the $s^{\pi_\ell}_\ell = 3/2^+$ states are defined using 7 independent parameters $\{\tau(1),\tau^\prime(1),\tau_1,\tau_2,\eta_1,\eta_2,\eta_3\}$. Here, we drop the superscript $c$ on $\chi$ and $\eta$ for simplicity. The resulting expressions for the $g_i$, $f_i$ and $k_i$  form factors are provided in appendix~\ref{App:ffhqet}.

By matching the respective currents, the relations between the form factors in the helicity and HQET bases are obtained. These relations can be found in appendix~\ref{App:ffrel}. Using eqs.~\ref{eq:appDs0s}-\ref{eq:appDs2s}, the SM decay rate is obtained in terms of HQET form factors. 

\subsection{Methodology for HQET Parameter Extraction}

Assuming that possible NP effects appear only in the tauonic mode, we perform a data-driven analysis to extract the form factor parameters from measurements of semileptonic decays with light leptons. At present, there are no available experimental data for the $B_s \to D_s^{**} \ell \nu_\ell$ modes. However, $SU(3)$ flavor symmetry can be used to relate the $B_s \to D_s^{**}$ transitions to the corresponding $B \to D^{**}$ decays where $D^{**} = \{D_{0}^*,D_{1}^*,D_{1},D_{2}^*\}$. Consequently, the transition form factors for $B \to D^{**}$, which are related to the Isgur-Wise functions, serve as essential inputs for obtaining information on the form factors in the $B_s \to D_s^{**}$ case. In our analysis, we perform a $\chi^2$ test to extract these parameters.  A similar analysis was done in ref.~\cite{Bernlochner:2016bci,Bernlochner:2017jxt}, but on account of recent experimental measurements~\cite{Belle:2022yzd}, an updated analysis is performed in this work. Measurements of branching fractions are available for all the modes, however these were obtained in the $D^{**} \to D^{(*)}\pi$ or the $D^{**} \to D\pi\pi$ channels, where they are determined as cascade decays of $\mathcal{B}(B \to D^{**}\ell\nu_\ell) \times \mathcal{B}(D^{**} \to D^{(*)}\pi)$ and $\mathcal{B}(B \to D^{**}\ell\nu_\ell) \times \mathcal{B}(D^{**} \to D\pi\pi) $~\cite{ParticleDataGroup:2024cfk}. To eliminate the branching fraction of the strong decay of the daughter particle, certain assumptions are made as given below \cite{Bernlochner:2016bci}
\begin{eqnarray}\label{eq:br_assum}
\mathcal{B}(\bar{D}_0^* \to \bar{D}\pi) &=& 1 \nonumber\, , \\
\mathcal{B}(\bar{D}_1^* \to \bar{D}^*\pi) &=& 1 \nonumber\, , \\
\mathcal{B}(\bar{D}_1 \to \bar{D}^*\pi) + \mathcal{B}(\bar{D}_1 \to \bar{D}\pi\pi) &=& 1\, , \nonumber \\
\mathcal{B}(\bar{D}_2^* \to \bar{D}\pi) + \mathcal{B}(\bar{D}_2^* \to \bar{D}^*\pi) &=& 1\, .
\end{eqnarray}
For the $\bar{D}_0^*$ and $\bar{D}_1^*$ particles, the dominant decay channels are $\bar{D}_0^* \to \bar{D}\pi$ and $\bar{D}_1^* \to \bar{D}^*\pi$. Decays to other daughter particles are heavily suppressed and hence the above assumptions are made \cite{Bernlochner:2017jxt,ParticleDataGroup:2024cfk}. No signal has been observed for $\bar{D}_1 \to \bar{D}\pi$, whereas the $\bar{D}_1 \to \bar{D}^*\pi\pi$ mode has been detected. However, based on the available inputs on the cascade decay mode from the particle data group (PDG) \cite{ParticleDataGroup:2024cfk}, it is inferred that the $\bar{D}_1 \to \bar{D}^*\pi\pi$ channel is constrained by an upper bound that is two orders of magnitude smaller than that for $\bar{D}_1 \to \bar{D}^*\pi$. Similarly, no signal has been reported for $\bar{D}_2^* \to \bar{D}^*\pi\pi$ and the non-resonant $\bar{D}_2^* \to \bar{D}\pi\pi$ decay only has an upper bound. These facts motivate the assumptions adopted in eq.~\ref{eq:br_assum}. To account for the missing isospin conjugate decay modes, the associated corrections are incorporated using Clebsch-Gordan coefficients as follows \cite{Bernlochner:2016bci}
\begin{eqnarray}~\label{eq:isospin_corr}
\frac{\mathcal{B}(D^{**} \to D^{(*)^0} \pi^-)}{\mathcal{B}(D^{**} \to D^{(*)}\pi)} = \frac{2}{3}\, ,\nonumber \\
\frac{\mathcal{B}(D^{**} \to D^{(*)-}\pi^+\pi^-)}{\mathcal{B}(D^{**}\to D^{(*)}\pi\pi)} = \frac{9}{16}\, .
\end{eqnarray}
Following the assumptions stated in eqs.~\ref{eq:br_assum} and \ref{eq:isospin_corr}, we extract the associated branching fractions for the relevant semileptonic decay modes, and these are presented in table~\ref{tab:corr_bran_frac}~\footnote{There is a discrepancy between the latest Belle measurement~\cite{Belle:2022yzd} and earlier measurements~\cite{BaBar:2008dar,Belle:2007uwr} for the branching fraction of $B^+ \to D_0^{*0}\ell^+\nu_\ell$ decay. Here, we consider only the latest Belle result and not the PDG average.}. We use these as inputs to extract the HQET parameters. 
\begin{table}
\begin{center}
	\begin{tabular}{|c|c|}
		\hline
		Decay & Branching Fraction \\
		\hline
		$B^+ \to D_0^{*0}\ell\nu_\ell$ & $0.810 (334)\times 10^{-3}$ \\
		$B^+ \to D_1^{*0}\ell\nu_\ell$ & $2.55 (90)\times 10^{-3}$ \\
		$B^0 \to D_1^{*-}\ell\nu_\ell$ & $3.75 (90)\times 10^{-3}$ \\
		$B^+ \to D_1^{0}\ell\nu_\ell$ & $6.127 (356)\times 10^{-3}$ \\
		$B^0 \to D_1^{-}\ell\nu_\ell$ & $6.088 (471)\times 10^{-3}$ \\
		$B^+ \to D_2^{*0}\ell\nu_\ell$ & $3.975 (309)\times 10^{-3}$ \\
		$B^0 \to D_2^{*-}\ell\nu_\ell$ & $3.105 (345)\times 10^{-3}$ \\
		\hline
	\end{tabular}
	\caption{Extracted branching fractions of $B \to D^{**}\ell\nu_\ell$ decays.}\label{tab:corr_bran_frac}
\end{center}
\end{table}
In addition to these branching fraction inputs, the available bin-wise measurements of the normalized differential decay rates are also used in our analysis. Specifically, the normalized differential decay rates have been measured in five bins for $B \to D_0^*\ell\nu_\ell$ and four bins for $B \to D_2^*\ell\nu_\ell$. These measurements are presented in table~\ref{tab:bin_wise}. Since the observables of table~\ref{tab:bin_wise} are normalized differential decay rates, all the bin results should add up to unity. Hence, one of the bins is not linearly independent. Therefore, four bins of $B \to D_0^*\ell\nu_\ell$ and three bins of $B \to D_2^*\ell\nu_\ell$ are used in the fitting. For $B^+ \to D_0^{*0}\ell\nu_\ell$, we drop the second bin ($1.08 - 1.16$), and for $B^+ \to D_2^{*0}\ell\nu_\ell$, we drop the first bin ($1.00 - 1.08$), as there are large associated errors compared to the other bins. Together, the $B \to D_0^*\ell\nu_\ell$ and $B \to D_1^*\ell\nu_\ell$ modes provide a total of seven observables, and also a total of seven observables with the $B \to D_1\ell\nu_\ell$ and $B \to D_2^*\ell\nu_\ell$ modes. There is a measurement of the branching ratio in the $B^0 \to D_0^{*-}\ell^+\nu_\ell$ decay, but it only has an upper bound $\mathcal{B}(B^0 \to D_0^{*-}\ell^+\nu_\ell) < 0.66 \times 10^{-3}$. Therefore, this result is not considered in our analysis.
\begin{table}
\begin{center}
	\begin{tabular}{|c|c|c|}
		\hline
		$w$ & $B^+ \to D_0^{*0}\ell\nu_\ell$ & $B^+ \to D_2^{*0}\ell\nu_\ell$ \\
		\hline
		1.00 - 1.08 & $0.05 (2)$ & $0.06 (2)$ \\
		1.08 - 1.16 & $0.02 (5)$ & $0.30 (5)$ \\
		1.16 - 1.24 & $0.30 (8)$ & $0.38 (3)$ \\
		1.24 - 1.32 & $0.30 (9)$ & $0.26 (6)$ \\
		1.32 - 1.40 & $0.33 (13)$ & -\\
		\hline
	\end{tabular}
	\caption{Bin-wise measurement of normalized differential decay rates.}\label{tab:bin_wise}
\end{center}
\end{table}

\subsection{HQET Form Factor Fits}

In this section, we extract the HQET from factor parameters relevant to the $B \to D^{**}$ decays, which are essential for estimating the shape of the form factors in $B_s \to D_s^{**}$ transitions. As we have seen, for transitions to the $s^{\pi_\ell}_\ell = 1/2^+$ states, the relevant form factors depend on five independent parameters $\{\zeta(1),\zeta^\prime(1),\zeta_1,\chi_1,\chi_2\}$. The transitions to the $s^{\pi_\ell}_\ell = 3/2^+$ states involve seven independent parameters $\{\tau(1),\tau^\prime(1),\tau_1,\tau_2,\eta_1,\eta_2,\eta_3\}$. Limited experimental data prevent a complete extraction of all HQET parameters for these channels, so certain approximations are adopted for a combined analysis and simultaneous extraction. The mass splitting $m_{D^*} - m_{D} \sim 140$ GeV is larger than that of $m_{D_1^*} - m_{D_0^*}$ and $m_{D_2^*} - m_{D_1}$, indicating that chromomagnetic corrections are suppressed~\cite{Leibovich:1997em,Leibovich:1997tu}. Thus, we drop chromomagnetic contributions to the form factors. In the absence of this operator, the parameters $\chi_{1,2}$ and $\eta_{1,2,3}$ are set to zero. The remaining parameter sets to be extracted simultaneously are therefore $\{\zeta(1),\zeta^\prime(1),\zeta_1\}$  for $s_\ell^{\pi_\ell}=1/2^+$ transitions and $\{\tau(1),\tau^\prime(1),\tau_1,\tau_2\}$ for $s_\ell^{\pi_\ell}=3/2^+$ transitions. It should be noted that in $B \to D^{(*)}$ decays, the leading order Isgur-Wise function is normalized to unity due to the flavor symmetry of the light degrees of freedom between the parent and daughter mesons. This normalization does not apply here, since the parent particle has $s_\ell^{\pi_\ell} = 1/2^-$ while the daughter particles carry $s_\ell^{\pi_\ell}= 1/2^+$ or $s_\ell^{\pi_\ell} = 3/2^+$. 

Performing a $\chi^2$ analysis, which corresponds to maximizing the log-likelihood function, we extract the relevant form factor parameters. The $\chi^2$ function is defined as
\begin{eqnarray}\label{eq:chi2}
\chi^2 = \sum_{i,j}(\mathcal{O}^{\text{theo}}_i - \mathcal{O}^{\text{data}}_i) \, \, Cov^{-1}_{i,j} \, (\mathcal{O}^{\text{theo}}_j - \mathcal{O}^{\text{data}}_j)^T \, ,
\end{eqnarray} 	
where $Cov$ is the covariance matrix, and $\mathcal{O}^{\text{theo}}$ and $\mathcal{O}^{\text{data}}$ represent the theoretical and experimental inputs of the observables, respectively. In our analysis, $\mathcal{O}^{\text{data}}$ are the branching ratios from tables~\ref{tab:corr_bran_frac} and \ref{tab:bin_wise}, whereas $\mathcal{O}^{\text{theo}}$ are the theoretical branching ratios obtained using the integrated differential decay rate of eq.~\ref{eq:sm_decay_rate}, with appropriate mass replacements. The numerical values of the input parameters used in this work are taken from \cite{ParticleDataGroup:2024cfk} and are presented in table \ref{tab:input}.

\begin{table}[t]
	\begin{center}
		\renewcommand*{\arraystretch}{1.13}
		\resizebox{1.02 \textwidth}{!}{
		\begin{tabular}{|c@{\hskip 60pt}|c@{\hskip 60pt}|c|}
			\hline
			$m_{D_{s0}^*} = 2.318$\,GeV & $m_{D_{s1}^*} = 2.460$\,GeV & $\tau_{B_s} = 1.516 \times 10^{-12}$\,s\\
			$m_{D_{s1}} = 2.535$\,GeV & $m_{D_{s2}^*} = 2.569$\,GeV & $\tau_{B^+} = 1.638 \times 10^{-12}$\,s\\
			%
			$m_{B_s} = 5.367$\,GeV & $m_{B^+} = 5.279$\,GeV & $\tau_{B^0} = 1.517 \times 10^{-12}$\,s\\
			%
			$m_{D^+} = 1.869$\,GeV & $m_{D_s} = 1.969$\,GeV & $V_{cb} = 41.1(12) \times 10^{-3}$\\
			%
			$m_{\tau} = 1.777$\,GeV & $m_{\mu} = 0.105$\,GeV & $G_F = 1.166 \times 10^{-5}$\,GeV$^{-2}$\\
			%
			$m_{b} = 4.183$\,GeV & $m_{c} = 1.273$\,GeV & \\
			\hline
		\end{tabular}
	}
		\caption{Input parameters used in this work.}\label{tab:input}
	\end{center}
\end{table}

\begin{table}[t]
	\begin{center}
		\renewcommand*{\arraystretch}{1.5}
		\resizebox{0.65 \textwidth}{!}{
			\begin{tabular}{|c|c|ccc|}
				\hline
				$\chi^2/dof = 4.2/4$ & \text{Fitted Value} & $\zeta(1)$ & $\zeta^\prime(1)$ & $\zeta_1$ \\
				\hline
				$\zeta(1)$ & $0.416 (116)$ & 1 & -0.9 & -0.364 \\
				$\zeta^\prime(1)$ & $0.570 (452)$ & & 1 & 0.627 \\
				$\zeta_1$ & $0.951 (154)$ & & & 1 \\
				\hline
			\end{tabular}
		}
		\caption{Best-fit results and correlations of the HQET parameters for $B \to D_0^*\ell\nu_\ell$ and $B \to D_1^*\ell\nu_\ell$.}~\label{tab:zeta_fit}
	\end{center}
\end{table}	
\begin{table}[t]
	\begin{center}
		\renewcommand*{\arraystretch}{1.5}
		\resizebox{0.65\textwidth}{!}{
			\begin{tabular}{|c|c|cccc|}
				\hline
				$\chi^2/dof = 2.27/3$ & \text{Fitted Value} & $\tau(1)$ & $\tau^\prime(1)$ & $\tau_1$ & $\tau_2$ \\
				\hline
				$\tau(1)$ & $0.809 (146)$ & 1 & -0.997 & -0.69 & 0.7 \\
				$\tau^\prime(1)$ & $-1.561 (749)$ & & 1 & 0.73 & -0.738 \\
				$\tau_1$ & $0.251 (251)$ & & & 1 & -0.988 \\
				$\tau_2$ & $-0.74 (154)$ & & & & 1 \\
				\hline
			\end{tabular}
		}
		\caption{Best-fit results and correlations of the HQET parameters for $B \to D_1\ell\nu_\ell$ and $B \to D_2^*\ell\nu_\ell$.}\label{tab:tau_fit}
	\end{center}
\end{table}	
The combined analysis of extracting the HQET parameters for the decay modes with $s_\ell^{\pi_\ell} = 1/2^+$ and $s_\ell^{\pi_\ell} = 3/2^+$ daughter states yields the parameter results presented in tables~\ref{tab:zeta_fit} and \ref{tab:tau_fit}. The fit results are consistent with those of \cite{Bernlochner:2016bci, Bernlochner:2017jxt} within $1 \sigma$ for all the considered decay modes. Our analysis of the parameter sets $\{\zeta(1),\zeta^\prime(1),\zeta_1\}$ and $\{\tau(1),\tau^\prime(1),\tau_1,\tau_2\}$ yields p-values of $38\%$ and $52\%$, respectively. In both cases, the fitted parameters exhibit strong correlations.


\section{Form Factors in the $z$-Parametrization}~\label{sec:form_factor}
Within the HQET framework, the IW parameters are Taylor-expanded around $\omega = 1$, as this expansion is more reliable near zero recoil ($q^2_{\text{max}}$). For a phenomenological analysis, we adopt a model-independent parametrization that satisfies both unitarity and analyticity across the entire $q^2$ region. In this regard, the $z$-parametrization provides a convenient representation of the form factors over the full physical kinematic region.

\subsection{BSZ parametrization}
The $z$-parametrization exploits analyticity and unitarity properties of the form factors to constrain the functional form and the expansion coefficients. A conformal transformation is used as 	
\begin{eqnarray}
z(q^2,t_0) = \frac{\sqrt{t_+-q^2} - \sqrt{t_+ - t_0}}{\sqrt{t_+-q^2} + \sqrt{t_+ - t_0}}\, ,
\end{eqnarray}
where $t_+ = (m_{B_s} + m_{D_s^{**}})^2$. The conformal variable $z$ is defined in such a way that the branch cut in the complex $q^2$ plane for $q^2>0$ maps onto the unit circle $|z| = 1$. The real axis below the branch point, $-\infty<q^2<t_+$ is mapped onto $z~\in~[-1,1]$. Hence, the entire $q^2$ region is confined to a unit disc $|z|<1$. The physical decay region in the $z$ plane is very small, and allows for a rapid convergence of the series. The parameter $t_0$ is a free parameter and is the center of the disc, that is, $z(t_0,t_0) = 0$. One of the simplest choices for $t_0$ is zero, but this does not always ensure rapid convergence. Hence, the optimal value for $t_0$ is $t_0 = t_+ \left(1 - \sqrt{1 - \frac{t_-}{t_+}}\right)$, where $t_- = (m_{B_s} - m_{D_s^{**}})^2$ is the maximum physical value of $q^2$. At this optimal point, $z(q^2 = 0) = -z(q^2 = t_-)$, so the entire physical region lies within $-|z_{\text{max}}| \leq z \leq |z_{\text{max}}|$. In this work, we employ the Bharucha-Straub-Zwicky (BSZ) parametrization~\cite{Bharucha:2015bzk} to describe the form factors. Within this scheme, the expansion takes the form
\begin{eqnarray}\label{eq:BSZ_exp}
F(q^2) = \frac{1}{1 - \frac{q^2}{m_{\text{pole}}^2}} \sum_{n=0}^N a_n \left(z(q^2,t_0) - z_0\right)^n \, ,
\end{eqnarray}	
where $F$ can be any of the (axial)-vector form factors of eqs.~\ref{eq:hadmat_sca},~\ref{eq:hadmat_axvec} and \ref{eq:hadmat_tens}, and $z_0 = z(0,t_0)$. To maintain the analyticity condition of form factors, possible poles arising from resonances need to be explicitly removed. Table~\ref{tab:pole_mass} lists the masses $m_{\text{pole}}$ of the lowest lying $B_c$ resonance states, along with their corresponding $J^P$ values. The shape of each form factor is then determined using the $z$-expansion, with coefficients $a_n$ extracted from a combined analysis. 
\begin{table}
\begin{center}
	\renewcommand*{\arraystretch}{1.4}
	\resizebox{0.55 \textwidth}{!}{
	\begin{tabular}{|c|c|c|c|c|}
		\hline
		$J^P$ & $0^-$ & $0^+$ & $1^-$ & $1^+$ \\
		\hline
		Pole mass~\cite{Gubernari:2018wyi} & 6.275 & 6.420 & 6.330 & 6.767 \\
		\hline
	\end{tabular}
}
\caption{Lowest lying $B_c$ meson masses.}\label{tab:pole_mass}
\end{center}
\end{table}
Given the limited data set, we truncate the form factor series expansion at $N=1$. The convergence of the $z$-series follows the hierarchy $a_0 > a_1 \left(z(q^2,t_0) - z_0 \right) > a_2 \left(z(q^2,t_0) - z_0\right)^2$. In this work, we truncate the BSZ expansion at linear order, that is, up to the first power of $z$. To account for the uncertainty arising from neglected higher order terms, we introduce an additional truncation error. Specifically, we estimate the possible contribution of the quadratic term while neglecting relatively small higher order terms beyond it, following the prescription of ~\cite{Bourrely:2008za, Dey:2025xjg}. This error is propagated as a systematic uncertainty in the form factors. The truncation uncertainty is defined as 
\begin{equation}\label{eq:trunc_errorP}
	\delta F_i(q^2)= \frac{a_{N+1}^{\text{max}} \left(z(q^2,t_0) - z_0\right)^{N+1}}{1-\frac{q^2}{m_{\text{pole}}^2}}\, ,
\end{equation}
where $a_{N+1}^\text{max}$ denotes the maximum value of the coefficient $a_{N+1}$. Since we truncate the series at $N=1$, the relevant parameter is $a_2^\text{max}$. The value of $a_2^{\text{max}}$ can be determined using unitarity bounds on the form factors. However, calculations of relevant outer functions are not trivial and are beyond the scope of our current work. The value of $a_2^{\text{max}}$ is evaluated such that the truncation error is $5\%$ of the leading order term for each of the form factors as the convergence of the $z$-series requires $a_1 \left(z(q^2,t_0) - z_0\right) >> a_2 \left(z(q^2,t_0) - z_0\right)^2$. In the following subsection, we perform a combine fit to extract the BSZ coefficients and estimate the coefficient $a_2$, which is used to evaluate the truncation error associated with each transition form factors. 
\subsection{Form Factor Shape}
In this subsection, we perform a $\chi^2$ analysis to extract $a_n$ using the necessary inputs already presented in the previous sections. The analysis is performed on synthetic data generated from the central values, uncertainties and correlations of the HQET parameters presented in tables~\ref{tab:zeta_fit} and ~\ref{tab:tau_fit}. The synthetic form factor data for the relevant decay modes are obtained from eqs.~\ref{eq:ff_Ds0s}-\ref{eq:ff_Ds2s}. The $\chi^2$ function defined in eq.~\ref{eq:chi2} is used, where $\mathcal{O}^{\text{theo}}$ here denotes the BSZ expansion of the form factors (eq.~\ref{eq:BSZ_exp}), with the expansion coefficients $a_n$ treated as free parameters, and $\mathcal{O}^{\text{data}}$ represents the corresponding synthetic data. In this analysis, each form factor is treated independently, and the synthetic data points are selected near the high-$q^2$ region, where HQET predictions are more reliable. The results obtained for each decay channel are presented below.

\begin{enumerate}
\item $B_s \to D_{s0}^*\ell\nu_\ell$

For the $B_s \to D_{s0}^*$ transition, the relevant matrix elements are provided in eq.~\ref{eq:hadmat_sca}. Two form factors are required to parameterize the (axial)-vector matrix elements. To determine the shape of these form factors and hence $a_n$, we perform a $\chi^2$-analysis using synthetic data summarized in table~\ref{tab:syn_dat_Ds0s}. We also incorporate available LCSR estimates at $q^2 = 0$, where $f_+(0) = f_0(0) = -0.24 (21)$~\cite{Gubernari:2023rfu}, as an additional input. The $\chi^2$-analysis for $a_n$ using only HQET synthetic data is referred to as the BSZ fit (HQET), whereas the analysis that includes both HQET and LCSR inputs is called the BSZ fit (HQET + LCSR). The form factors $f_+(q^2)$ and $f_0(q^2)$ are parameterized with two BSZ coefficients, truncated at $N=1$. The kinematic constraint $f_+(0) = f_0(0)$ implies $a_0^{f_+} = a_0^{f_0}$, thereby reducing the number of free parameters to three. Since the maximum value of $z$ is $\sim 0.021$, the rapid convergence of the $z$-series ensures that a small number of parameters is sufficient to parameterize the form factors. 
\begin{table}[t]
	\begin{center}
		\renewcommand*{\arraystretch}{1.4}
		\resizebox{0.55 \textwidth}{!}{
		\begin{tabular}{|c|c|c|c|}
			\hline
			Form Factors & \multicolumn{3}{c|}{$q^2$ values (GeV$^2$)} \\
			\cline{2-4}
			& 7.3 & 8.3 & $q^2_{\text{max}}= 9.3$ \\
			\hline
			$f_+(q^2)$ & $0.233 (63)$ & $0.252 (69)$ & $0.271 (75)$ \\
			$f_0(q^2)$ & $0.077 (33)$ & $0.079 (29)$ & $0.082 (23)$\\
			\hline
		\end{tabular}
	}
		\caption{Generated synthetic data points for $B_s \to D_{s0}^*$ form factors.} 
		\label{tab:syn_dat_Ds0s}
	\end{center}
\end{table}
The fitted BSZ parameters are presented in table~\ref{Tab:Ds0sBSZfit}. The resulting $q^2$-dependence of the form factors, with $1\sigma$ bands, is displayed in figure~\ref{fig:ffDs0s}. 
\begin{table}[t]
	\begin{center}
		\renewcommand*{\arraystretch}{1.6}
		\resizebox{0.95 \textwidth}{!}{
		\begin{tabular}{|c|c|c|c|c|}
			\hline
			Fit & $\chi^2/dof$ & $a_0^{f_+} \equiv a_0^{f_0}$ & $a_1^{f_+}$ & $a_1^{f_0}$ \\
			\hline
			\text{BSZ fit (HQET)} & $1.20/2$ & $0.102 (34)$ & $-1.67 (47)$ & $0.66 (62)$ \\
			\hline
			\text{BSZ fit (HQET + LCSR)} & $3.59/3$ & $0.102 (29)$ & $-2.2 (12)$ & $0.64 (59)$ \\
			\hline
		\end{tabular}
	}
	\caption{Fit results for the BSZ coefficients $a_n$ in the $B_s \to D_{s0}^*$ transition.} \label{Tab:Ds0sBSZfit}
	\end{center}
\end{table}	

\begin{figure}[t]
	\begin{center}
		\includegraphics[width=0.48\textwidth]{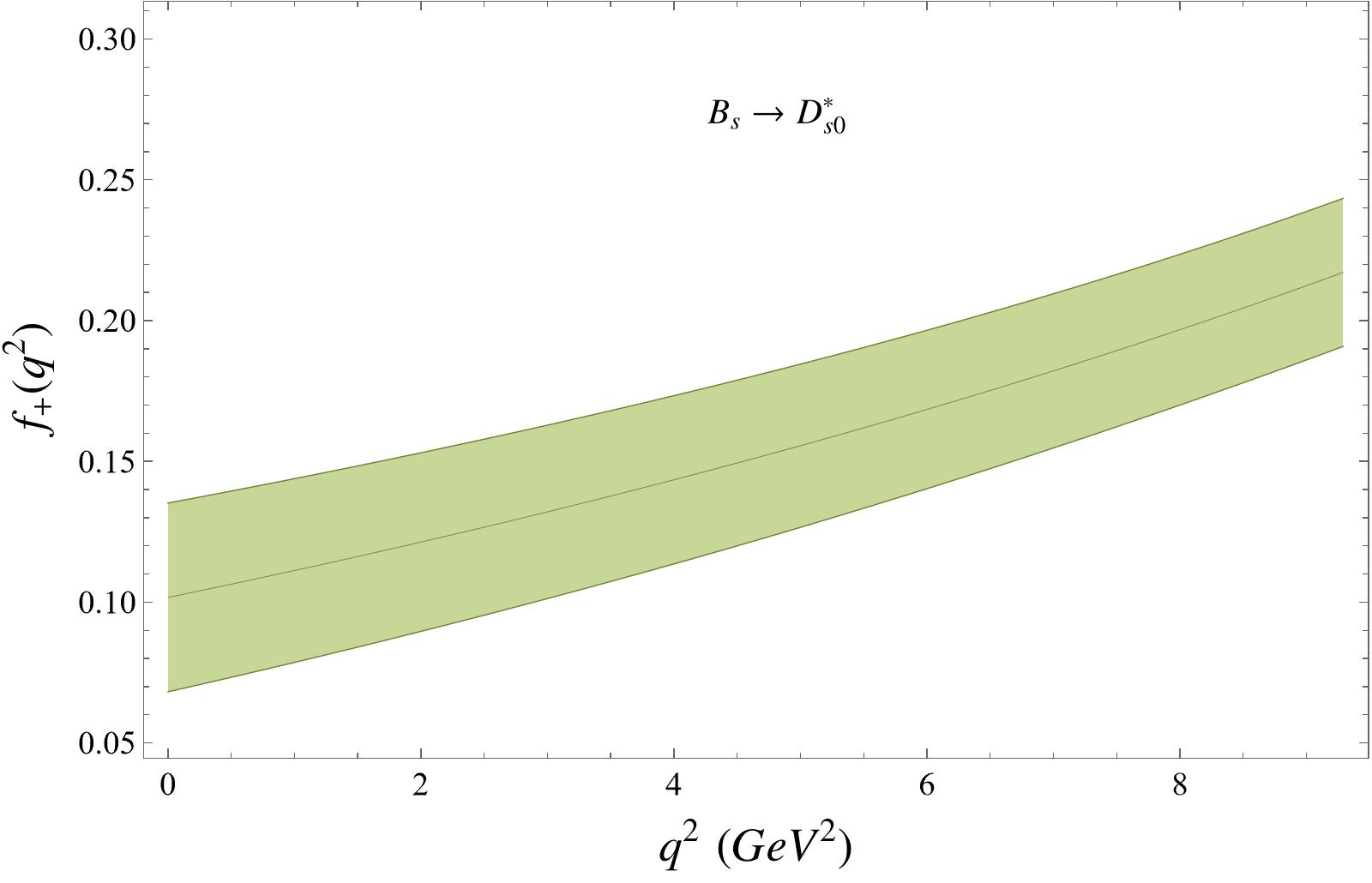} \hspace{0.01\textwidth}
		\includegraphics[width=0.48\textwidth]{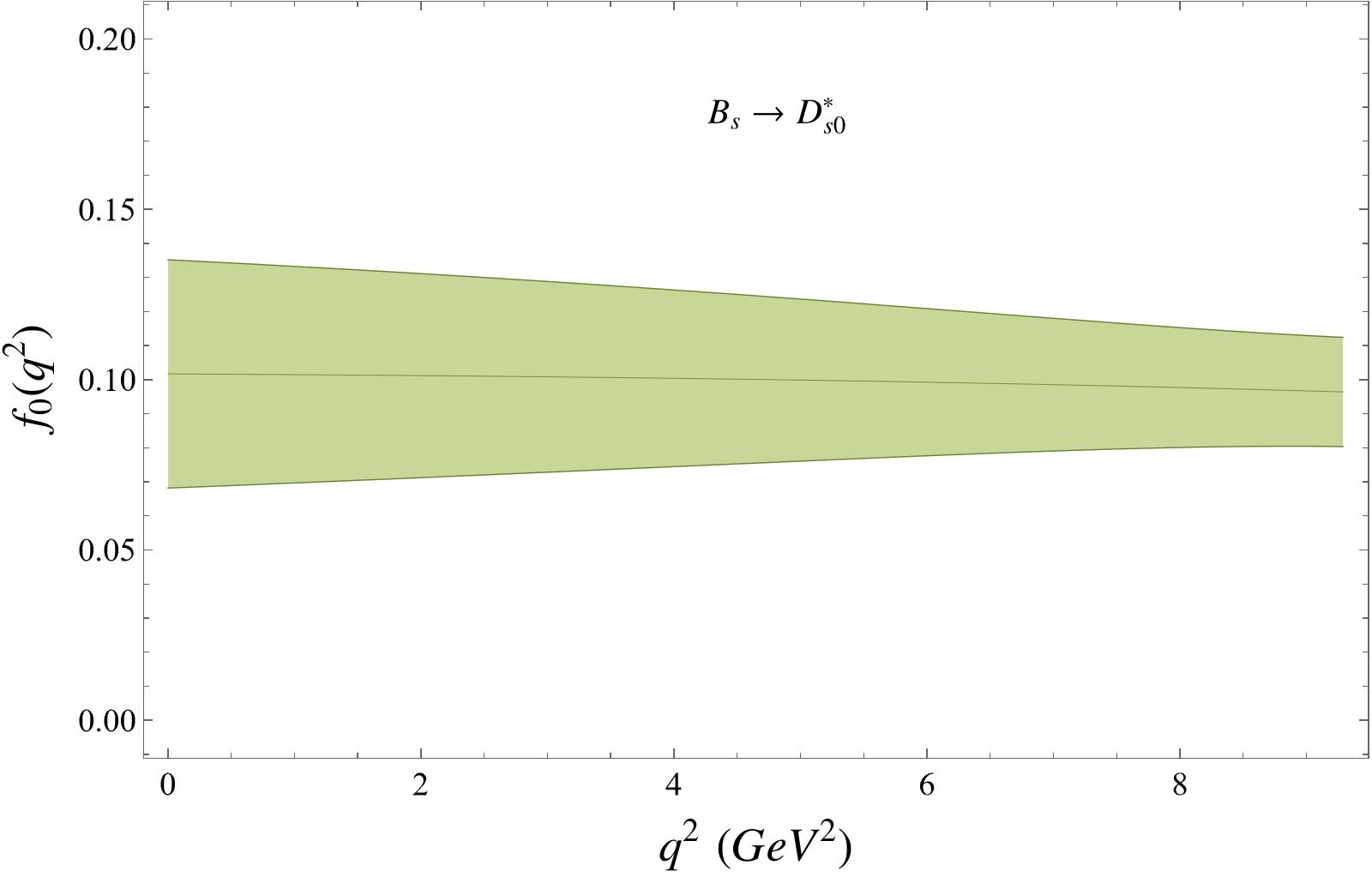}
	\end{center}
	\caption{$q^2$-distribution for $B_s \to D_{s0}^*$ form factors computed using the BSZ fit (HQET)  given in table~\ref{Tab:Ds0sBSZfit}.}\label{fig:ffDs0s}
\end{figure}

We also obtain form factor estimates at $q^2 = 0$ using both HQET inputs and the BSZ parametrization.  Our estimates for the $B_s \to D_{s0}^{*}$ transition form factors at $q^2 = 0$ are summarized below
\begin{align}~\label{eq:fq20prd_comb}
	f_+(0) = f_{0}(0) = \begin{cases} 
		0.107 (43) ~~ \text{HQET} \\ 
		0.102 (34) ~~ \text{BSZ fit (HQET)} \\
		 0.102 (29) ~~ \text{BSZ fit (HQET + LCSR)} \end{cases}\, .
\end{align}
Using the HQET parametrization, we find that the form factor uncertainty at $q^2=0$ is approximately $40\%$. The BSZ parametrization reduces this uncertainty to about $33\%$. When LCSR inputs are incorporated into the analysis, the uncertainty further decreases to $28\%$.
\item $B_s \to D_{s1}^*\ell\nu_\ell$

\begin{table}[ht]
	\begin{center}
		\renewcommand*{\arraystretch}{1.6}
		\resizebox{0.6 \textwidth}{!}{
			\begin{tabular}{|c|c|c|c|}
				\hline
				Form Factors & \multicolumn{3}{c|}{$q^2$ values (GeV$^2$)} \\
				\cline{2-4}
				& 5.5 & 7.0 & $q^2_{\text{max}}= 8.45$ \\
				\hline
				$A(q^2)$ & $0.536 (78)$ & $0.484 (98)$ & $0.43 (12)$ \\
				$V_0(q^2)$ & $-0.252 (29)$ & $-0.209 (36)$ & $-0.168 (47)$ \\
				$V_1(q^2)$ & $0.058 (6)$ & $0.028 (3)$ & $0.002 (1)$ \\
				$V_2(q^2)$ & $1.01 (10)$ & $0.96 (11)$ & $0.90 (12)$ \\
				\hline
			\end{tabular}
		}
			\caption{Generated synthetic data for $B_s \to D_{s1}^*$ form factors.}\label{tab:syn_dat_Ds1s}
	\end{center}
\end{table}

The relevant matrix elements for the $B_s \to D_{s1}^*$ transition are given in eqs.~\ref{eq:hadmat_axvec}. Four form factors parameterize the (axial)-vector matrix elements. A $\chi^2$-analysis is conducted to determine the shapes of these form factors within the BSZ parametrization.  The analysis takes into account the synthetic data generated from eq.~\ref{eq:ff_Ds1s} using the HQET parameter values listed in table~\ref{tab:zeta_fit}, including their correlations. We generate the synthetic data points at $q^2= 5.5$ GeV$^2$, $7.0$ GeV$^2$ and $8.45$ GeV$^2$ and summarise them in table~\ref{tab:syn_dat_Ds1s}. Since LCSR inputs for these modes are not yet available, the analysis relies solely on synthetic data generated using HQET parameters. With the $z$-series truncated at $N = 1$, the BSZ parametrization comprises eight coefficients, two per form factor, for the four form factors. The maximum value of $z$ here is $\sim 0.019$, and the rapid convergence of the $z$-series ensures that only a limited number of parameters is needed to parameterise the form factors. However, the kinematic constraint at $q^2=0$ (eq.~\ref{eq:kin_const_vec}) implies that one leading-order coefficient is not independent. This reduces the number of free parameters to seven. In table~\ref{Tab:Ds1sBSZ}, we present the fitted BSZ parameters \footnote{To differentiate between $B_s \to D_{s1}^*$ and $B_s \to D_{s1}$ transition form factors, the form factor parameter names here are marked with an asterisk.}. Upon obtaining the full set of form factor parameters, we extract the form factors over the full kinematic region. The $1\sigma$ bands obtained from the fit are shown in figure~\ref{fig:ffDs1s}.
\begin{table}[t]
	\begin{center}
		\renewcommand*{\arraystretch}{1.4}
		\resizebox{1.02 \textwidth}{!}{
			\begin{tabular}{|c|c|c|c|c|c|c|c|}
				\hline
				$\chi^2/dof = 1.99/1$ & $a_0^{A^*}$ & $a_1^{A^*}$ & $a_1^{V^*_0}$ & $a_0^{V^*_1}$ & $a_1^{V^*_1}$ & $a_0^{V^*_2}$ & $a_1^{V^*_2}$ \\
				\hline
				Fit Results & $0.660 (61)$ & $8.2 (29)$ & $-6.7 (14)$ & $0.126 (14)$ & $3.36 (36)$ & $0.980 (64)$ & $8.5(26)$ \\
				\hline
			\end{tabular}
		}
		\caption{Fit results for the BSZ coefficients $a_n$ in the $B_s \to D_{s1}^*$ transition.}\label{Tab:Ds1sBSZ}
	\end{center}
\end{table} 
\begin{figure}[t]
	\begin{center}
		\includegraphics[width=0.48\textwidth]{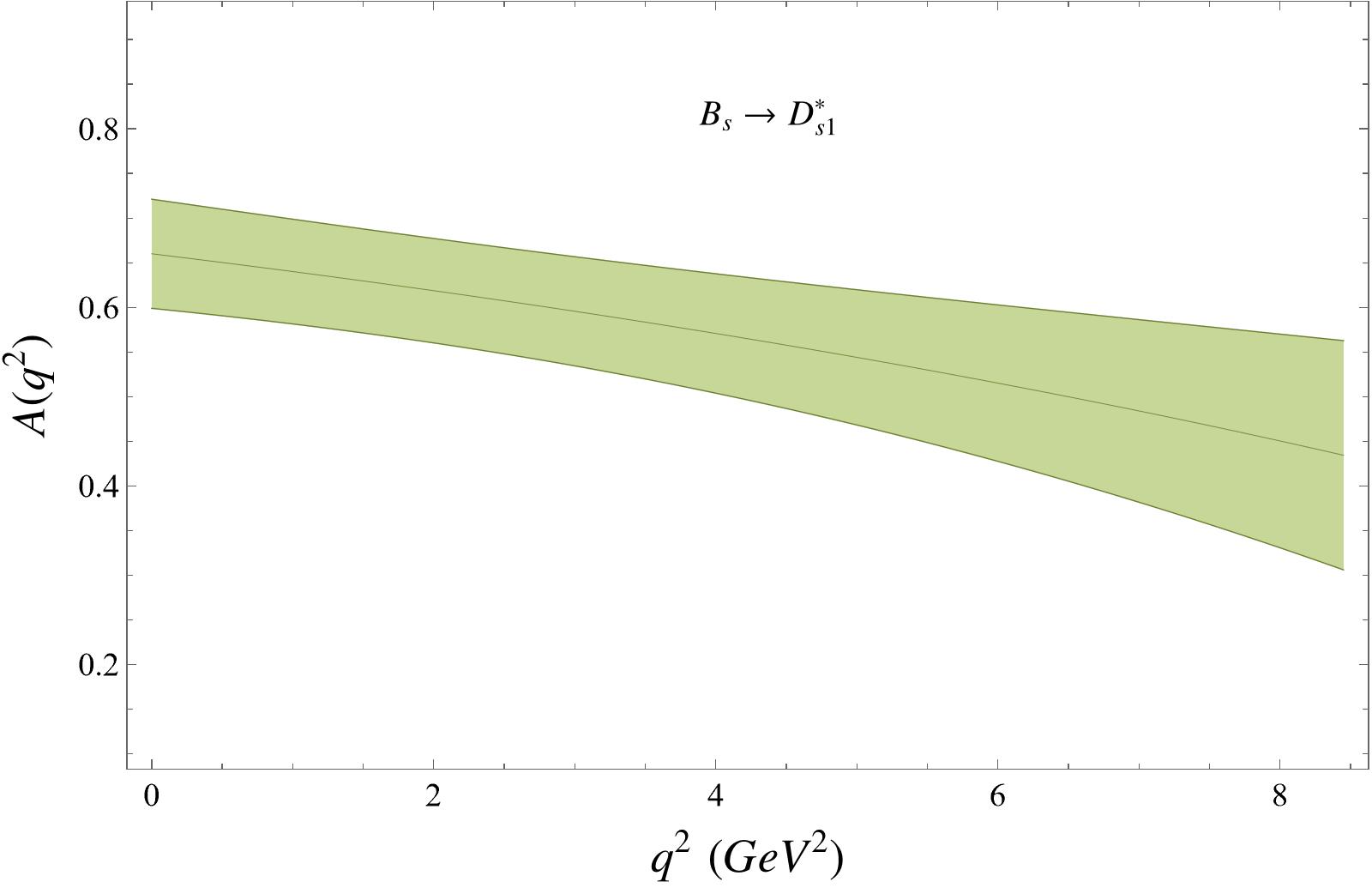}
		\hspace{0.01\textwidth}
		\includegraphics[width=0.48\textwidth]{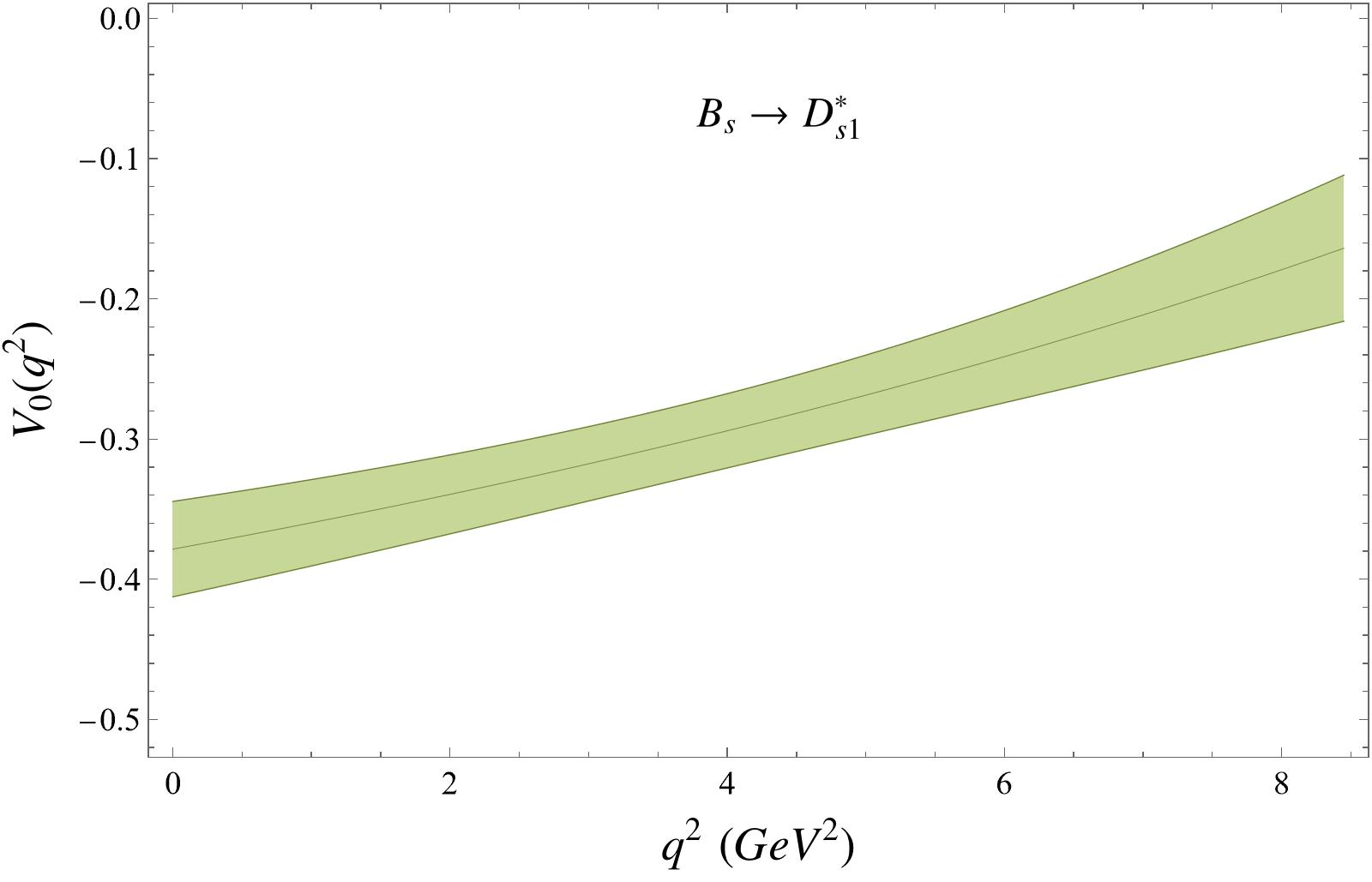}
		
		\vspace{0.3cm}
		
		\includegraphics[width=0.48\textwidth]{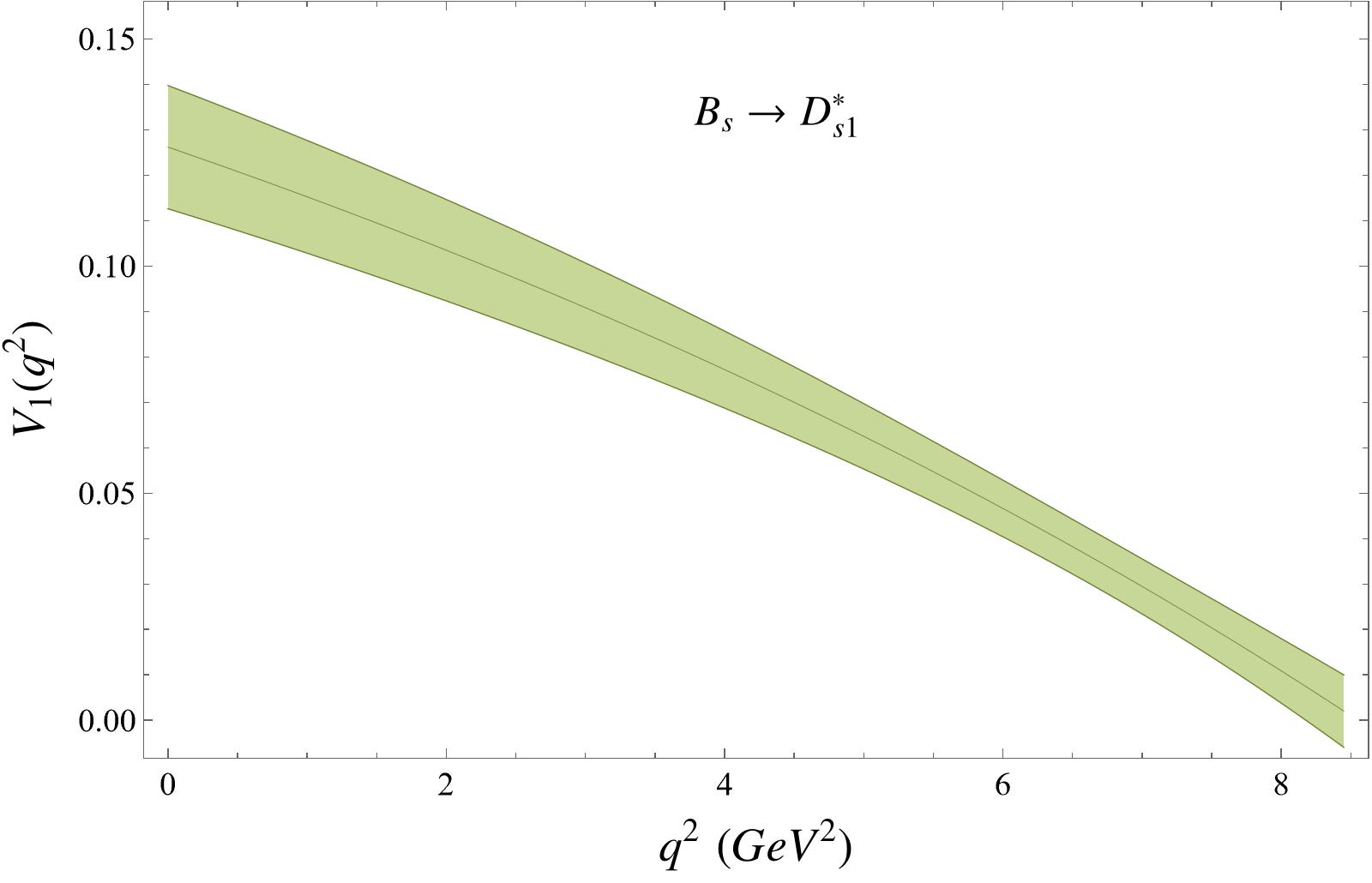}
		\hspace{0.01\textwidth}
		\includegraphics[width=0.48\textwidth]{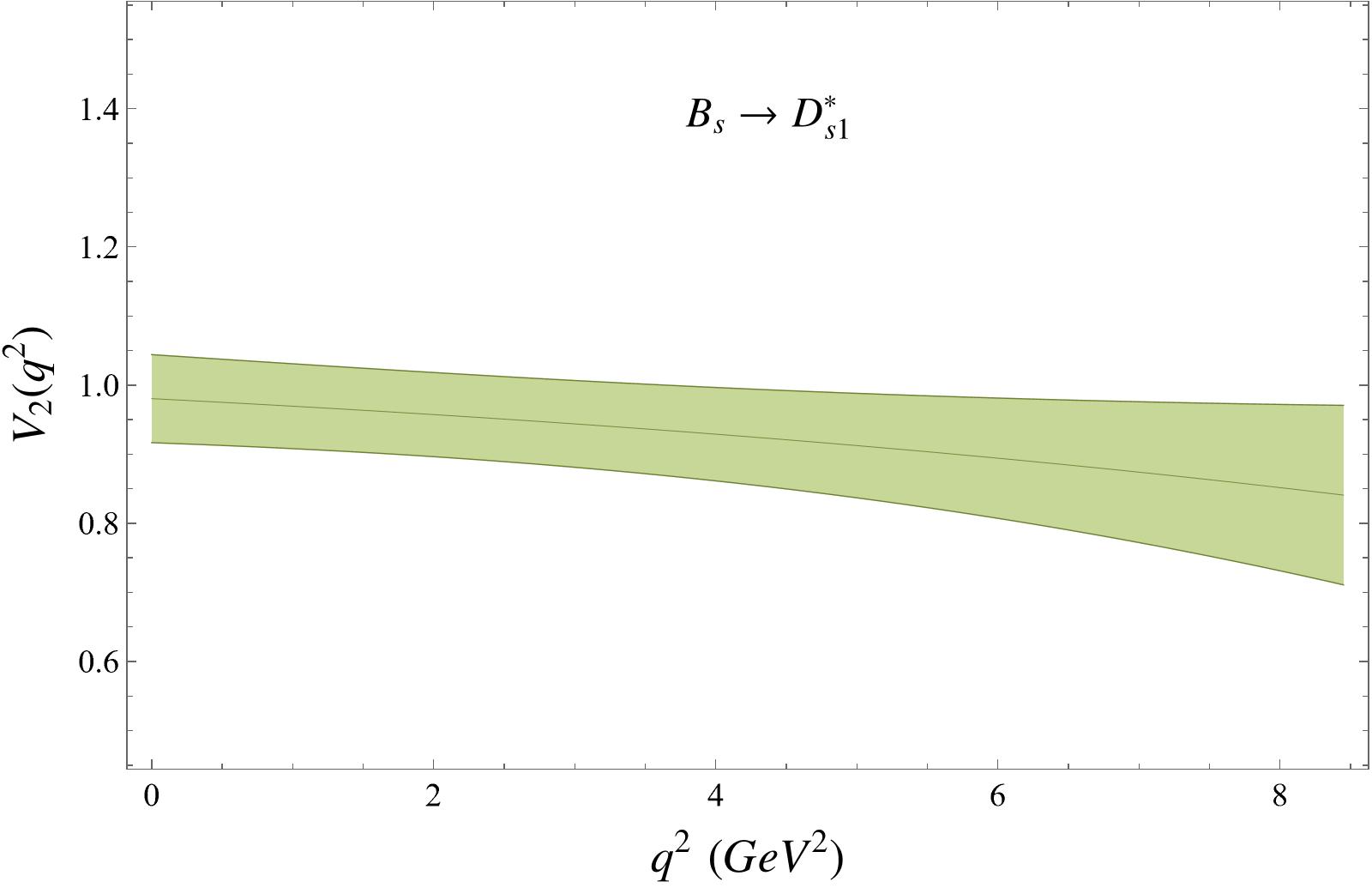}
	\end{center}
	\caption{$q^2$-distribution for $B_s \to D_{s1}^*$ form factors computed with the fit results in table~\ref{Tab:Ds1sBSZ}.}\label{fig:ffDs1s}
\end{figure}

Using both HQET and the BSZ parametrization, we also obtain predictions for the form factors at $q^2 = 0$. For the $B_s \to D_{s1}^{*}$ transition, the predicted form factors at $q^2 = 0$ are
\begin{align}~\label{eq:Ds1stffprd1}
	A(0) = 
	\begin{cases} 
		0.727 (80) ~~ \text{HQET} \\
		0.660 (61) ~~ \text{BSZ} 
	\end{cases}\, ,
	\hspace{1.5cm}
	V_0(0) = 
	\begin{cases}
		-0.407 (50) ~~ \text{HQET} \\
		-0.379 (34) ~~ \text{BSZ}
	\end{cases}\, ,\nonumber
\end{align}
\begin{align}
	V_1(0) = 
	\begin{cases} 
		0.193 (23) ~~ \text{HQET} \\
		0.126 (14) ~~ \text{BSZ} 
	\end{cases}\, ,
	\hspace{1.5cm}
	V_2(0) = 
	\begin{cases}
		1.21 (15) ~~ \text{HQET} \\
		0.980 (64) ~~ \text{BSZ}
	\end{cases}\, .
\end{align}
The HQET parametrization yields a form factor uncertainty of approximately $(11-15)\%$ at $q^2=0$, whereas the BSZ parametrization reduces this uncertainty to about $(7-11)\%$. 

\item $B_s \to D_{s1}\ell\nu_\ell$

The analysis of this mode follows a methodology similar to that of the $B_s \to D_{s1}^*\ell\nu_\ell$ mode, as the $B_s \to D_{s1}$ form factors are also defined by eqs.~\ref{eq:hadmat_axvec}. Here, the synthetic data is generated at $q^2= 4.0$ GeV$^2$, $6.0$ GeV$^2$ and $8.02$ GeV$^2$ using the HQET parameters in table~\ref{tab:tau_fit} and eq.~\ref{eq:ff_Ds1}. The results are listed in table~\ref{tab:syn_dat_Ds1}. The maximum value of $z$ here is $\sim 0.017$ and the kinematic constraint at $q^2=0$ (eq.~\ref{eq:kin_const_vec}) implies that there are only seven free parameters, with $a_0^{V_1}$ removed from the free parameter space. The fitted BSZ coefficients are presented in table~\ref{Tab:Ds1BSZ}. There are no available LCSR inputs yet for the $B_s \to D_{s1}$ transition, and thus the analysis depends only on the synthetic data generated using HQET parameters. The resulting form factors across the full kinematic region are displayed in figure~\ref{fig:ffDs1}, with the green band indicating the $1\sigma$ region.

\begin{table}[t]
	\begin{center}
		\renewcommand*{\arraystretch}{1.4}
		\resizebox{0.65 \textwidth}{!}{
		\begin{tabular}{|c|c|c|c|}
			\hline
			Form Factors & \multicolumn{3}{c|}{$q^2$ values (GeV$^2$)} \\
			\cline{2-4}
			& 4.0 & 6.0 & $q^2_{\text{max}}= 8.02$ \\
			\hline
			$A(q^2)$ & $-0.67 (16)$ & $-0.80 (16)$ & $-0.91 (16)$ \\
			$V_0(q^2)$ & $-0.599 (32)$ & $-0.679 (80)$ & $-0.74 (13)$ \\
			$V_1(q^2)$ & $-0.17 (13)$ & $-0.196 (83)$ & $-0.199 (36)$ \\
			$V_2(q^2)$ & $0.63 (68)$ & $0.71 (59)$ & $0.82 (50)$ \\
			\hline
		\end{tabular}
	}
	\caption{Generated synthetic data for $B_s \to D_{s1}$ form factors.}\label{tab:syn_dat_Ds1}
	\end{center}
\end{table}

\begin{table}[t]
	\begin{center}
		\renewcommand*{\arraystretch}{1.4}
		\resizebox{1.02 \textwidth}{!}{
			\begin{tabular}{|c|c|c|c|c|c|c|c|}
				\hline
				$\chi^2/dof = 8.69/4$ & $a_0^{V_0}$ & $a_1^{V_0}$ & $a_1^{V_1}$ & $a_0^{V_2}$ & $a_1^{V_2}$ & $a_0^{A}$ & $a_1^{A}$ \\
				\hline
				Fit Results & $-0.611 (37)$ & $-5.73 (77)$  & $-8.5 (18)$ & $-0.14 (21)$ & $-13.7 (48)$ & $-0.49 (19)$ & $7.6 (44)$ \\
				\hline
			\end{tabular}
		}
			\caption{Fit results for the BSZ coefficients $a_n$ in the $B_s \to D_{s1}$ transition.}\label{Tab:Ds1BSZ}
	\end{center}
\end{table}
At $q^2 = 0$, the estimated $B_s \to D_{s1}$ form factors, obtained using both HQET inputs and the BSZ parametrization, are given below

\begin{align}\label{eq:Ds1ffprd1}
	A(0) = 
	\begin{cases} 
		-0.36 (19) ~~ \text{HQET} \\
		-0.49 (19) ~~ \text{BSZ} 
	\end{cases}\, ,
	\hspace{1.5cm}
	V_0(0) = 
	\begin{cases}
		-0.39 (13) ~~ \text{HQET} \\
		-0.611 (37) ~~ \text{BSZ}
	\end{cases}\, ,\nonumber
\end{align}
\begin{align}
	V_1(0) = 
	\begin{cases} 
		-0.065 (226) ~~ \text{HQET} \\
		-0.442 (76) ~~ \text{BSZ} 
	\end{cases}\, ,
	\hspace{1.5cm}
	V_2(0) = 
	\begin{cases}
		0.52 (83) ~~ \text{HQET} \\
		-0.14 (21) ~~ \text{BSZ}
	\end{cases}\, .
\end{align}

\begin{figure}[t]
	\begin{center}
		\includegraphics[width=0.48\textwidth]{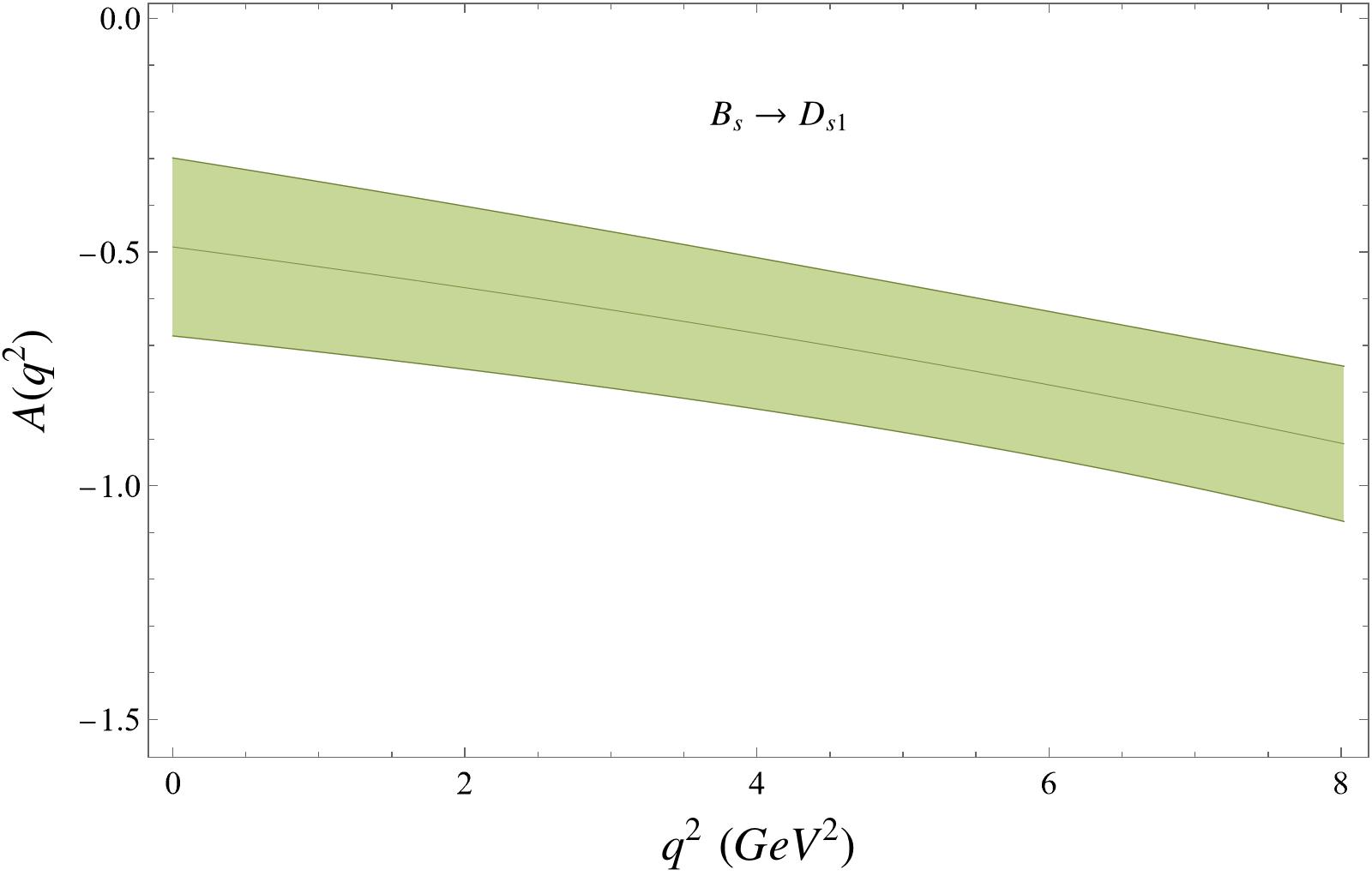}
		\hspace{0.01\textwidth}
		\includegraphics[width=0.48\textwidth]{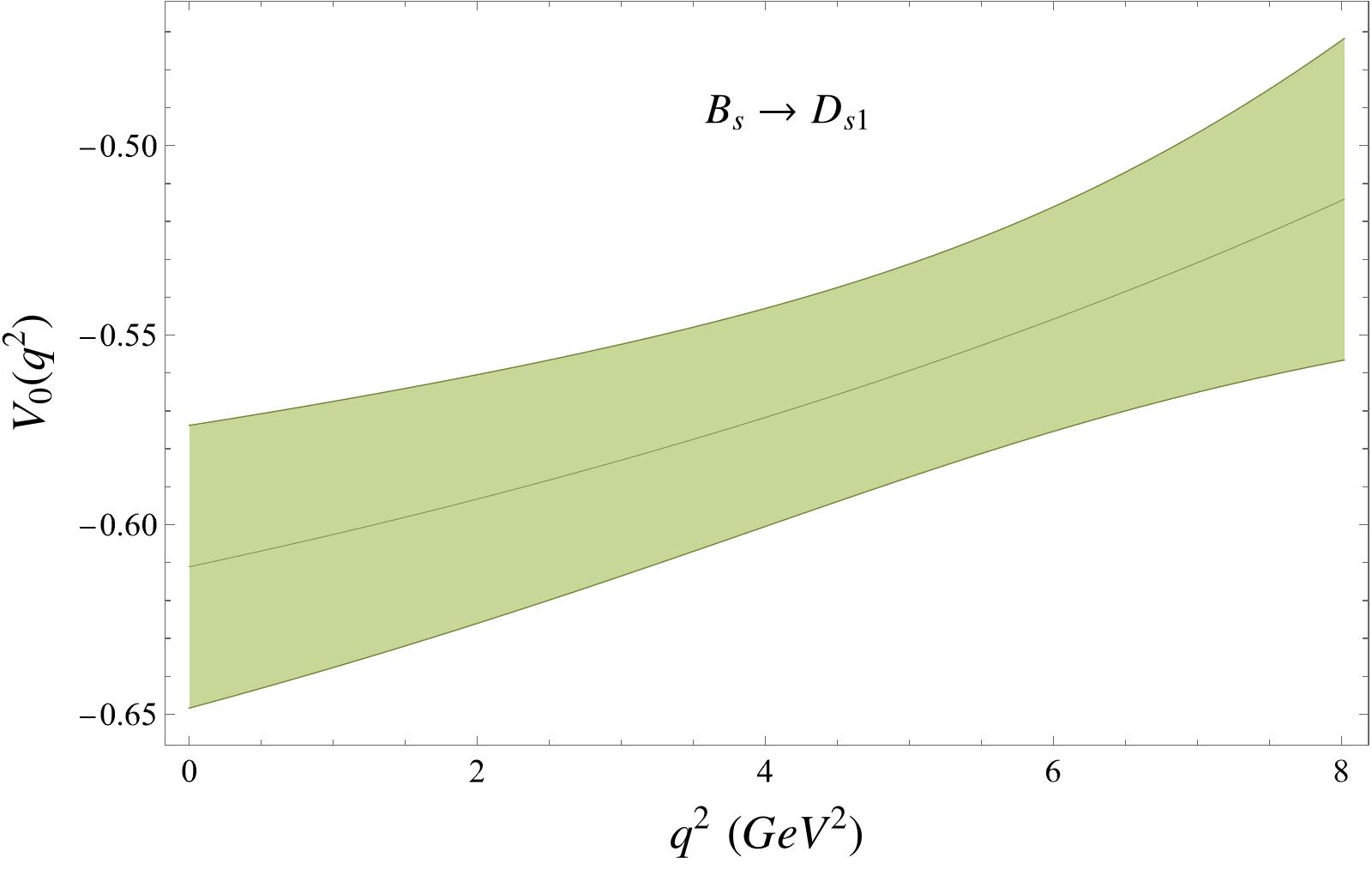}
		
		\vspace{0.3cm}
		
		\includegraphics[width=0.48\textwidth]{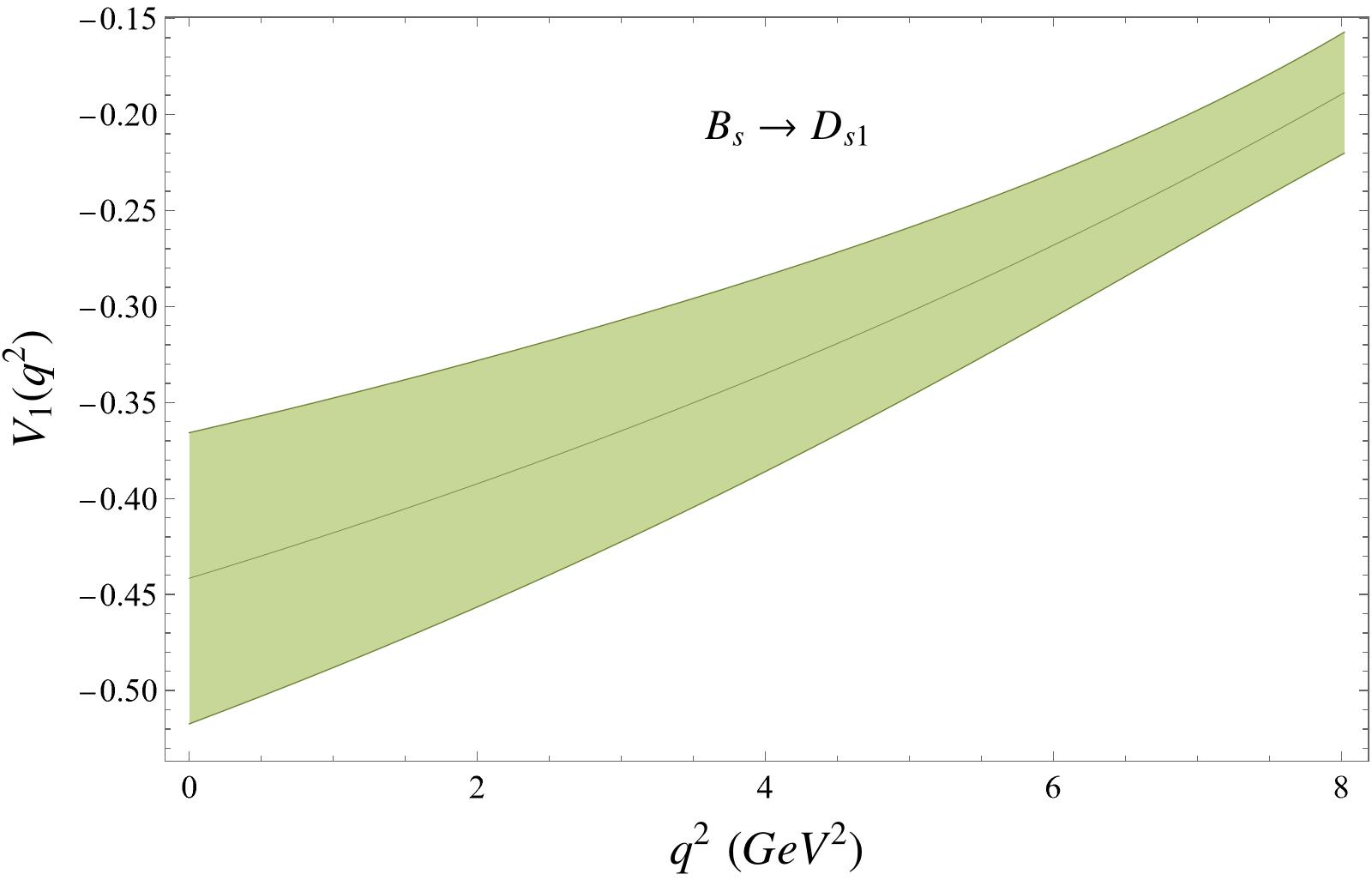}
		\hspace{0.01\textwidth}
		\includegraphics[width=0.48\textwidth]{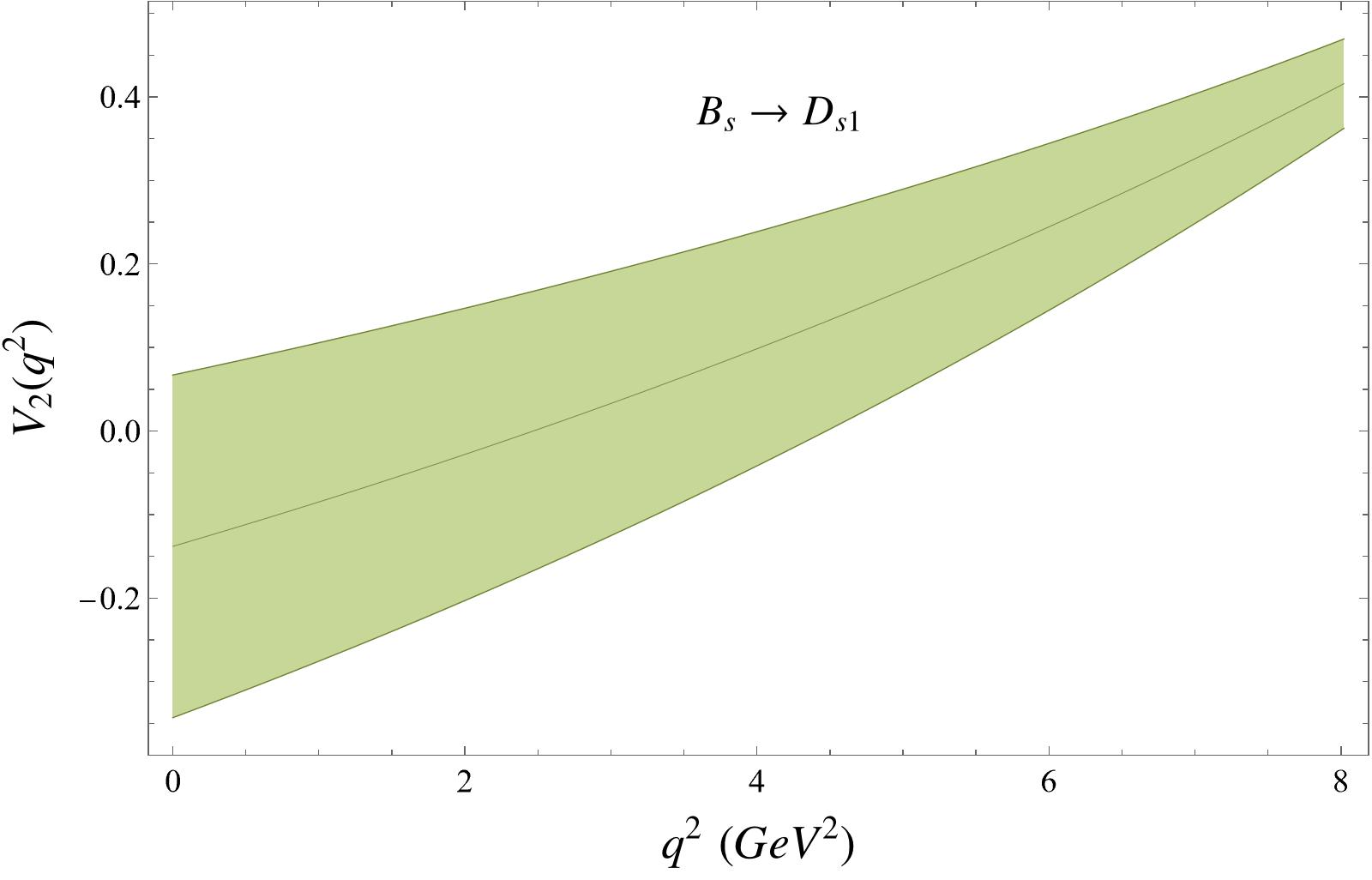}
	\end{center}
	\caption{$q^2$-distribution for $B_s \to D_{s1}$ form factors computed with the fit results in table~\ref{Tab:Ds1BSZ}.}\label{fig:ffDs1}
\end{figure}
\item $B_s \to D_{s2}^*\ell\nu_\ell$

The $B_s \to D_{s2}^*$ transition is described by four form factors $V (q^2)$, $A_0 (q^2)$, $A_1 (q^2)$ and $A_2 (q^2)$ as given in eqs.~\ref{eq:hadmat_tens}. Using eq.~\ref{eq:ff_Ds2s} with HQET inputs from table~\ref{tab:tau_fit}, synthetic data are generated at $q^2= 4.0$ GeV$^2$, $6.0$  GeV$^2$ and 7.83 GeV$^2$, which are summarized in table~\ref{tab:syn_dat_Ds2s}.  For $N=1$, the BSZ parametrization introduces eight coefficients, two per form factor. Since $z_\text{\text{max}} \sim 0.017$, the expansion converges quickly, requiring only a few parameters. Similar to the previous cases, the kinematic constraint of eq.~\ref{eq:kin_const_vec} eliminates $a_0^{A_1}$, leaving seven free parameters, reported in table~\ref{Tab:Ds2sBSZ}. Since LCSR inputs for these modes are unavailable, the analysis is performed solely using synthetic data generated from HQET parameters. The resulting form factors over the full kinematic region are shown in figure~\ref{fig:ffDs2s}, with the green band representing the $1\sigma$ region.

\begin{table}[t]
	\begin{center}
		\renewcommand*{\arraystretch}{1.4}
		\resizebox{0.65 \textwidth}{!}{
		\begin{tabular}{|c|c|c|c|}
			\hline
			Form Factors & \multicolumn{3}{c|}{$q^2$ values (GeV$^2$)} \\
			\cline{2-4}
			& 4.0 & 6.0 & $q^2_{\text{max}}= 7.83$ \\
			\hline
			$V(q^2)$ & $-1.19 (81)$ & $-1.34 (77)$ & $-1.49 (73)$ \\
			$A_0(q^2)$ & $-0.60 (17)$ & $-0.71 (23)$ & $-0.81 (28) $ \\
			$A_1(q^2)$ & $-0.607 (43) $ & $-0.675 (76)$ & $-0.73 (13)$ \\
			$A_2(q^2)$ & $-0.80 (51)$ & $-0.94 (47)$ & $-1.07 (45)$ \\
			\hline
		\end{tabular}
	}
	\caption{Generated synthetic data for $B_s \to D_{s2}^*$ form factors.}\label{tab:syn_dat_Ds2s}
	\end{center}
\end{table}


\begin{table}[t]
	\begin{center}
		\renewcommand*{\arraystretch}{1.6}
		\resizebox{0.95 \textwidth}{!}{
		\begin{tabular}{|c|c|c|c|c|c|c|c|}
			\hline
			$\chi^2/dof = 0.003/1$ & $a_1^{A_0}$ & $a_0^{A_1}$ & $a_1^{A_1}$ & $a_0^{A_2}$& $a_1^{A_2}$ & $a_0^{V}$ & $a_1^{V}$ \\
			\hline
			Fit Results & $5.5 (41)$ & $-0.528 (94)$ & $2.4 (54)$ & $-0.64 (30)$ & $7.8 (49)$ & $-0.98 (87)$ & $6.4 (95)$ \\
			\hline
		\end{tabular}
	     }
		\caption{Fit results for the BSZ coefficients $a_n^i$ for the $B_s \to D_{s2}^*$ transition.}\label{Tab:Ds2sBSZ}
	\end{center}
\end{table}

\begin{figure}[t]
	\begin{center}
		\includegraphics[width=0.48\textwidth]{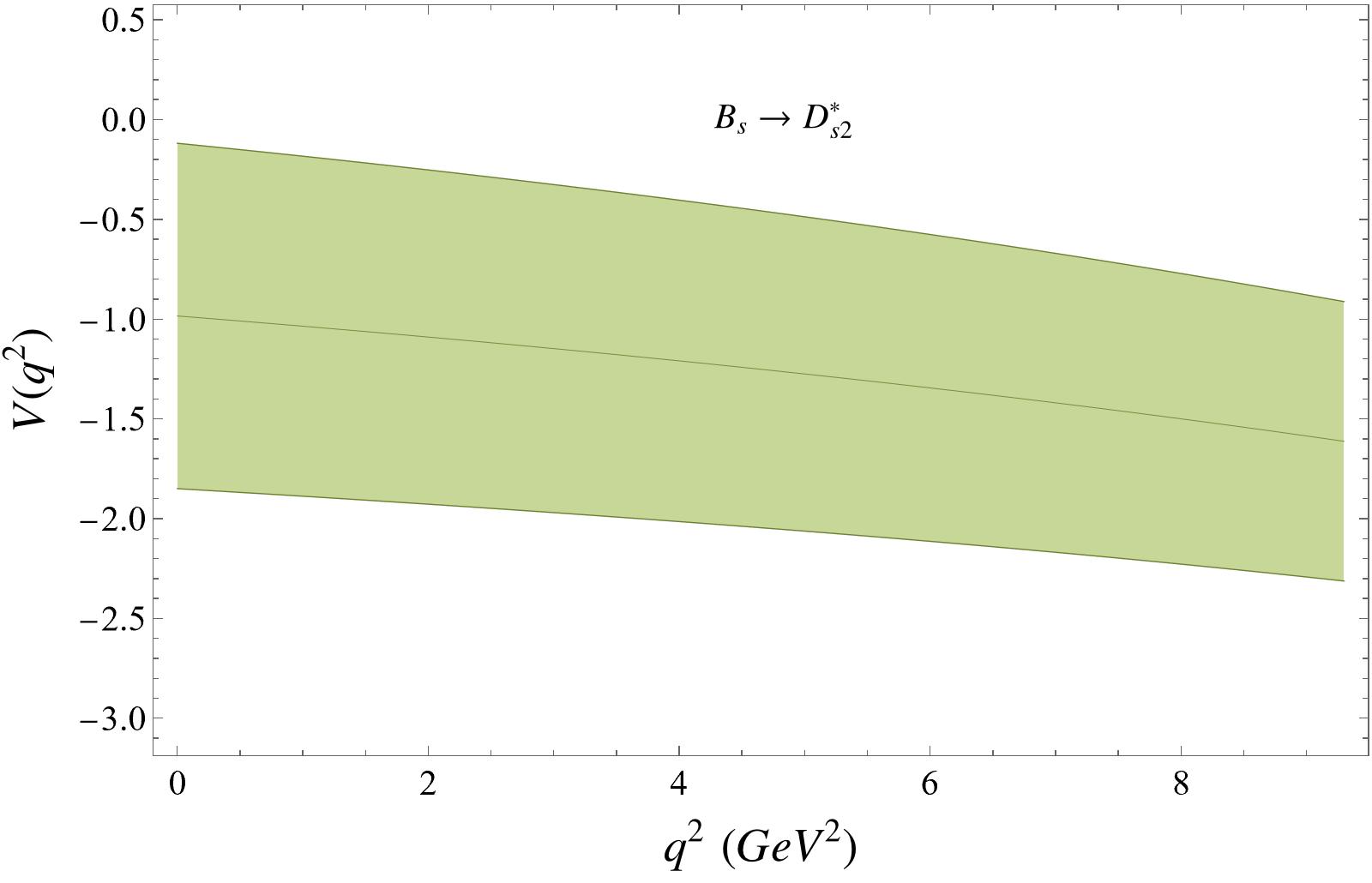}
		\hspace{0.01\textwidth}
		\includegraphics[width=0.48\textwidth]{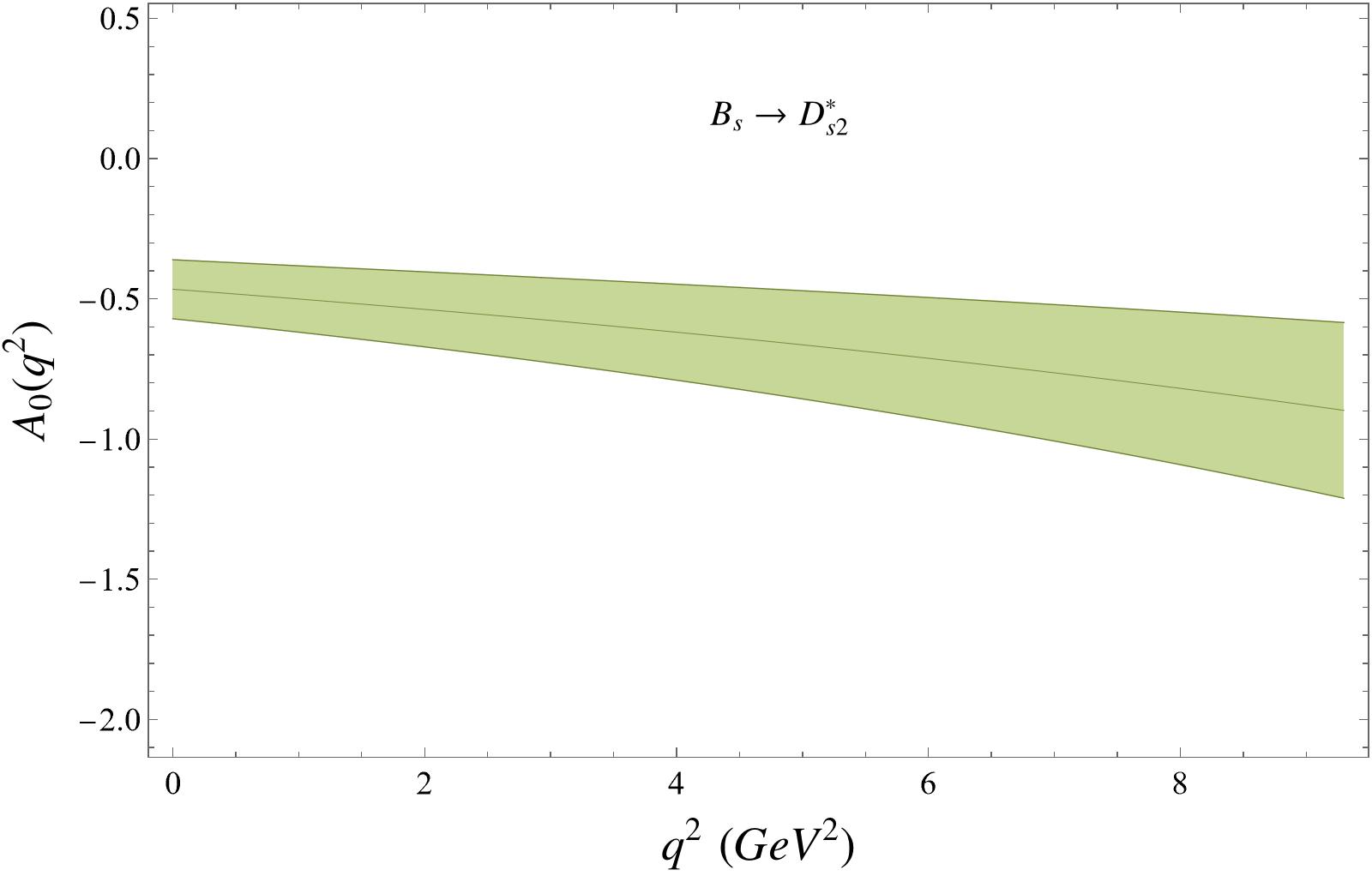}
		
		\vspace{0.3cm}
		
		\includegraphics[width=0.48\textwidth]{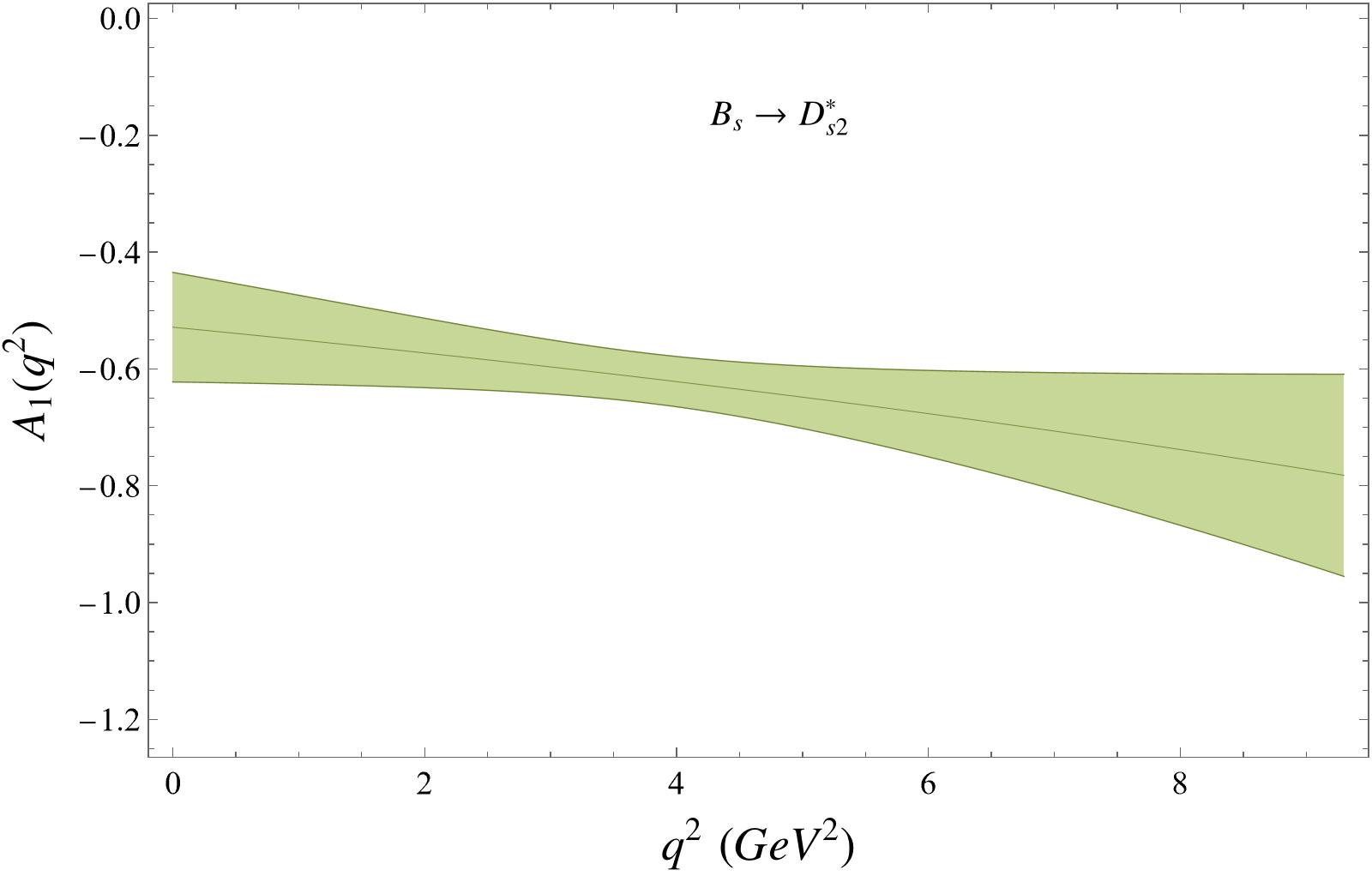}
		\hspace{0.01\textwidth}
		\includegraphics[width=0.48\textwidth]{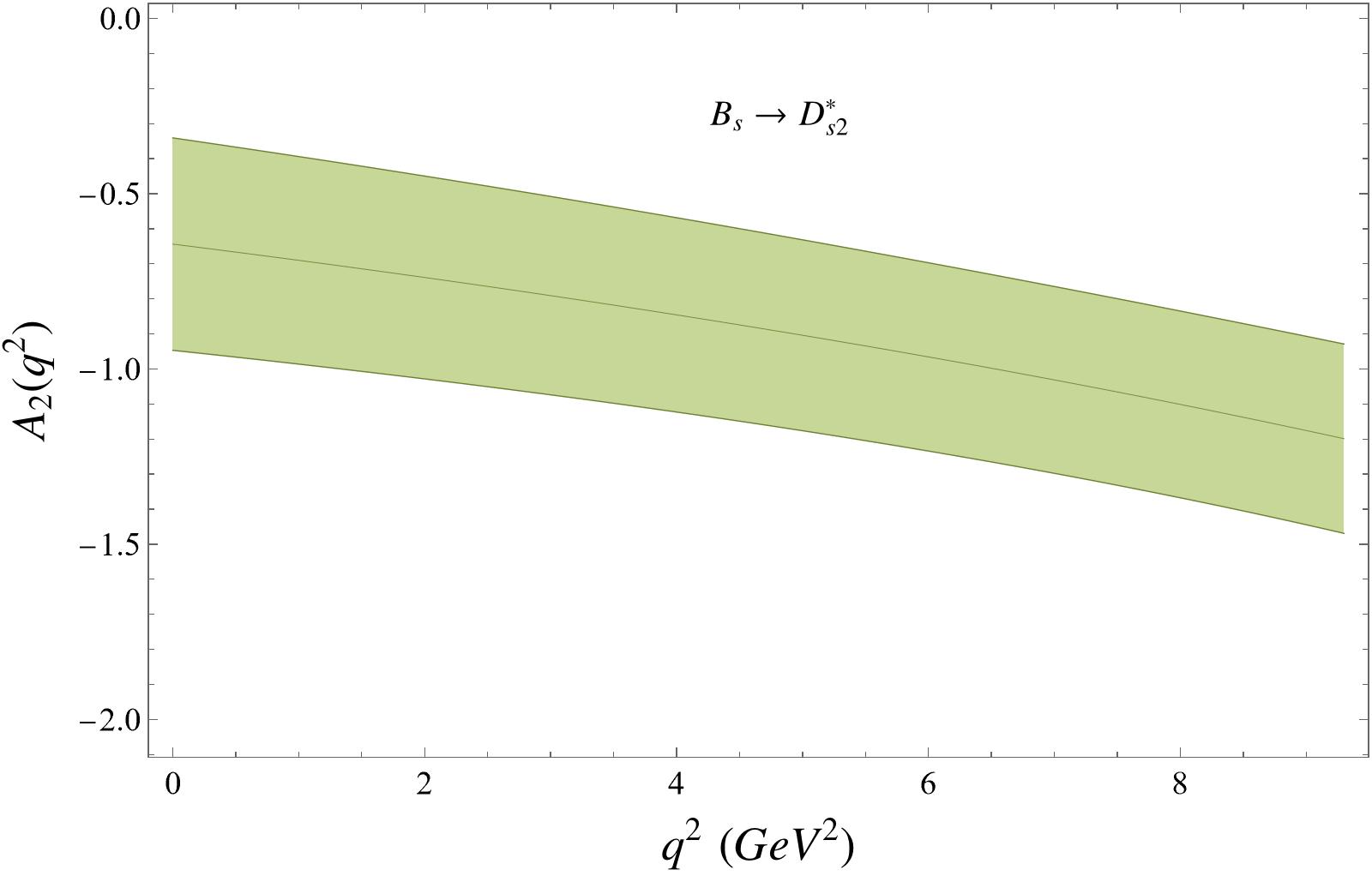}
	\end{center}
		\caption{$q^2$-distribution for $B_s \to D_{s2}^*$ form factors computed with the fit results in table~\ref{Tab:Ds2sBSZ}.}~\label{fig:ffDs2s}
\end{figure}

Our form factor estimates at $q^2 = 0$, obtained using both HQET inputs and the BSZ parametrization, are 
\begin{align}~\label{eq:Ds2stffprd1}
	V(0) = 
	\begin{cases} 
		-0.88 (91) ~~ \text{HQET} \\
		-0.98 (87) ~~ \text{BSZ} 
	\end{cases}\, ,
	\hspace{1.5cm}
	A_0(0) = 
	\begin{cases}
		-0.392 (89) ~~ \text{HQET} \\
		-0.47 (11) ~~ \text{BSZ}
	\end{cases}\, ,\nonumber
\end{align}
\begin{align}
	A_1(0) = 
	\begin{cases} 
		-0.44 (17) ~~ \text{HQET} \\
		-0.53 (9) ~~ \text{BSZ}
	\end{cases}\, ,
	\hspace{1.5cm}
	A_2(0) = 
	\begin{cases}
		-0.53 (60) ~~ \text{HQET} \\
		-0.64 (30) ~~ \text{BSZ}
	\end{cases}\, .
\end{align}
\end{enumerate}

The form factor estimates at $q^2=0$ for each of the considered transitions, as given in eqs.~\ref{eq:fq20prd_comb} $-$ \ref{eq:Ds2stffprd1}, serve as benchmarks for future lattice QCD determinations or other model-dependent estimates. The shapes of the transition form factors across all modes can be further constrained once lattice QCD and/or LCSR inputs are available. In addition, precise experimental measurements will be essential to establish their behavior over the full kinematic region. 
\section{Observables Predictions within the Standard Model }\label{sec:SM_pred}
With the form factor estimates available across the full kinematic region, we now proceed to make predictions for several observables defined in subsection~\ref{sec:obs} for both the muon and tau decay modes. 

\subsection{$q^2$-distribution of the Observables}
\begin{enumerate}
\item $B_s \to D_{s0}^*\ell\nu_\ell$

\begin{figure}[t]
	\begin{center}
		\includegraphics[width=.48\textwidth]{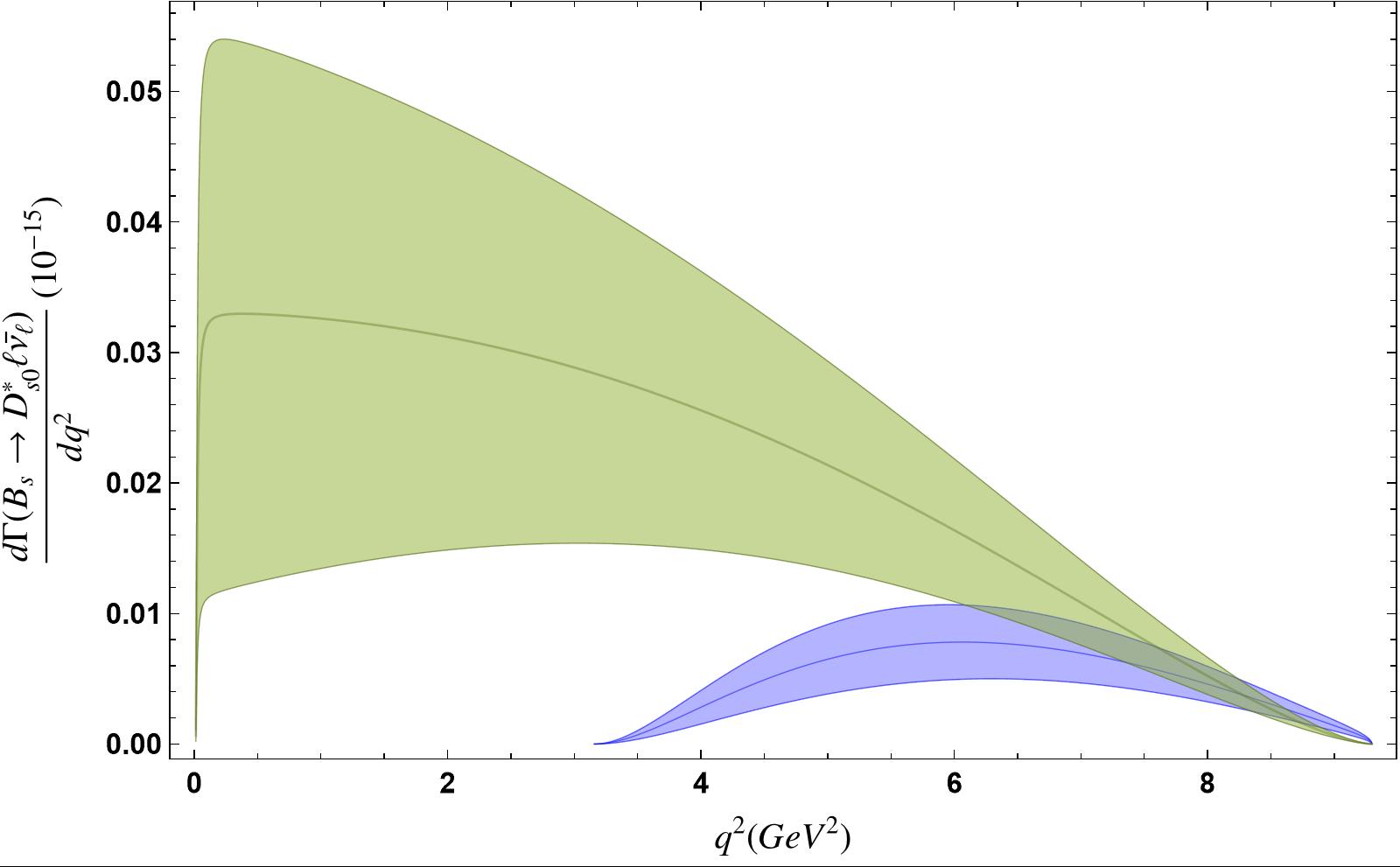}
		\hspace{0.01\textwidth}
		\includegraphics[width=.48\textwidth]{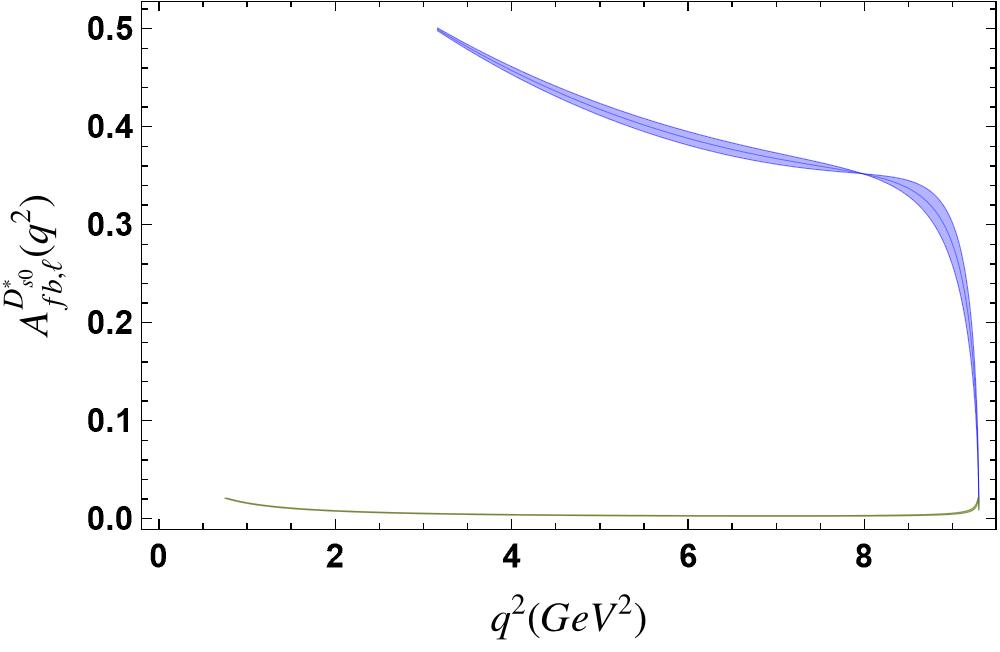}
		
		\vspace{0.3cm}
		
		\includegraphics[width=0.48\textwidth]{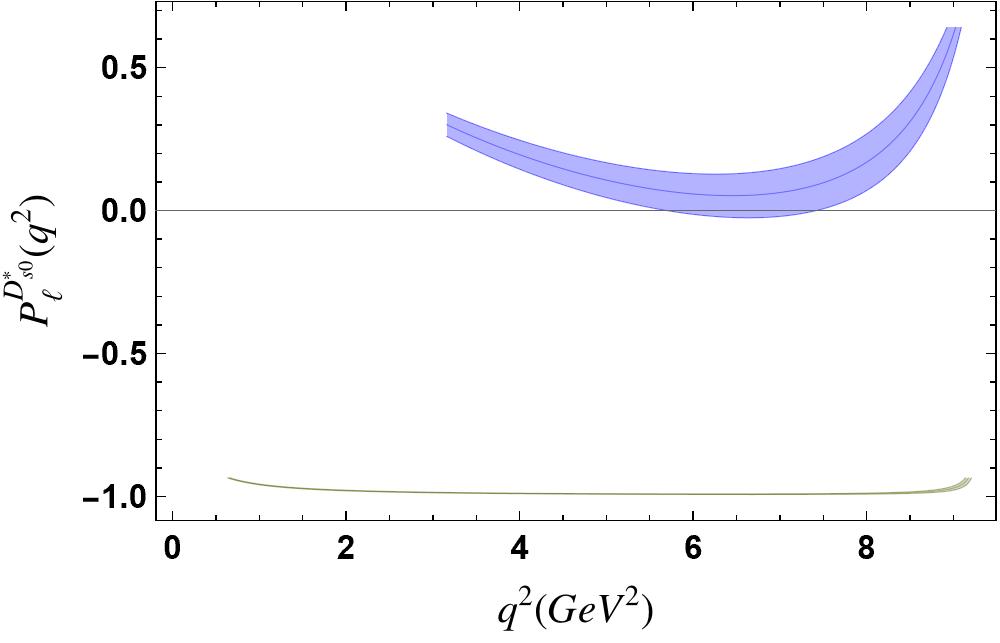}
		\hspace{0.01\textwidth}
		\includegraphics[width=0.48\textwidth]{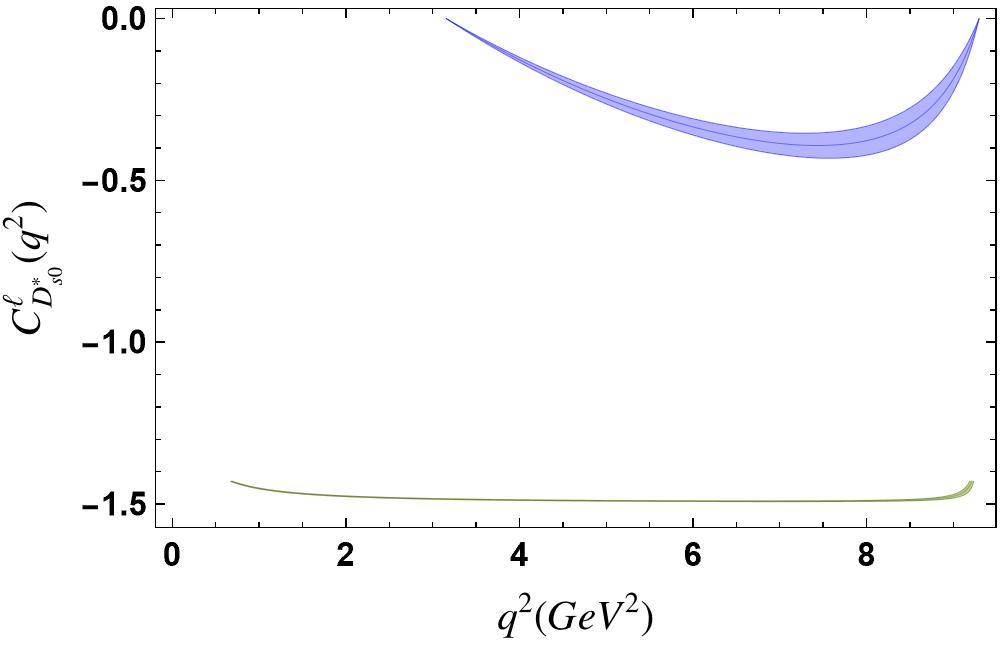}
		
		\includegraphics{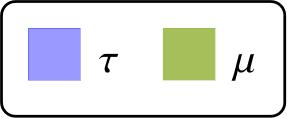}
	\end{center}
	\caption{$q^2$-distribution of $B_s \to D_{s0}^*\ell \nu_\ell$ observables. The bands indicate the $1\sigma$ regions.	}\label{fig:SMObsDs0s}
\end{figure}

The $q^2$-distribution of the observables associated with the $B_s \to D_{s0}^*\ell \nu$ decay  is shown in figure~\ref{fig:SMObsDs0s}. From this figure, it is evident that both $C_{D_{s0}^*}^\mu(q^2)$ and $P_\mu^{D_{s0}^*}(q^2)$ exhibit negligible uncertainties across the full $q^2$ kinematic range. The observables of $B_s \to D_{s0}^*\ell\nu_\ell$ depend solely on two hadronic helicity amplitudes, $H_0(q^2)$ and $H_t(q^2)$, which constitute the primary sources of uncertainty. In the SM, $H_t(q^2)$ is strongly suppressed for light leptons, while $H_0(q^2)$ cancels in the ratio-based observables as $m_\ell^2 \to 0$. As a result, $C_{D_{s0}^*}^\mu(q^2)$ and $P_\mu^{D_{s0}^*}(q^2)$ have marginal uncertainties, and are also flat throughout the $q^2$ region. Furthermore, eq.~\ref{eq:abc_Ds0s} shows that the angular coefficient $b_{\theta_\ell}$ is $m_\ell^2$-suppressed, leading to $A_{fb,\mu}^{D_{s0}^*}(q^2)$ values being very close to zero. Across the full kinematic region, $A_{fb,\ell}^{D_{s0}^*}(q^2)$ remains positive, while $C_{D_{s0}^*}^\ell(q^2)$ remains negative. We also find that Across the full kinematic region, $A_{fb,\ell}^{D_{s0}^*}(q^2)$ remains positive, while $C_{D_{s0}^*}^\ell(q^2)$ remains negative. We also find that $P_\mu^{D_{s0}^*}(q^2)$ is negative and $P_\tau^{D_{s0}^*}(q^2)$ is positive throughout the entire $q^2$ region.

\item $B_s \to D_{s1}^*\ell\nu_\ell$

\begin{figure}[t]
	\centering
	
	\begin{minipage}[c]{0.31\textwidth}
		\centering
		\includegraphics[width=\linewidth]{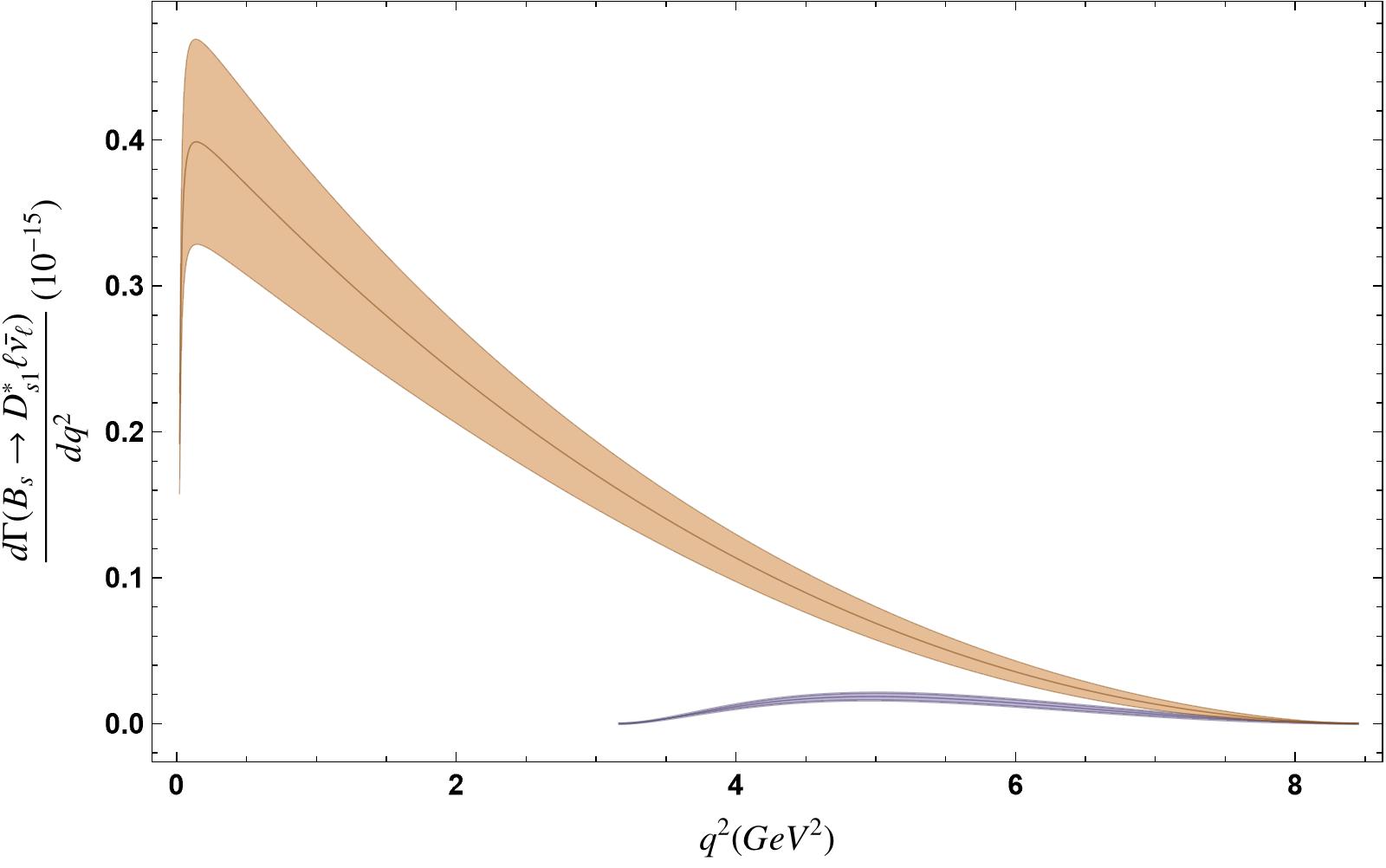}
	\end{minipage}
	\hspace{0.01\textwidth}
	\begin{minipage}[c]{0.31\textwidth}
		\centering
		\includegraphics[width=\linewidth]{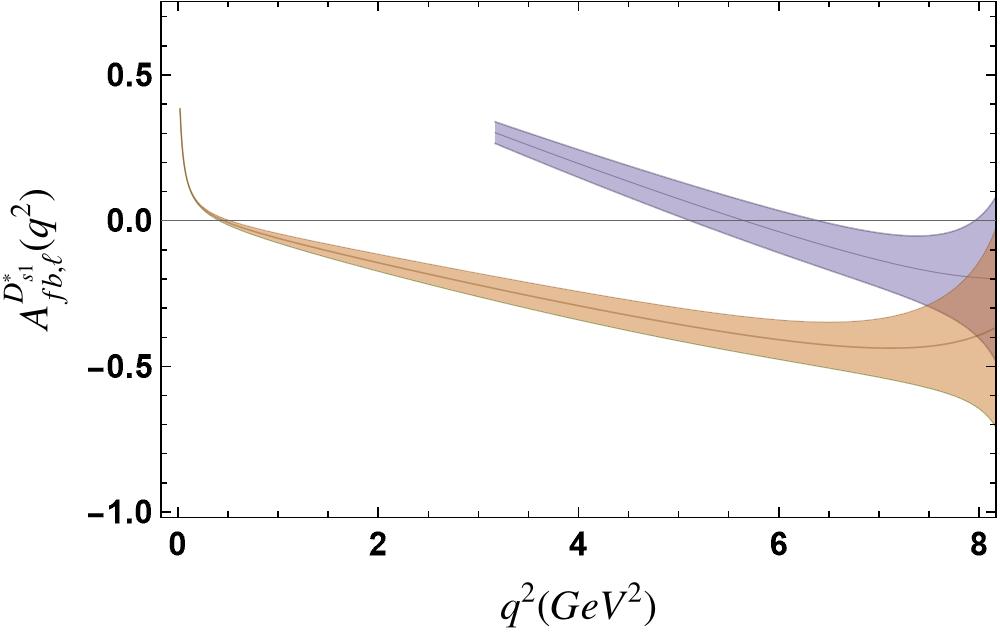}
	\end{minipage}
	\hspace{0.01\textwidth}
	\begin{minipage}[c]{0.31\textwidth}
		\centering
		\includegraphics[width=\linewidth]{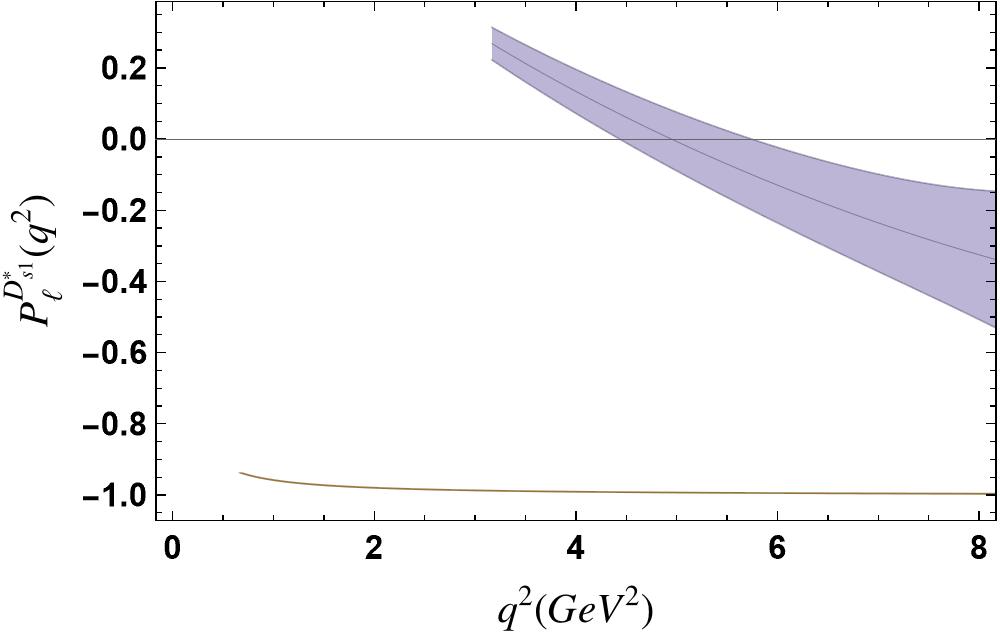}
	\end{minipage}
	
	\vspace{0.3cm}
	
	\begin{minipage}[c]{0.31\textwidth}
		\centering
		\includegraphics[width=\linewidth]{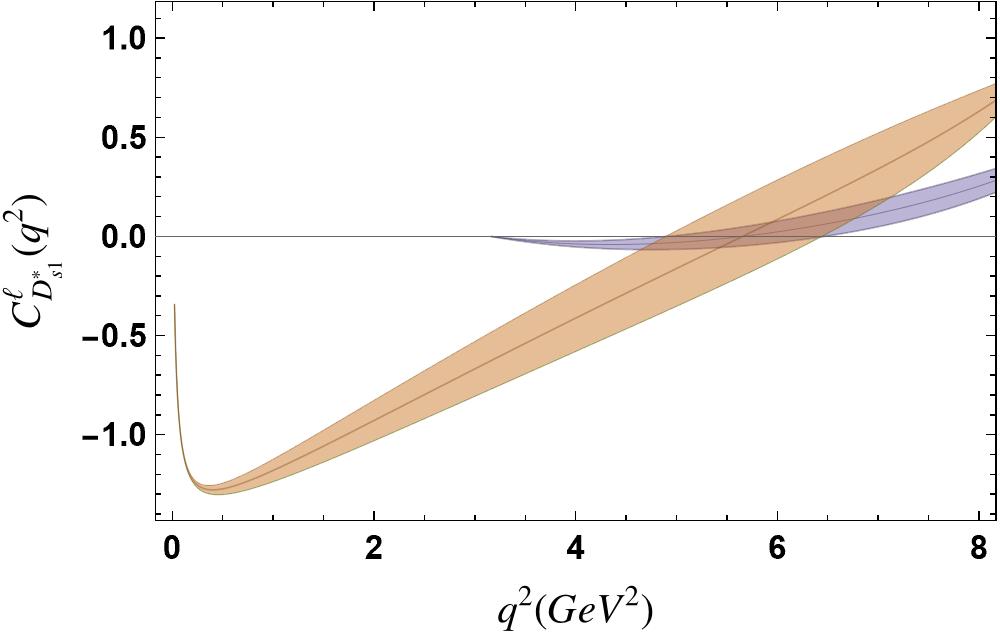}
	\end{minipage}
	\hspace{0.01\textwidth}
	\begin{minipage}[c]{0.31\textwidth}
		\centering
		\includegraphics[width=\linewidth]{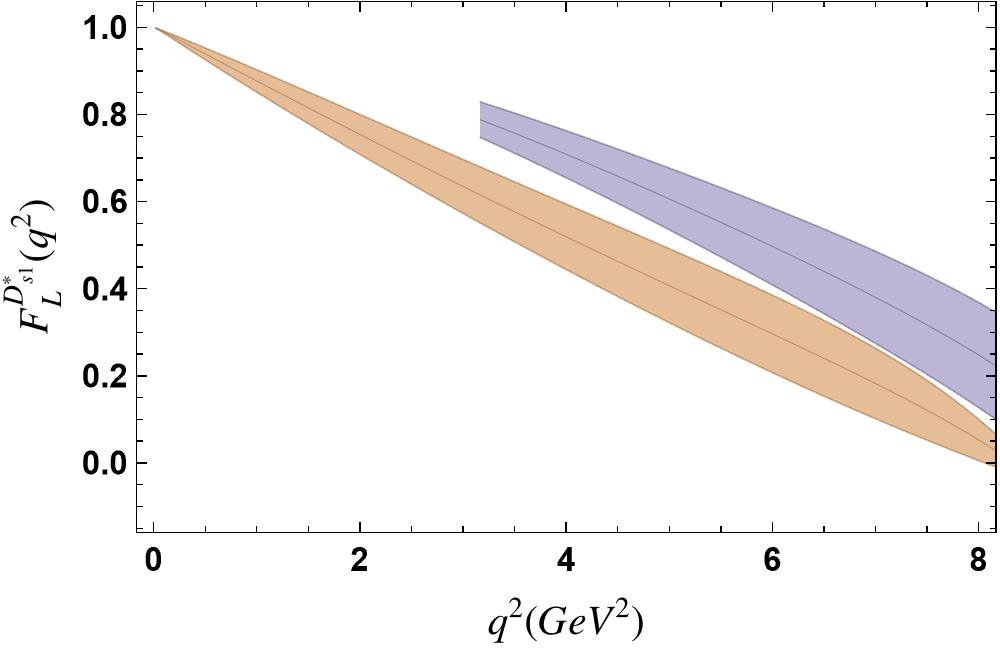}
	\end{minipage}
	\hspace{0.01\textwidth}
	\begin{minipage}[c]{0.31\textwidth}
		\centering
		\raisebox{0.25\height}{
			\includegraphics[width=0.34\linewidth]{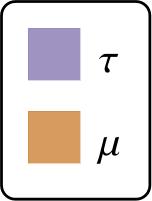}
		}
	\end{minipage}
	\caption{$q^2$-distribution of $ B_s \to D_{s1}^*\ell \nu_\ell$ observables. The bands indicate the $1\sigma$ regions.}\label{fig:SMObsDs1s}
\end{figure}

For this mode, the $q^2$-distribution is shown in figure~\ref{fig:SMObsDs1s}. Except for the differential decay rate, the $q^2$ spectrum for all other observables is shown only up to $q^2 = 8$ GeV$^2$. This is because the uncertainty grows significantly near $q^2_{\text{max}}$ due to the long tail of the decay rate, which is consolidated around zero. The observables here depend on four hadronic helicity amplitudes, and thus no trivial cancellation occurs in the ratio-based observables as $m_\ell^2 \to 0$. Consequently, there are residual uncertainties. However, these uncertainties are negligible for $P_\mu^{D_{s1}^*}(q^2)$, as this observable is defined as the difference between the decay rates corresponding to different helicity states of the charged lepton, normalized by the total decay rate. As a result, the hadronic helicity amplitudes cancel, leaving only the leptonic helicity contributions. 

Zero-crossings are observed for $A_{fb,\ell}^{D_{s1}^*}(q^2)$ and $C_{D_{s1}^*}^\ell(q^2)$ in both the muon and tau channels, whereas $P_\ell^{D_{s1}^*}(q^2)$ exhibits a zero-crossing only in the tau channel. There is a notable difference in the zero-crossings of  $A_{fb,\mu}^{D_{s1}^*}(q^2)$  and $A_{fb,\tau}^{D_{s1}^*}(q^2)$ . The zero-crossing of $A_{fb,\mu}^{D_{s1}^*}(q^2)$ occurs at $q^2 = 0.458(57)$ GeV$^2$, whereas for $A_{fb,\tau}^{D_{s1}^*}(q^2)$, it is at $q^2 = 5.67(67)$ GeV$^2$. For $C_{D_{s1}^*}^{\mu}(q^2)$ and $C_{D_{s1}^*}^\tau(q^2)$, the zero-crossings coincide at $q^2 = 5.67(83)$ GeV$^2$. The zero-crossing for $P_\tau^{D_{s1}^*}(q^2)$ occurs at $q^2 = 4.94(67)$ GeV$^2$, in the neighbourhood of the central kinematic region. The longitudinal polarization $F_{L}^{D_{s1}^{*}}(q^2)$ remains positive throughout the entire physical region for both the lepton channels.

\item $B_s \to D_{s1}\ell\nu_\ell$

The $q^2$-distribution is shown in figure~\ref{fig:SMObsDs1}. As for the  $ B_s \to D_{s1}^*\ell \nu_\ell$ mode, the lepton polarization asymmetry $P_\mu^{D_{s1}}(q^2)$ exhibits negligible errors. In this mode, a zero-crossing is observed only for $A_{fb,\ell}^{D_{s1}}(q^2)$  and $P_\tau^{D_{s1}}(q^2)$. A striking difference in the zero-crossing pattern of $A_{fb,\ell}^{D_{s1}}(q^2)$ can be observed - it occurs at $q^2 = 0.455(95)$ GeV$^2$ for the muon channel and at $q^2 = 5.69(43)$ GeV$^2$ for the tau channel. Also, $P_\mu^{D_{s1}}(q^2)$ has no zero crossing, whereas $ P_\tau^ {D _ {s1}} (q^2) $ crosses zero at $q^2 = 3.78(27)$ GeV$^2$ in the central $q^2$ region.  Throughout the physical region, the convexity  $C_{D_{s1}}^\ell(q^2)$ is negative, whereas the longitudinal polarization $F_{L}^{D_{s1}}(q^2)$ is positive.

\begin{figure}[t]
	\begin{minipage}[c]{0.31\textwidth}
		\centering
		\includegraphics[width=\textwidth]{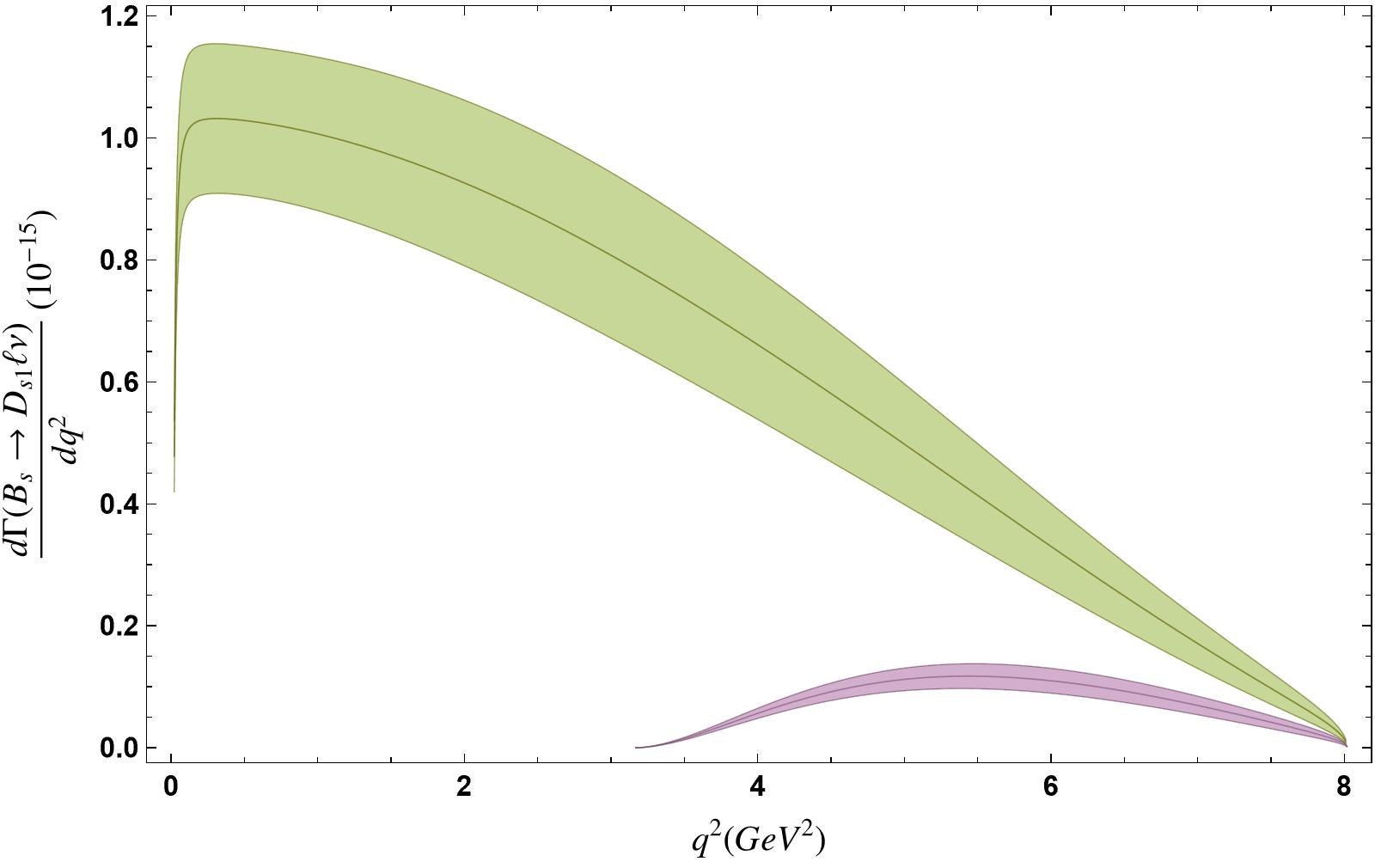}
	\end{minipage}
	\hspace{0.01\textwidth}
	\begin{minipage}[c]{0.31\textwidth}
		\centering
		\includegraphics[width=\textwidth]{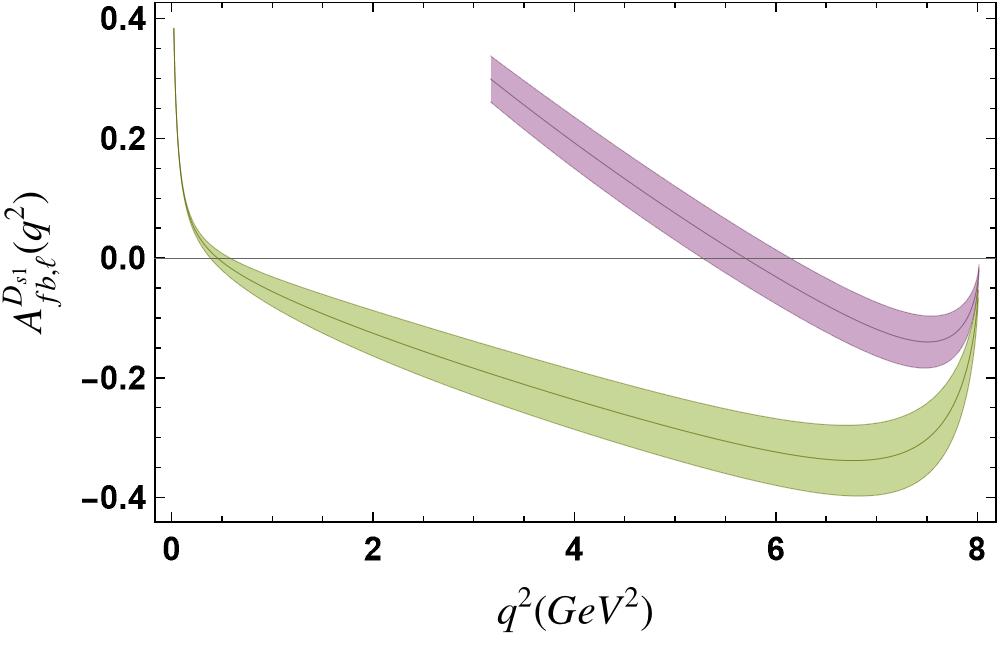}
	\end{minipage}
	\hspace{0.01\textwidth}
	\begin{minipage}[c]{0.31\textwidth}
		\centering
		\includegraphics[width=\textwidth]{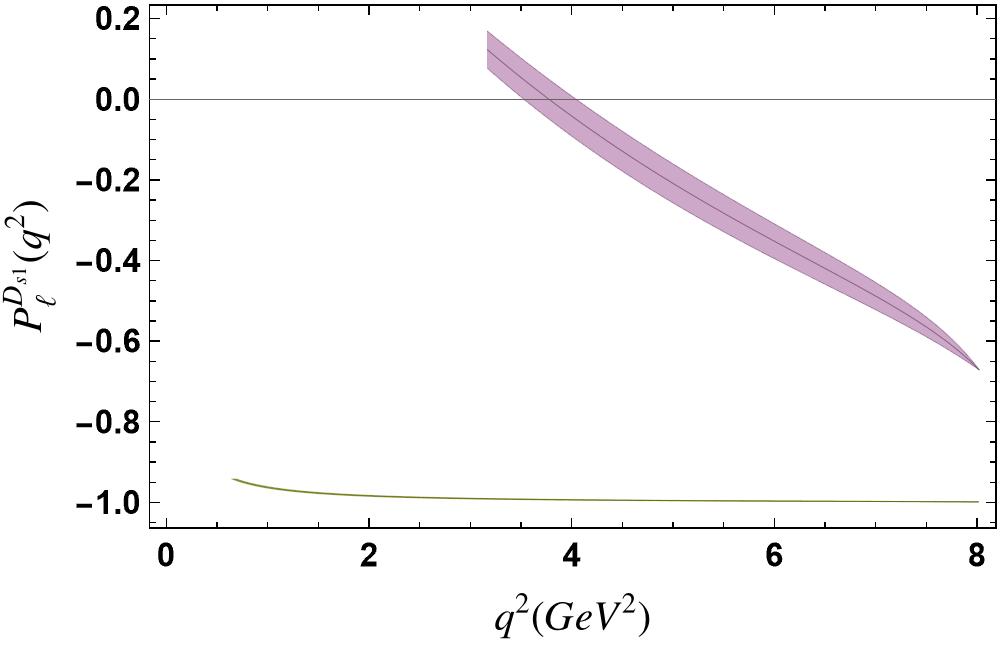}
	\end{minipage}
	
	\vspace{0.3cm}
	
	\begin{minipage}[c]{0.31\textwidth}
		\centering
		\includegraphics[width=\textwidth]{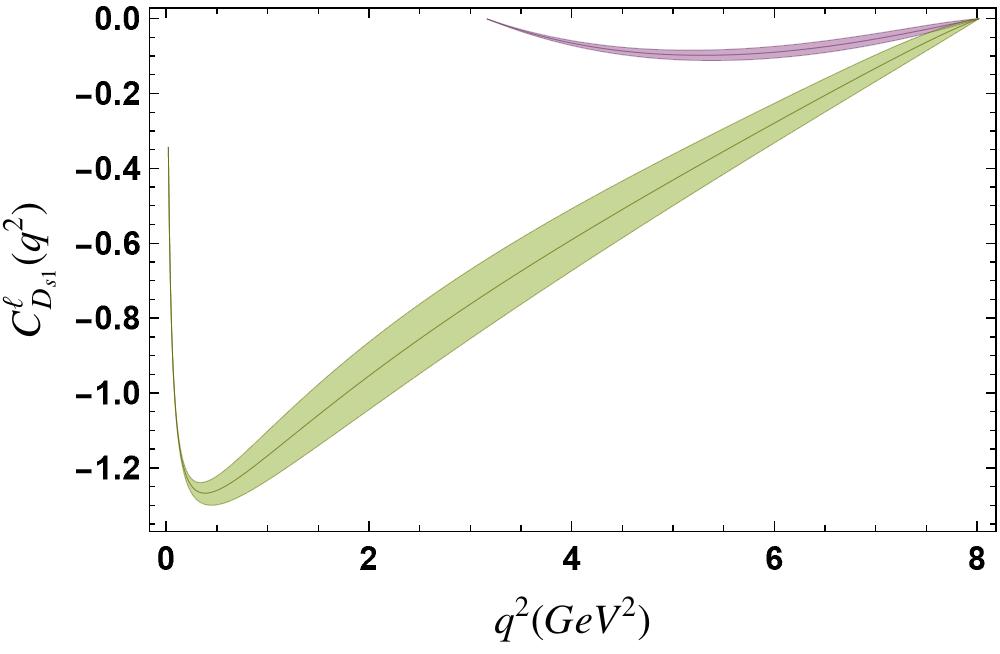}
	\end{minipage}
	\hspace{0.01\textwidth}
	\begin{minipage}[c]{0.31\textwidth}
		\centering
		\includegraphics[width=\textwidth]{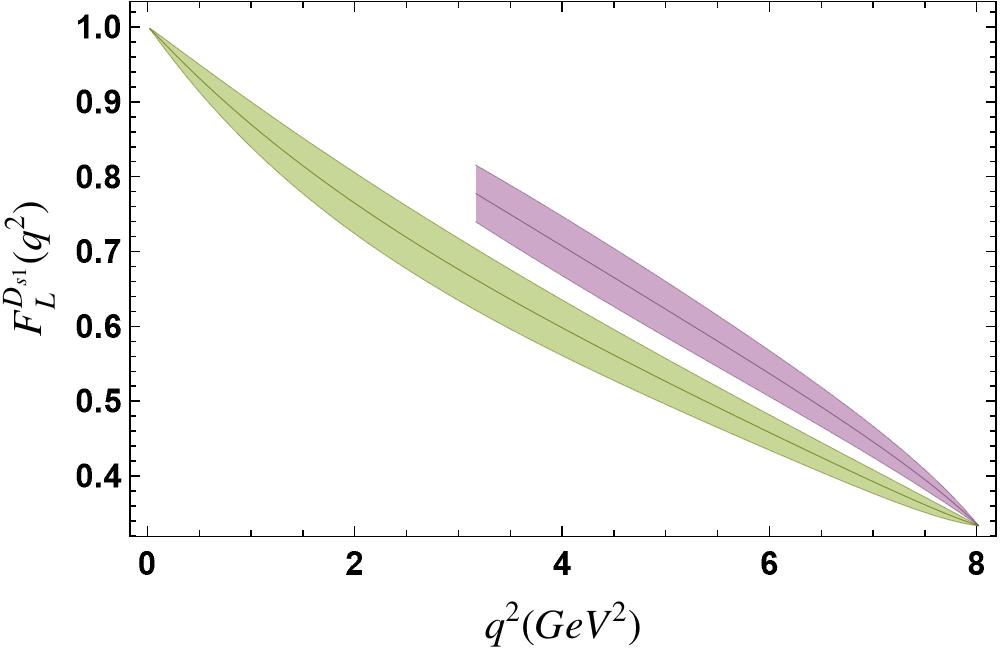}
	\end{minipage}
	\hspace{0.01\textwidth}
	\begin{minipage}[c]{0.31\textwidth}
		\centering
		\raisebox{0.25\height}{
			\includegraphics[width=0.34\linewidth]{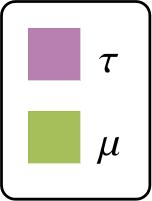}
		}
	\end{minipage}
	\caption{$q^2$-distribution of $B_s \to D_{s1} \ell \nu_\ell$ observables. The bands indicate the $1\sigma$ regions.}\label{fig:SMObsDs1}
\end{figure}

\item $B_s \to D_{s2}^*\ell\nu_\ell$

The $q^2$-distribution is shown in figure~\ref{fig:SMObsDs2s}. Similar to the previous cases, only $P_\mu^{D_{s2}^*}(q^2)$ shows negligible errors. In this mode, the zero-crossing occurs only for $A_{fb,\ell}^{D_{s2}^*}(q^2)$. It occurs at $q^2 = 0.27(24)$ GeV$^2$ in the muon channel and at $q^2 = 4.8(20)$ GeV$^2$ near the central $q^2$ region in the tau channel. The convexity $C_{D_{s2}^*}^\ell(q^2)$ and the lepton polarization asymmetry $P_\ell^{D_{s2}^*}(q^2)$ are negative throughout the physical region, whereas the longitudinal polarization of the daughter meson $F_{L}^{D_{s2}^*}(q^2)$ is positive.

\begin{figure}[t]
	\begin{minipage}[c]{0.31\textwidth}
		\centering
		\includegraphics[width=\textwidth]{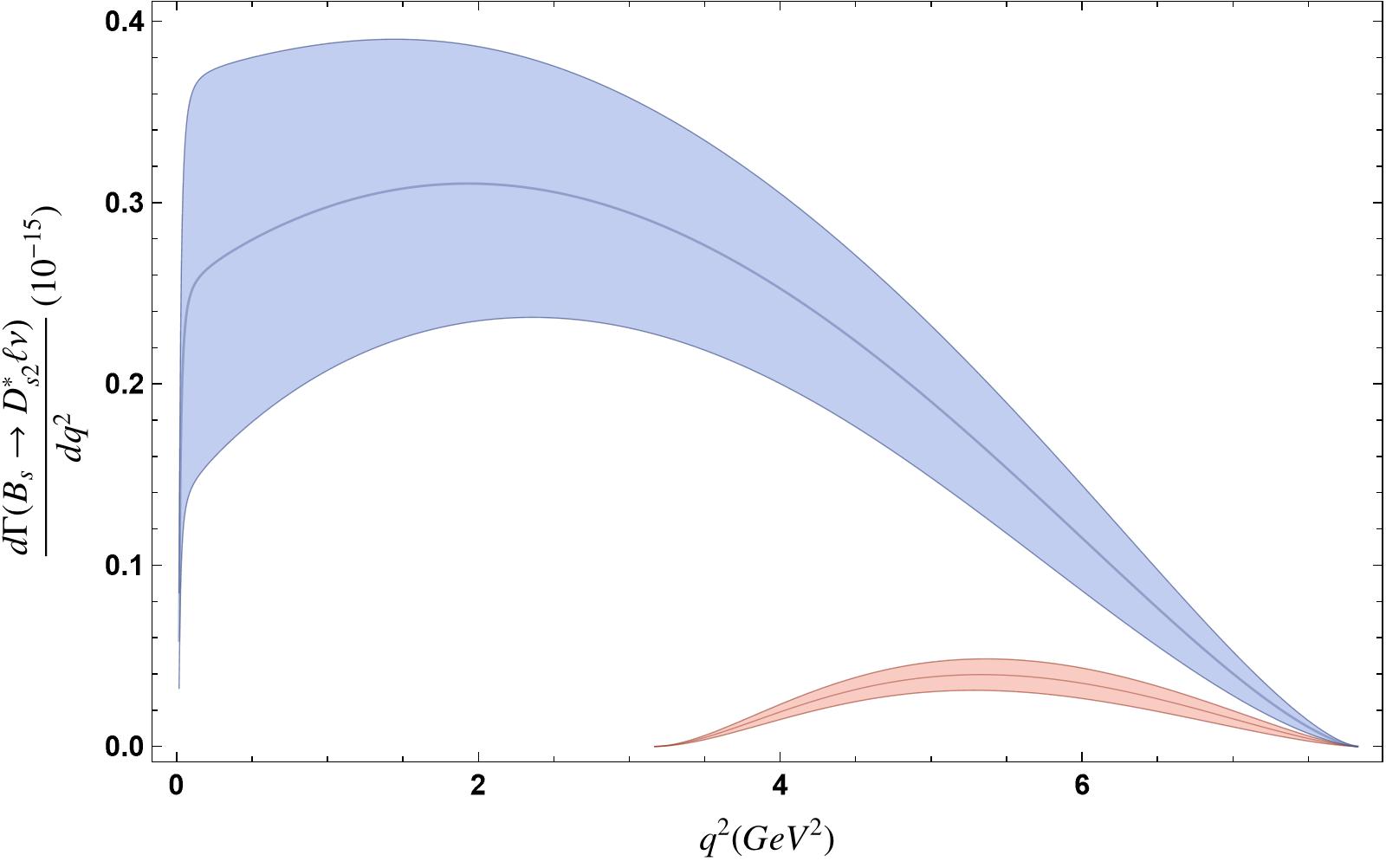}
	\end{minipage}
	\hspace{0.01\textwidth}
	\begin{minipage}[c]{0.31\textwidth}
		\centering
		\includegraphics[width=\textwidth]{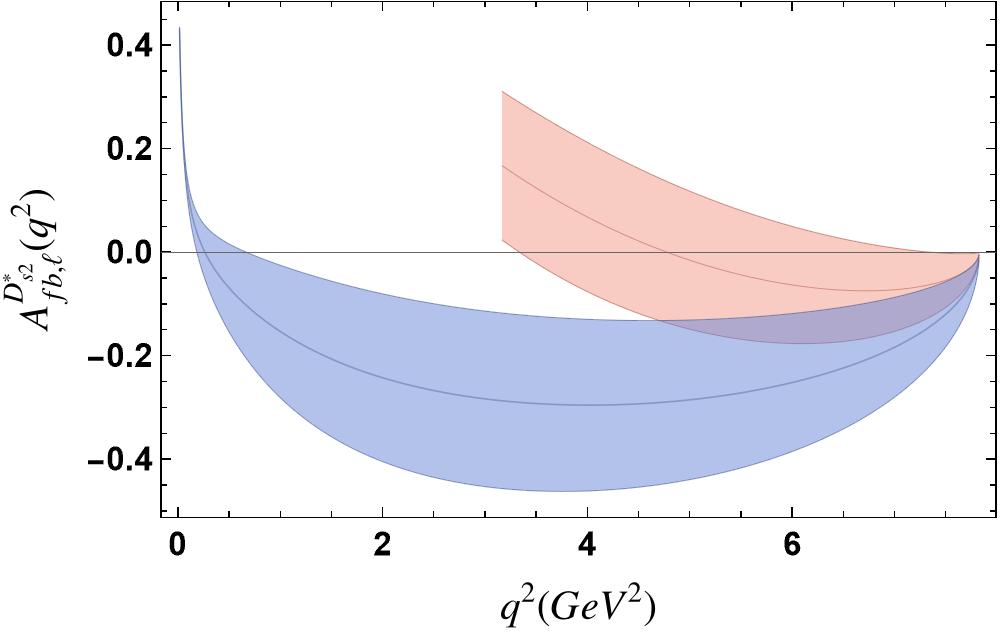}
	\end{minipage}
	\hspace{0.01\textwidth}
	\begin{minipage}[c]{0.31\textwidth}
		\centering
		\includegraphics[width=\textwidth]{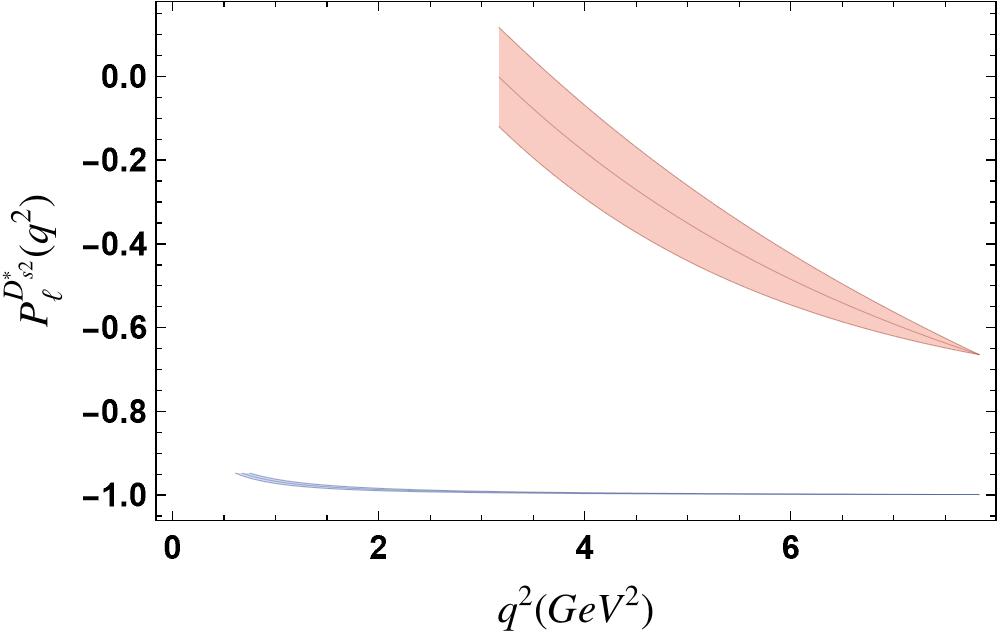}
	\end{minipage}
	
	\vspace{0.3cm}
	
	\begin{minipage}[c]{0.31\textwidth}
		\centering
		\includegraphics[width=\textwidth]{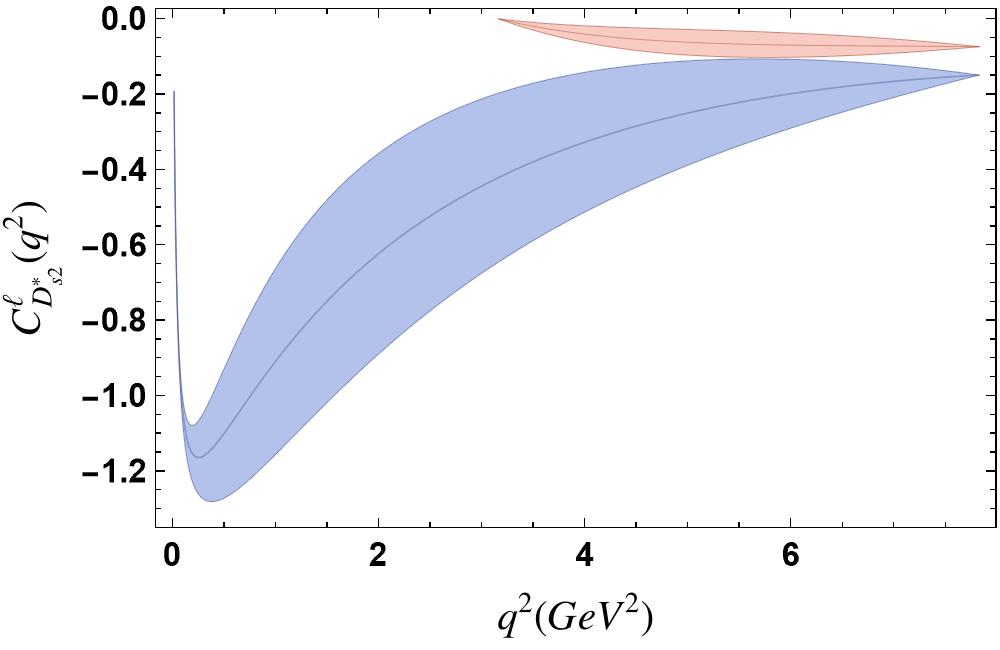}
	\end{minipage}
	\hspace{0.01\textwidth}
	\begin{minipage}[c]{0.31\textwidth}
		\centering
		\includegraphics[width=\textwidth]{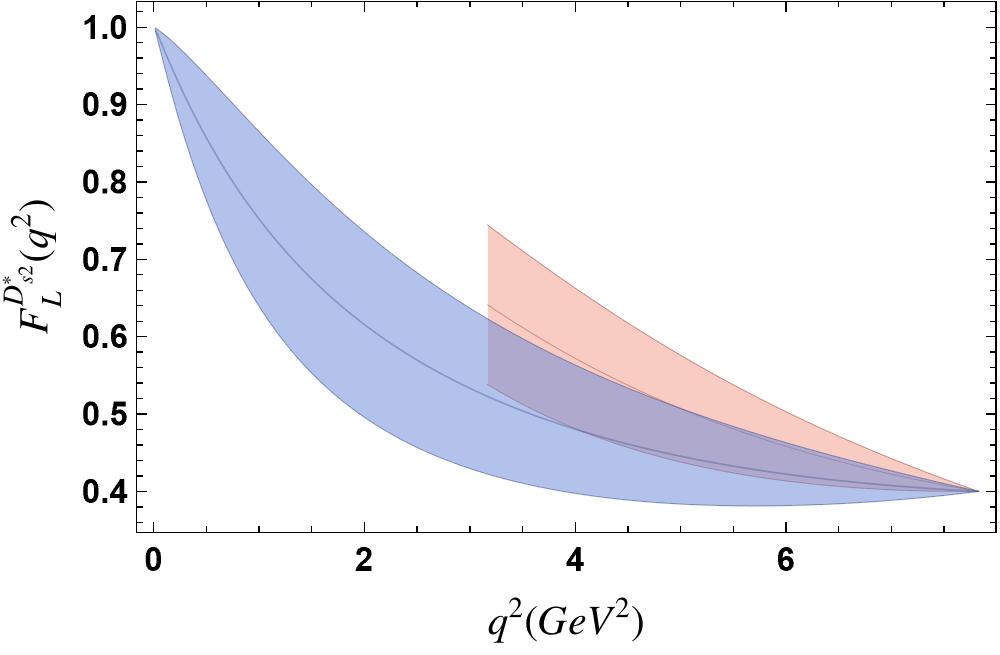}
	\end{minipage}
	\hspace{0.01\textwidth}
	\begin{minipage}[c]{0.31\textwidth}
		\centering
		\raisebox{0.25\height}{
			\includegraphics[width=0.34\linewidth]{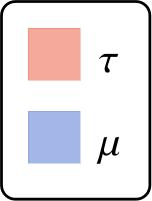}
		}
	\end{minipage}
	\caption{$q^2$-distribution of $B_s \to D_{s2}^*\ell \nu_\ell$ observables.  The bands indicate the $1\sigma$ regions.}\label{fig:SMObsDs2s}
\end{figure}

\end{enumerate}

\subsection{Predictions for Integrated Observables}
The predicted integrated observables for the muon and tau channels are presented in table~\ref{tab:int_tau}. In general, the uncertainties in ratio-based observables are reduced due to the positive correlation between the numerator and denominator terms.

\begin{table}
	\begin{center}
		\renewcommand*{\arraystretch}{1.6}
		\resizebox{1.01 \textwidth}{!}{
		\begin{tabular}{|c|c|c|c|c|c|c|}
			\hline
			Decay Channel & Mode & $\mathcal{B}\times 10^{-3}$ & $\langle A^{D_s^{**}}_{fb,\ell} \rangle$ & $\langle P^{D_s^{**}}_\ell \rangle$ & $\langle C^\ell_{D_s^{**}} \rangle$ & $\langle F^{D_s^{**}}_L \rangle$ \\
			\hline
			\multirow{2}{*}{$B_s \to D_{s0}^*\ell \nu_\ell$}&$\mu$ & $0.43(20)$ &$0.014(2)$&$-0.963(6)$&$-1.458(6)$& $-$\\
			\cline{2-7}
			&$\tau$ & $0.069(25)$ &$0.387(5)$&$0.120(75)$&$-0.31(3)$& $-$ \\
			\hline
			\multirow{2}{*}{$B_s \to D_{s1}^*\ell \nu_\ell$}&$\mu$ & $2.67(37)$ &$-0.134(31)$&$-0.940(5)$&$-0.86(11)$& $0.742(51)$\\
			\cline{2-7}
			&$\tau$ & $0.120(18)$ &$0.039(68)$&$-0.045(98)$&$-0.01(4)$&$0.563(84)$\\
			\hline
			\multirow{2}{*}{$B_s \to D_{s1}\ell \nu_\ell$}&$\mu$ & $11.3(17)$ &$-0.153(39)$&$-0.962(4)$&$-0.808(85)$&$0.711(40)$\\
			\cline{2-7}
			&$\tau$ & $0.82(15)$ &$0.017(44)$&$-0.295(47)$&$-0.079(12)$&$0.567(33)$\\
			\hline
			\multirow{2}{*}{$B_s \to D_{s2}^*\ell \nu_\ell$}&$\mu$ & $3.73(83)$ &$-0.22(15)$&$-0.971(9)$&$-0.138(51)$&$0.591(95)$\\
			\cline{2-7}
			&$\tau$ & $0.247(55)$ &$-0.025(121)$&$-0.399(83)$&$-0.016(7)$& $0.490(58)$\\
			\hline
		\end{tabular}
	}
	\caption{SM predictions for the integrated observables of $B_s \to D_s^{**} \ell\nu_\ell$ decays.}~\label{tab:int_tau}
	\end{center}
\end{table}
\begin{enumerate}
	\item[$\bullet$] The errors in the branching ratios of $B_s \to D_{s0}^* \mu \nu_\mu$ and $B_s \to D_{s0}^*\tau \nu_\tau$ are $47\%$ and $37\%$, respectively. As discussed earlier, $\langle P^{D_{s0}^*}_\mu \rangle$ and $\langle C_{D_{s0}^*}^\mu \rangle$ have negligible errors. The angular coefficients $b$ and $c$ given in eq.~\ref{eq:abc_Ds0s} are strongly correlated with the decay rate. Hence, in the ratios $\langle A_{fb,\ell}^{D_{s0}^*} \rangle$ and $\langle C_{D_{s0}^*}^\tau \rangle$, the errors are reduced. The observable $\langle P^{D_{s0}^*}_\tau \rangle $ has an error of $62\%$, as evident from its definition.
	\item[$\bullet$] The errors in the branching ratios of $B_s \to D_{s1}^*\mu\nu_\mu$ and $B_s \to D_{s1}^*\tau\nu_\tau$ are $14\%$ and $15\%$, respectively. The observable $\langle P^{D_{s1}^*}_\mu \rangle$ has a negligible error, as discussed in the previous section. All other observables, except $F_L^{D_{s1}^*}(q^2)$, have zero-crossings. The cancellations between positive and negative contributions yield small integrated central values for these observables. Hence, the calculated percent errors are larger than those of the decay rates in the table. The longitudinal polarization of the daughter has a smaller error than that of the decay rate because of the correlations between the terms in the numerator and denominator.
	\item[$\bullet$] The errors in the branching ratios of $B_s \to D_{s1}\mu\nu_\mu$ and $B_s \to D_{s1}\tau\nu_\tau$ are $15\%$ and $18\%$, respectively. Here, $\langle P^{D_{s1}}_\mu \rangle$ has a negligible error, as in the previous cases. The convexity $\langle C^\ell_{D_{s1}} \rangle$ and the longitudinal polarization $\langle F^{D_{s1}}_L \rangle$ have smaller errors than the decay rate due to the correlations. The observables $\langle A_{fb,\ell}^{D_{s1}} \rangle$ and $\langle P^{D_{s1}}_\tau \rangle$ have slightly larger errors than the decay rate due to the zero-crossing.
	\item[$\bullet$] The branching ratios of $B_s \to D_{s2}^*\mu\nu_\mu$ and $B_s \to D_{s2}^*\tau\nu_\tau$ have errors of $22\%$ each. Again, $\langle P^{D_{s2}^*}_\mu \rangle$ has a negligible error, as in the other modes. The angular coefficients $a(q^2)$, $b(q^2)$, $c(q^2)$ for this mode are defined in eq.~\ref{eq:abc_Ds2s}. It is found that the coefficients $b$ and $c$ have larger errors than $a$, and $|a| >> |c|$. Hence, the branching ratio has the same error as $a$. The ratios $\langle A_{fb,\ell}^{D_{s2}^*} \rangle $ and $\langle C_{D_{s2}^*}^\ell \rangle$ have larger errors than the decay rate due to $b$ and $c$. The errors in $\langle A_{fb,\ell}^{D_{s2}^*}\rangle $ are even larger because of the zero-crossing. The observable $\langle P^{D_{s2}^*}_\tau \rangle$ and the longitudinal polarization $\langle F^{D_{s2}^*}_L \rangle$ have smaller errors than the decay rate.
\end{enumerate}
In addition to the observables in table~\ref{tab:int_tau}, we provide quantitative estimates of the LFU ratio $R_{D_s^{**}}$ defined as
\begin{eqnarray}
	R_{D_s^{**}} = \frac{\mathcal{B}(B\to D_s^{**} \tau \nu) }{\mathcal{B}(B\to D_s^{**} \mu \nu)} \, \,.
\end{eqnarray}
The estimates of the LFU ratios are key results of this analysis and serve as benchmarks for testing the SM and for probing new physics beyond it. Our results for the LFU ratios are summarized in table~\ref{tab:R_comp}. 
\begin{table}[t]
	\begin{center}
		\renewcommand*{\arraystretch}{1.6}
		\resizebox{0.85\textwidth}{!}{
			\begin{tabular}{|c|c|c|c|c|}
				\hline
				  & $R_{D_{s0}^*}$ & $R_{D_{s1}^*}$ & $R_{D_{s1}}$ & $R_{D_{s2}^*}$ \\
				\hline
				Our work (HQET + BSZ) & $0.158(20)$ & $0.045(5)$ & $0.073(4)$ & $0.066(9)$ \\
				\hline
				\text{HQET} \cite{Bernlochner:2016bci} & $0.09(4)$ & $0.07(3)$ & $0.09(2)$ & $0.07(1)$ \\
				\hline
				\text{LCSR} \cite{Gubernari:2023rfu} & $0.14^{+0.07}_{-0.02}$ & - & - & - \\
				\hline
			\end{tabular}
		}
	\end{center}
	\caption{SM estimates of LFU observables in $B_s \to D_s^{**} \ell\nu_\ell$ decays.}\label{tab:R_comp}
\end{table}
We find that the errors for the LFU ratios $R_{D_{s0}^*}$, $R_{D_{s1}^*}$, $R_{D_{s1}}$ and $R_{D_{s2}^*}$ are $13\%$, $11\%$, $5\%$ and $14\%$, respectively. We explicitly verify that, due to the high positive correlation between the terms in the numerator and denominator, the errors in the LFU ratios are reduced. We also compare our results with those of other works ~\cite{Bernlochner:2016bci, Gubernari:2023rfu}. The comparison is shown in table~\ref{tab:R_comp}. Our predictions for the LFU ratios are within $1\sigma$ agreement with the results of \cite{Bernlochner:2016bci}, except for the ratio $R_{D_{s0}^*}$. Within the $2\sigma$ confidence interval, our estimate of $R_{D_{s0}^*}$ is consistent with that of \cite{Bernlochner:2016bci}. Our prediction for $R_{D_{s0}^*}$ is consistent with the predictions of \cite{Gubernari:2023rfu} within $1\sigma$.
	
\section{New Physics Sensitivity Test}\label{sec:NP}

With the precise SM determinations of the $B_s \to D_s^{**}$ form factors in the previous sections, we proceed to analyze new physics sensitivity of the $B_s \to D_s^{**}\ell\nu_\ell$ modes, testing all possible contributing Lorentz interactions within model-independent and model-dependent frameworks. The WET, SMEFT and 2HDM frameworks serve as illustrations in this work. For the Wilson coefficients, certain benchmark values are considered which are obtained by performing a $\chi^2$ fit to the available experimental observables relevant to $b \to c\tau\nu_\tau$ transitions. The observables considered in the $\chi^2$ fit are $R(D^{(*)})$, $R(\Lambda_c)$, $\mathcal{B}(\Lambda_b \to \Lambda_c\tau\nu_\tau)$ and $F_L(D^*)$ \cite{HFLAV,LHCb:2022piu,Belle:2019ewo,LHCb:2023ssl}. We do not include $P_\tau(D^*)$ in the fit due to its large experimental uncertainties \cite{Belle:2016dyj}. Similarly, $R(J/\psi)$ is not considered, as recent CMS measurements exhibit significant uncertainties \cite{CMS:2024seh,CMS:2024uyo}. The theoretical predictions for the observables used in the fit are obtained using the following expressions 

\begin{table}[t]
	\begin{center}
		\renewcommand*{\arraystretch}{1.45}
		\resizebox{0.75 \textwidth}{!}{
			\begin{tabular}{|c|c|c|}
				\hline
				Observable & Experimental Value & Theoretical Prediction \\
				\hline
				$R(D)$ & $0.358(24)$ & $0.296(4)$ \cite{HFLAV}\\
				$R(D^*)$ & $0.281(11)$ & $0.254(5)$ \cite{HFLAV}\\
				%
				$F_L(D^*)$ & $0.410(70)$ & $0.424(9)$\\
				%
				$R(\Lambda_c)$ & $0.242(71)_{\textrm{Syst.}}(26)_{\textrm{Stat.}}$ & $0.333(10)$\\
				$\mathcal{B}(\Lambda_b \to \Lambda_c\tau\nu_\tau)$ & $0.015 (4)$ & $0.0178(8)$ \\
				\hline
			\end{tabular}
		}
	\end{center}
\caption{Observables considered in the fitting of the Wilson coefficients. }\label{tab:expt_res}
\end{table}

\begin{eqnarray}
\frac{R(D)}{R(D)_{\text{SM}}} &=& \vert 1 + C_{V_1} + C_{V_2} \vert^2 + 1.3440 \, \vert C_{S_1} + C_{S_2} \vert^2 + 0.8271 \, \text{Re}[(1+C_{V_1}+ C_{V_2})C_T^*] \nonumber \\ && + 1.7126 \, \text{Re}[(1+C_{V_1} + C_{V_2})(C_{S_1}^* + C_{S_2}^*)] + 0.4798 \, \vert C_T \vert^2 \, , \nonumber \\
\frac{R(D^*)}{R(D^*)_{\text{SM}}} &=& \vert 1 + C_{V_1} \vert^2 + \vert C_{V_2} \vert^2 - 1.7932 \,  \text{Re}[(1+C_{V_1})C_{V_2}^*] + 0.0358 \vert C_{S_1} - C_{S_2} \vert^2 \nonumber \\ && + 12.4506 \vert C_T \vert^2 - 4.4536 \, \text{Re}[(1+C_{V_1})C_T^*] + 5.8859 \, \text{Re} [C_{V_2}C_T^*] \nonumber \\ && + 0.1016 \, \text{Re} [(1+C_{V_1}-C_{V_2})(C^*_{S_1} - C^*_{S_2})]\, , \nonumber \\
\frac{F_L(D^*)}{F_L(D^*)_{\text{SM}}} &=& \frac{R(D^*)_{\text{SM}}}{R(D^*)}\bigg\{ \vert 1 + C_{V_1} - C_{V_2} \vert^2 + 0.2398 \, \text{Re}[(1+C_{V_1} - C_{V_2})(C_{S_1}^* - C_{S_2}^*)]  \nonumber \\ && - 3.8129 \, \text{Re}[(1+C_{V_1}- C_{V_2})C_T^*] + 0.0845 \, \vert C_{S_1} - C_{S_2} \vert^2 + 5.3176 \, \vert C_T \vert^2 \bigg\}\, . \nonumber \\
\end{eqnarray}
In the above equations, the expressions for $R(D)$ are obtained by performing a combined fit to lattice QCD data from HPQCD \cite{Na:2015kha} and Fermilab lattice/MILC \cite{FermilabLattice:2015ilb}. Similarly, the expressions for $R(D^*)$ and $F_L(D^*)$ are obtained using lattice inputs from HPQCD \cite{Harrison:2023dzh}, Fermilab lattice/MILC \cite{FermilabLattice:2021cdg} and JLQCD \cite{Aoki:2023qpa}. Using lattice inputs from refs.~\cite{Detmold:2015aaa,Datta:2017aue}, we obtain expressions for $R(\Lambda_c)$ and $\mathcal{B}(\Lambda_b \to \Lambda_c\tau\nu_\tau)$\footnote{The detailed expressions for the form factors used in this analysis are lengthy and are not presented here.}. The experimental values and the computed SM values of the observables are presented in table~\ref{tab:expt_res}. The measured values of $R(D)$ and $R(D^*)$ have a correlation of $-0.374$ \cite{HFLAV}. The fit procedure followed here is similar to that in \cite{Nandi:2024aia}, where the form factor information for $\Lambda_b \to \Lambda_c$ is used in the fit.  For the SM, we obtain $\chi^2_{\text{SM}} = 16.32$, which for 5 degrees of freedom (dof) corresponds to a p-value of $6.04\times10^{-3}$, indicating that the observations deviate from the SM predictions by $2.75\sigma$. 

\subsection{WET Framework}	

\begin{figure}[ht]
\begin{center}
	\includegraphics[width=0.31\textwidth]{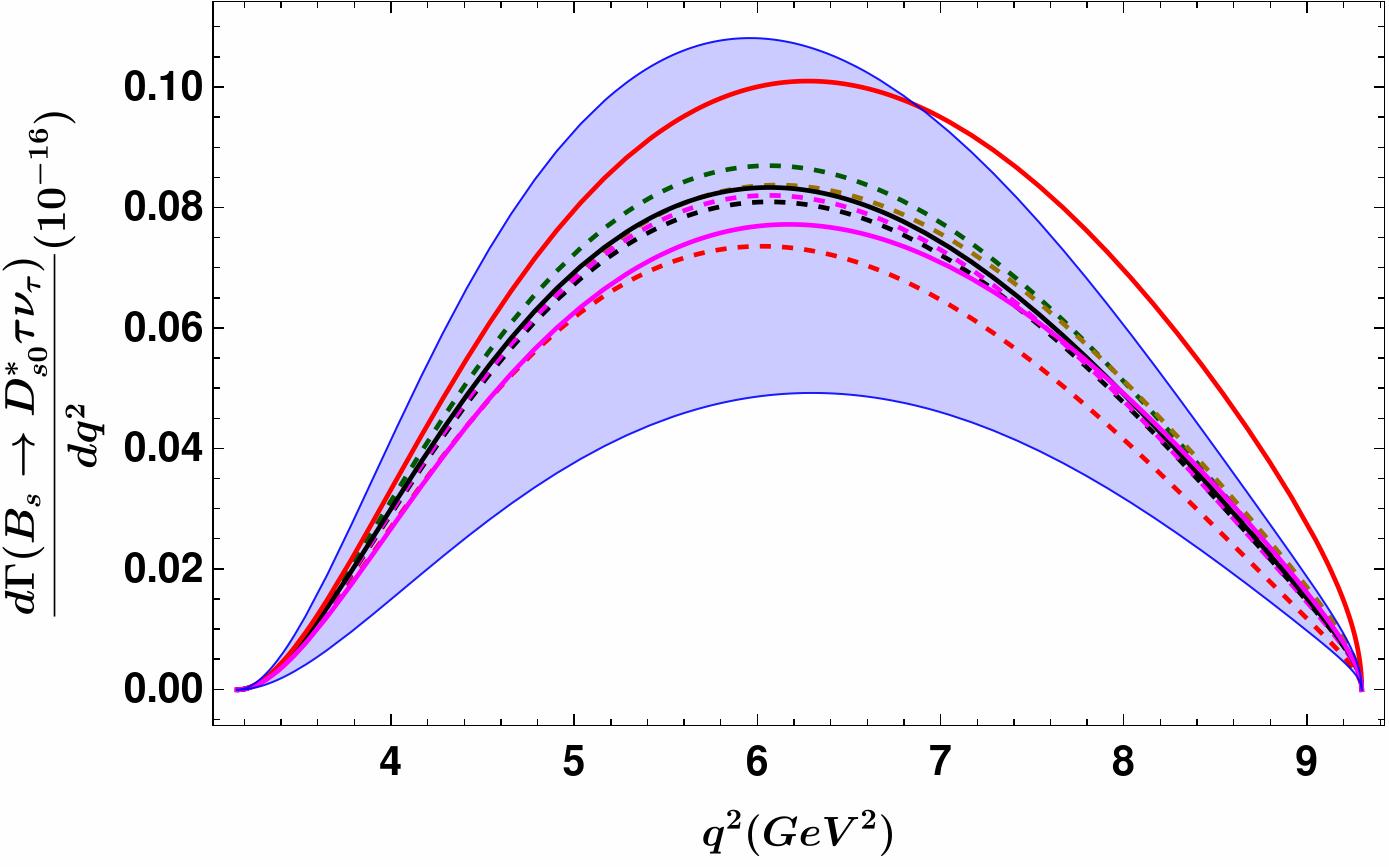}
	\hspace{0.01\textwidth}
	\includegraphics[width=0.31\textwidth]{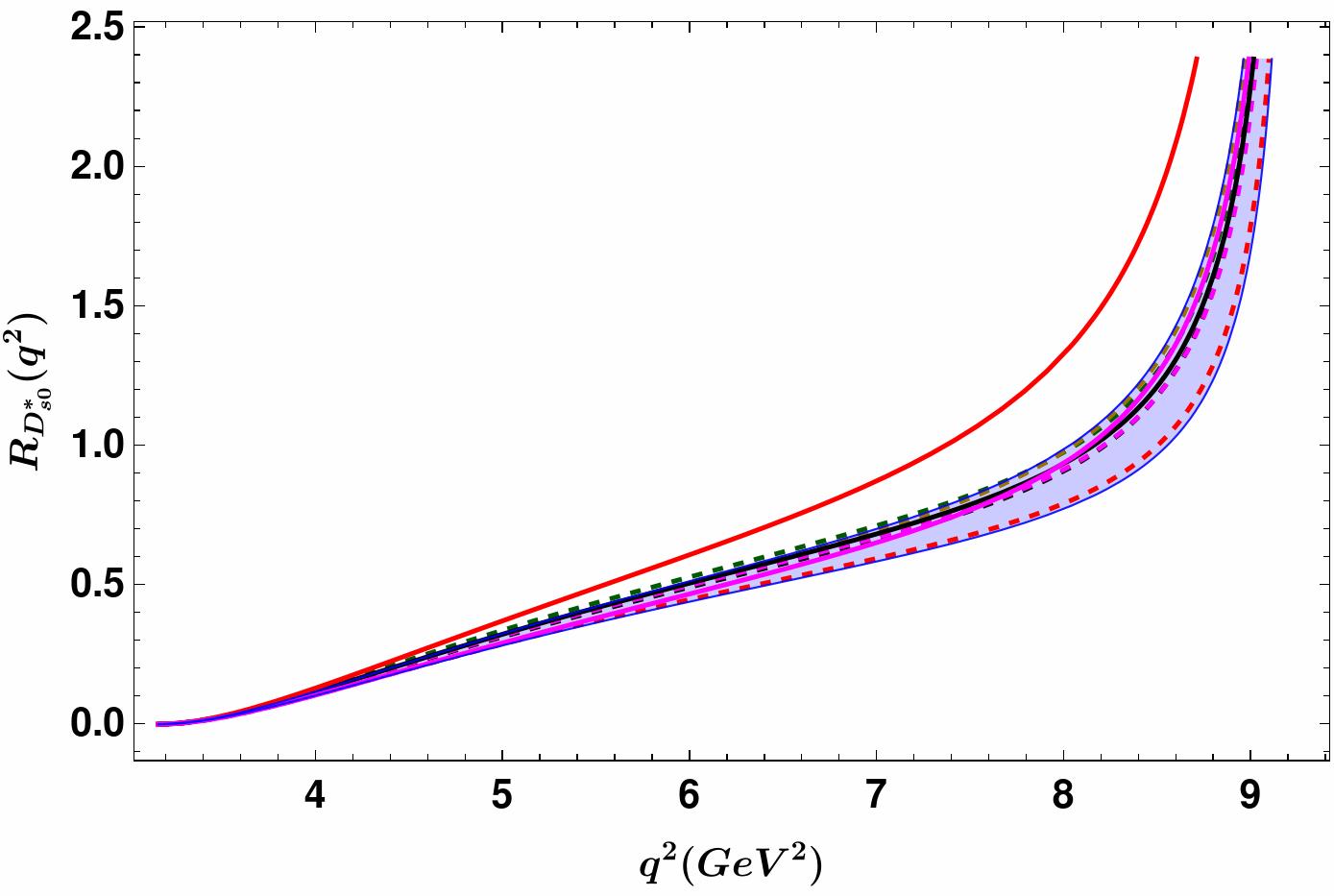}
	\hspace{0.01\textwidth}
	\includegraphics[width=0.31\textwidth]{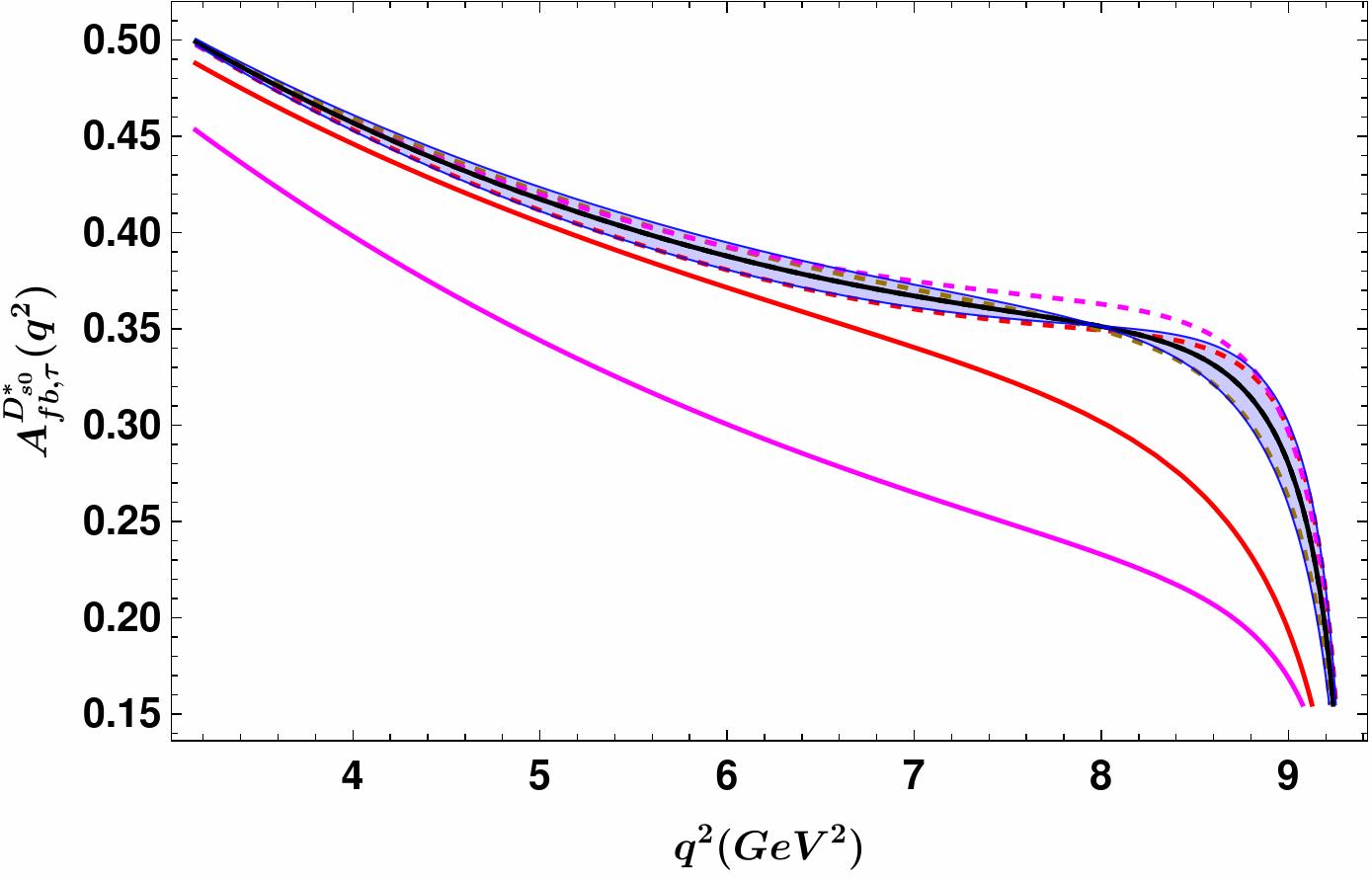}
	
	\vspace{0.3cm}
	
	\includegraphics[width=0.31\textwidth]{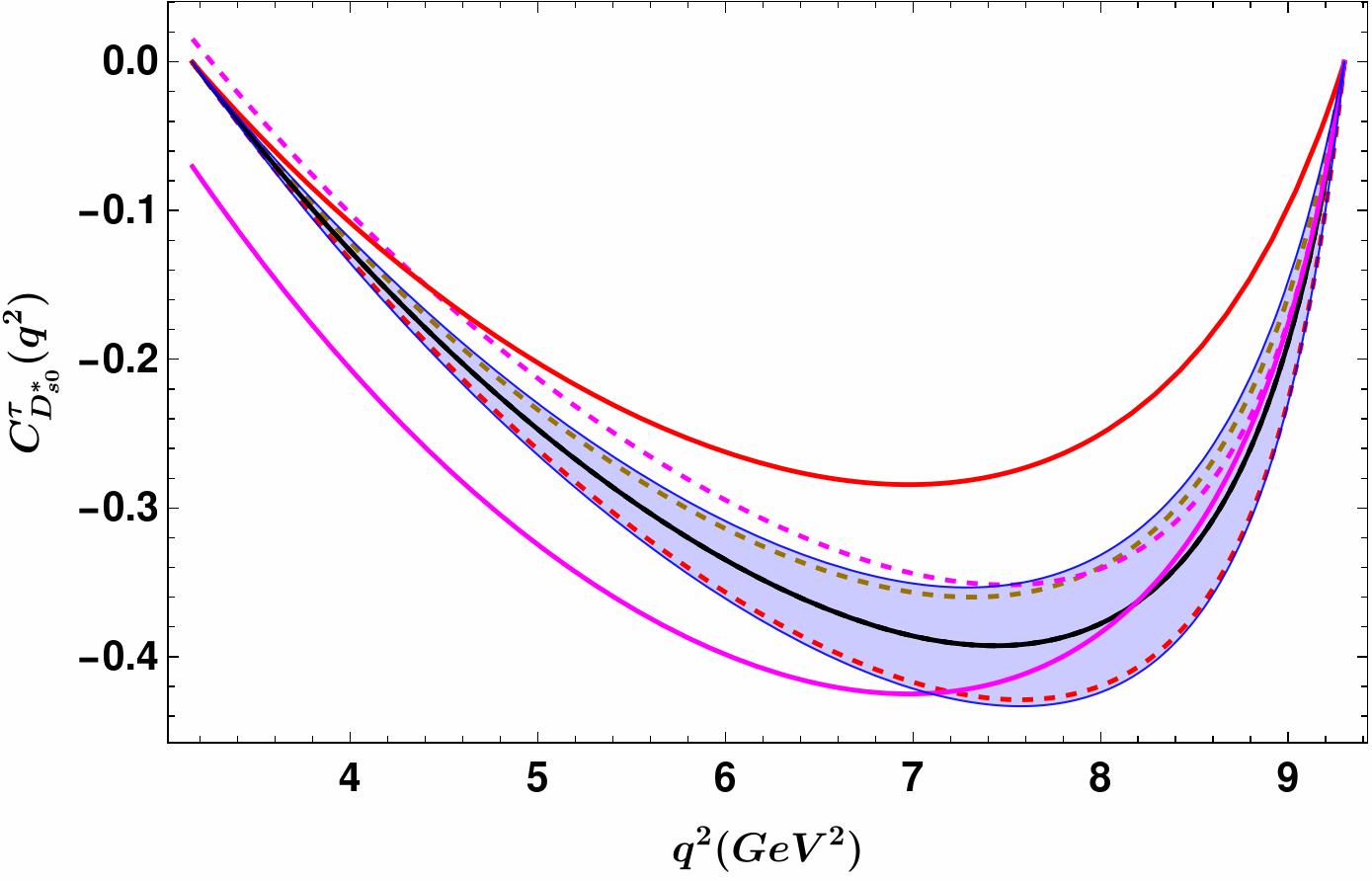}
	\hspace{0.01\textwidth}
	\includegraphics[width=0.31\textwidth]{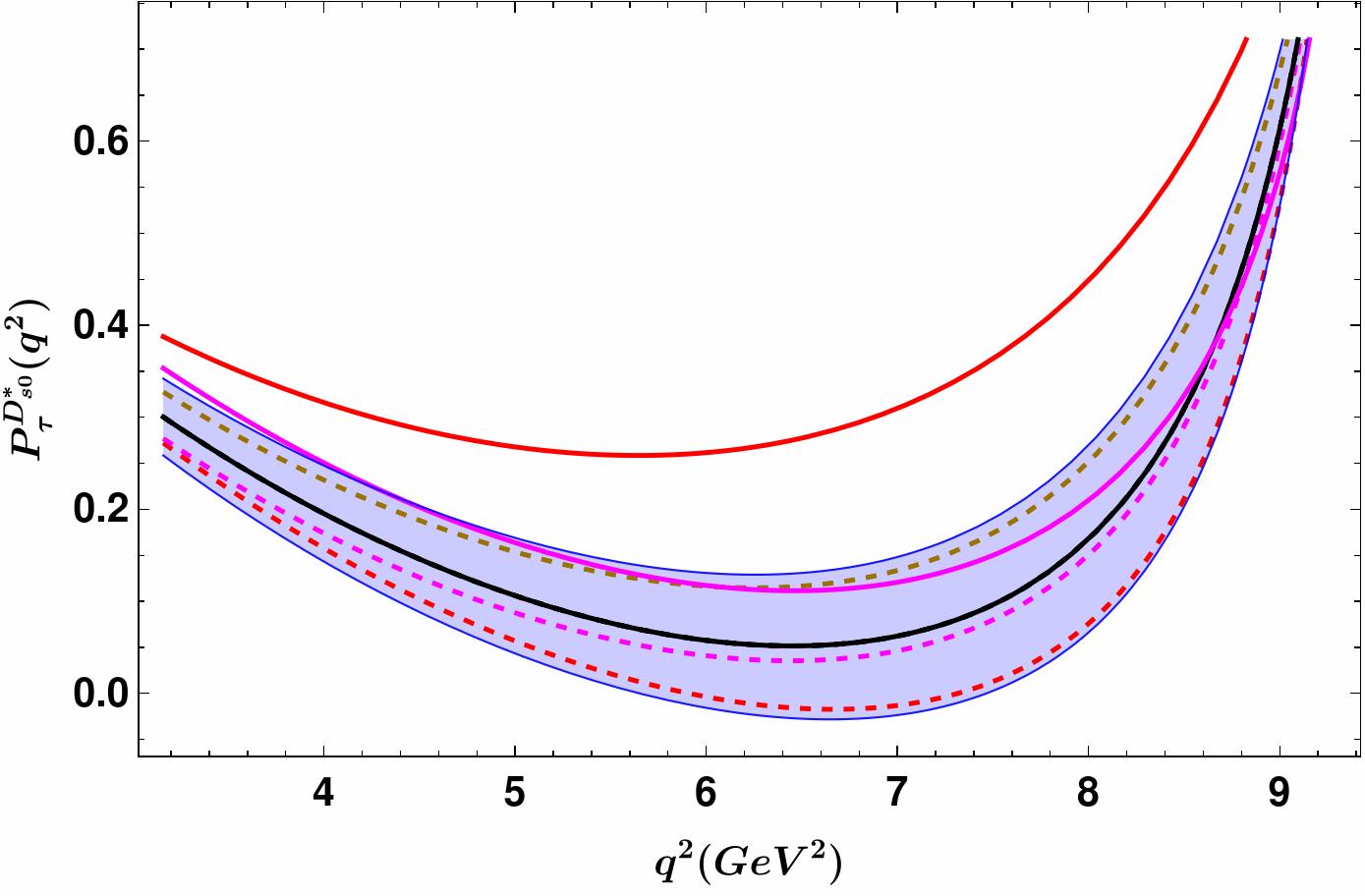}
	
	\vspace{0.3cm}
	
	\includegraphics[scale=0.4]{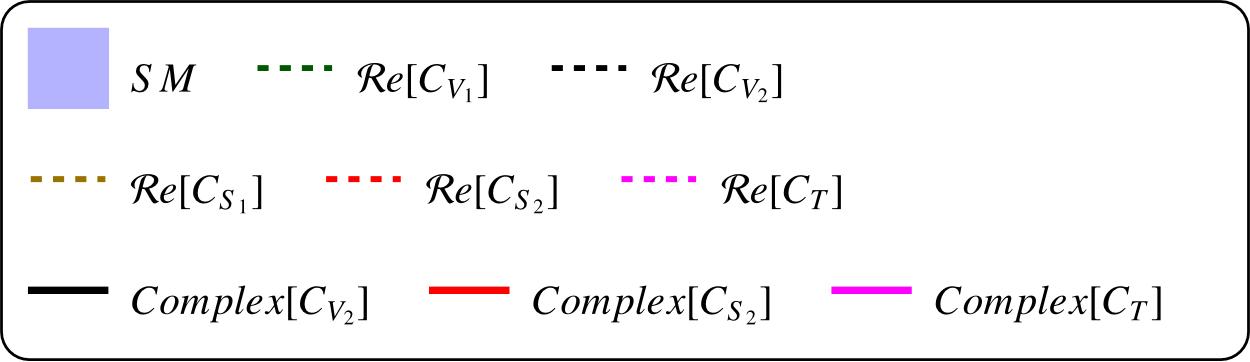}
\end{center}
\caption{$q^2$-distribution of $B_s \to D_{s0}^*\tau\nu_\tau$ observables in WET.}\label{fig:1d_comp_Ds0s}
\end{figure}

\begin{figure}[ht]
\begin{center}
	\includegraphics[width=0.31\textwidth]{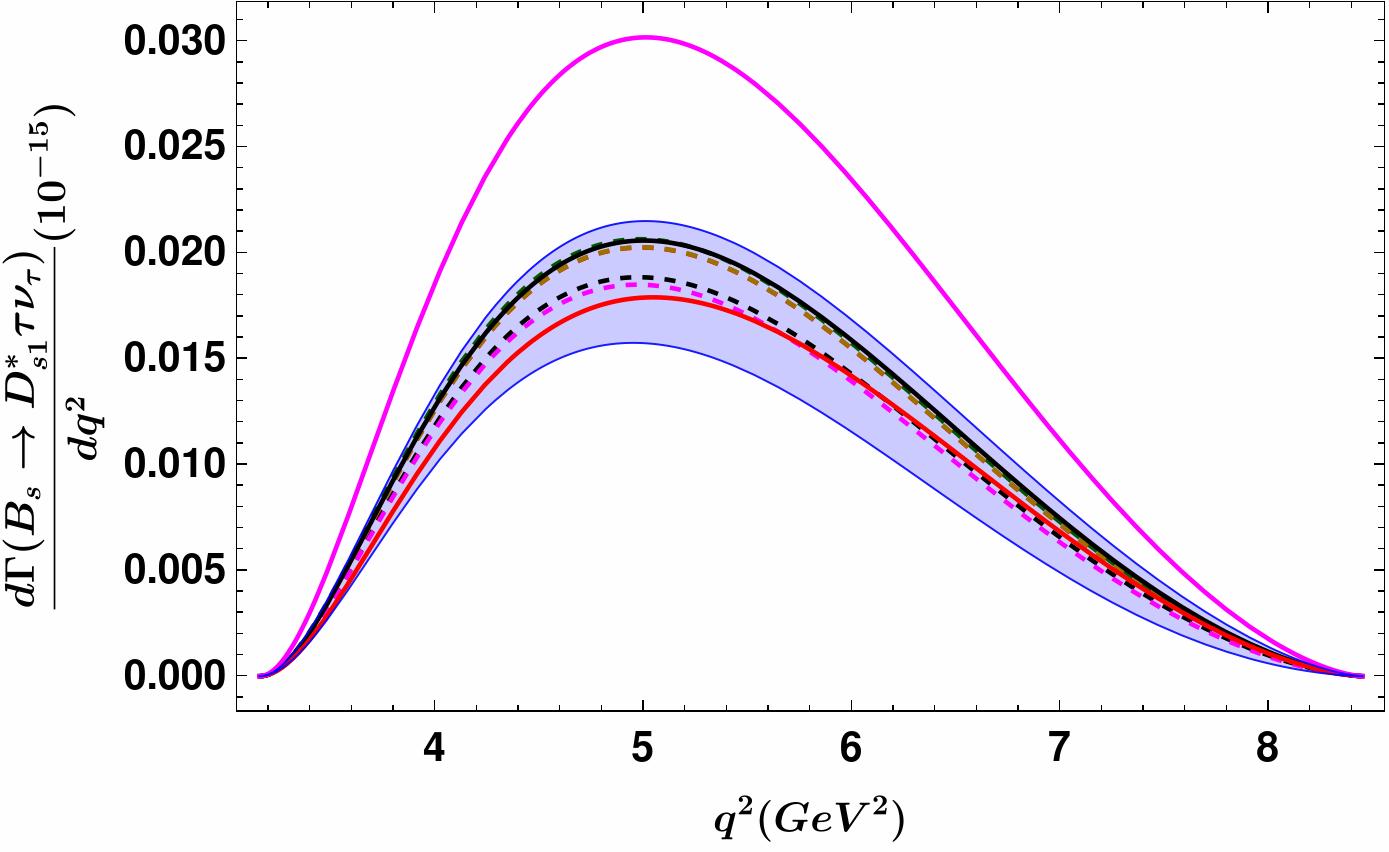}
	\hspace{0.01\textwidth}
	\includegraphics[width=0.31\textwidth]{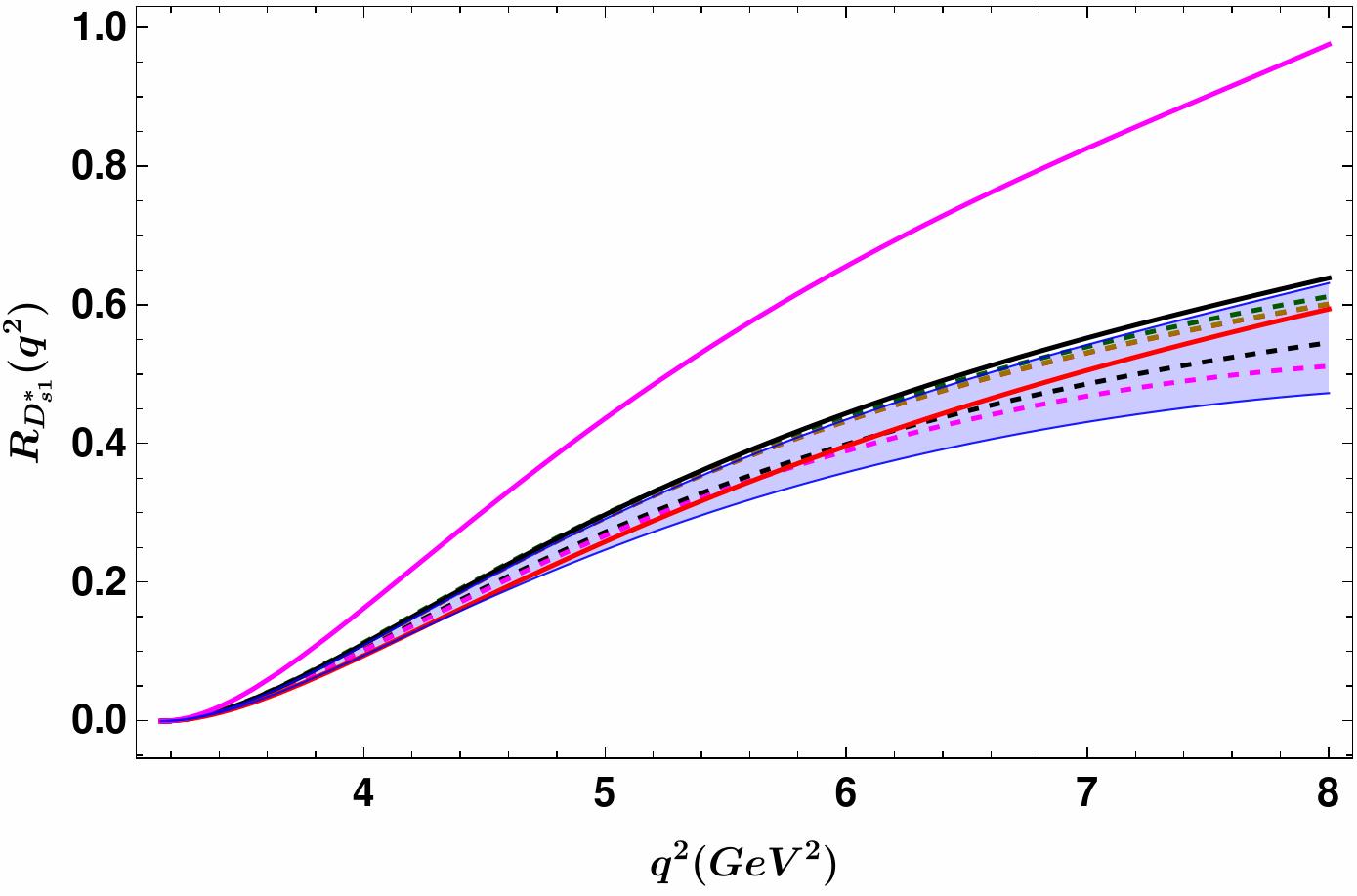}
	\hspace{0.01\textwidth}
	\includegraphics[width=0.31\textwidth]{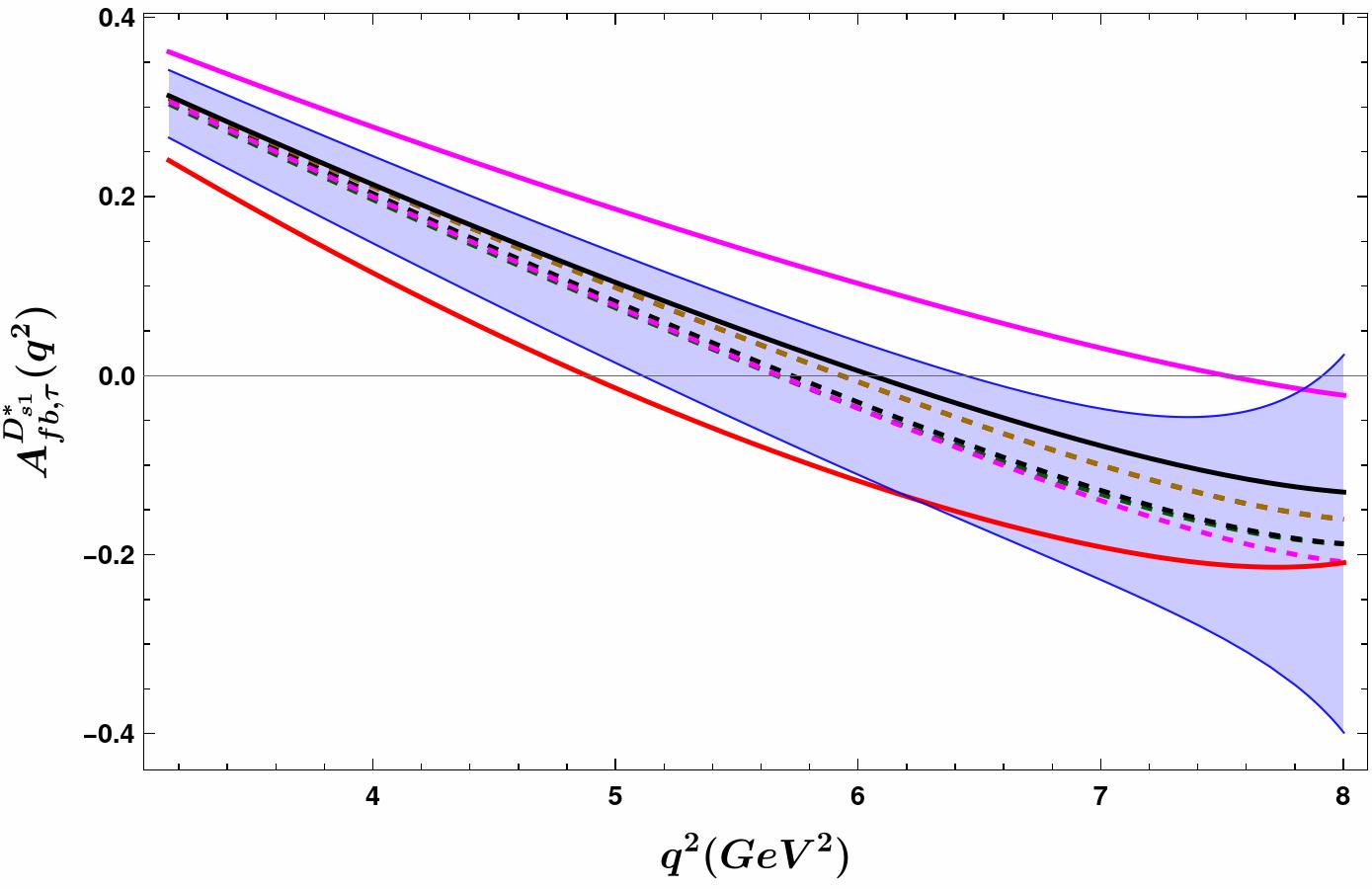}
	
	\vspace{0.3cm}
	
	\includegraphics[width=0.31\textwidth]{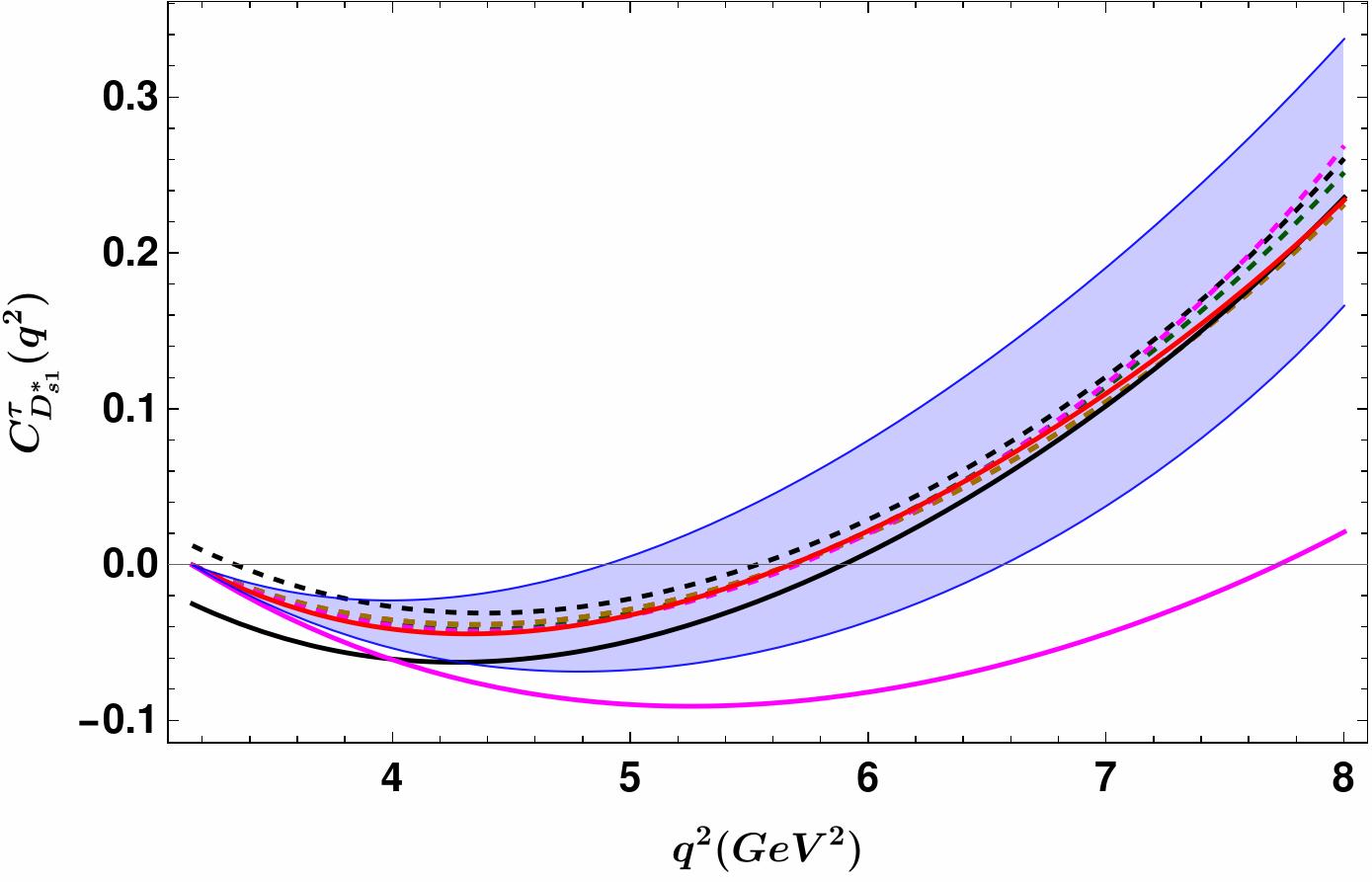}
	\hspace{0.01\textwidth}
	\includegraphics[width=0.31\textwidth]{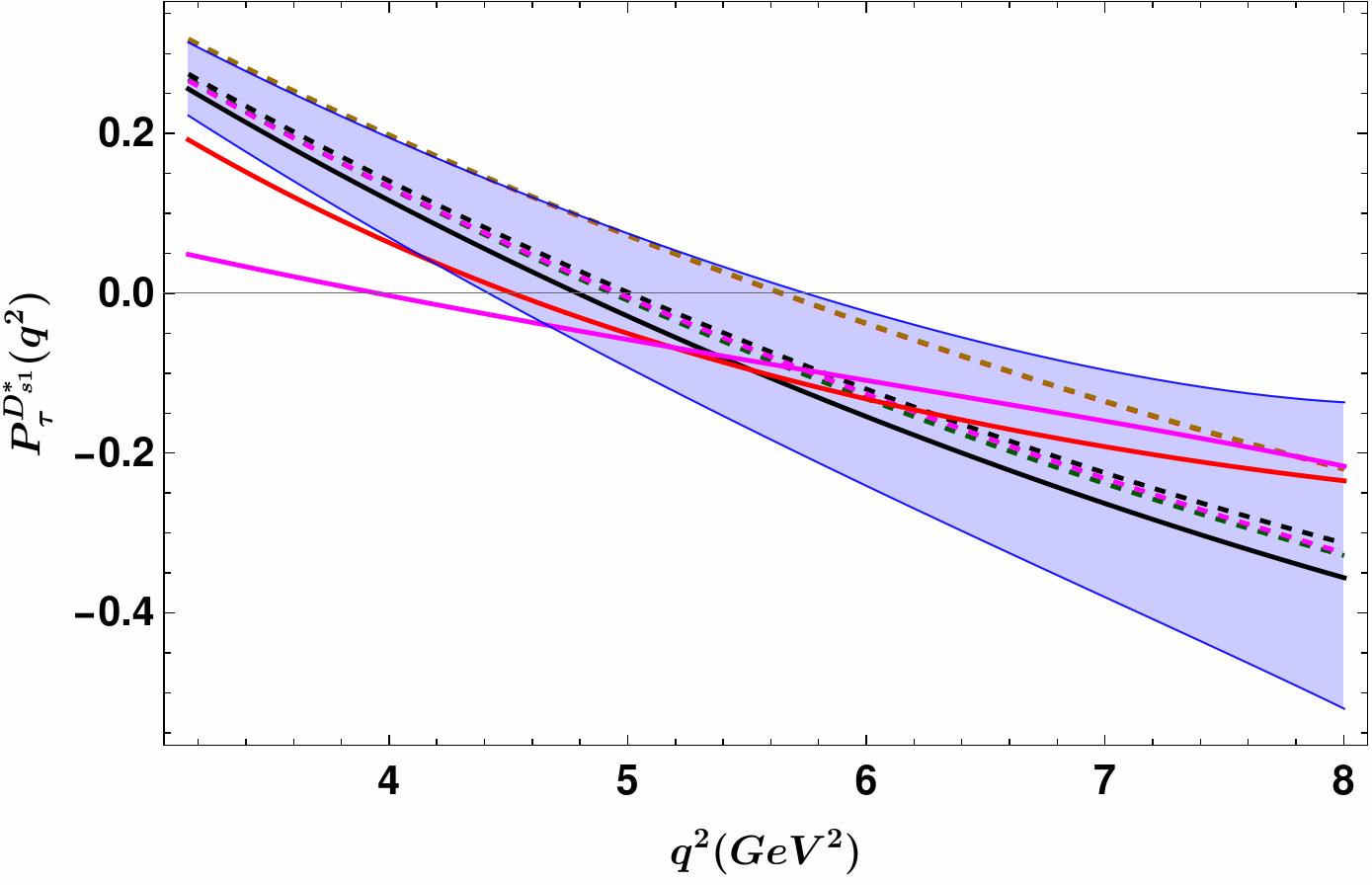}
	\hspace{0.01\textwidth}
	\includegraphics[width=0.31\textwidth]{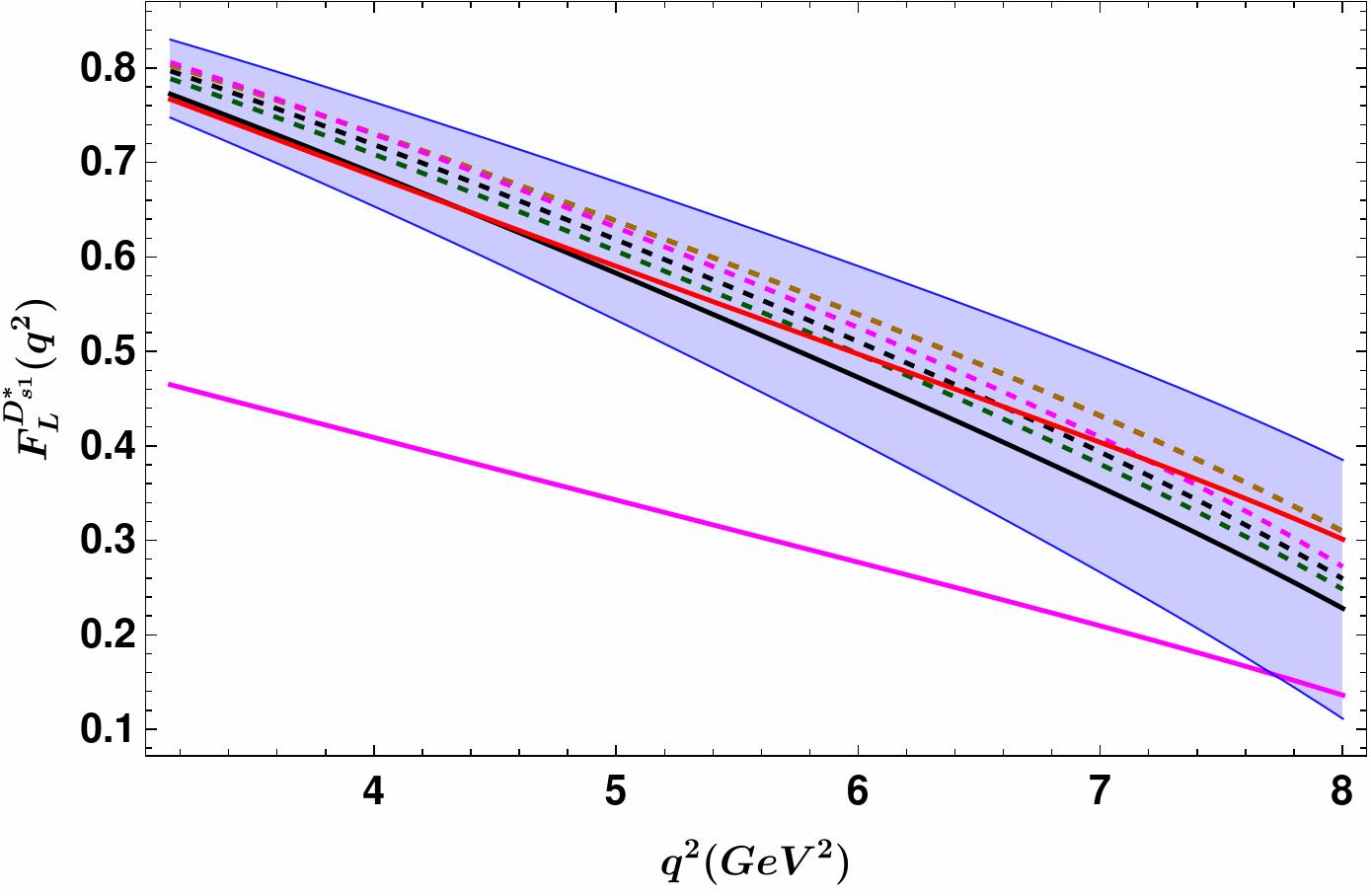}
	
	\vspace{0.3cm}
	
	\includegraphics[scale=0.4]{DsstNPscnWEFTleg.jpg}
\end{center}
\caption{$q^2$-distribution of $B_s \to D_{s1}^*\tau\nu_\tau$ observables in WET.}\label{fig:1d_comp_Ds1s}
\end{figure}

\begin{figure}[!htb]	
\begin{center}
	\includegraphics[width=0.31\textwidth]{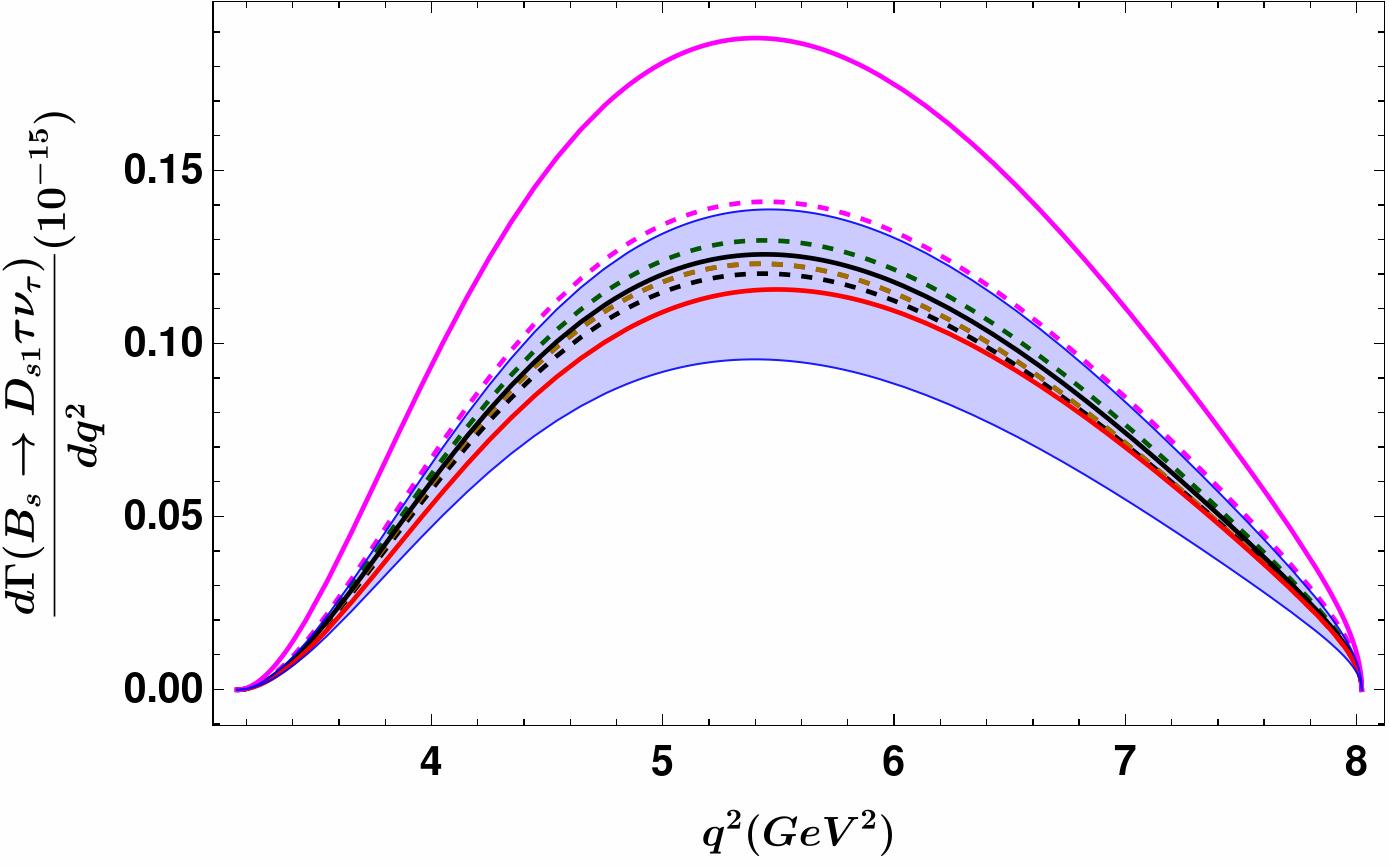}
	\hspace{0.01\textwidth}
	\includegraphics[width=0.31\textwidth]{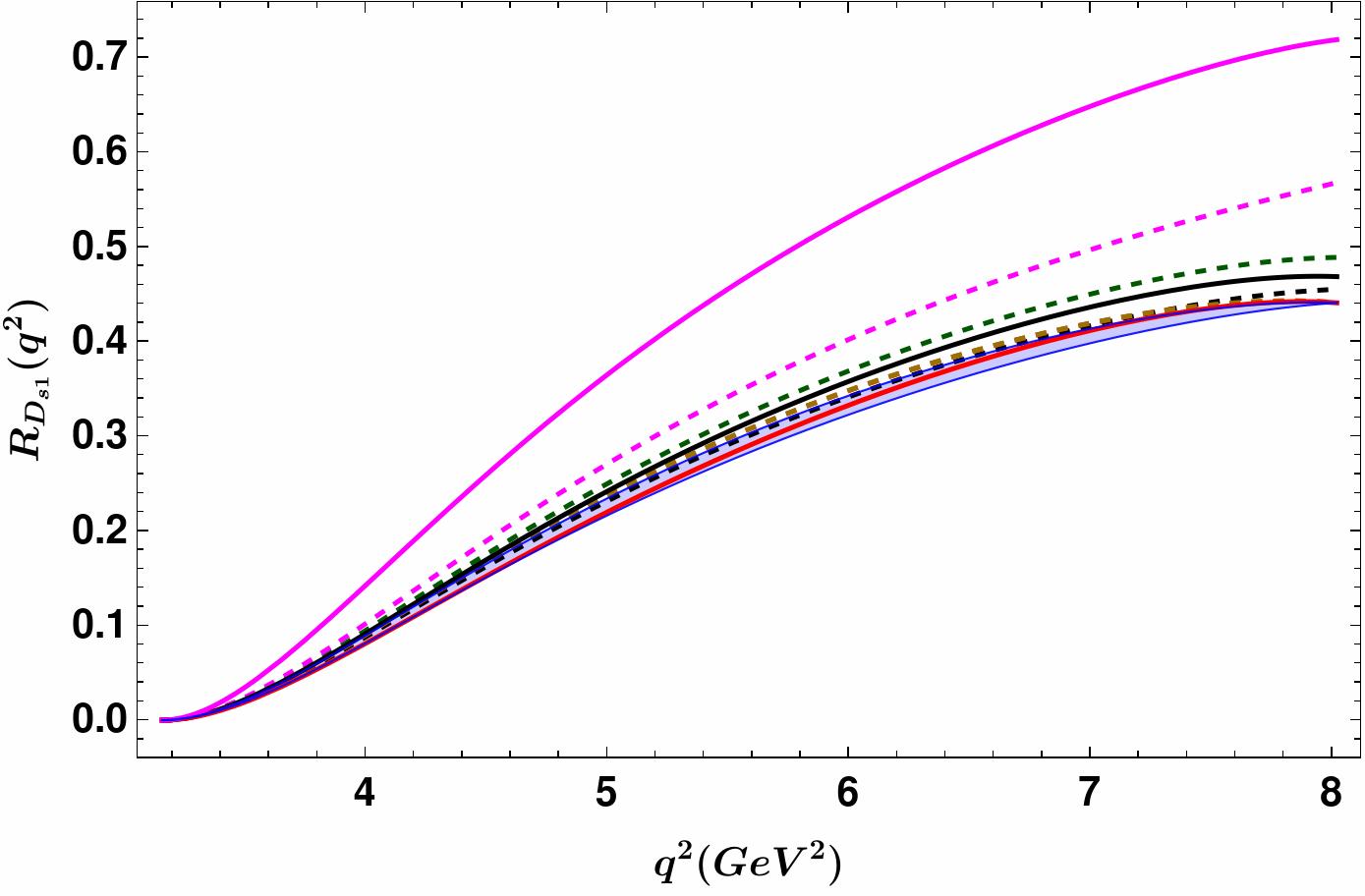}
	\hspace{0.01\textwidth}
	\includegraphics[width=0.31\textwidth]{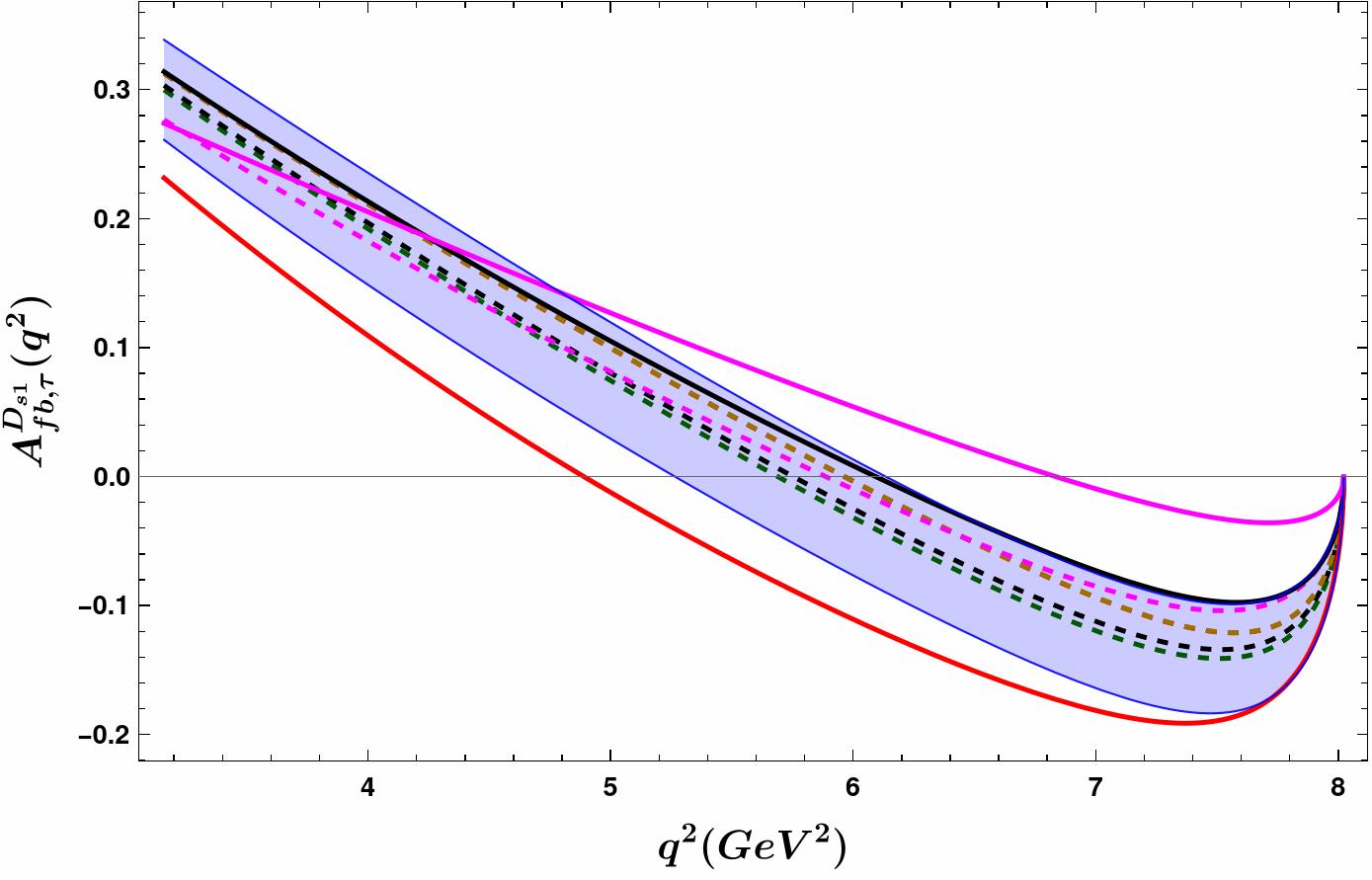}
	
	\vspace{0.3cm}
	
	\includegraphics[width=0.31\textwidth]{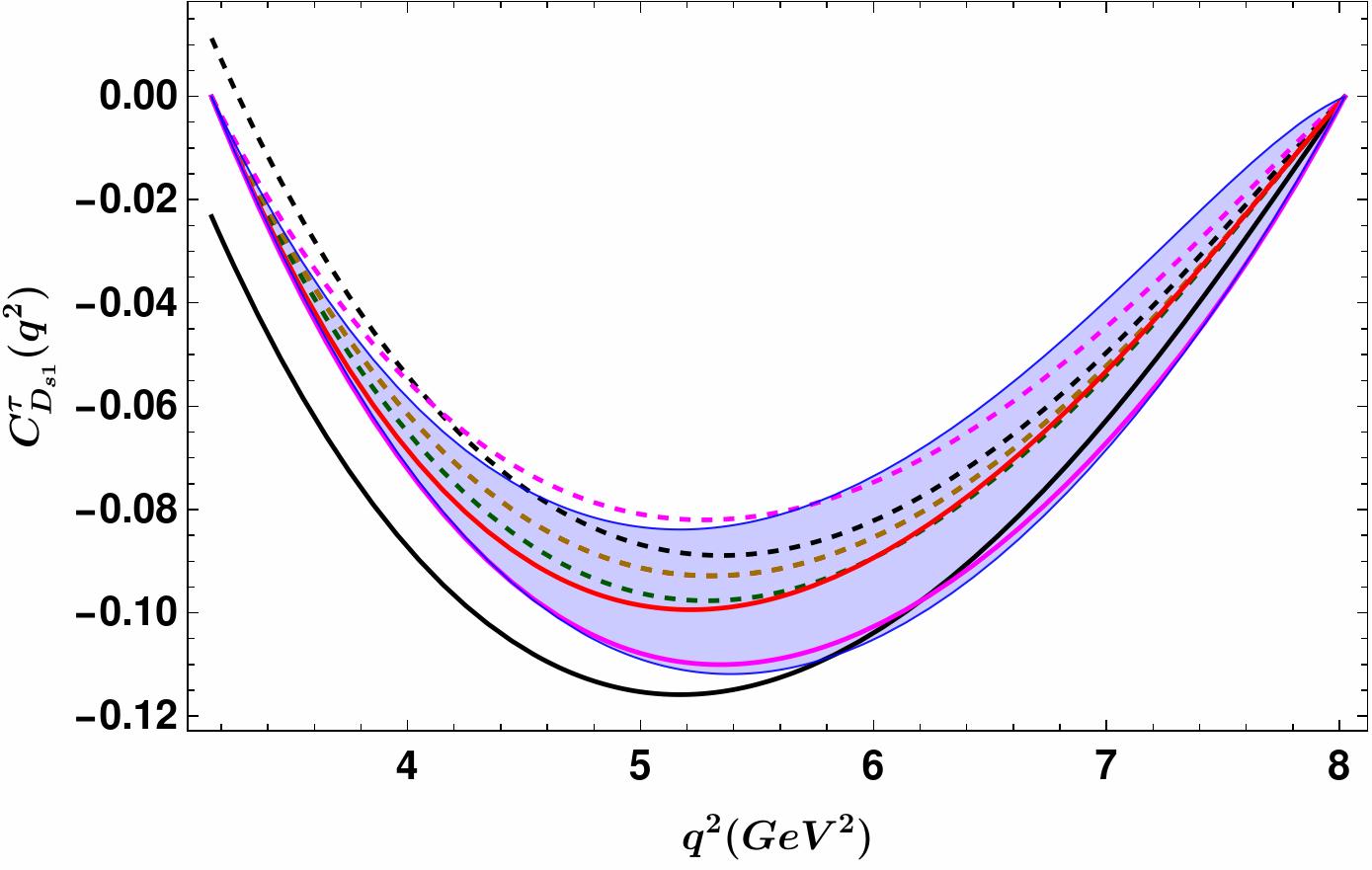}
	\hspace{0.01\textwidth}
	\includegraphics[width=0.31\textwidth]{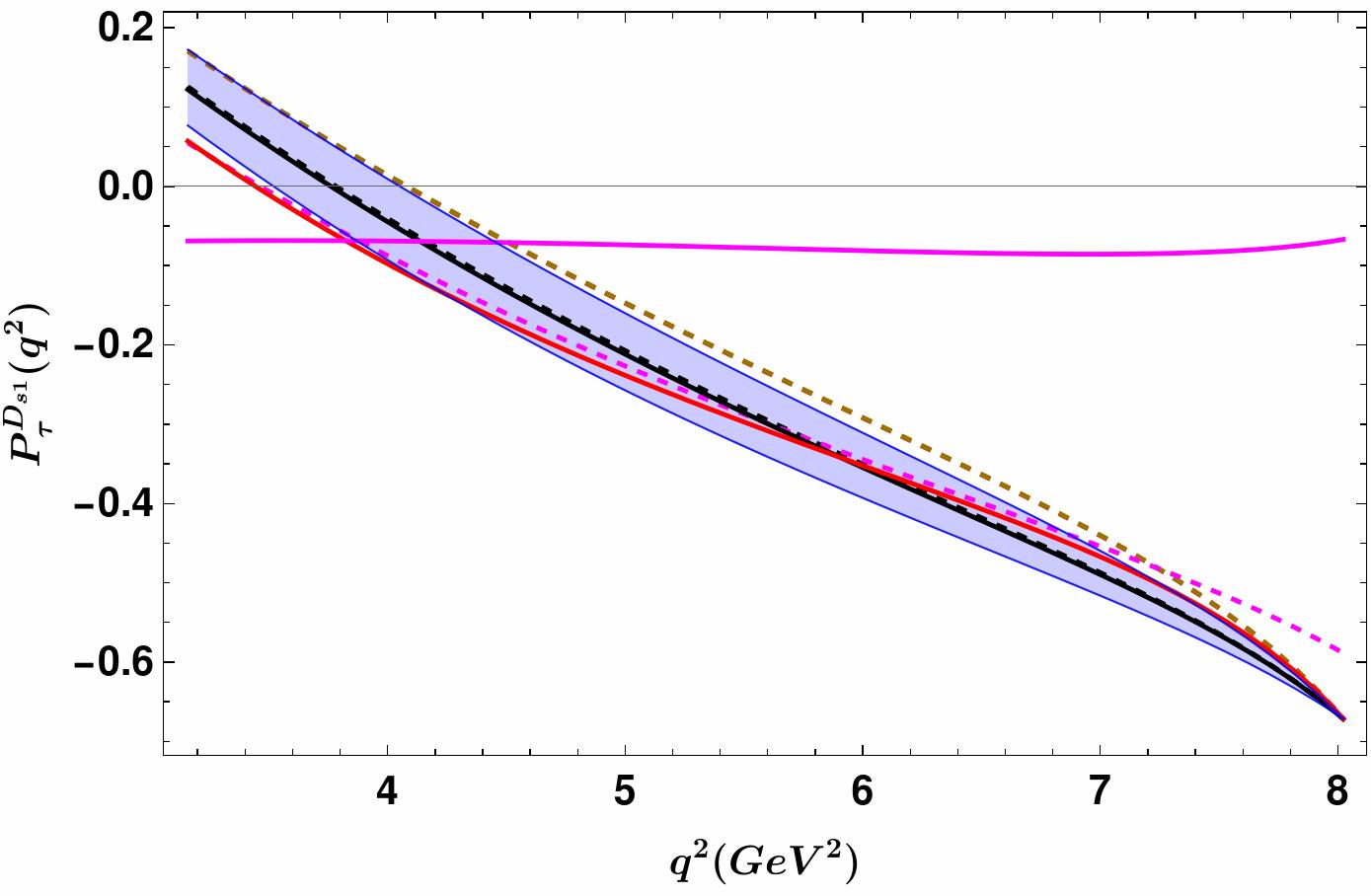}
	\hspace{0.01\textwidth}
	\includegraphics[width=0.31\textwidth]{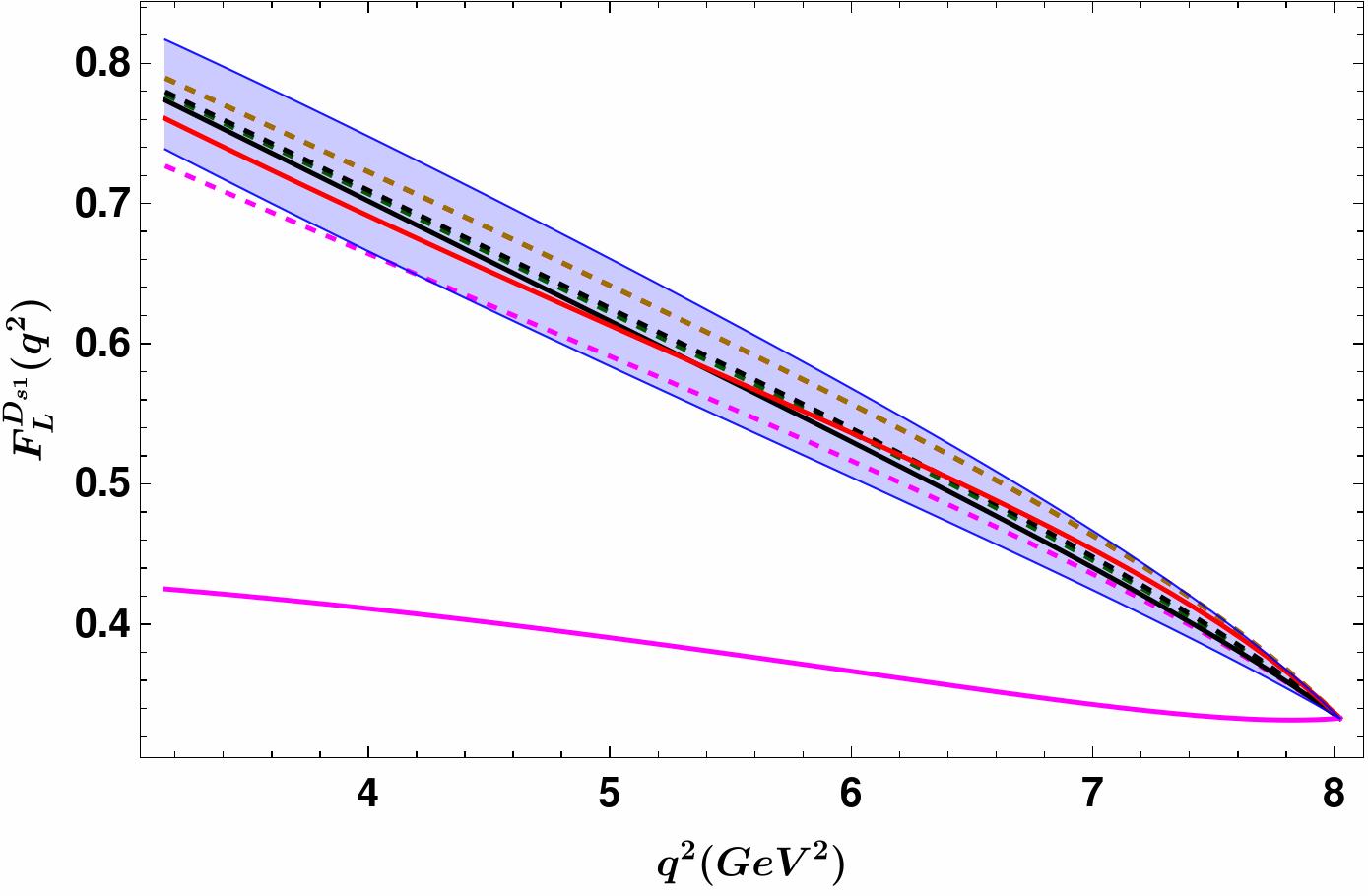}
	
	\vspace{0.3cm}
	
	\includegraphics[scale=0.4]{DsstNPscnWEFTleg.jpg}
\end{center}
\caption{$q^2$-distribution of $B_s \to D_{s1}\tau\nu_\tau$ observables in WET.}\label{fig:1d_comp_Ds1}
\end{figure}

\begin{figure}[ht]
\begin{center}
	\includegraphics[width=0.31\textwidth]{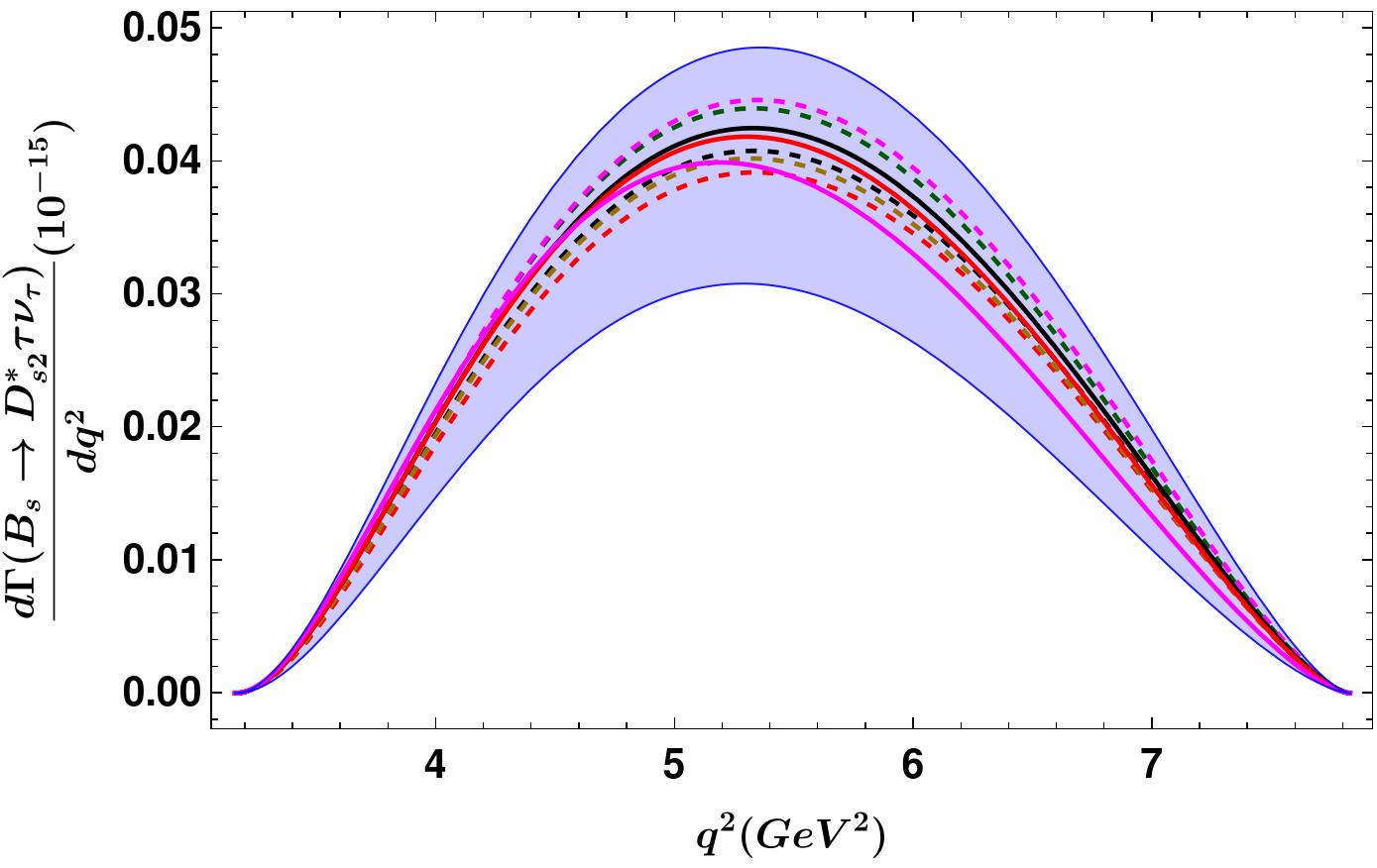}
	\hspace{0.01\textwidth}
	\includegraphics[width=0.31\textwidth]{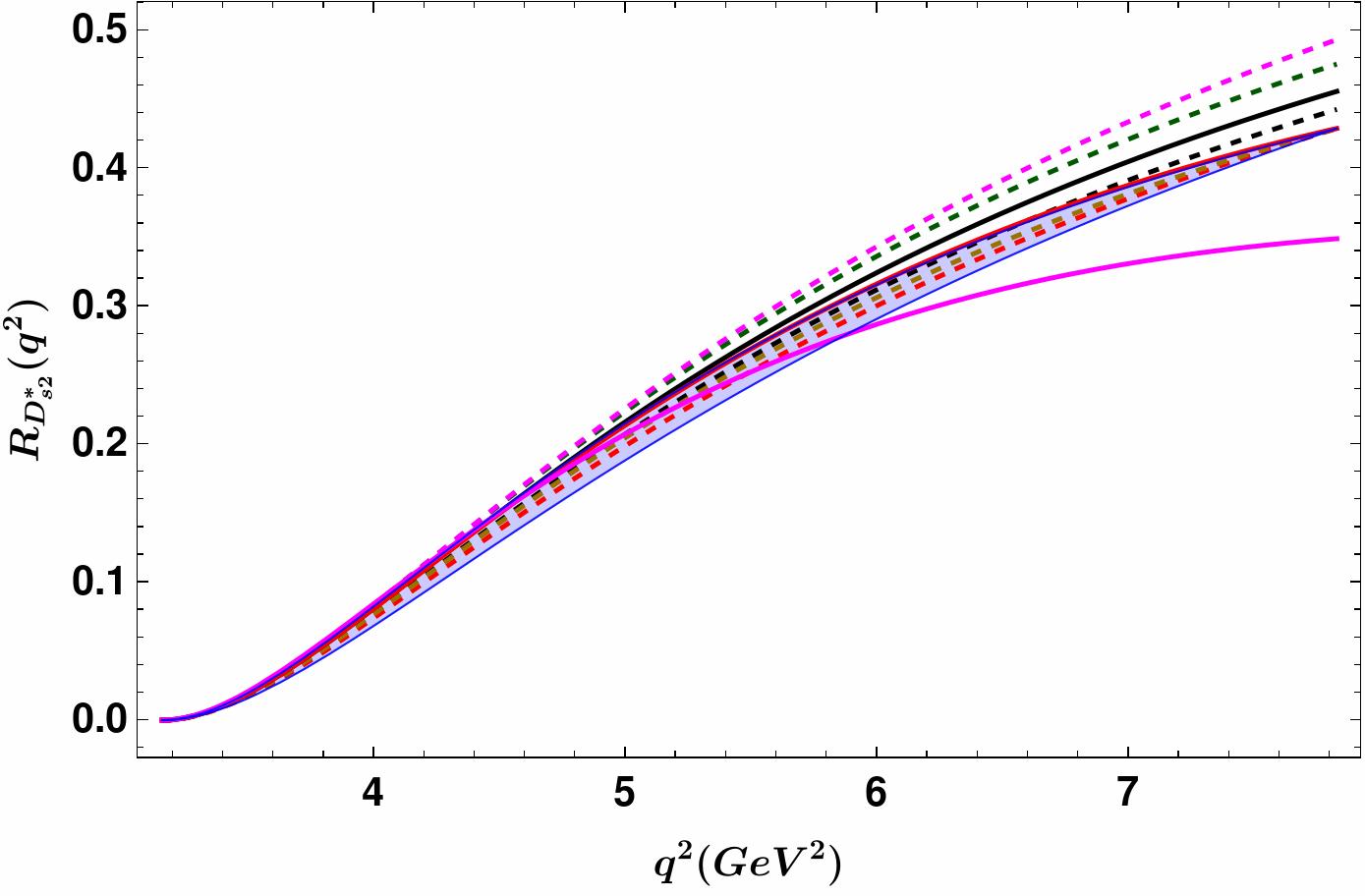}
	\hspace{0.01\textwidth}
	\includegraphics[width=0.31\textwidth]{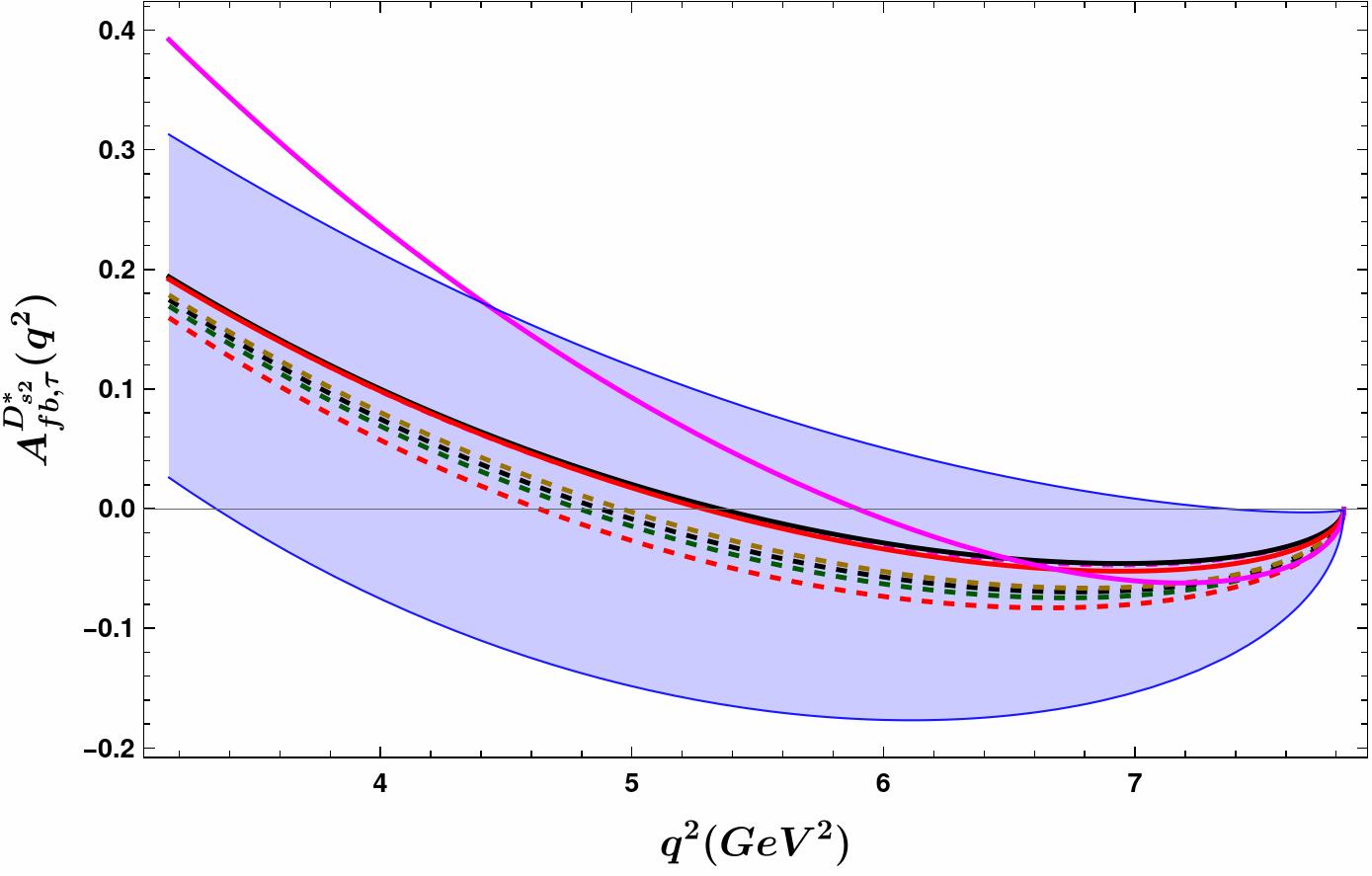}
	
	\vspace{0.3cm}
	
	\includegraphics[width=0.31\textwidth]{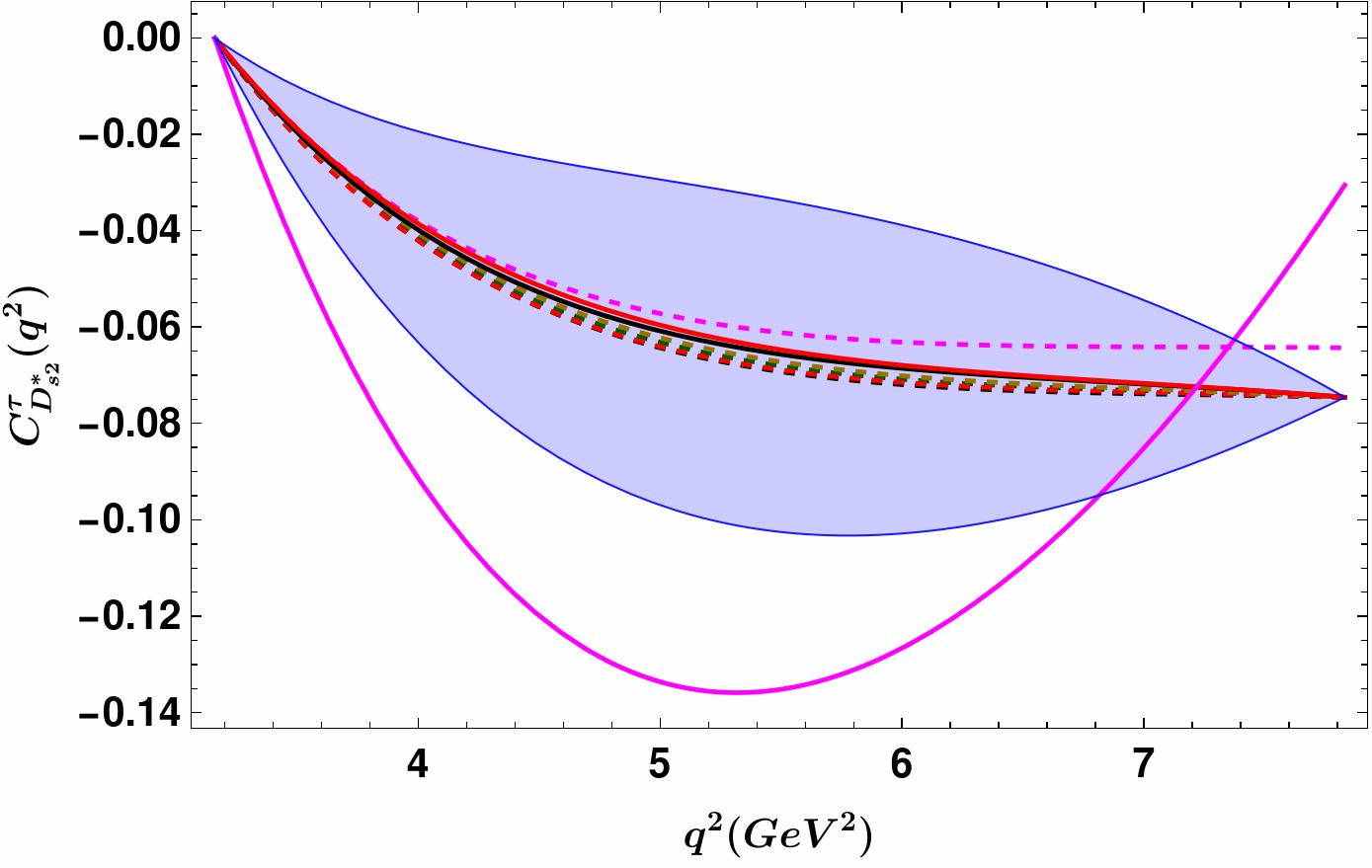}
	\hspace{0.01\textwidth}
	\includegraphics[width=0.31\textwidth]{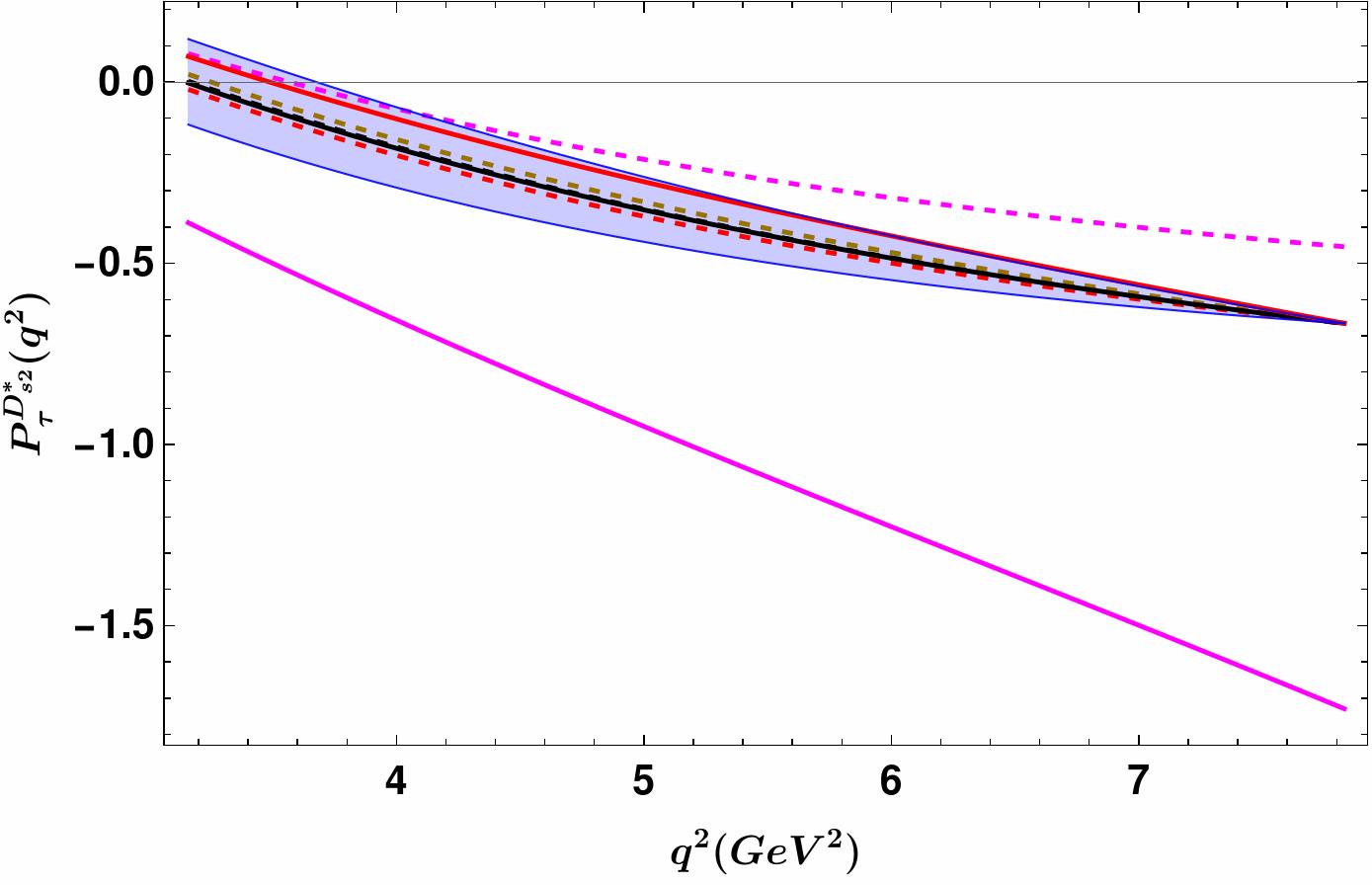}
	\hspace{0.01\textwidth}
	\includegraphics[width=0.31\textwidth]{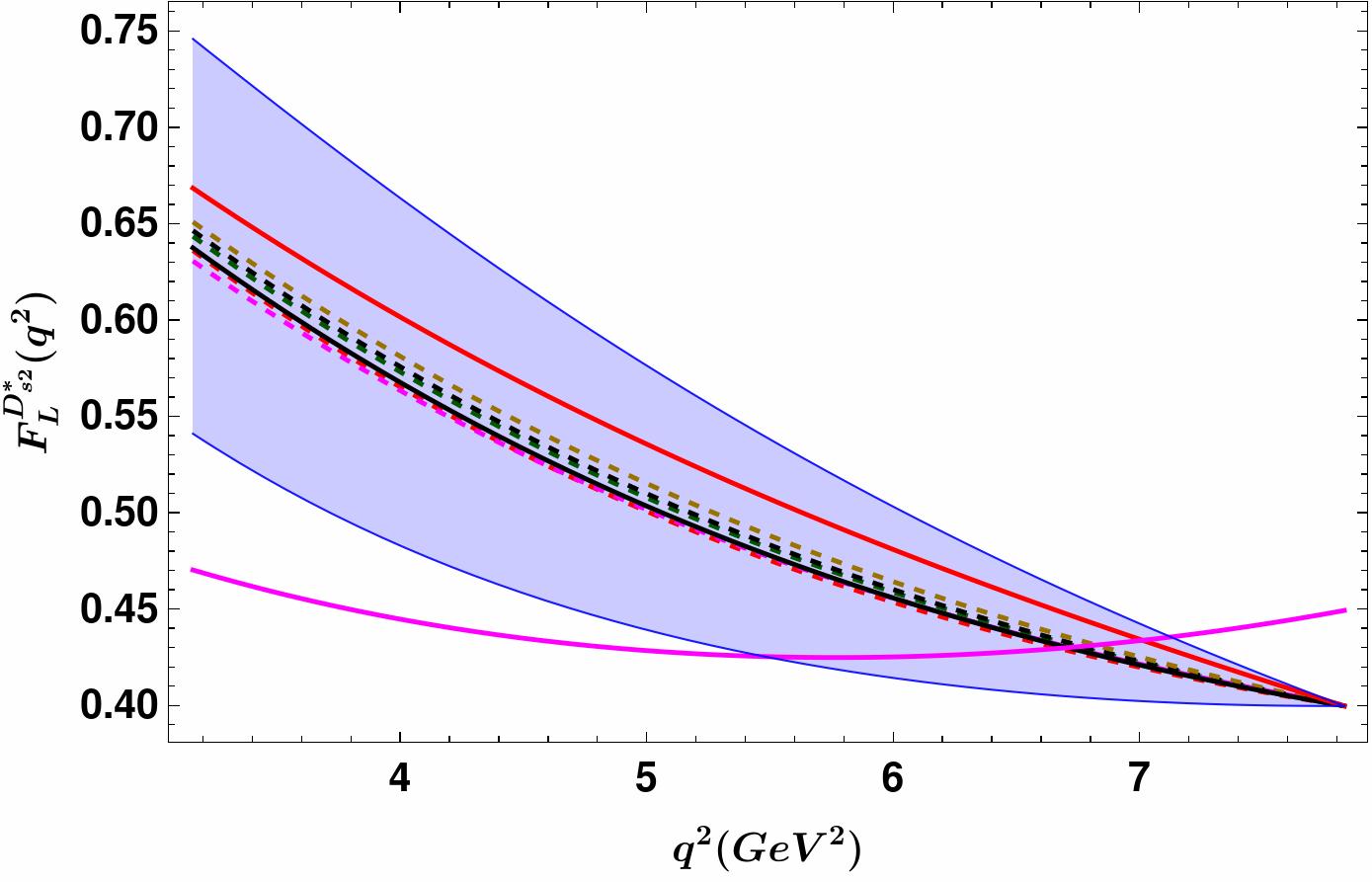}
	
	\vspace{0.3cm}
	
	\includegraphics[scale=0.4]{DsstNPscnWEFTleg.jpg}
\end{center}
\caption{$q^2$-distribution of $B_s \to D_{s2}^*\tau\nu_\tau$ observables in WET.}\label{fig:1d_comp_Ds2s}
\end{figure}

In this framework, NP sensitivity is examined by considering one operator at a time, while fixing all others to zero. This is done for both real and complex WCs. The obtained best-fit points for the NP couplings are presented in table~\ref{tab:1d_real}. These values are then used as benchmarks to test NP sensitivity. Figures~\ref{fig:1d_comp_Ds0s}-\ref{fig:1d_comp_Ds2s} display the $q^2$-dependence of various observables for the modes considered, in the SM and in various NP scenarios. To complement the information provided by the plots, we calculate the tension between the SM and NP predictions using the expression given below
\begin{eqnarray}\label{eq:tension_formula}
\sigma_{\textrm{dev}} = \left\vert \frac{\langle \mathcal{O}_{\text{SM}} \rangle - \langle \mathcal{O}_{\text{NP}} \rangle}{\sqrt{\sigma_{\text{SM}}^2 + \sigma_{\text{NP}}^2}} \right\vert \, ,
\end{eqnarray}
where in the above equation, $\langle \mathcal{O}_{\text{SM}} \rangle$ and $\langle \mathcal{O}_{\text{NP}} \rangle$ are the integrated values of the observables in the presence of SM and NP, respectively, with $\sigma_{\text{SM}}$ and $\sigma_{\text{NP}}$ denoting their respective errors. The tension calculation highlights the observables that can potentially reveal NP distinguishable from the SM predictions. For the $B_s \to D_{s0}^*\tau\nu_\tau$ and $B_s \to D_{s1}^*\tau\nu_\tau$ modes, the computed tension between the SM and the WET observables is given in table~\ref{tab:tens_wet_zeta}. Table~\ref{tab:tens_wet_tau} presents the computed tension for the $B_s \to D_{s1}\tau\nu_\tau$ and $B_s \to D_{s2}^*\tau\nu_\tau$ modes.

\begin{table}[t]
		\begin{center}
			\renewcommand*{\arraystretch}{1.6}
			\resizebox{1.02\textwidth}{!}{
		\begin{tabular}{|c|c|c|c|c|c|c|c|c|}
			\hline
			Scenario & \multicolumn{5}{c|}{Real} & \multicolumn{3}{c|}{Complex} \\
			\hline
			WCs & $C_{V_L}$ & $C_{V_R}$ & $C_{S_L}$ & $C_{S_R}$ & $C_T$ & $C_{V_R}$ & $C_{S_L}$ & $C_T$ \\
			\hline
			Best-fit values & $0.051(20)$ & $-0.007(31)$ &  $0.127(47)$ & $0.128(45)$  & $-0.019(11)$    &  $0.039(34)$ & $-0.28(36)$ & $0.071(69)$ \\
			& & & & & & $ + 0.360(64)i$& $+0.66(18)i$ & $+ 0.162(44)i$  \\
			\hline
			p-value & $0.25$ & $0.02$ & $0.22$ & $0.31$ & $0.06$ & 0.21 & 0.13 & 0.13 \\
			\hline
		\end{tabular}
	}
\end{center}
\caption{Fit results for one-operator scenario in WET.}\label{tab:1d_real}
\end{table}	


As shown in figure~\ref{fig:1d_comp_Ds0s}, for the $B_s \to D_{s0}^*\tau\nu_\tau$ mode, the NP sensitivity of various observables is highest in the presence of the complex $C_{S_2}$ WC, except for the differential branching fraction. The deviations from the SM are at most $2\sigma$, except for $A^{D_{s0}^*}_{fb,\tau}$, as reinforced by the computed tension in table~\ref{tab:tens_wet_zeta}. The complex $C_T$ coupling also induces a deviation from the SM for $A^{D_{s0}^*}_{fb,\tau}$, however, the deviation is within $2\sigma$. The predictions for $B_s \to D_{s1}^*\tau\nu_\tau$ are shown in figure~\ref{fig:1d_comp_Ds1s}.  The observables are most sensitive to the complex $C_T$ coupling, with deviations less than $2\sigma$ from the SM. Additionally, the observables $C^\tau_{D_{s1}^*}$ and $A^{D_{s1}^*}_{fb,\tau}$ exhibit zero-crossings at $q^2 = 7.7(15)$ GeV$^2$ and $q^2 = 7.5(14)$ GeV$^2$, respectively, in the presence of complex $C_T$. From figure~\ref{fig:1d_comp_Ds1}, it is evident that the observables of $B_s \to D_{s1}\tau\nu_\tau$ are mainly receptive to the complex $C_T$ coupling. However, as shown in table~\ref{tab:tens_wet_tau}, only $R_{D_{s1}}$ and $F_L^{D_{s1}}$ demonstrate more than $2\sigma$ deviations from the SM prediction. Due to the larger errors in the fitted NP couplings, the other observables show tensions below $2\sigma$. In the presence of complex $C_T$, the observable $A^{D_{s1}}_{fb,\tau}$ exhibits a zero-crossing at $q^2 = 6.83(80)$ GeV$^2$, which lies $1.26\sigma$ away from the SM. 
\begin{table}[t]
\begin{center}
	\renewcommand*{\arraystretch}{1.6}
	\resizebox{1.02\textwidth}{!}{
	\begin{tabular}{|c|c|c|c|c|c|c|c|c|c|c|c|c|}
		\hline
		& & \multicolumn{5}{c|}{$B_s \to D_{s0}^*\tau\nu_\tau$} & \multicolumn{6}{c|}{$ B_s \to D_{s1}^*\tau\nu_\tau$} \\
		\hline
		& WCs & $\Gamma$ & $\langle R_{D_{s0}^*} \rangle$ & $\langle A^{D_{s0}^*}_{fb,\tau} \rangle$ & $\langle C^\tau_{D_{s0}^*} \rangle$ & $\langle P^{D_{s0}^*}_\tau \rangle$ & $\Gamma$ & $\langle R_{D_{s1}^*} \rangle$ & $\langle A^{D_{s1}^*}_{fb,\tau} \rangle$ & $\langle C^\tau_{D_{s1}^*} \rangle$ & $\langle P^{D_{s1}^*}_\tau \rangle$ & $\langle F_L^{D_{s1}^*} \rangle$ \\
		\hline
		Real & $C_{V_1}$ & 0.21 & 0.60 & 0.0 & 0.0 & 0.0 & 0.45 & 0.64 & 0.0 & 0.0 & 0.0 & 0.0 \\
		& $C_{V_2}$ & 0.06 & 0.17 & 0.0 & 0.0 & 0.0 & 0.04 & 0.06 & 0.07 & 0.13 & 0.08 & 0.10 \\
		& $C_{S_1}$ & 0.15 & 0.44 & 0.14 & 0.53 & 0.58 & 0.36 & 0.52 & 0.24 & 0.01 & 0.60 & 0.29 \\
		& $C_{S_2}$ & 0.13 & 0.40 & 0.50 & 0.56 & 0.61 & 0.36 & 0.52 & 0.24 & 0.01 & 0.60 & 0.29 \\
		& $C_{T}$ & 0.08 & 0.24 & 0.78 & 0.82 & 0.17 & 0.06 & 0.08 & 0.02 & 0.02 & 0.04 & 0.22 \\
		\hline
		Complex & $C_{V_2}$ & 0.12 & 0.34 & 0.0 & 0.0 & 0.0 & 0.46 & 0.66 & 0.33 & 0.25 & 0.16 & 0.20\\
		& $C_{S_2}$ & 0.58 & 1.10 & 2.26 & 1.29 & 1.35 & 0.10 & 0.13 & 0.69 & 0.0 & 0.16 & 0.09\\
		& $C_{T}$ & 0.0 & 0.0 & 1.58 & 0.75 & 0.46 & 1.45 & 1.58 & 1.12 & 1.13 & 0.22 & 1.62\\
		\hline
	\end{tabular}
}
\end{center}
\caption{Computed tensions for observables in $B_s \to D_{s0}^*\tau\nu_\tau$ and $B_s \to D_{s1}^*\tau\nu_\tau$ within WET (in units of $\sigma$).}\label{tab:tens_wet_zeta}
\end{table}
\begin{table}[t]
\begin{center}
	\renewcommand*{\arraystretch}{1.6}
	\resizebox{1.02\textwidth}{!}{
	\begin{tabular}{|c|c|c|c|c|c|c|c|c|c|c|c|c|c|}
		\hline
		& & \multicolumn{6}{c|}{$B_s \to D_{s1}\tau\nu_\tau$} & \multicolumn{6}{c|}{$B_s \to D_{s2}^*\tau\nu_\tau$} \\
		\hline
		& WCs & $\Gamma$ & $\langle R_{D_{s1}} \rangle$ & $\langle A^{D_{s1}}_{fb,\tau} \rangle$ & $\langle C^\tau_{D_{s1}} \rangle$ & $\langle P^{D_{s1}}_\tau \rangle$ & $\langle F_L^{D_{s1}} \rangle$ & $\Gamma$ & $\langle R_{D_{s2}^*} \rangle$ & $\langle A^{D_{s2}^*}_{fb,\tau} \rangle$ & $\langle C^\tau_{D_{s2}^*} \rangle$ & $\langle P^{D_{s2}^*}_\tau \rangle$ & $\langle F_L^{D_{s2}^*} \rangle$ \\
		\hline
		Real & $C_{V_1}$ & 0.41 & 1.52 & 0.0 & 0.0 & 0.0 & 0.0 & 0.34 & 0.56 & 0.0 & 0.0 & 0.0 & 0.0\\
		& $C_{V_2}$ & 0.10 & 0.35 & 0.10 & 0.36 & 0.02 & 0.06 & 0.09 & 0.15 & 0.03 & 0.02 & 0.0 & 0.02\\
		& $C_{S_1}$ & 0.18 & 0.72 & 0.42 & 0.20 & 0.89 & 0.40 & 0.04 & 0.07 & 0.06 & 0.02 & 0.15 & 0.08\\
		& $C_{S_2}$ & 0.18 & 0.71 & 0.41 & 0.19 & 0.88 & 0.40 & 0.04 & 0.06 & 0.06 & 0.02 & 0.14 & 0.08\\
		& $C_{T}$ & 0.75 & 1.86 & 0.23 & 0.70 & 0.05 & 0.52 & 0.38 & 0.63 & 0.18 & 0.15 & 1.15 & 0.07\\
		\hline
		Complex & $C_{V_2}$ & 0.28 & 0.96 & 0.56 & 0.70 & 0.04 & 0.12 & 0.22 & 0.36 & 0.19 & 0.05 & 0.01 & 0.04\\
		& $C_{S_2}$ & 0.05 & 0.17 & 0.86 & 0.05 & 0.20 & 0.11 & 0.16 & 0.26 & 0.17 & 0.07 & 0.53 & 0.30\\
		& $C_{T}$ & 1.71 & 2.41 & 1.06 & 0.55 & 1.93 & 2.26 & 0.02 & 0.02 & 0.54 & 0.95 & 1.22 & 0.68\\
		\hline
	\end{tabular}
}
\end{center}
\caption{Computed tensions for observables in $B_s \to D_{s1}\tau\nu_\tau$ and $B_s \to D_{s2}^*\tau\nu_\tau$ within WET (in units of $\sigma$).}\label{tab:tens_wet_tau}
\end{table}
In the SM, $P^{D_{s1}}_\tau$ exhibits a zero-crossing, whereas this is absent in the presence of complex $C_T$. This is a notable observation for $P^{D_{s1}}_\tau$ that can be confronted experimentally. The predictions for $B_s \to D_{s2}^*\tau\nu_\tau$ observables are shown in figure~\ref{fig:1d_comp_Ds2s}. It can be seen that $P^{D_{s2}^*}_\tau$ is largely sensitive to the complex $C_T$ coupling, but due to the large errors in the NP couplings, it is in agreement with the SM at $2\sigma$.

\subsection{SMEFT Framework}

Having analyzed the one-operator WET scenario, we now proceed to analyze the NP in the SMEFT approach. The SMEFT Lagrangian is given as \cite{Grzadkowski:2010es}
\begin{eqnarray}\label{eq:smeft_lag}
\mathcal{L}_{\text{SMEFT}} = \mathcal{L}_{\text{SM}} + \frac{1}{\Lambda^2}\sum_i C^{(6)}_i(\Lambda)Q_i\, ,
\end{eqnarray}
where $\mathcal{L}_{\text{SM}}$ is the dimension-4 SM Lagrangian and $\Lambda = 1$~TeV is the NP scale considered in this work. The dimension-6 operators allowed by SM gauge symmetry that can mediate semileptonic $b\to c\ell\nu_\ell$ transitions are given by
\begin{align}\label{eq:smeft_ops}
&Q^{(3)}_{\ell q} = (\bar{\ell}\gamma_\mu \tau^I \ell)(\bar{q}\gamma^\mu \tau^I q) \,, & 
&Q_{\ell edq} = (\bar{\ell}^j e)(\bar{d}q^j) \,, \nonumber \\[0.5em]
&Q^{(1)}_{\ell equ} = (\bar{\ell}^j e)\epsilon_{jk} (\bar{q}^k u) \,, & 
&Q^{(3)}_{\ell equ} = (\bar{\ell}^j \sigma_{\mu\nu} e) \epsilon_{jk} (\bar{q}^k \sigma^{\mu\nu} u) \,.
\end{align} 
In the above, $q$ ($\ell$) are the left-handed quark (lepton) doublets, $u (d)$ are the right-handed up (down)-type quark singlets, $e$ is the right-handed charged lepton singlet, and $\tau^I$ represents a Pauli matrix. The corresponding Wilson coefficients of the above operators are given by $C_i = (C^{(3)}_{\ell q}, C_{\ell edq}, C^{(1)}_{\ell equ}, C^{(3)}_{\ell equ})$.

To analyze the effects of SMEFT on $b\to c\ell\nu_\ell$ modes, we first evolve the Wilson coefficients from the scale $\Lambda = 1$~TeV to the electroweak scale using the Renormalization Group Evolution (RGE) technique. At the EW scale, we perform a tree-level matching between the SMEFT and WET operators. The resulting Wilson coefficients are then evolved to the $m_b$ scale. It is to be noted that the right-handed vector operator $O_{V_2}$ in WET does not match any dimension-6 SMEFT operator. It can be matched with an SMEFT operator only at dimension-8, which, however, suppresses the corresponding Wilson coefficient by a factor of $1/\Lambda^4$. The RGE equation for the SMEFT couplings to evolve from the scale $\Lambda = 1$ TeV to the scale $\mu = m_Z$ is given by \cite{Gonzalez-Alonso:2017iyc}
\begin{eqnarray}\label{eq:smeft_run}
\begin{pmatrix}
	C^{(3)}_{\ell q} \\ C_{\ell edq} \\ C_{\ell equ}^{(1)} \\ C_{\ell equ}^{(3)} \end{pmatrix}_{(\mu = m_Z)} = \begin{pmatrix}
	1 & 0 & 0 & 0 \\ 0 & 1.19 & 0 & 0 \\ 0 & 0 & 1.20 & -0.185 \\ 0 & 0 & -0.00381 & 0.959 
\end{pmatrix} \begin{pmatrix}
	C^{(3)}_{\ell q} \\ C_{\ell edq} \\ C_{\ell equ}^{(1)} \\ C_{\ell equ}^{(3)}
\end{pmatrix}_{(\mu = 1\text{TeV})}\, .
\end{eqnarray}
As can be seen from the above equation, RGE introduces a mixing between the $C_{\ell equ}^{(1)}$ and $C_{\ell equ}^{(3)}$ couplings. 

The SMEFT Lagrangian of eq.~\ref{eq:smeft_lag} is defined in the flavor (weak) basis. We therefore perform a unitary transformation from the weak basis to the mass basis. Following the transformation relations given in ref.\,\cite{Aebischer:2015fzz}, the tree-level matching of the SMEFT operators given in eq.~\ref{eq:smeft_ops} with the WET operators given in eq.~\ref{eq:wet_ops} yields the following relations between the respective couplings
\begin{align}
&C_{V_1} = -\frac{1}{\sqrt{2}G_F}\frac{1}{\Lambda^2} \sum_n^3  \frac{V_{cn}}{V_{cb}}\left[\tilde{C}^{(3)33n3}_{\ell q}\right] \,,&
&C_{S_1} = -\frac{1}{2\sqrt{2}G_F}\frac{1}{\Lambda^2} \sum_n^3 \frac{V_{cn}}{V_{cb}}\left[\tilde{C}^{333n}_{\ell edq}\right]^* \,, \nonumber \\[0.5em]
&C_{S_2} = - \frac{1}{2\sqrt{2}G_F}\frac{1}{\Lambda^2} \sum_n^3 \frac{V_{nb}}{V_{cb}}\left[\tilde{C}^{(1)33n2}_{\ell equ}\right]^* \,, &
&C_{T} =- \frac{1}{2\sqrt{2}G_F}\frac{1}{\Lambda^2} \sum_n^3 \frac{V_{nb}}{V_{cb}}\left[\tilde{C}^{(3)33n2}_{\ell equ}\right]^* \,,  \label{eq:match_eqn}
\end{align} where $\tilde{C}^{(3)33n3}_{\ell q}, \tilde{C}^{333n}_{\ell edq}, \tilde{C}^{(1)33n2}_{\ell equ}, \tilde{C}^{(3)33n2}_{\ell equ}$ are the SMEFT couplings in the mass basis. The first two superscript indices of these couplings represent lepton generation and the last two represent quark generation. In the above equation, $V_{ij}$ represent CKM elements and
\begin{align}
&\left[\tilde{C}_{\ell q}^{(3)\ell\ell ij}\right] = \left[C_{\ell q}^{(3)\ell\ell ab}\right] S_{Lia}^{d\dagger} S^d_{Lbj} \,,&
&\left[\tilde{C}_{\ell edq}^{\ell\ell ji}\right]^* = \left[C_{\ell edq}^{\ell\ell ab}\right]^* S^{d\dagger}_{Lib} S^{d}_{Raj} \,, \nonumber \\[0.5em]
&\left[\tilde{C}_{\ell equ}^{(1)\ell\ell ji}\right]^* = \left[C_{\ell equ}^{(1)\ell\ell ab} \right]^* S^{u\dagger}_{Rib} S^d_{Laj} \,,&
&\left[\tilde{C}_{\ell equ}^{(3)\ell\ell ji}\right]^* = \left[C_{\ell equ}^{(3)\ell\ell ab} \right]^* S^{u\dagger}_{Rib} S^d_{Laj} \,, \label{eq:tilde_eqn}
\end{align} 
where $S^u_{L(R)}$ and $S^d_{L(R)}$ are the unitary matrices that transform the left(right)-handed up-type and down-type quarks, respectively. The CKM matrix is defined as $V_{CKM} = S^{u\dagger}_L~S^d_L$. For simplicity, we drop the tilde symbol in the rest of this paper.

Now evolving the WET couplings from the $m_Z$ scale to the $m_b$ scale, the running is given in ref.\,\cite{Gonzalez-Alonso:2017iyc} as
\begin{eqnarray}\label{eq:wet_run}
\begin{pmatrix}
	C_{V_1} \\ C_{S_1} \\ C_{S_2} \\ C_{T}
\end{pmatrix}_{(\mu=m_b)} = \begin{pmatrix}
	1 & 0 & 0 & 0 \\ 0 & 1.46 & 1.45 \times 10^{-6} & -0.0177 \\ 0 & 1.45 \times 10^{-6} & 1.46 & -0.0177 \\ 0 & -1.72 \times 10^{-4} & -1.72 \times 10^{-4} & 0.878
\end{pmatrix} \begin{pmatrix}
	C_{V_1} \\ C_{S_1} \\ C_{S_2} \\ C_{T}
\end{pmatrix}_{(\mu=m_Z)}\, .
\end{eqnarray}
By performing the RGE and matching using eqs.~\ref{eq:smeft_run}-\ref{eq:wet_run}, we obtain the following relations 
\begin{eqnarray} \label{eqn:bc_match_num}
(C_{V_1})_{b \to c} &=& - 1.439 \left[C_{\ell q}^{(3)3323}\right]\, , \nonumber \\
(C_{S_1})_{b \to c} &=& - 1.25 \left[C_{\ell edq}^{3332}\right]^*\, , \nonumber \\
(C_{S_2})_{b \to c} &=& - 1.305 \left[C_{\ell equ}^{(1)3332}\right]^* + 0.214 \left[C_{\ell equ}^{(3)3332}\right]^* \, ,\nonumber \\
(C_T)_{b \to c} &=& 2.8 \times 10^{-3} \left[C_{\ell equ}^{(1)3332}\right]^* - 0.627 \left[C_{\ell equ}^{(3)3332}\right]^*\, ,
\end{eqnarray} where the above couplings represent only CKM-favored couplings.

\begin{figure}[t]
\begin{center}
	\includegraphics[width=0.31\textwidth]{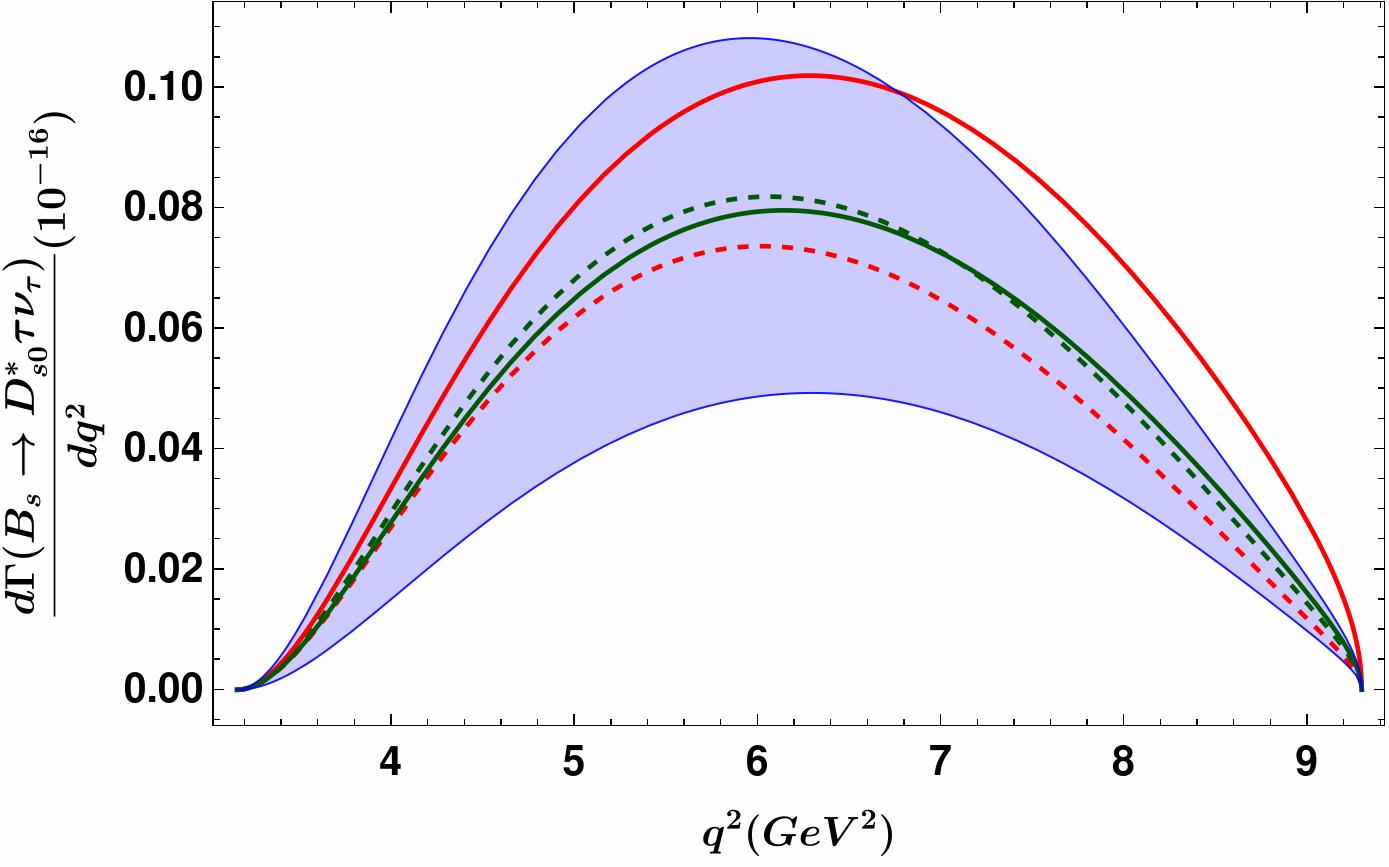}
	\hspace{0.01\textwidth}
	\includegraphics[width=0.31\textwidth]{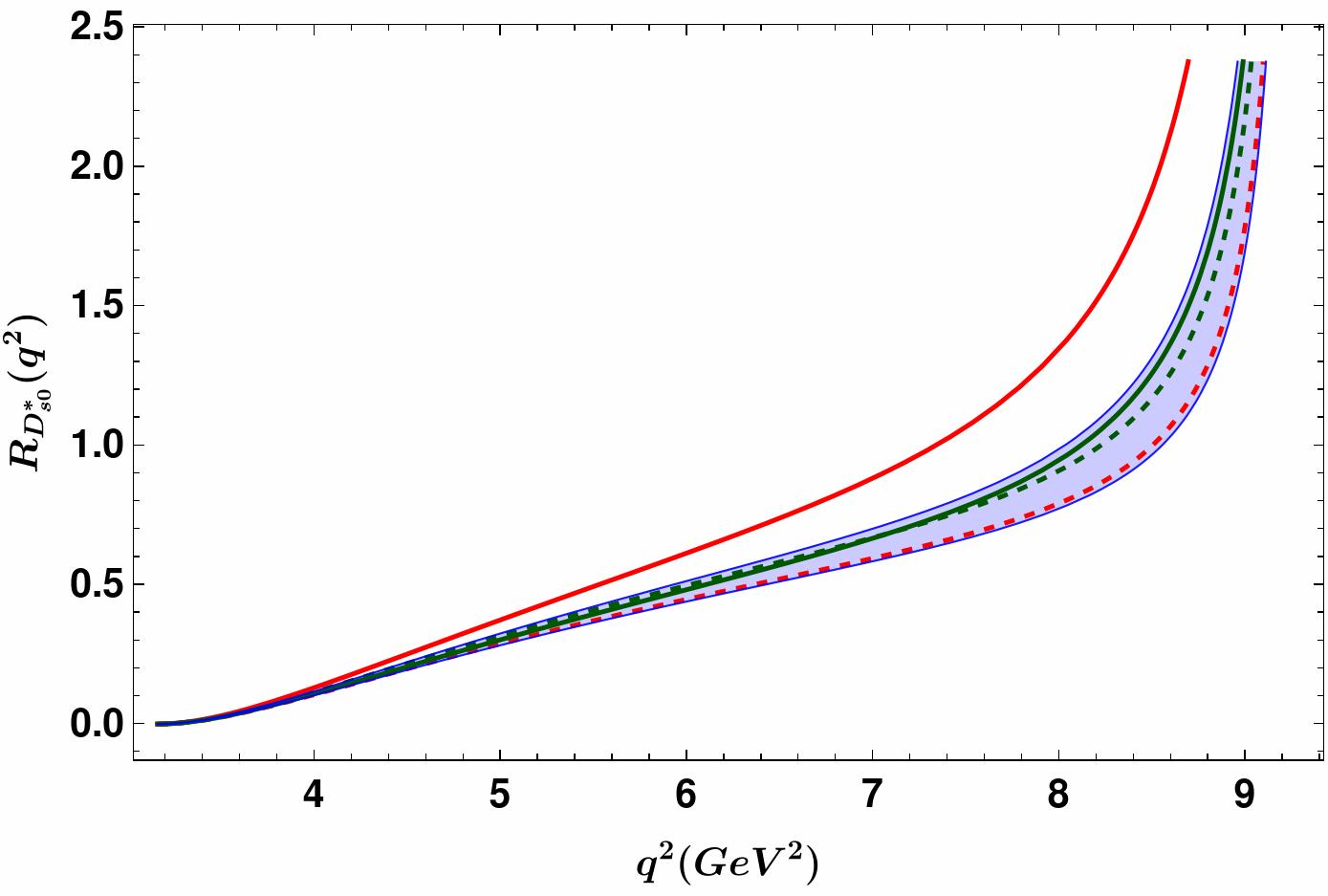}
	\hspace{0.01\textwidth}
	\includegraphics[width=0.31\textwidth]{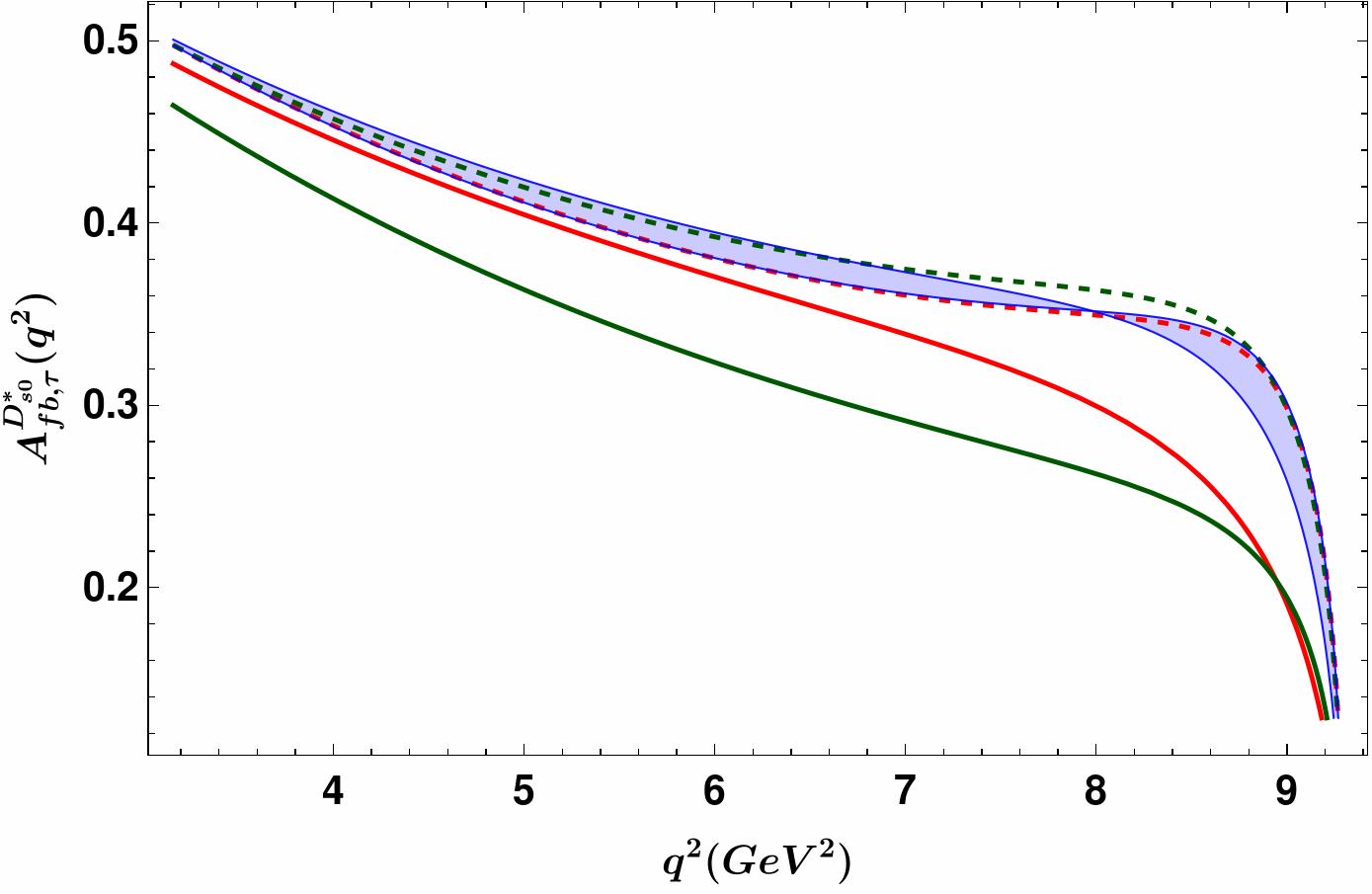}
	
	\vspace{0.3cm}
	
	\includegraphics[width=0.31\textwidth]{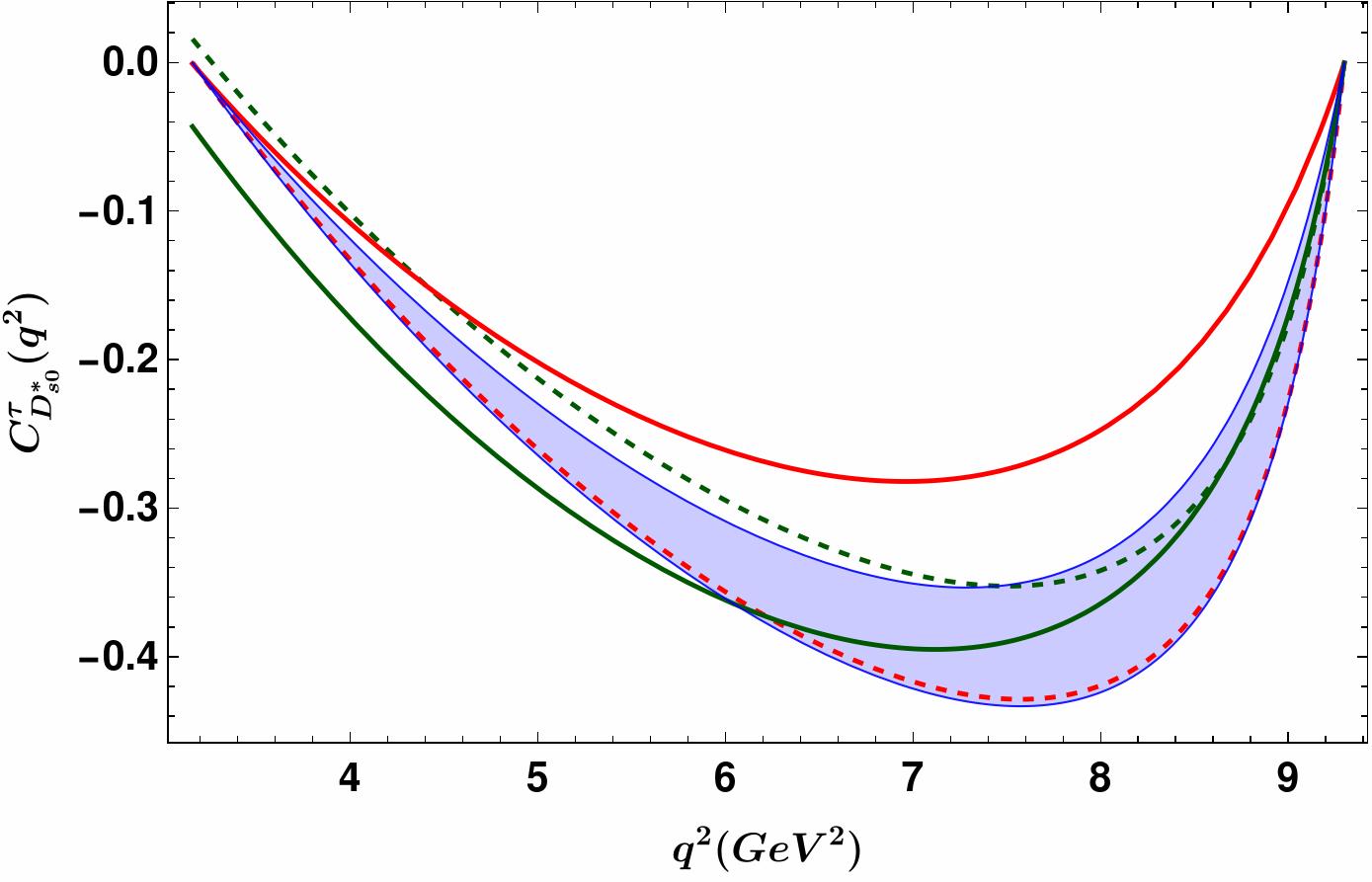}
	\hspace{0.01\textwidth}
	\includegraphics[width=0.31\textwidth]{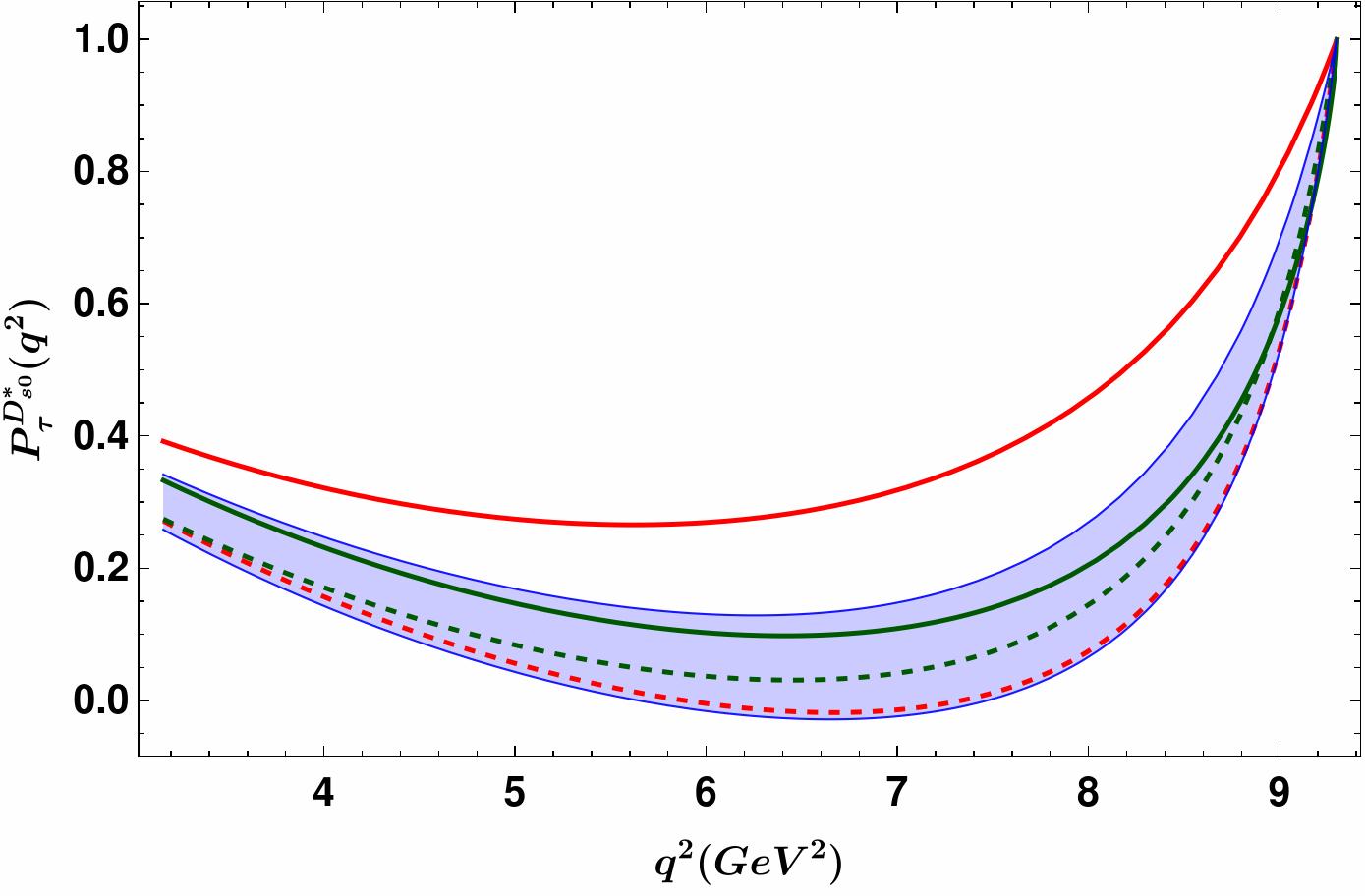}
	
	\vspace{0.3cm}
	
	\includegraphics[scale=0.4]{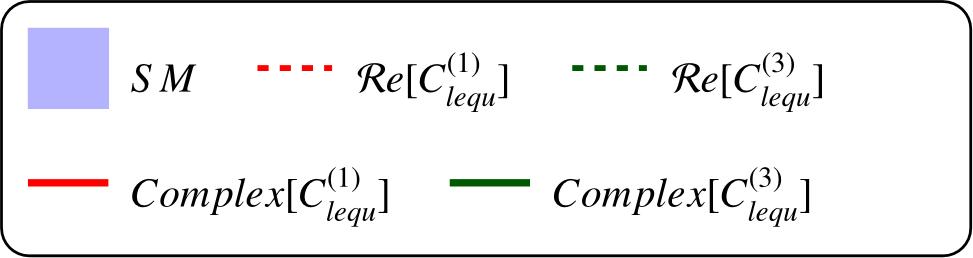}
\end{center}
\caption{$q^2$-distribution of $B_s \to D_{s0}^*\tau\nu_\tau$ observables in SMEFT.}\label{fig:comp_smeft_Ds0s}
\end{figure}

\begin{figure}[t]
\begin{center}
	\includegraphics[width=0.31\textwidth]{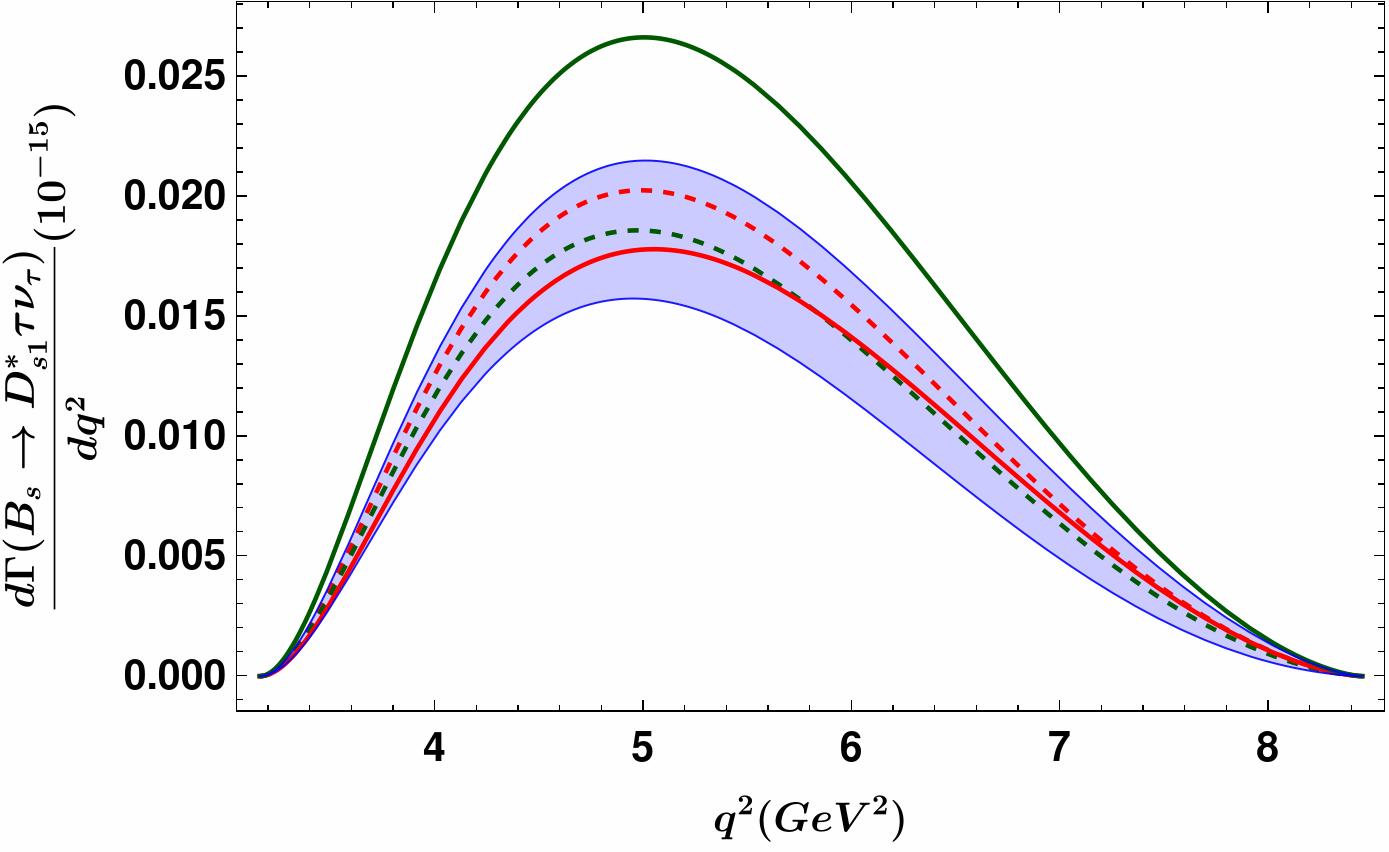}
	\hspace{0.01\textwidth}
	\includegraphics[width=0.31\textwidth]{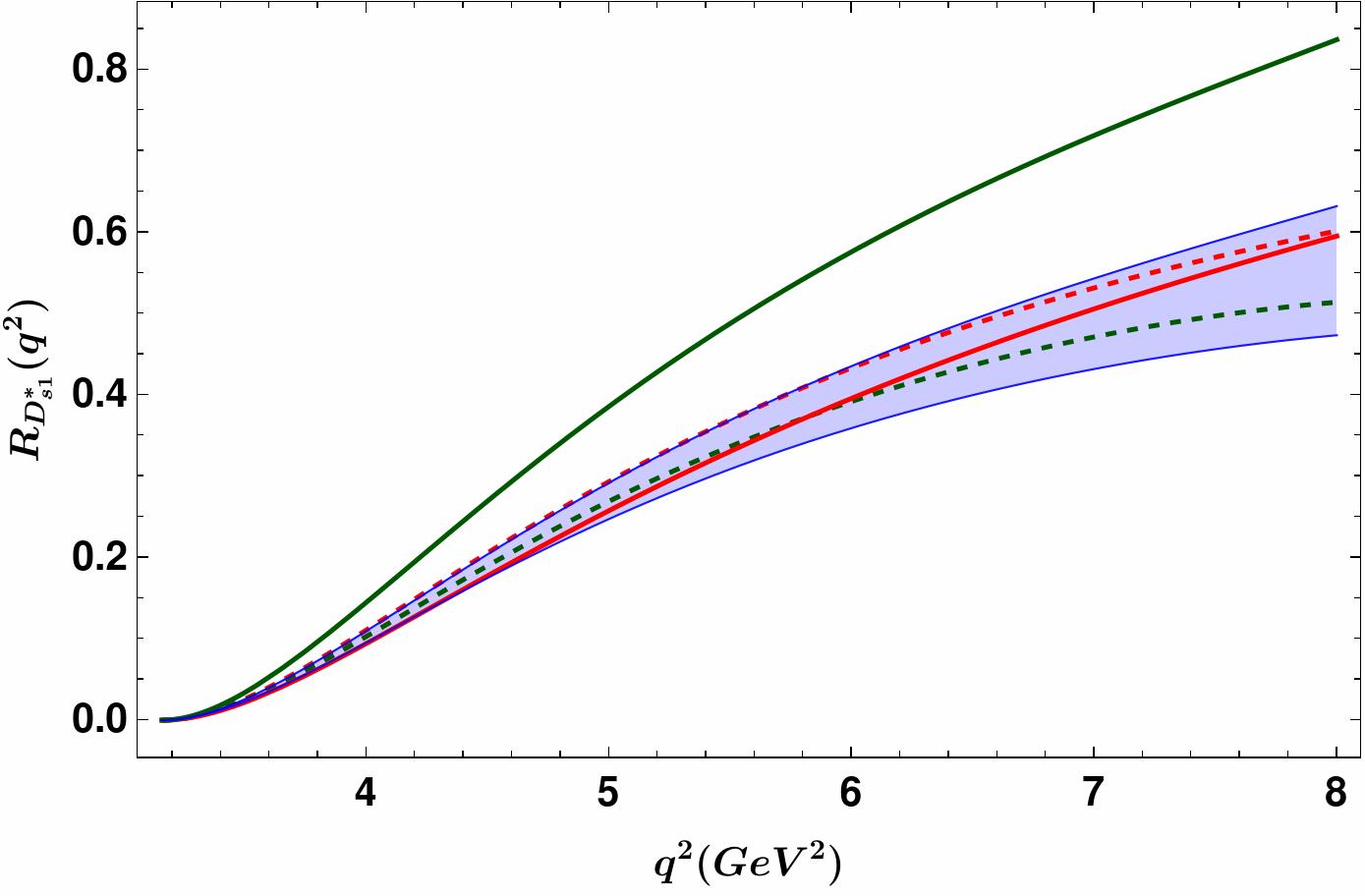}
	\hspace{0.01\textwidth}
	\includegraphics[width=0.31\textwidth]{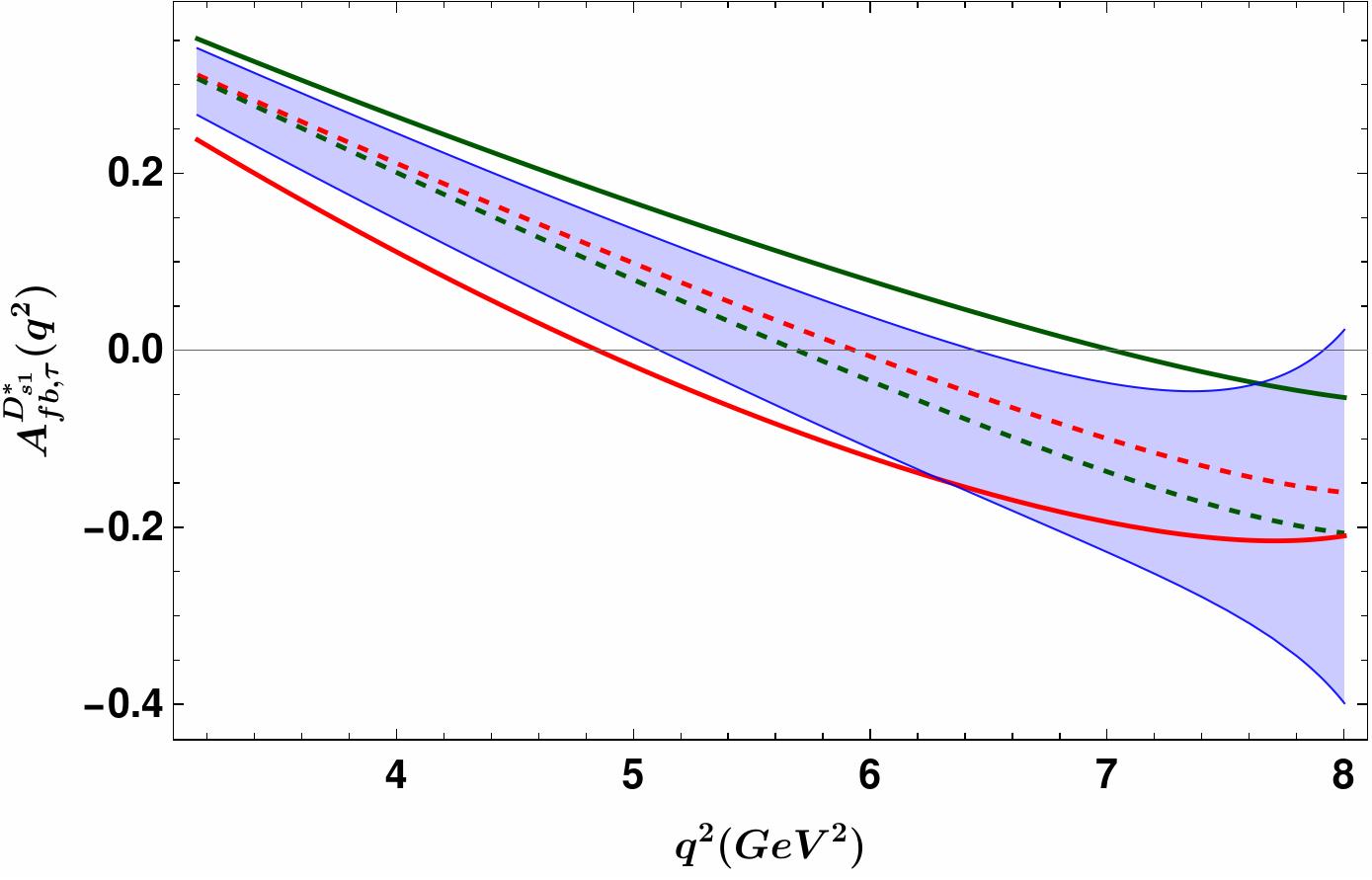}
	
	\vspace{0.3cm}
	
	\includegraphics[width=0.31\textwidth]{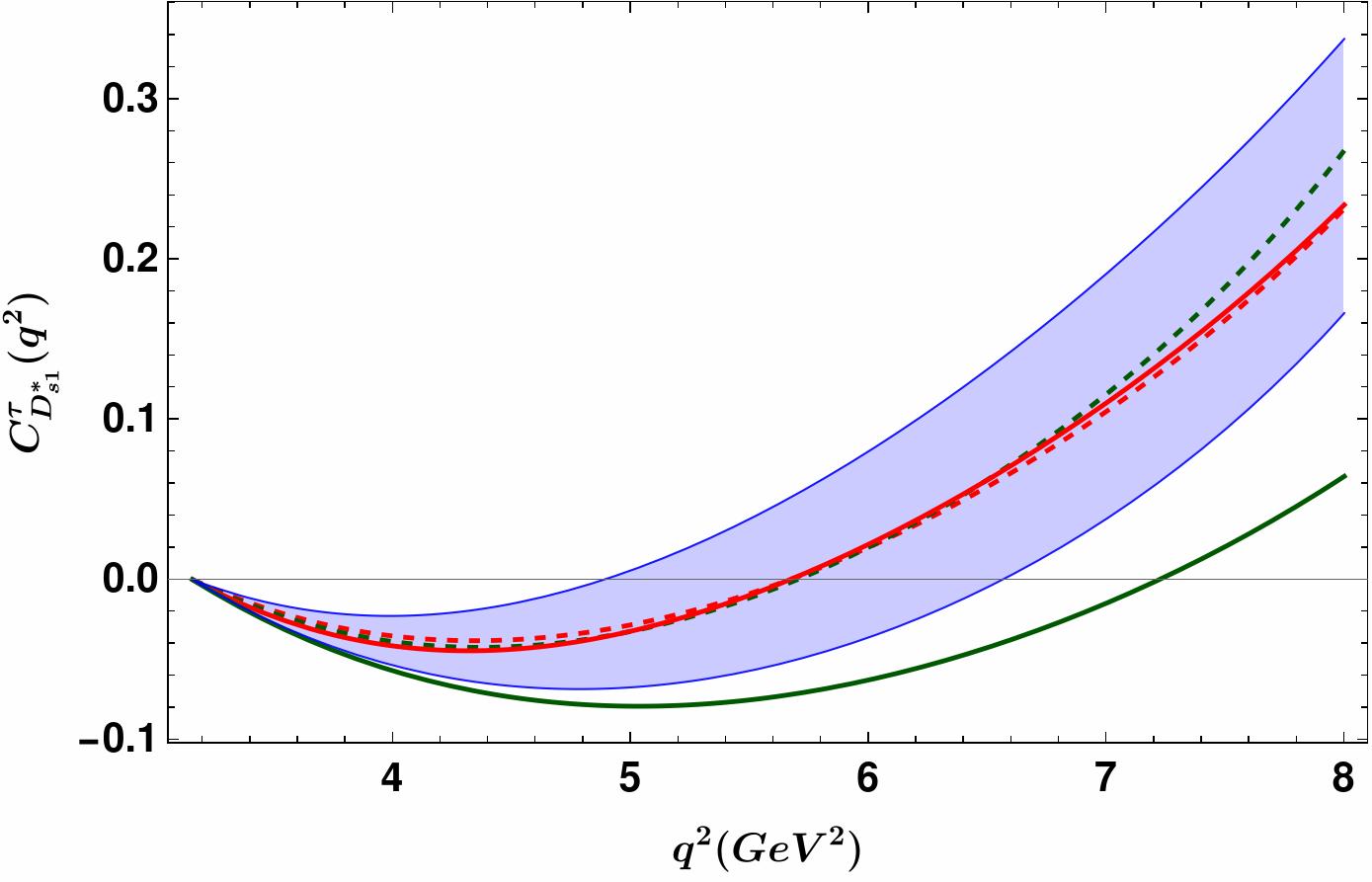}
	\hspace{0.01\textwidth}
	\includegraphics[width=0.31\textwidth]{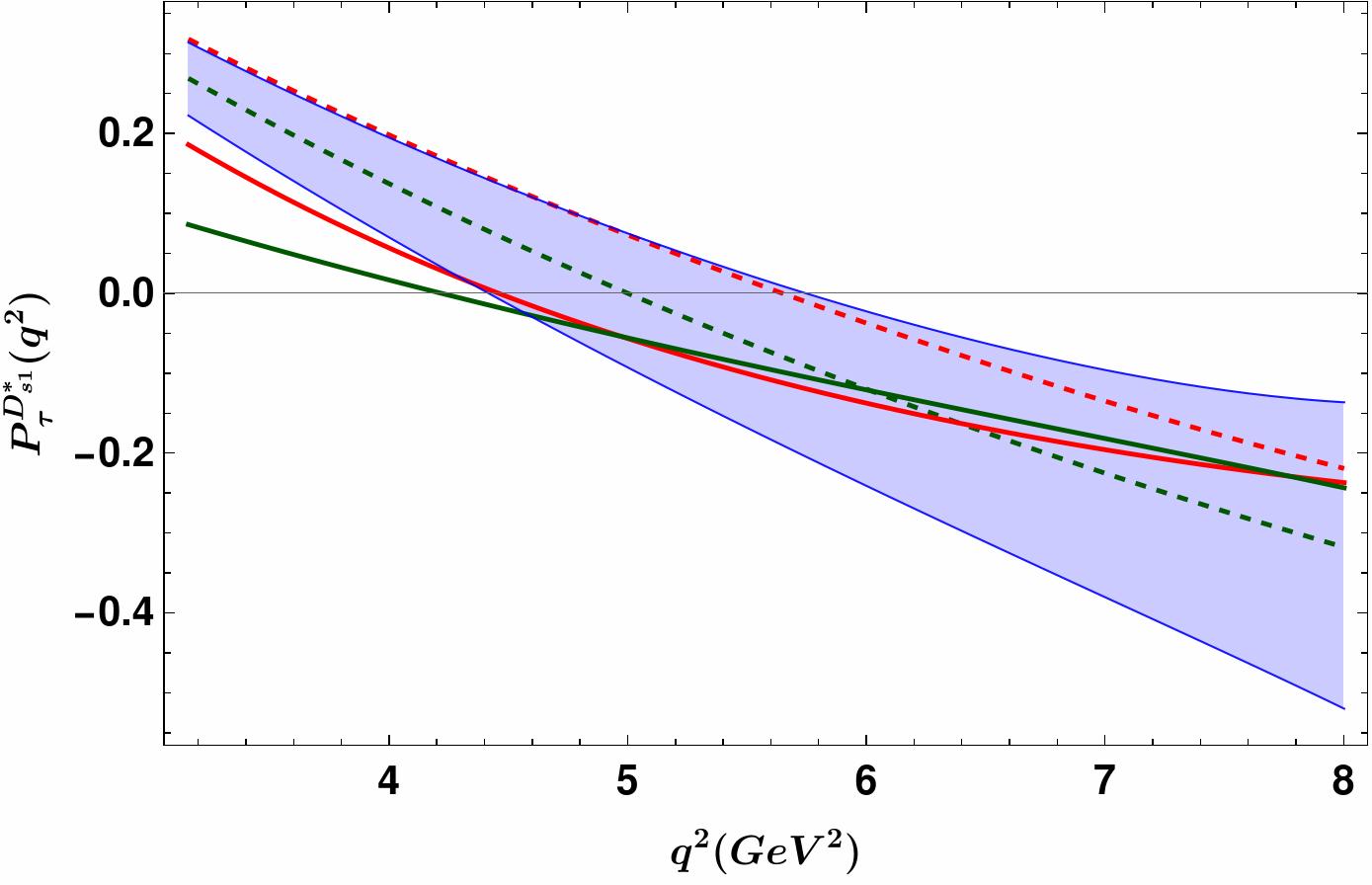}
	\hspace{0.01\textwidth}
	\includegraphics[width=0.31\textwidth]{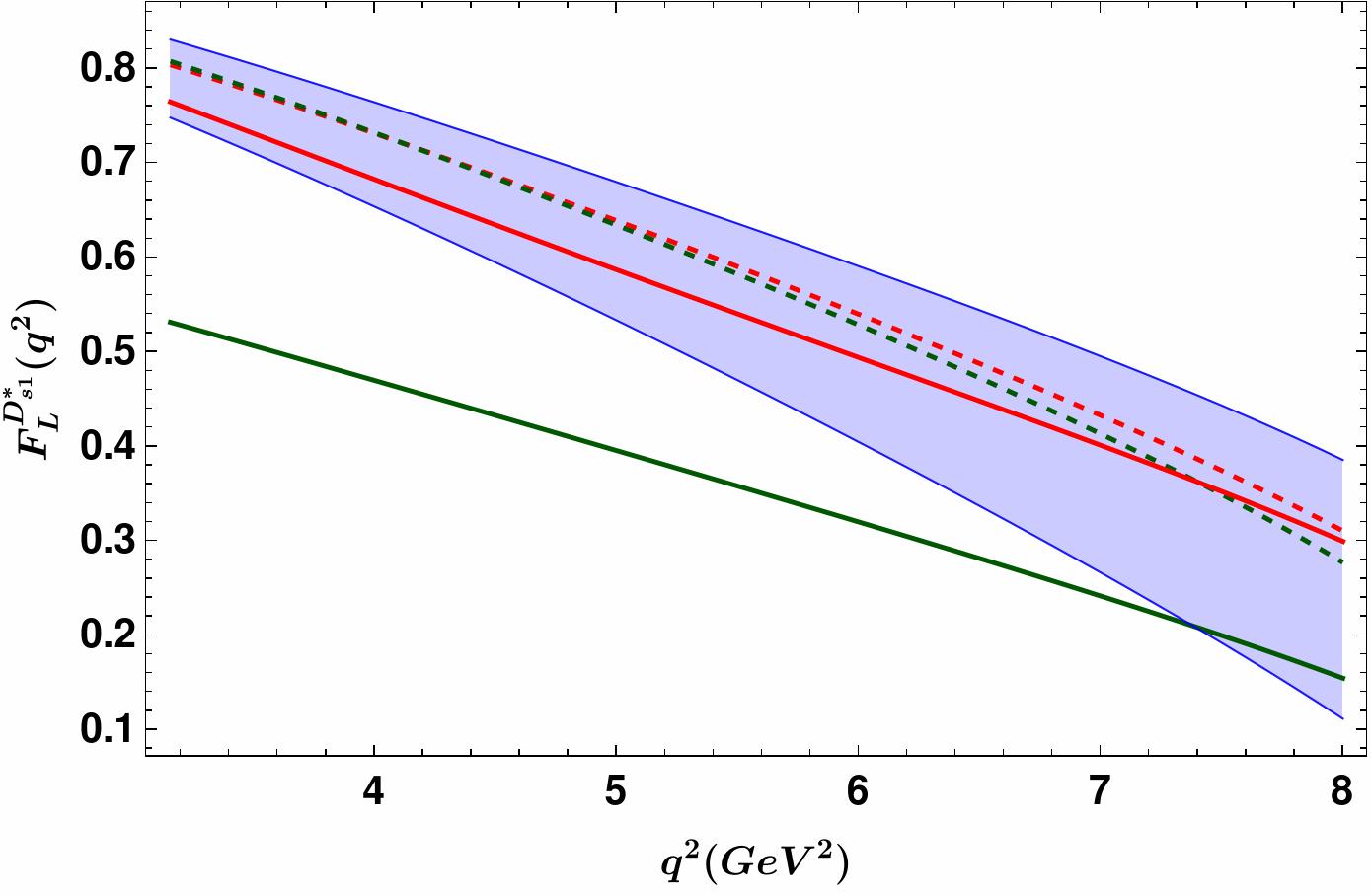}
	
	\vspace{0.3cm}
	
	\includegraphics[scale=0.4]{DsstNPscnSMEFTleg.jpg}
\end{center}
\caption{$q^2$-distribution of $B_s \to D_{s1}^*\tau\nu_\tau$ observables in SMEFT.}\label{fig:comp_smeft_Ds1s}
\end{figure}

\begin{figure}[t]
	\begin{center}
		\includegraphics[width=0.31\textwidth]{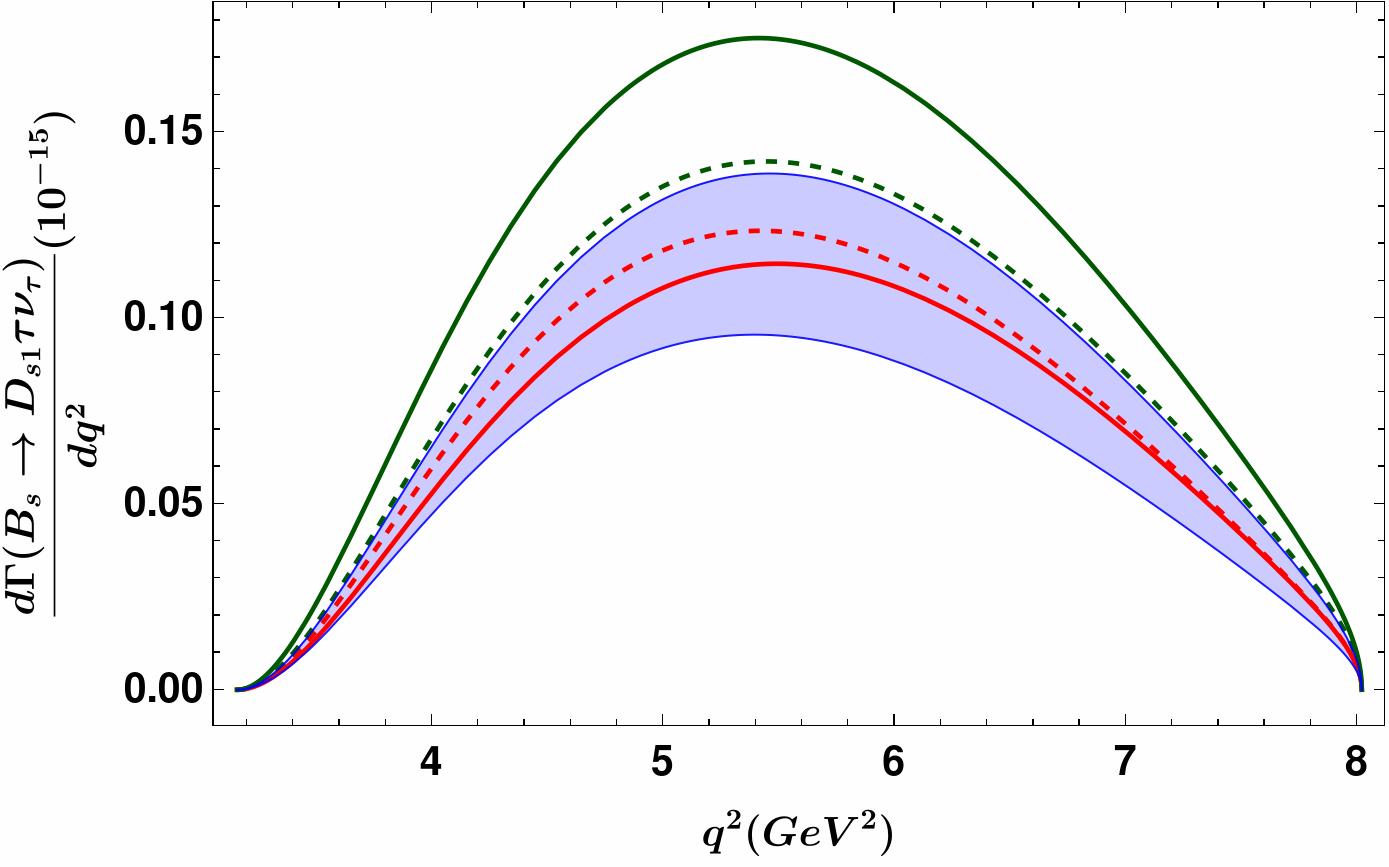}
		\hspace{0.01\textwidth}
		\includegraphics[width=0.31\textwidth]{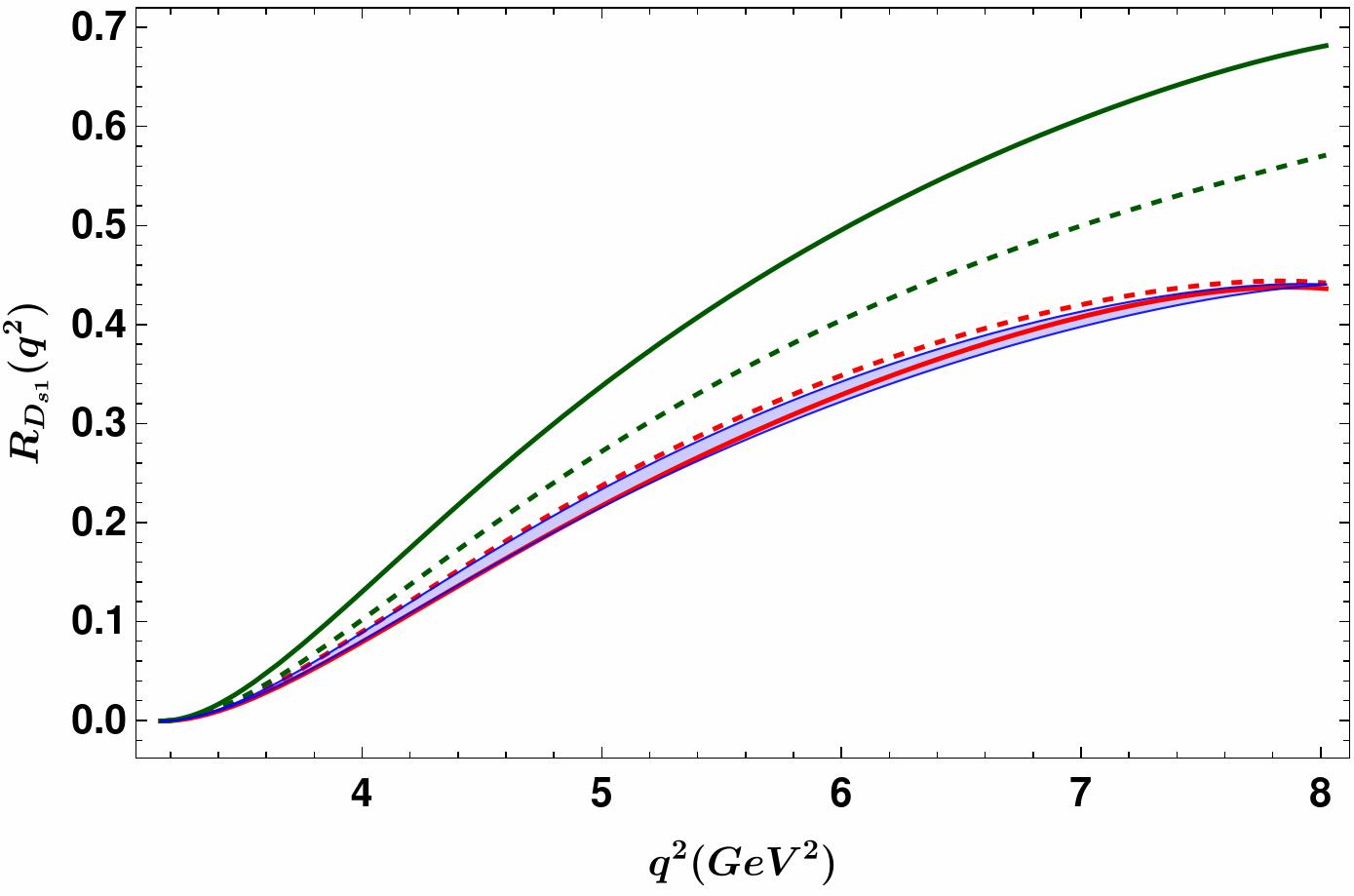}
		\hspace{0.01\textwidth}
		\includegraphics[width=0.31\textwidth]{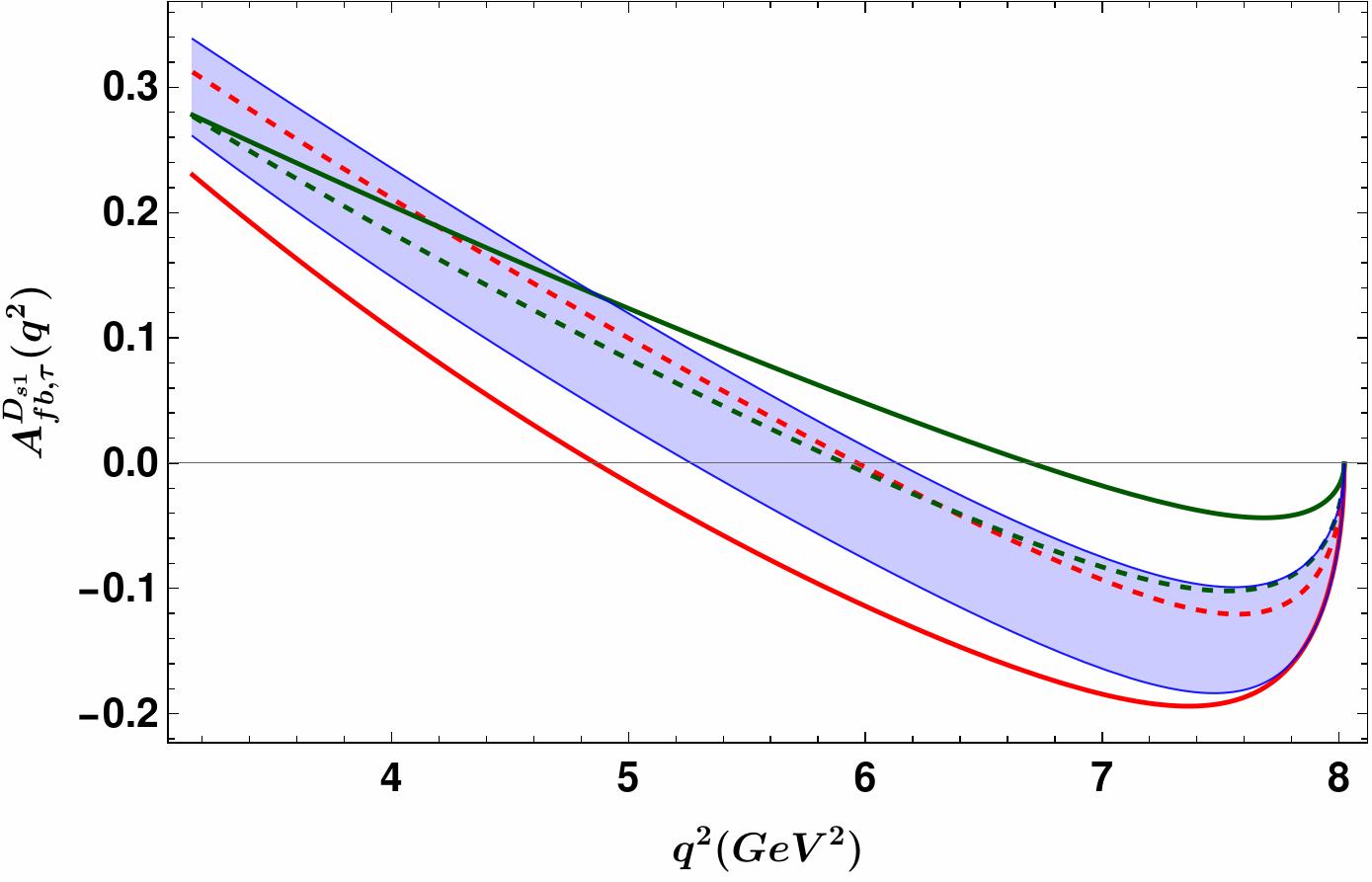}
		
		\vspace{0.3cm}
		
		\includegraphics[width=0.31\textwidth]{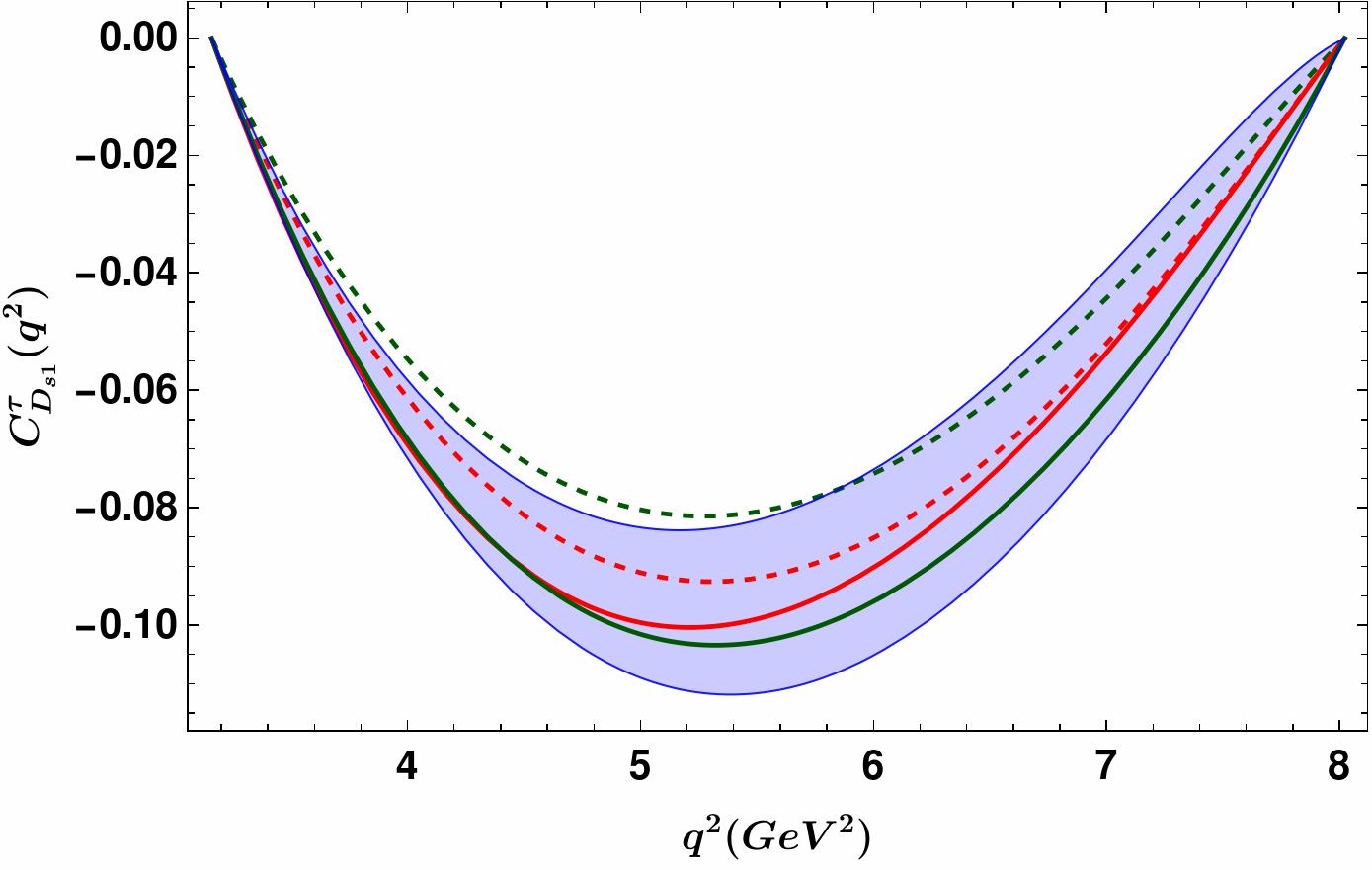}
		\hspace{0.01\textwidth}
		\includegraphics[width=0.31\textwidth]{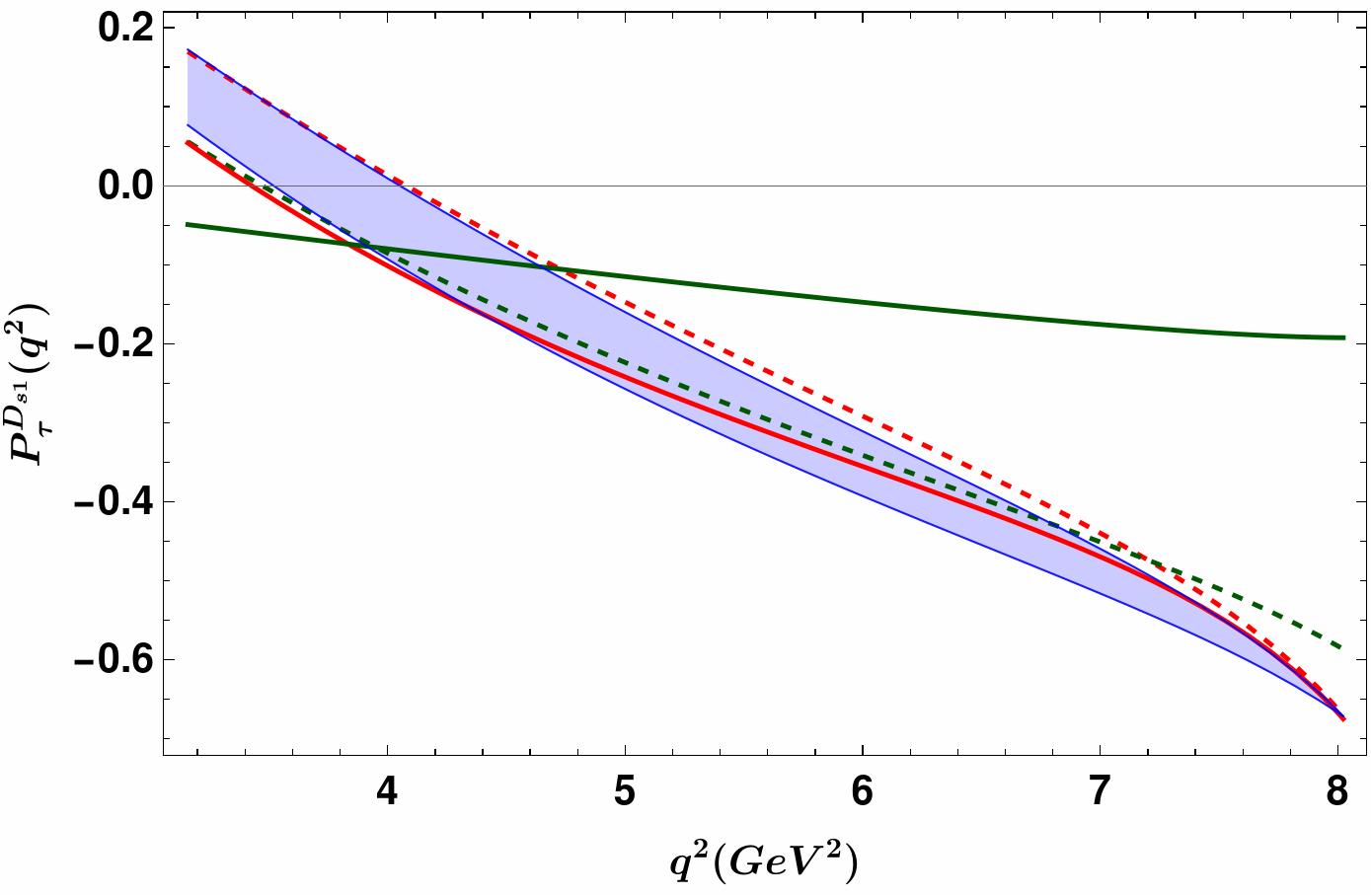}
		\hspace{0.01\textwidth}
		\includegraphics[width=0.31\textwidth]{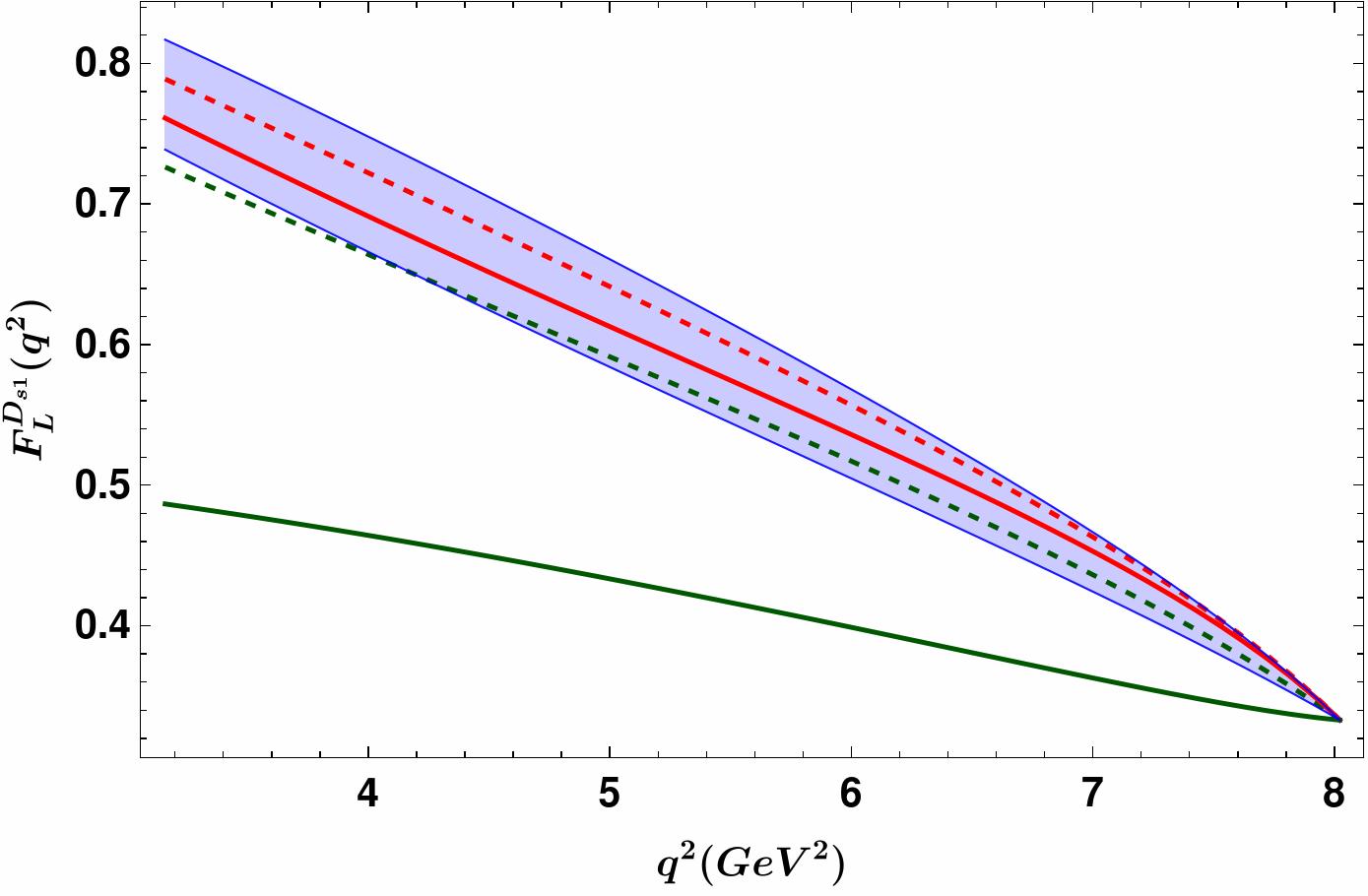}
		
		\vspace{0.3cm}
		
		\includegraphics[scale=0.4]{DsstNPscnSMEFTleg.jpg}
	\end{center}
	\caption{$q^2$-distribution of $B_s \to D_{s1}\tau\nu_\tau$ observables in SMEFT.}\label{fig:comp_smeft_Ds1}
\end{figure}

\begin{figure}[t]
	\begin{center}
		\includegraphics[width=0.31\textwidth]{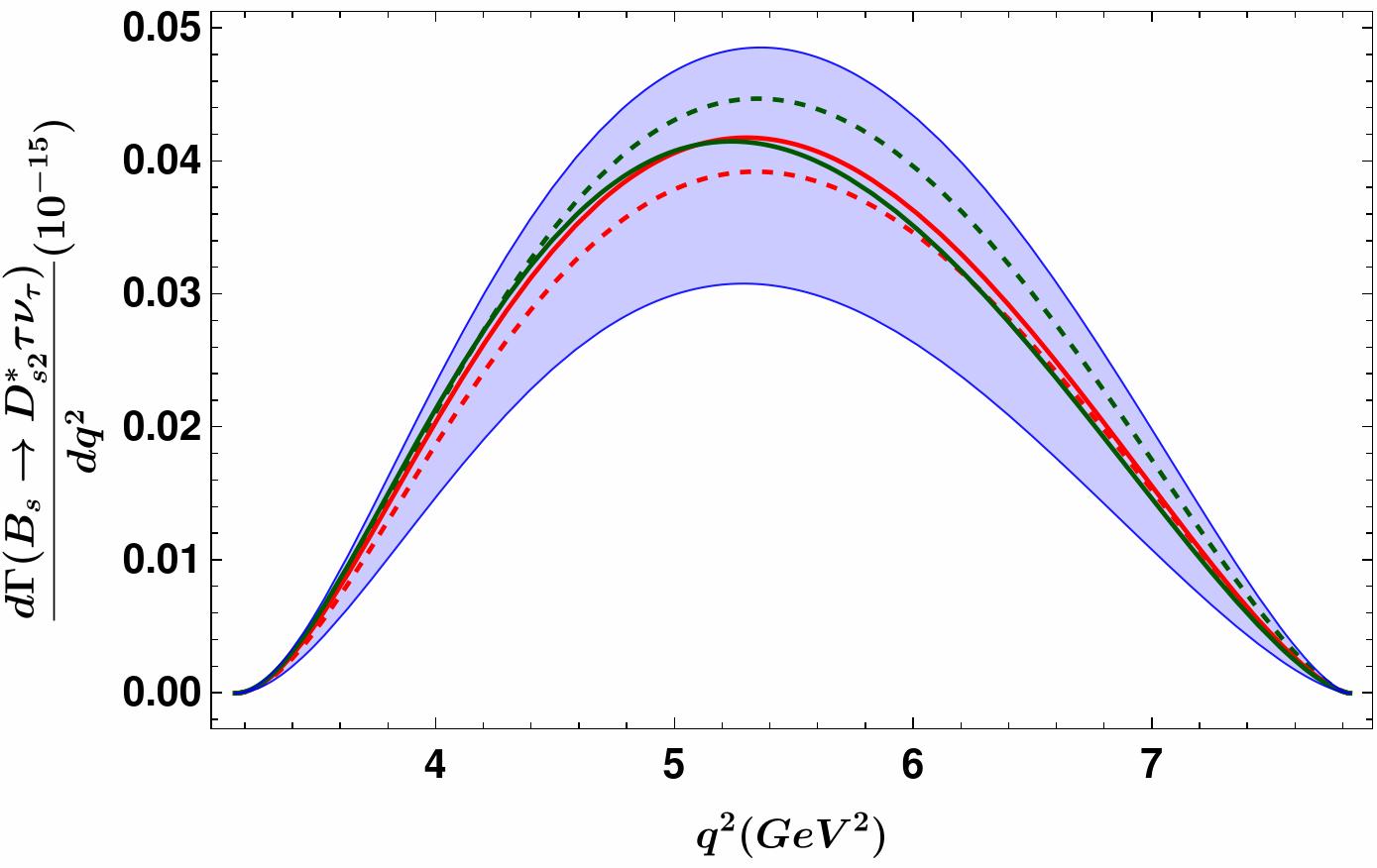}
		\hspace{0.01\textwidth}
		\includegraphics[width=0.31\textwidth]{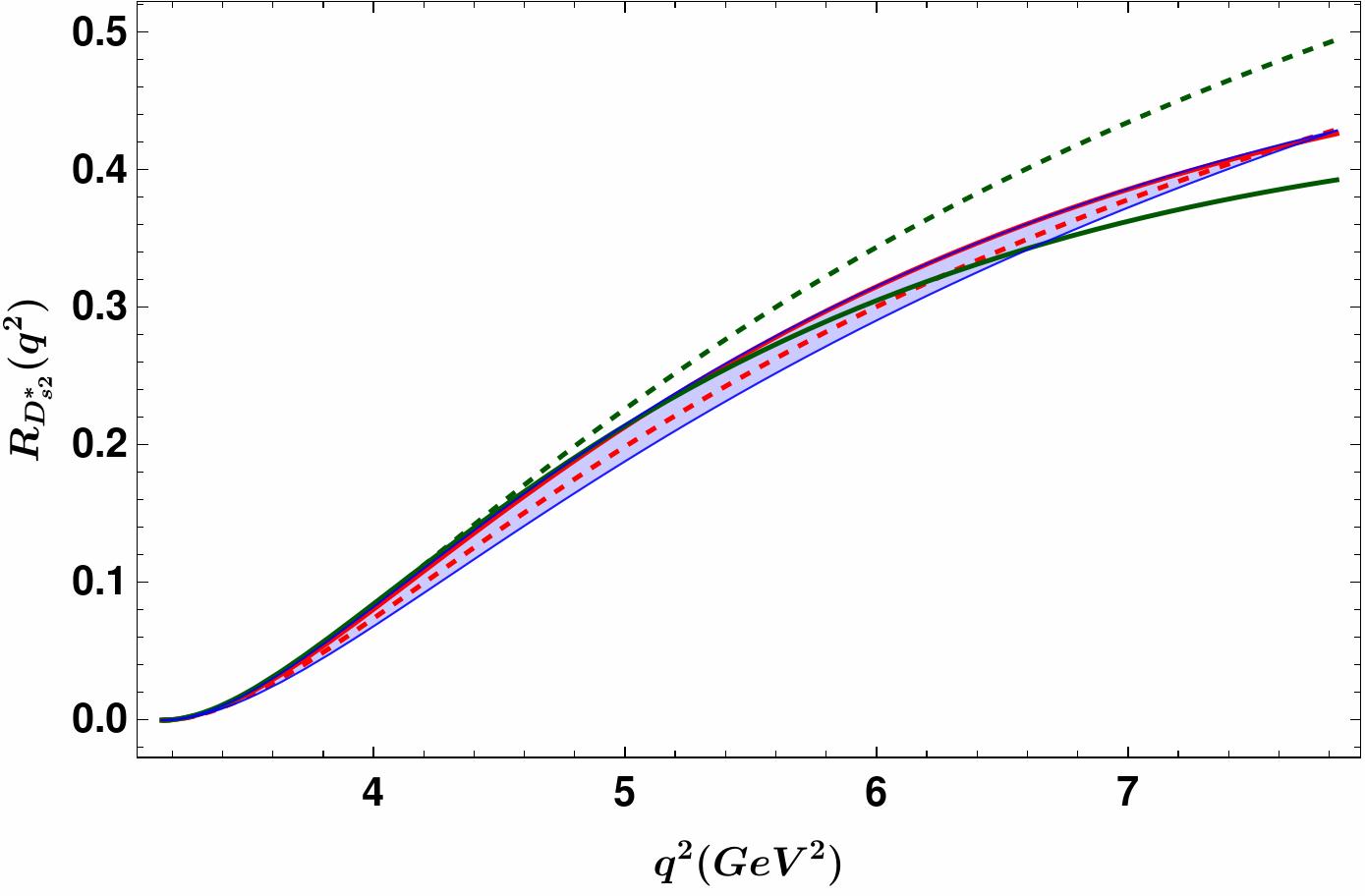}
		\hspace{0.01\textwidth}
		\includegraphics[width=0.31\textwidth]{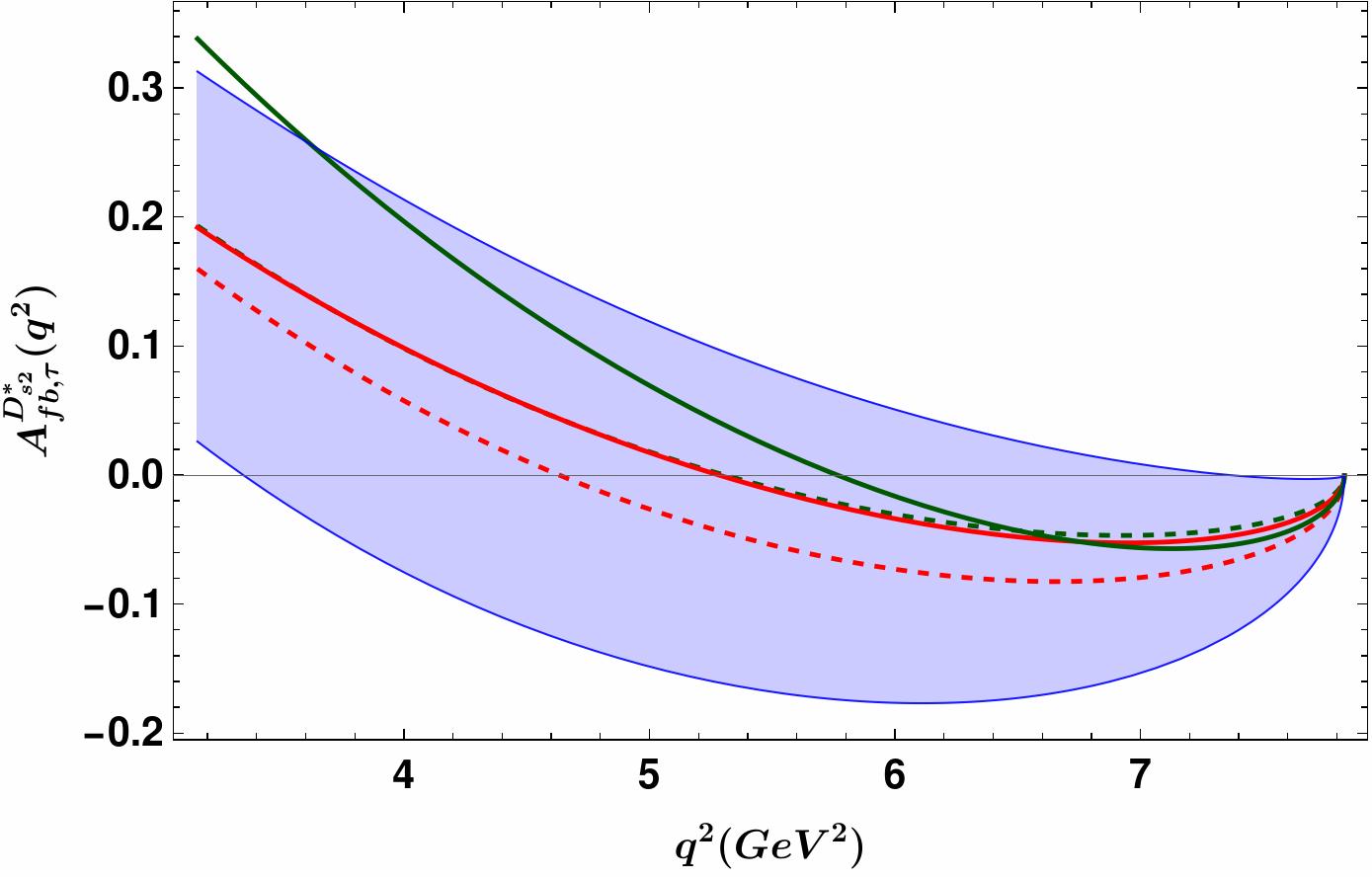}
		
		\vspace{0.3cm}
		
		\includegraphics[width=0.31\textwidth]{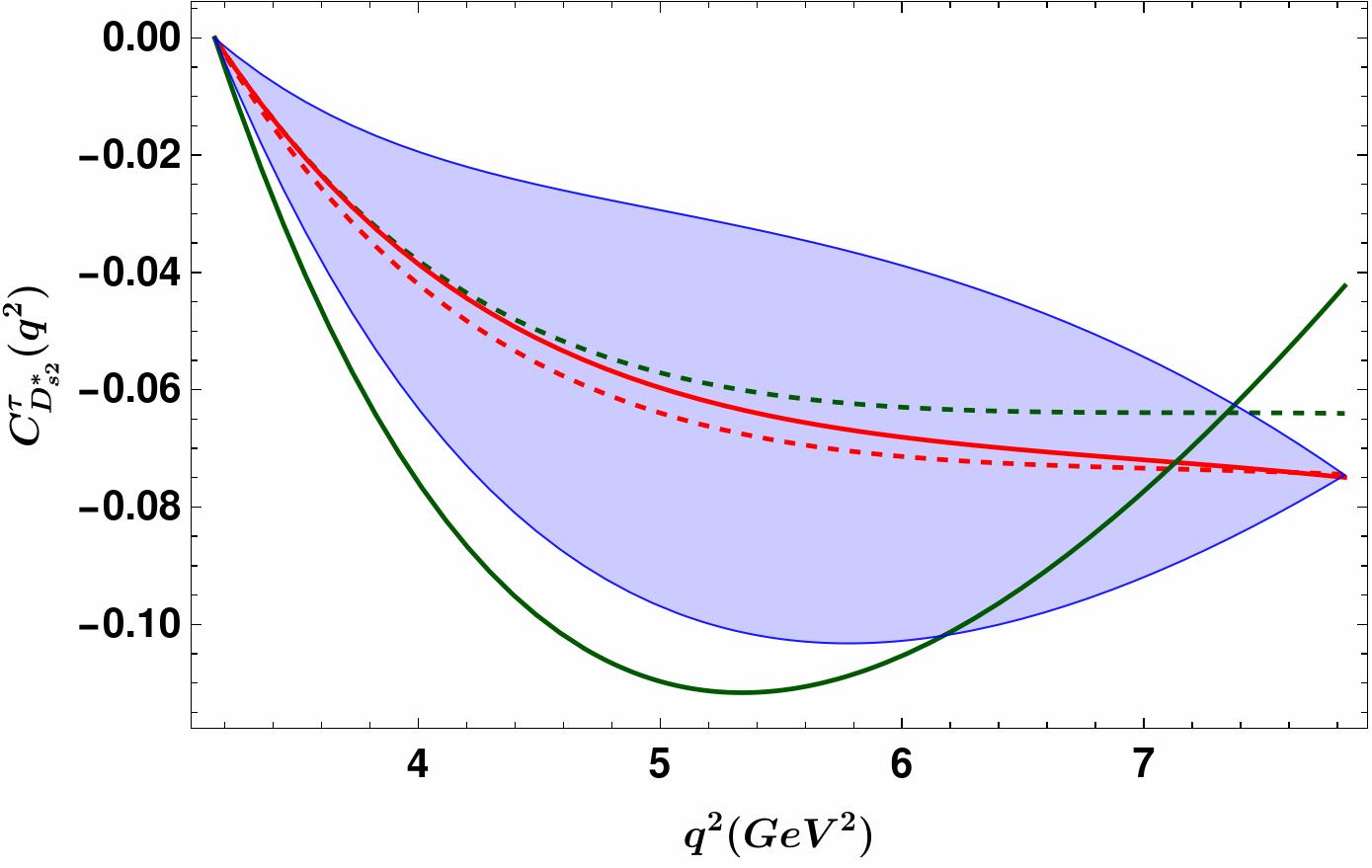}
		\hspace{0.01\textwidth}
		\includegraphics[width=0.31\textwidth]{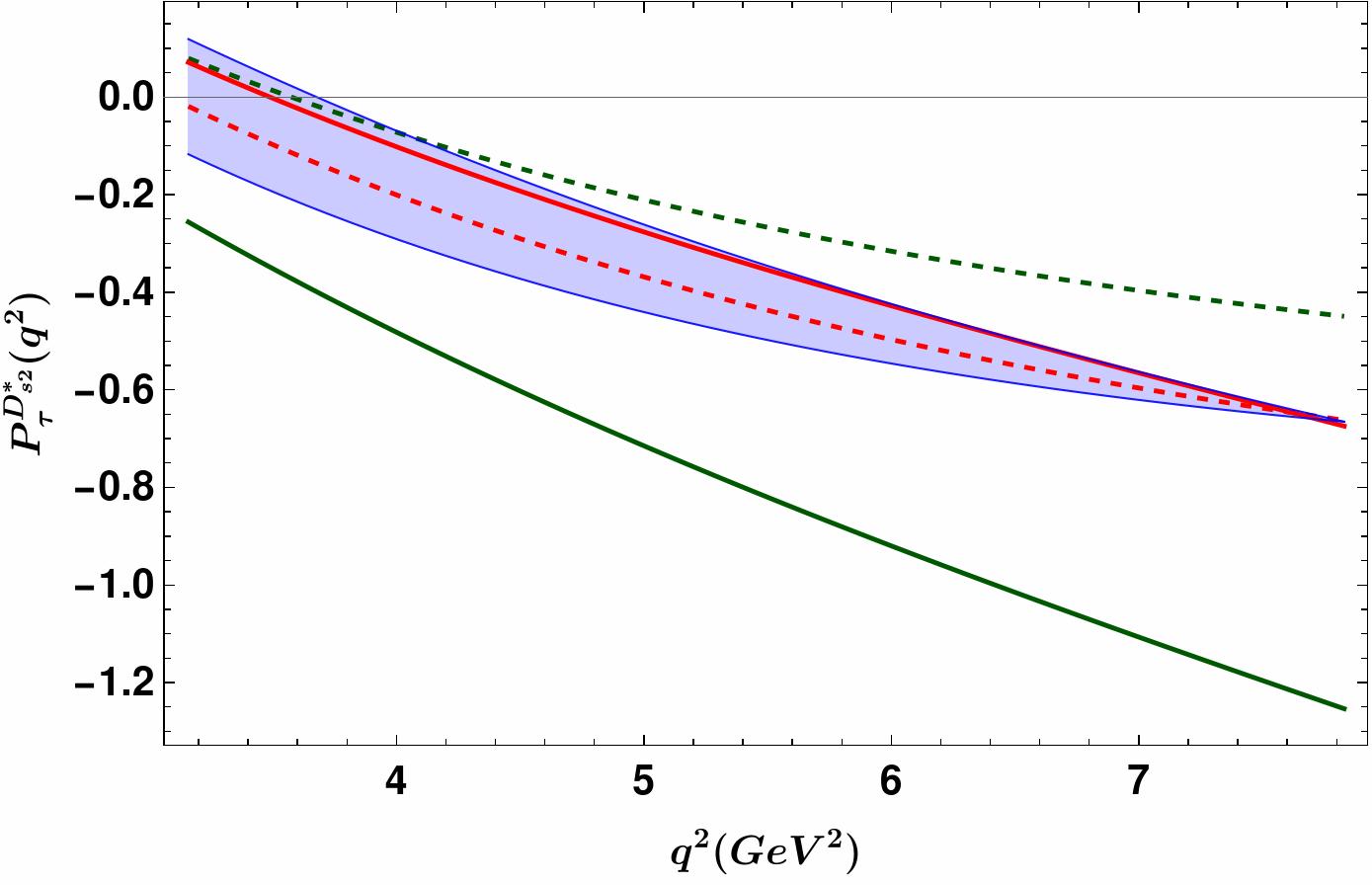}
		\hspace{0.01\textwidth}
		\includegraphics[width=0.31\textwidth]{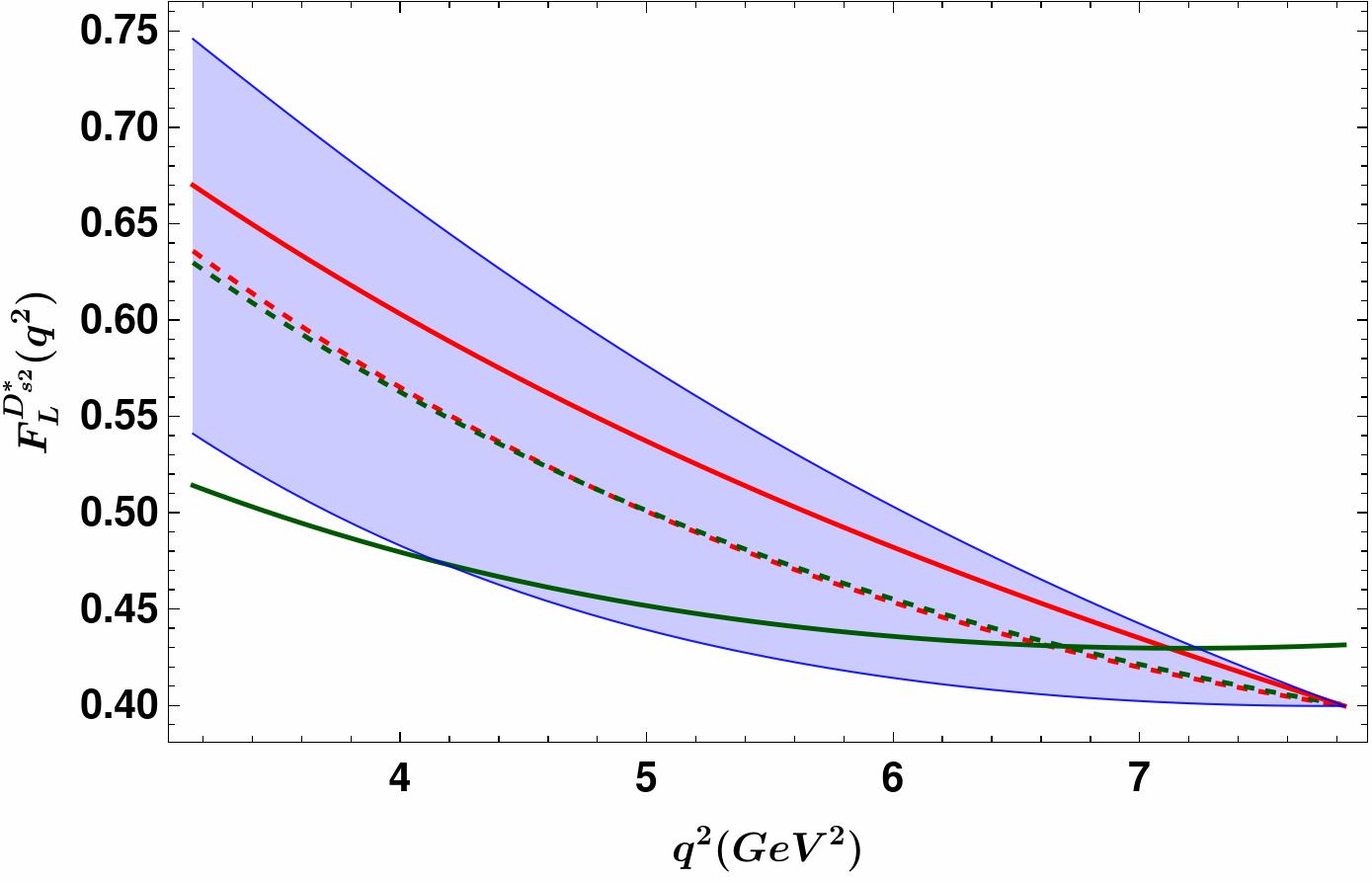}
		
		\vspace{0.3cm}
		
		\includegraphics[scale=0.4]{DsstNPscnSMEFTleg.jpg}
	\end{center}
	\caption{$q^2$-distribution of $B_s \to D_{s2}^*\tau\nu_\tau$ observables in SMEFT.}\label{fig:comp_smeft_Ds2s}
\end{figure}

Next, we test the NP sensitivity of the $B_s \to D_s^{**}\tau \nu_\tau$ observables in the presence of one SMEFT operator at a time. We consider scenarios in which the couplings can be either real or complex. As shown in eq.~\ref{eqn:bc_match_num}, $C_{lq}^{(3)3323}$ and $C_{ledq}^{3332}$ differ from $C_{V_L}$ and $C_{S_R}$ by a multiplicative factor only. Therefore, the one-operator NP scenarios involving $C_{lq}^{(3)3323}$ and $C_{ledq}^{3332}$ are similar to those of $C_{V_1}$ and $C_{S_1}$. Hence, we do not display the results for $C_{lq}^{(3)3323}$ and $C_{ledq}^{3332}$ couplings as this would be repetitive. We probe NP sensitivity only with the $C_{lequ}^{(1)3332}$ and $C_{lequ}^{(3)3332}$ couplings, which are admixtures of the $C_{S_2}$ and $C_T$ couplings. From here on, we drop the generation indices '3332' from $C_{lequ}^{(1)}$ and $C_{lequ}^{(3)}$ for the remainder of the paper. The obtained best-fit values of $C_{lequ}^{(1)}$ and $C_{lequ}^{(3)}$ are presented in table~\ref{tab:real_smeft} and these are used as benchmark values for the NP sensitivity test. The resulting $q^2$-distributions are presented in figures~\ref{fig:comp_smeft_Ds0s}-\ref{fig:comp_smeft_Ds2s}. As in the WET framework, we also compute the tension between the SM and NP predictions using eq.~\ref{eq:tension_formula}. The results are presented in table~\ref{tab:tens_smeft_zeta} for the $B_s \to D_{s0}^*\tau\nu_\tau$ and $B_s \to D_{s1}^*\tau\nu_\tau$ modes, and in table~\ref{tab:tens_smeft_tau} for the $B_s \to D_{s1}\tau\nu_\tau$ and $B_s \to D_{s2}^*\tau\nu_\tau$ modes. From figure~\ref{fig:comp_smeft_Ds0s}, it is seen that most of the observables of $B_s \to D_{s0}^*\tau\nu_\tau$ are sensitive to both complex $C_{lequ}^{(1)}$ and complex $C_{lequ}^{(3)}$. Table~\ref{tab:tens_smeft_zeta} shows that in the presence of $C_{lequ}^{(1)}$, the observable $A^{D_{s0}^*}_{fb,\tau}$ deviates from the SM by more than $2\sigma$. All other observables deviate by less than $2\sigma$. Figures~\ref{fig:comp_smeft_Ds1s}, \ref{fig:comp_smeft_Ds1} and \ref{fig:comp_smeft_Ds2s} respectively demonstrate that the $B_s \to D_{s1}^*\tau\nu_\tau$, $B_s \to D_{s1}\tau\nu_\tau$ and $B_s \to D_{s2}^*\tau\nu_\tau$ observables are most sensitive to complex $C_{lequ}^{(3)}$. For $B_s \to D_{s1}^*\tau\nu_\tau$, the large errors render these observables consistent with the SM at $1\sigma$, as shown in table~\ref{tab:tens_smeft_zeta}. For $B_s \to D_{s1}\tau\nu_\tau$, the tensions are all under $2\sigma$ as seen in table~\ref{tab:tens_smeft_tau}. It is observed that $P_\tau^{D_{s1}}$ has a zero-crossing in the SM, which is absent in the presence of complex $C_{lequ}^{(3)}$. The observables of $B_s \to D_{s2}^*\tau\nu_\tau$ are consistent with the SM predictions within $1\sigma$.

\begin{table}
\begin{center}
	\renewcommand*{\arraystretch}{1.6}
	\resizebox{0.8\textwidth}{!}{
	\begin{tabular}{|c|c|c|c|c|}
		\hline
		   & \multicolumn{2}{c|}{Real} & \multicolumn{2}{c|}{Complex} \\
		\hline
		Coupling & $C_{lequ}^{(1)}$ & $C_{lequ}^{(3)}$  & $C_{lequ}^{(1)}$ & $C_{lequ}^{(3)}$\\
		\hline
		Best fit value & $-0.098(29)$ & $0.036(14)$  & $0.29(25) + 0.54(8)i$ & $-0.09(14) - 0.25(11)$\\
		\hline
		p-value & $0.13$ & $0.04$ & $0.21$ & $0.02$\\
		\hline
	\end{tabular}
  }
\end{center}
\caption{Fit results for one-operator scenario in SMEFT.}\label{tab:real_smeft}
\end{table}			

\begin{table}
	\begin{center}
		\renewcommand*{\arraystretch}{1.6}
		\resizebox{1.02\textwidth}{!}{
			\begin{tabular}{|c|c|c|c|c|c|c|c|c|c|c|c|c|}
				\hline
				& & \multicolumn{5}{c|}{$ B_s \to D_{s0}^*\tau\nu_\tau$} & \multicolumn{6}{c|}{$ B_s \to D_{s1}^*\tau\nu_\tau$} \\
				\hline
				& WCs & $\Gamma$ & $\langle R_{D_{s0}^*} \rangle$ & $\langle A^{D_{s0}^*}_{fb,\tau} \rangle$ & $\langle C^\tau_{D_{s0}^*} \rangle$ & $\langle P^{D_{s0}^*}_\tau \rangle$ & $\Gamma$ & $\langle R_{D_{s1}^*} \rangle$ & $\langle A^{D_{s1}^*}_{fb,\tau} \rangle$ & $\langle C^\tau_{D_{s1}^*} \rangle$ & $\langle P^{D_{s1}^*}_\tau \rangle$ & $\langle F_L^{D_{s1}^*} \rangle$ \\
				\hline
				Real & $C_{lequ}^{(1)}$ & 0.13 & 0.40 & 0.50 & 0.55 & 0.61 & 0.36 & 0.52 & 0.24 & 0.01 & 0.60 & 0.30 \\
				& $C_{lequ}^{(3)}$ & 0.08 & 0.22 & 0.77 & 0.82 & 0.20 & 0.04 & 0.05 & 0.04 & 0.03 & 0.08 & 0.25\\
				\hline
				Complex & $C_{lequ}^{(1)}$ & 0.59 & 1.09 & 2.27 & 1.29 & 1.35 & 0.11 & 0.15 & 0.69 & 0.0 & 0.19 & 0.11\\
				& $C_{lequ}^{(3)}$ & 0.05 & 0.15 & 0.80 & 0.26 & 0.32 & 0.77 & 0.81 & 0.79 & 0.79 & 0.22 & 0.92\\
				\hline
			\end{tabular}
		}
	\end{center}
\caption{Computed tensions for observables of $B_s \to D_{s0}^*\tau\nu_\tau$ and $B_s \to D_{s1}^*\tau\nu_\tau$ within SMEFT (in units of $\sigma$). }\label{tab:tens_smeft_zeta}
\end{table}

\begin{table}
	\begin{center}
		\renewcommand*{\arraystretch}{1.6}
		\resizebox{1.02\textwidth}{!}{
			\begin{tabular}{|c|c|c|c|c|c|c|c|c|c|c|c|c|c|}
				\hline
				& & \multicolumn{6}{c|}{$ B_s \to D_{s1}\tau\nu_\tau$} & \multicolumn{6}{c|}{$ B_s \to D_{s2}^*\tau\nu_\tau$} \\
				\hline
				& WCs & $\Gamma$ & $\langle R_{D_{s1}} \rangle$ & $\langle A^{D_{s1}}_{fb,\tau} \rangle$ & $\langle C^\tau_{D_{s1}} \rangle$ & $\langle P^{D_{s1}}_\tau \rangle$ & $\langle F_L^{D_{s1}} \rangle$ & $\Gamma$ & $\langle R_{D_{s2}^*} \rangle$ & $\langle A^{D_{s2}^*}_{fb,\tau} \rangle$ & $\langle C^\tau_{D_{s2}^*} \rangle$ & $\langle P^{D_{s2}^*}_\tau \rangle$ & $\langle F_L^{D_{s2}^*} \rangle$ \\
				\hline
				Real & $C_{lequ}^{(1)}$ & 0.18 & 0.75 & 0.42 & 0.21 & 0.88 & 0.40 & 0.03 & 0.06 & 0.06 & 0.02 & 0.13 & 0.08\\
				& $C_{lequ}^{(3)}$ & 0.78 & 1.95 & 0.26 & 0.72 & 0.00 & 0.52 & 0.39 & 0.65 & 0.18 & 0.16 & 1.18 & 0.07\\
				\hline
				Complex & $C_{lequ}^{(1)}$ & 0.08 & 0.27 & 0.85 & 0.10 & 0.23 & 0.11 & 0.15 & 0.25 & 0.17 & 0.07 & 0.51 & 0.31\\
				& $C_{lequ}^{(3)}$ & 1.10 & 1.32 & 0.84 & 0.24 & 1.01 & 1.19 & 0.12 & 0.19 & 0.41 & 0.53 & 0.59 & 0.45\\
				\hline
			\end{tabular}
		}
	\end{center}
\caption{Computed tensions for observables of $B_s \to D_{s1}\tau\nu_\tau$ and $B_s \to D_{s2}^*\tau\nu_\tau$ within SMEFT (in units of $\sigma$).}\label{tab:tens_smeft_tau}
\end{table}

\subsection{2HDM Framework}
In this section, we analyze NP sensitivity of the  $B_s \to D_s^{**}\tau \nu_\tau$ modes within the general 2HDM. The most general gauge-invariant scalar potential with two Higgs doublets is given by \cite{Athron:2024rir}
\begin{eqnarray}
V(\Phi_1,\Phi_2) &=& m_{11}^2 (\Phi_1^\dagger \Phi_1) + m_{22}^2 (\Phi_2^\dagger \Phi_2) - m_{12}^2 (\Phi_1^\dagger \Phi_2 + \Phi_2^\dagger \Phi_1) + \frac{1}{2} \lambda_1 (\Phi_1^\dagger \Phi_1)^2 + \frac{1}{2} \lambda_2 (\Phi_2^\dagger \Phi_2)^2 \nonumber \\ && + \lambda_3 (\Phi_1^\dagger \Phi_1)(\Phi_2^\dagger \Phi_2) + \lambda_4 (\Phi_1^\dagger \Phi_2)(\Phi_2^\dagger \Phi_1) + \bigg( \frac{1}{2} \lambda_5 (\Phi_1^\dagger \Phi_2)^2 + \big(\lambda_6 (\Phi_1^\dagger \Phi_1) \nonumber \\ && + \lambda_7 (\Phi_2^\dagger \Phi_2)\big) (\Phi_1^\dagger \Phi_2) + \rm{h.c.} \bigg)\, .
\end{eqnarray} 
The Higgs doublets are defined as
\begin{eqnarray}
\Phi_i = \begin{pmatrix}
	\phi_i^+ \\ \frac{1}{\sqrt{2}} (v_i + \rho_i + i\eta_i) 
\end{pmatrix}\, ,~~ i = 1,2~\, ,
\end{eqnarray}
where $v_i$ are the vacuum expectation values (vevs) of the two Higgs fields. They satisfy $v_1^2 + v_2^2 = v^2$, with $v$ being the SM vev. The linear combinations of the fields $\rho_i$, $\eta_i$ and $\phi^\pm_i$ form mass eigenstates given by
\begin{eqnarray}
\begin{pmatrix}
	H \\ h 
\end{pmatrix} = R_\alpha \begin{pmatrix}
	\rho_1 \\ \rho_2 \end{pmatrix}
~\, ,~
\begin{pmatrix}
	G_Z \\ A
\end{pmatrix} = R_\beta \begin{pmatrix}
	\eta_1 \\ \eta_2
\end{pmatrix}
~\, ,~
\begin{pmatrix}
	G_{W^\pm} \\ H^\pm 
\end{pmatrix} = R_\beta \begin{pmatrix}
	\phi_1^\pm \\ \phi_2^\pm 
\end{pmatrix}\, ,
\end{eqnarray}
where $\phi^-_i$ is the complex conjugate of $\phi^+_i$. In the above, $h$, $H$, $A$, and $H^\pm$ denote the Higgs bosons, and $G_{W^\pm}$ and $G_Z$ denote the Goldstone bosons that become the polarization states of the $W^\pm$ and $Z$ bosons, respectively. The rotation matrices $R_{\alpha(\beta)}$ have the form
\begin{eqnarray}
R_\theta = \begin{pmatrix}
	\cos\theta & \sin\theta \\ -\sin\theta & \cos\theta
\end{pmatrix}\, ,
\end{eqnarray}
with $\theta = \alpha,\beta$. The angle $\alpha$ is the mixing angle for the CP-even states $\rho_i$, while $\beta$ is the mixing angle for the CP-odd states $\eta_i$ and $\phi_i$. The angle $\beta$ is determined from the relation $\tan\beta = v_2/v_1$.

The most general Yukawa Lagrangian in the $\{\Phi_1,\Phi_2\}$ basis is given by 
\begin{align}
-\mathcal{L}_{\rm{Yukawa}} = \bar{Q}^0 (Y_u^1 \bar{\Phi}_1 + Y_u^2 \bar{\Phi}_2) u^0_R + \bar{Q}^0 (Y_d^1 \Phi_1 + Y_d^2 \Phi_2) d^0_R + \bar{L}^0 (Y_\ell^1 \Phi_1 + Y_\ell^2 \Phi_2) \ell^0_R + \text{h.c}.\, ,
\end{align}
where $\bar{\Phi}_j = i\sigma_2\Phi_j^*$ and $Y_f$ ($f = u,d,\ell$) denote the Yukawa couplings. The doublets $Q^0$ and $L^0$ represent left-handed quark and lepton doublets, respectively, in the flavor basis. The singlets $u_R^0$, $d_R^0$ and $\ell_R^0$ are right-handed up-type, down-type and lepton singlets, respectively, in the flavor basis. The fermion mass matrix is obtained as 
\begin{eqnarray}
M_f = \frac{1}{2} (v_1 Y_f^1 + v_2 Y_f^2)\, . 
\end{eqnarray}
Solving for $Y_f^1$, we obtain the relation
\begin{eqnarray}
Y^{1,ij}_f = \frac{\sqrt{2}}{v\cos\beta} M_f^{ij} - \tan \beta~Y_f^{2,ij}\, .
\end{eqnarray}
In the mass basis, the Yukawa Lagrangian becomes 
\begin{eqnarray}
-\mathcal{L}_{\rm{Yukawa}} &=& \bar{u}_i (V_{ik}\rho_d^{kj}P_R - V_{kj}\rho^{ki*}_u P_L)d_j H^+ + \bar{\nu}_i\rho^{ij}_\ell P_R \ell_j H^+ + \text{h.c}. \, ,
\end{eqnarray} 
with
\begin{eqnarray}
\rho^{ij}_f \equiv \frac{Y_f^{2,ij}}{\cos\beta} - \frac{\sqrt{2}\tan\beta\overline{M}_f^{ij}}{v}\, .
\end{eqnarray} 
For $b\to c\ell\nu_\ell$ decays, we have
\begin{eqnarray}
-\mathcal{L}_{\rm{Yukawa}} &=& \bar{c} (V_{ci}\rho_d^{i3}P_R - V_{ib}\rho^{i2*}_u P_L)b H^+ + \bar{\nu}\rho^{33}_\ell P_R \tau H^+ + {\rm{h.c.}}\, . 
\end{eqnarray}
On matching the above Lagrangian with the WET Lagrangian of eq. {\ref{Leff}}, the matching relations between the respective couplings are obtained as
\begin{eqnarray}
C_{S_1} = -F_{\rm{RG}}\frac{\sqrt{2}}{4 G_F V_{cb}} \frac{V_{cs} \rho_d^{23}  \rho_\ell^{33*} }{m_{H^+}^2}~~\, ,~~~~~ C_{S_2} = F_{\rm{RG}}\frac{\sqrt{2}}{4 G_F V_{cb}} \frac{V_{tb} \rho_u^{32*} \rho_\ell^{33*}}{m_{H^+}^2}\, , 
\end{eqnarray}
where $F_{\rm{RG}} \approx 1.5$ is the RGE running factor to match the 2HDM scale with the $m_b$ scale \cite{Crivellin:2023sig}.
Using $m_{H^+} =  130 $ GeV \cite{Athron:2024rir}, we obtain
\begin{eqnarray}\label{eqn:2HDM_match}
C_{S_1} = -63.844\rho^{23}_d\rho^{33*}_\ell ~~\, ,~~~~~C_{S_2} = 66.15\rho^{32*}_u\rho^{33*}_\ell\, . \label{eqn:cslrrho}
\end{eqnarray}

\begin{table}
\begin{center}
	\renewcommand*{\arraystretch}{1.2}
	\resizebox{0.6\textwidth}{!}{
	\begin{tabular}{|c|c|}
	 	\hline
		Fit result & $\rho_d^{23}\rho^{33*}_\ell = 0.007(3)$, $\rho_u^{32}\rho^{33*}_\ell = -0.014(2)$ \\
		\hline
		p-value & $0.22$ \\
		\hline
	\end{tabular}
}
\end{center}
\caption{Fit results for 2HDM couplings.}\label{tab:2HDM}
\end{table}	

\begin{figure}
\begin{center}
	\includegraphics[width=0.31\textwidth]{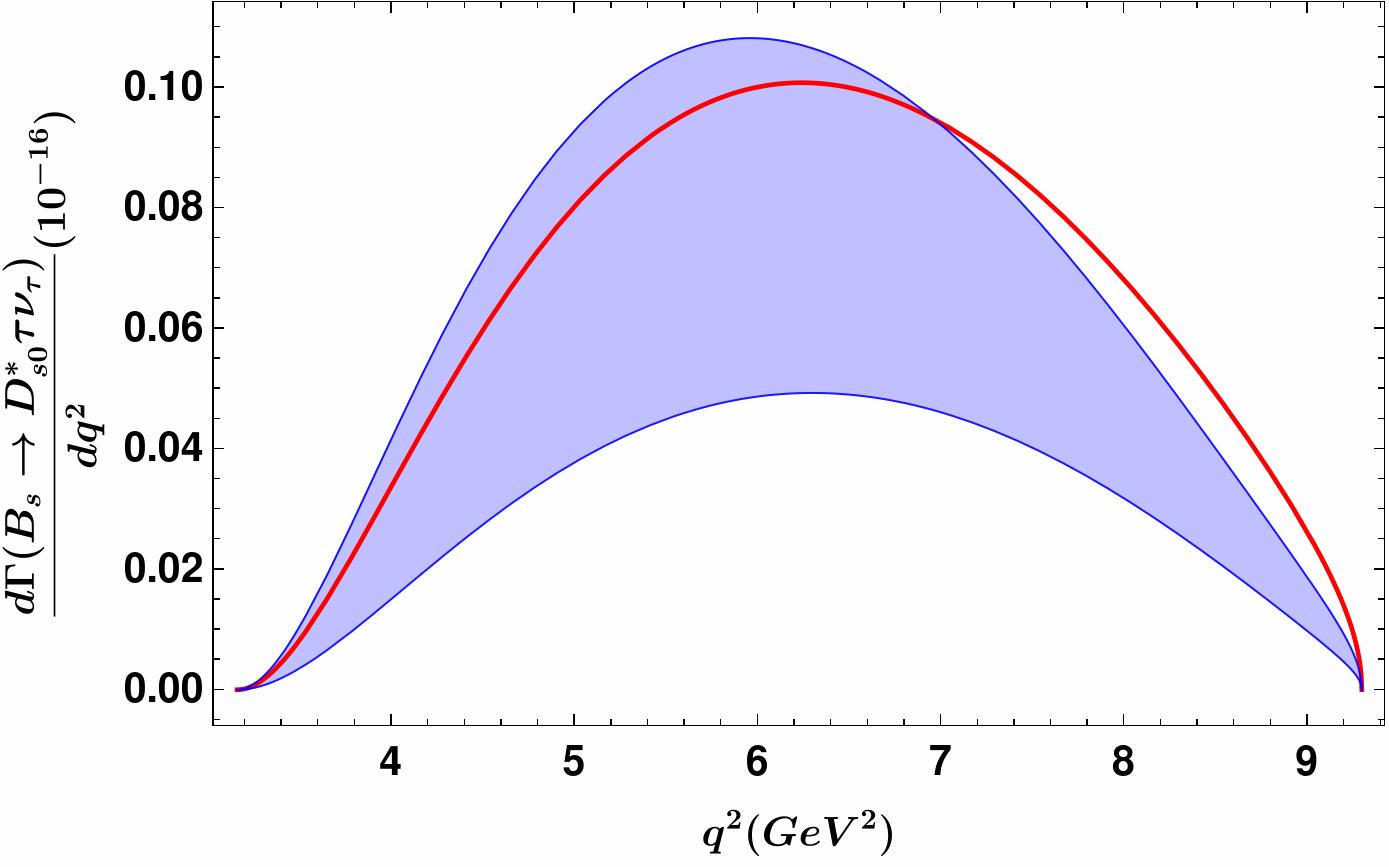}
	\hspace{0.01\textwidth}
	\includegraphics[width=0.31\textwidth]{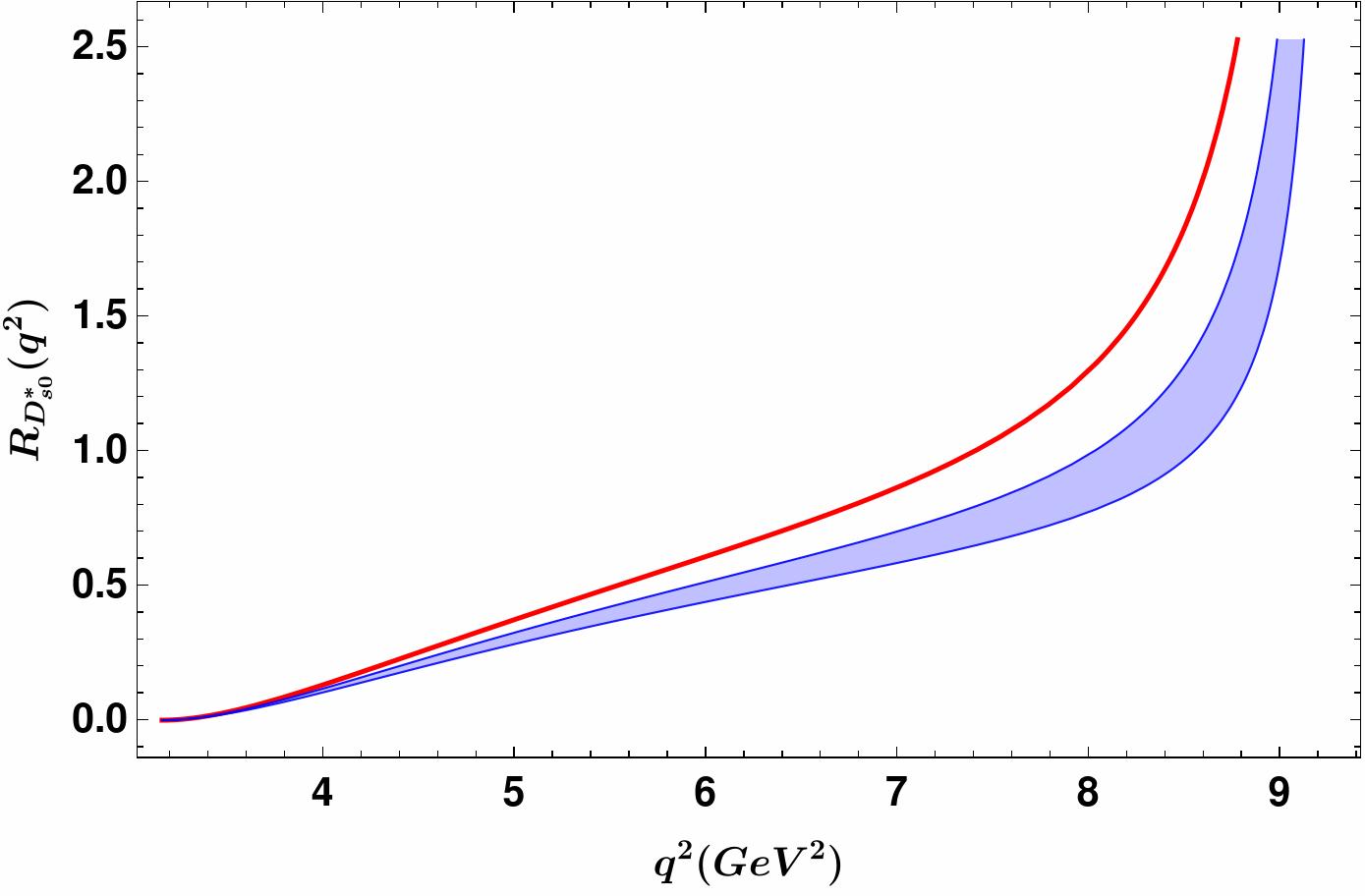}
	\hspace{0.01\textwidth}
	\includegraphics[width=0.31\textwidth]{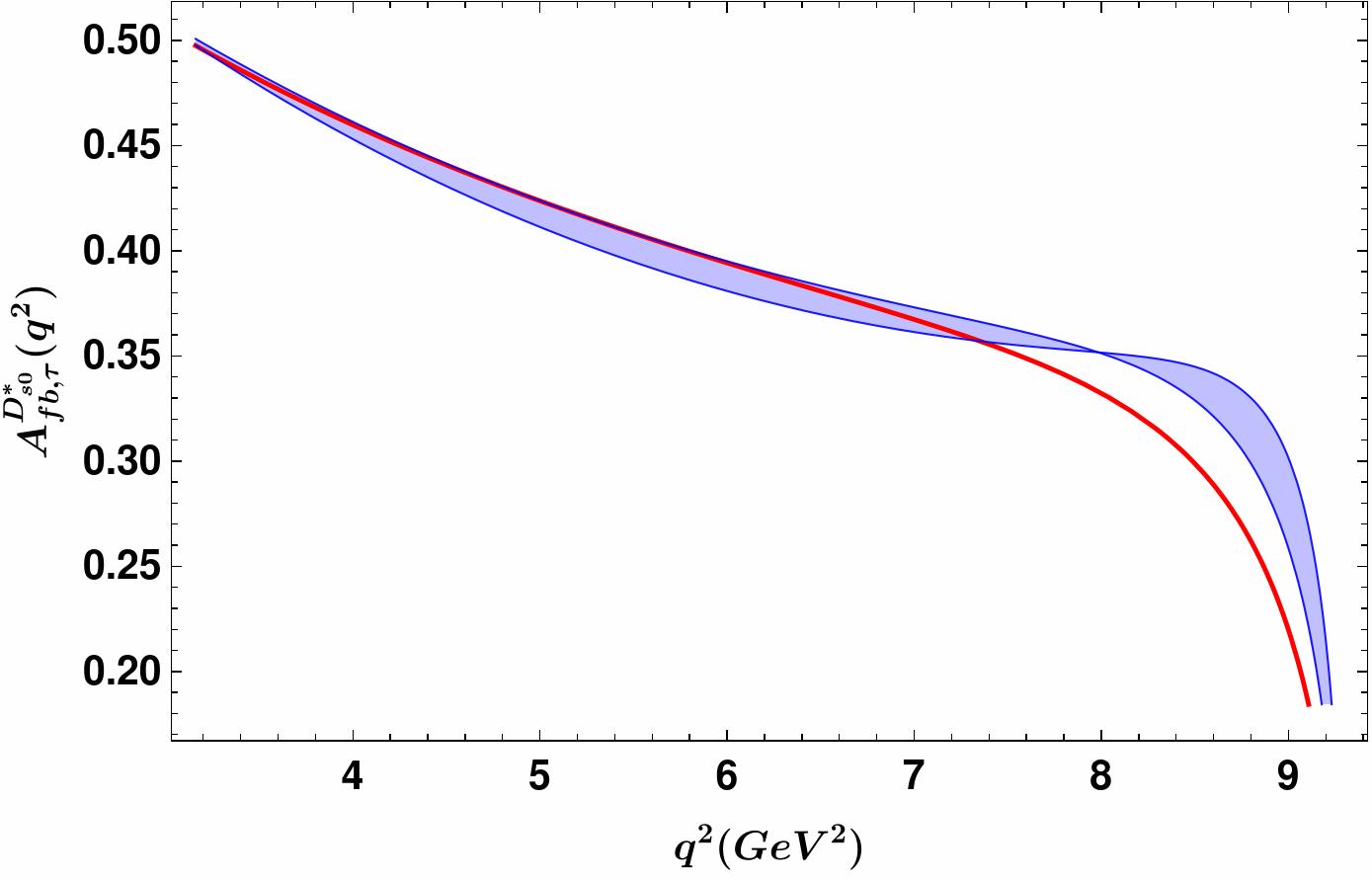}
	
	\vspace{0.3cm}
	
	\includegraphics[width=0.31\textwidth]{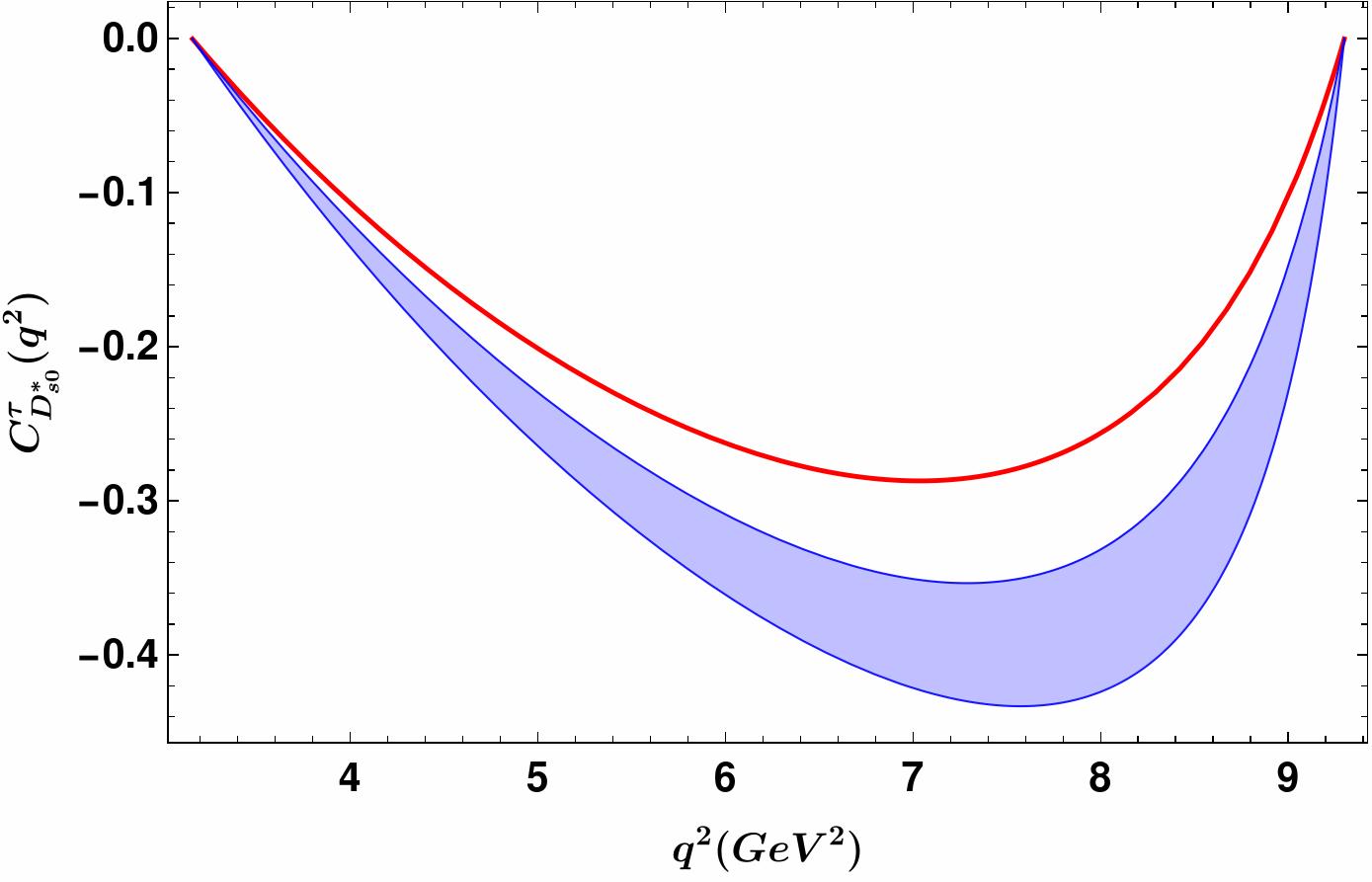}
	\hspace{0.01\textwidth}
	\includegraphics[width=0.31\textwidth]{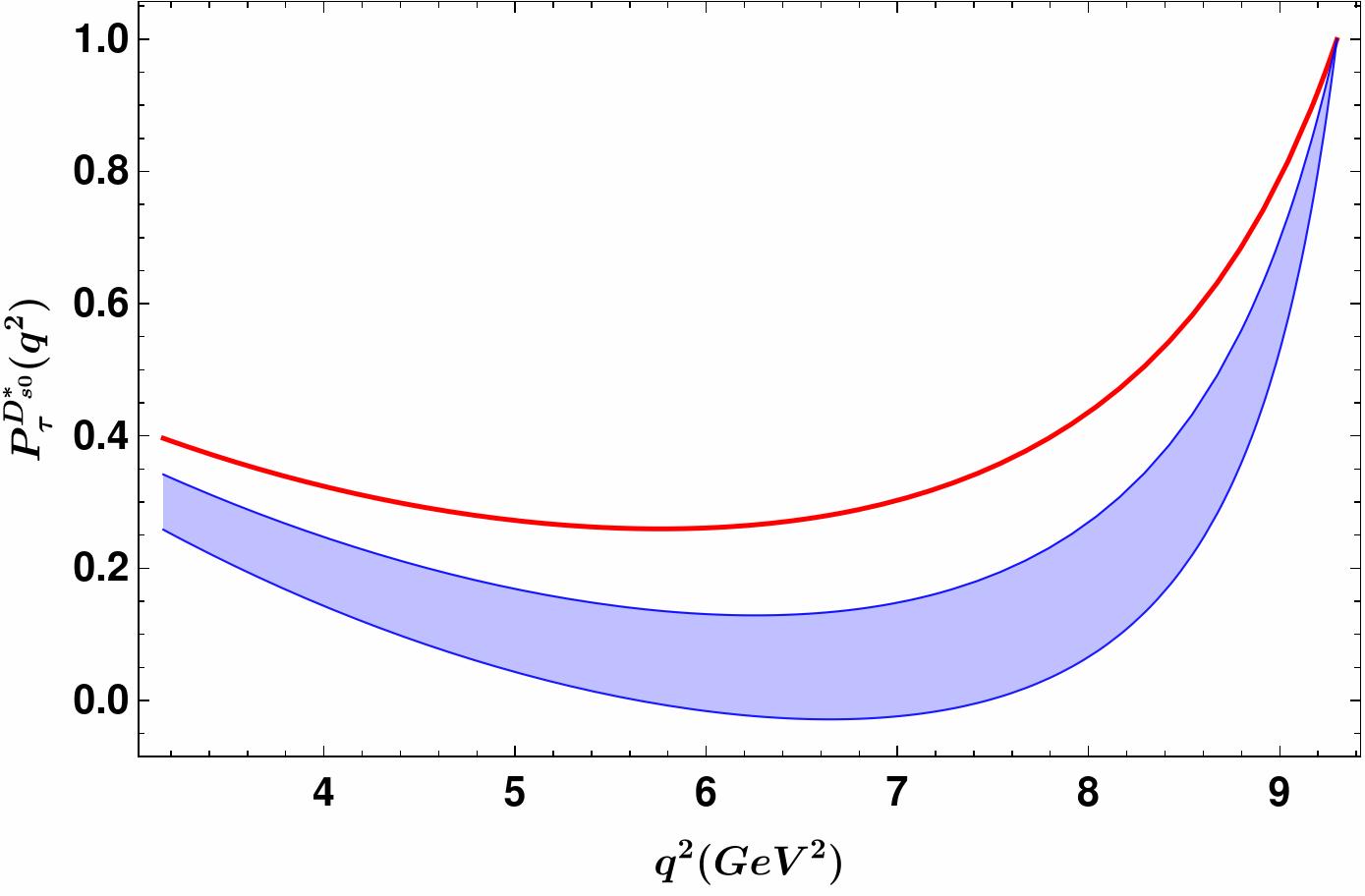}
	
	\vspace{0.3cm}
	
	\includegraphics[scale=0.34]{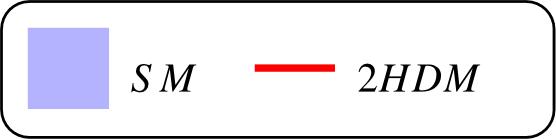}
\end{center}
\caption{$q^2$-distribution of $B_s \to D_{s0}^*\tau\nu_\tau$ observables in 2HDM.}\label{2HDM_Ds0s}
\end{figure}

\begin{figure}
\begin{center}
	\includegraphics[width=0.31\textwidth]{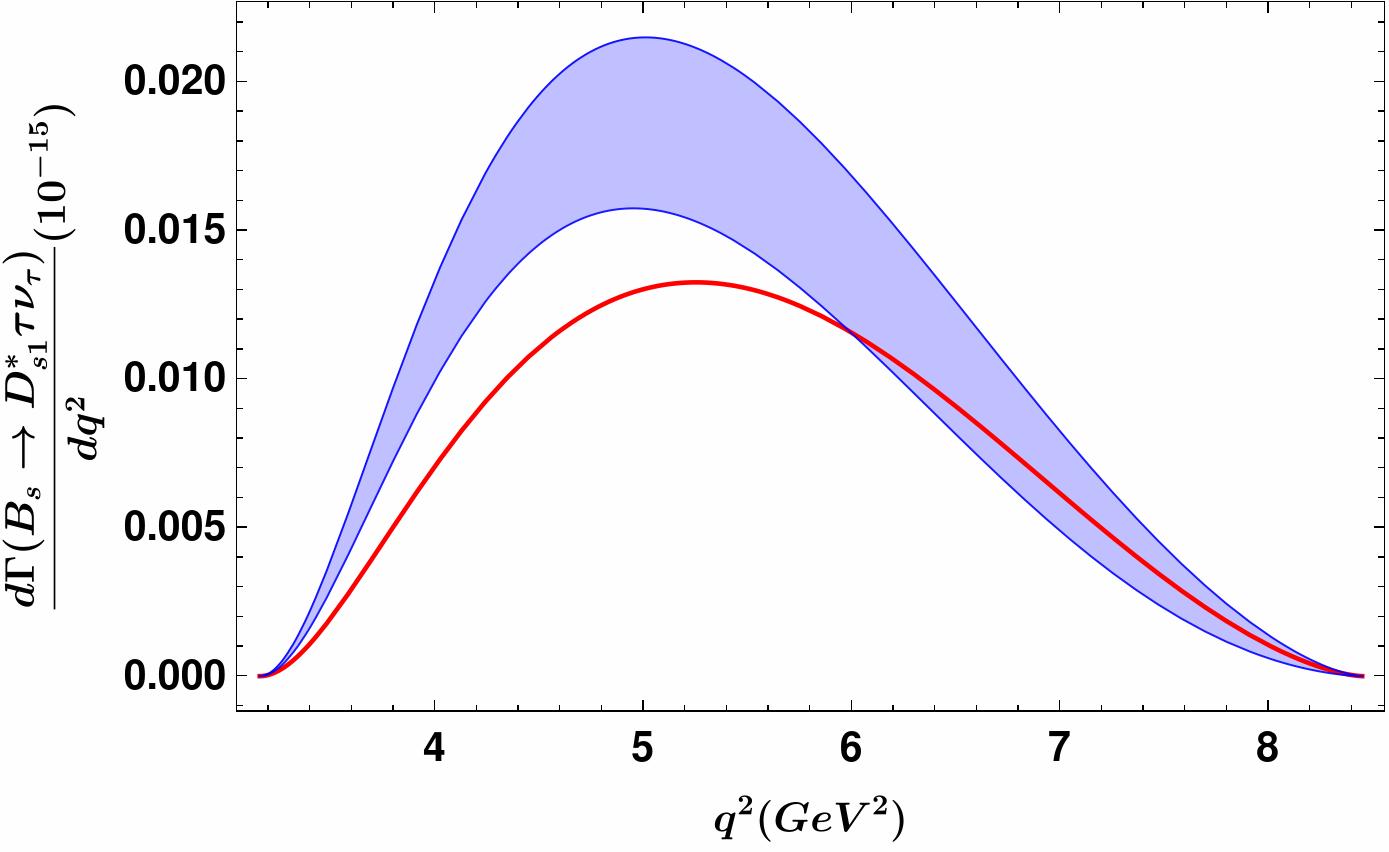}
	\hspace{0.01\textwidth}
	\includegraphics[width=0.31\textwidth]{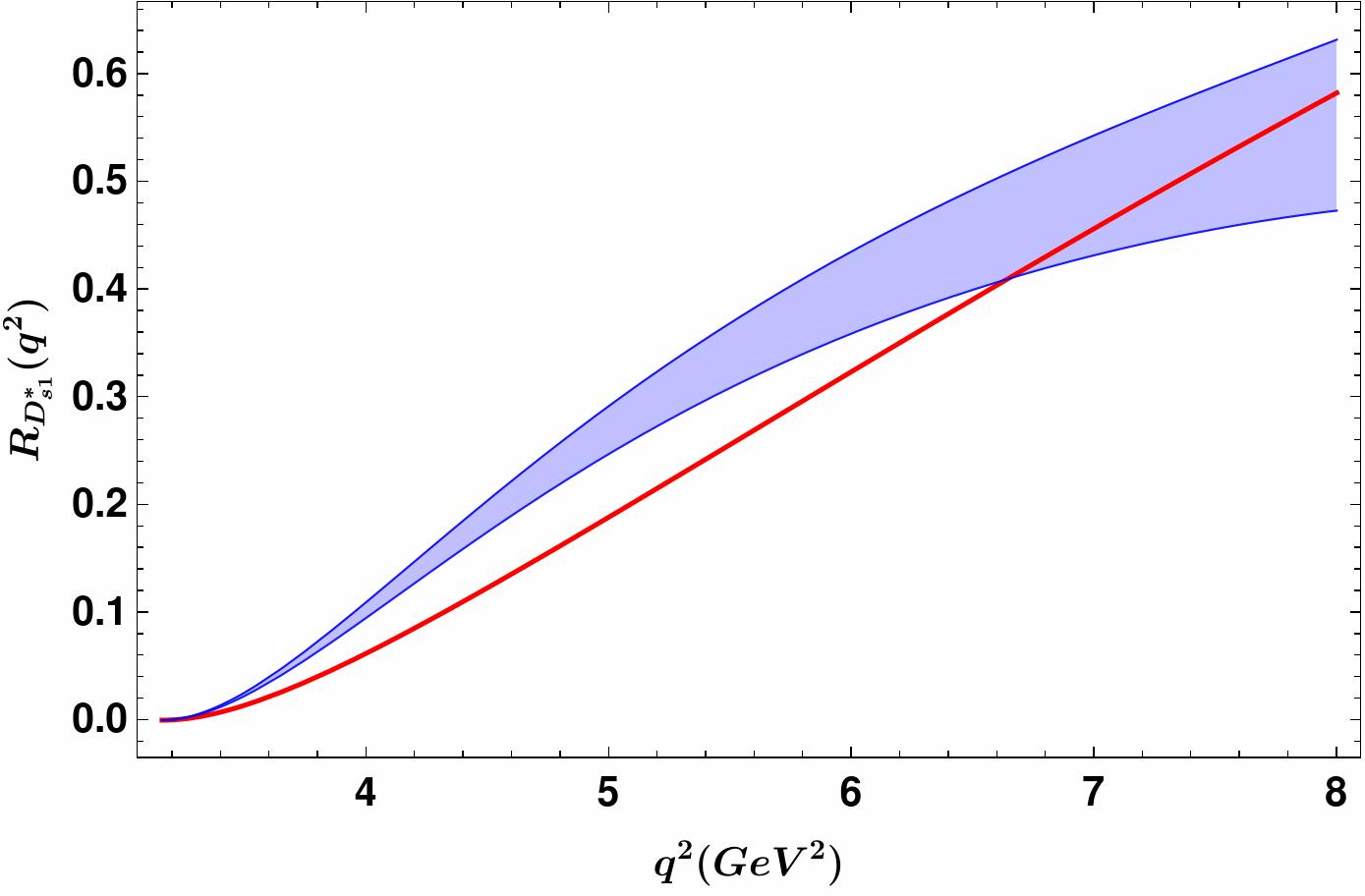}
	\hspace{0.01\textwidth}
	\includegraphics[width=0.31\textwidth]{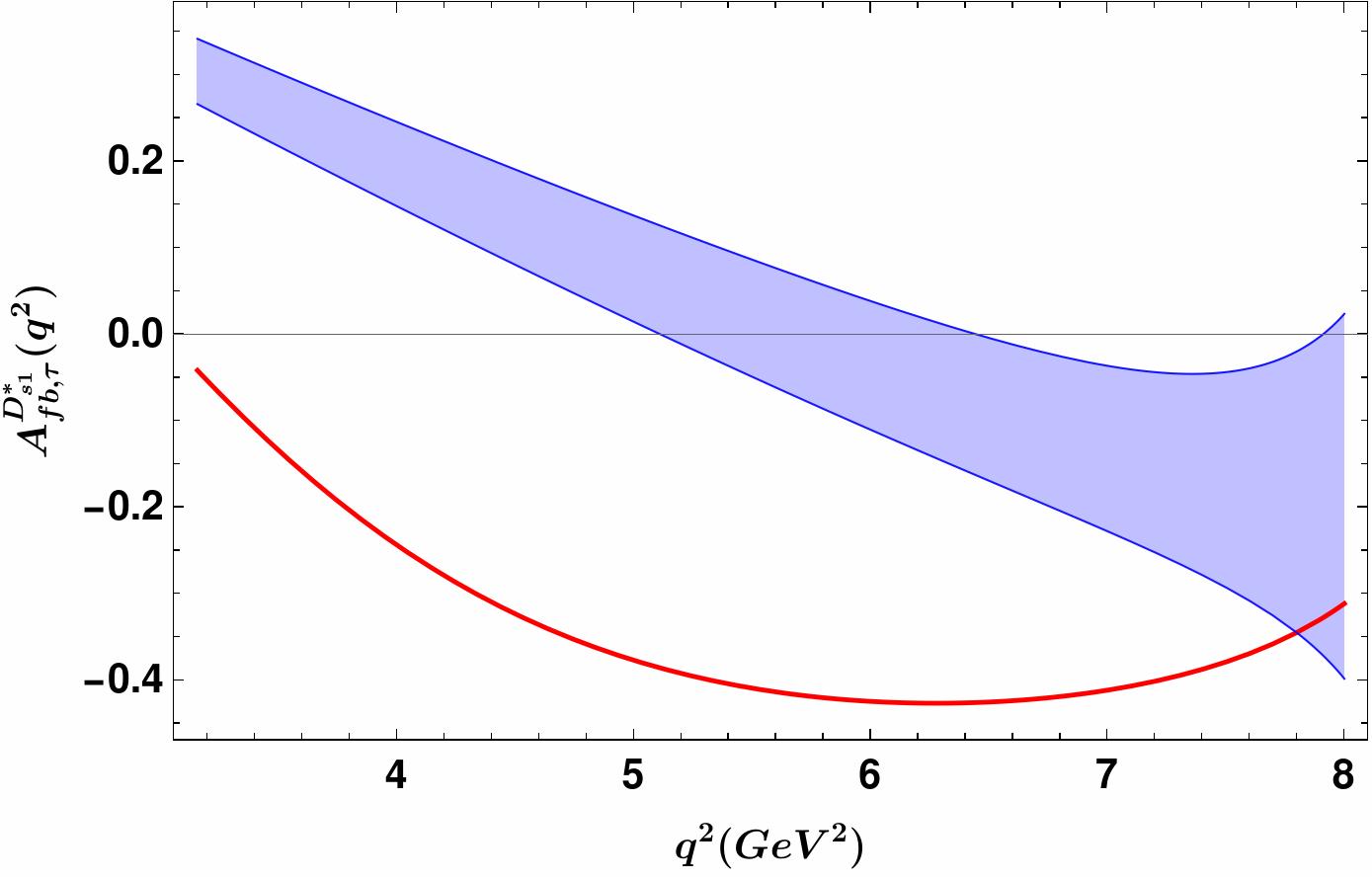}
	
	\vspace{0.3cm}
	
	\includegraphics[width=0.31\textwidth]{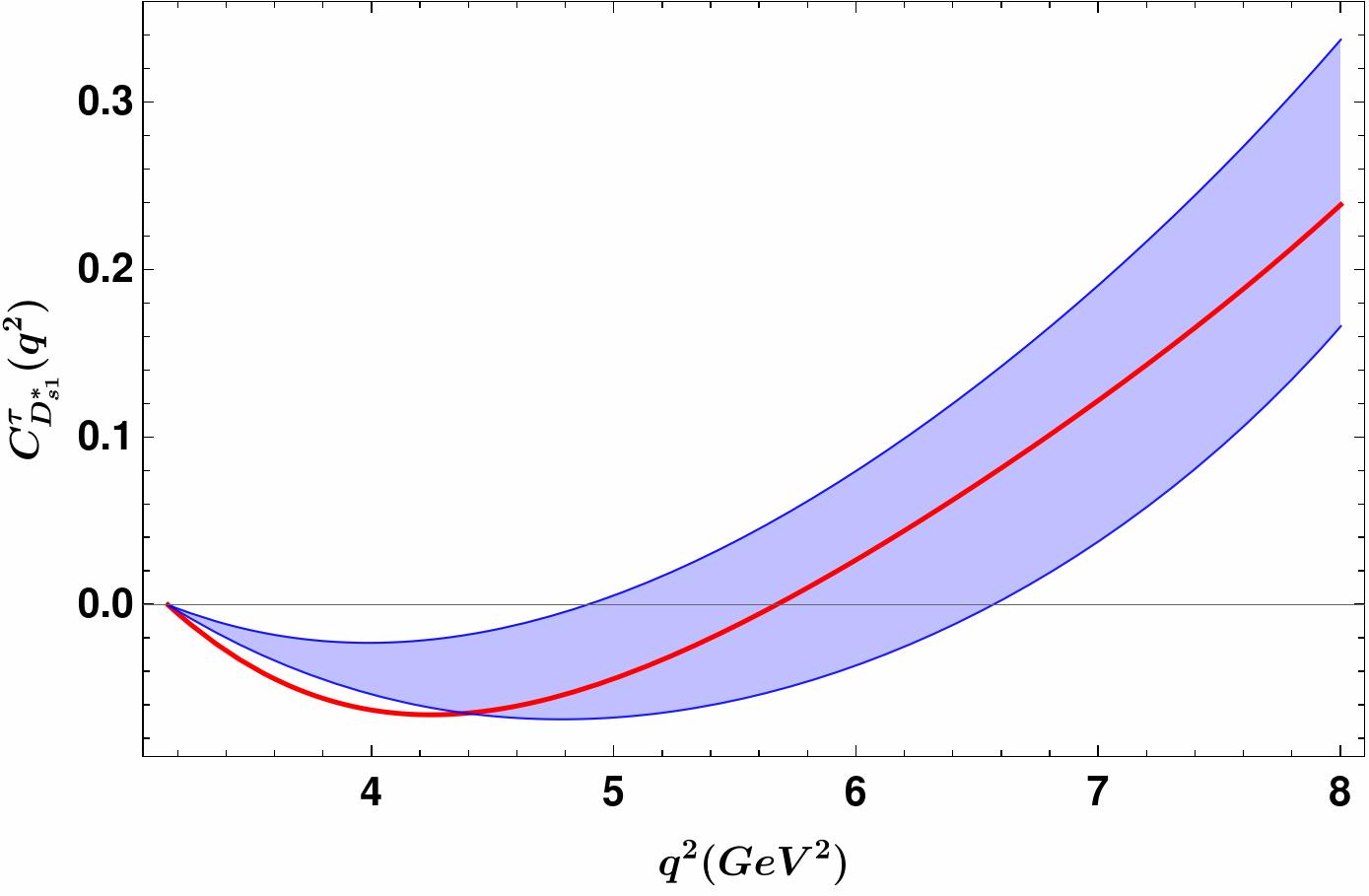}
	\hspace{0.01\textwidth}
	\includegraphics[width=0.31\textwidth]{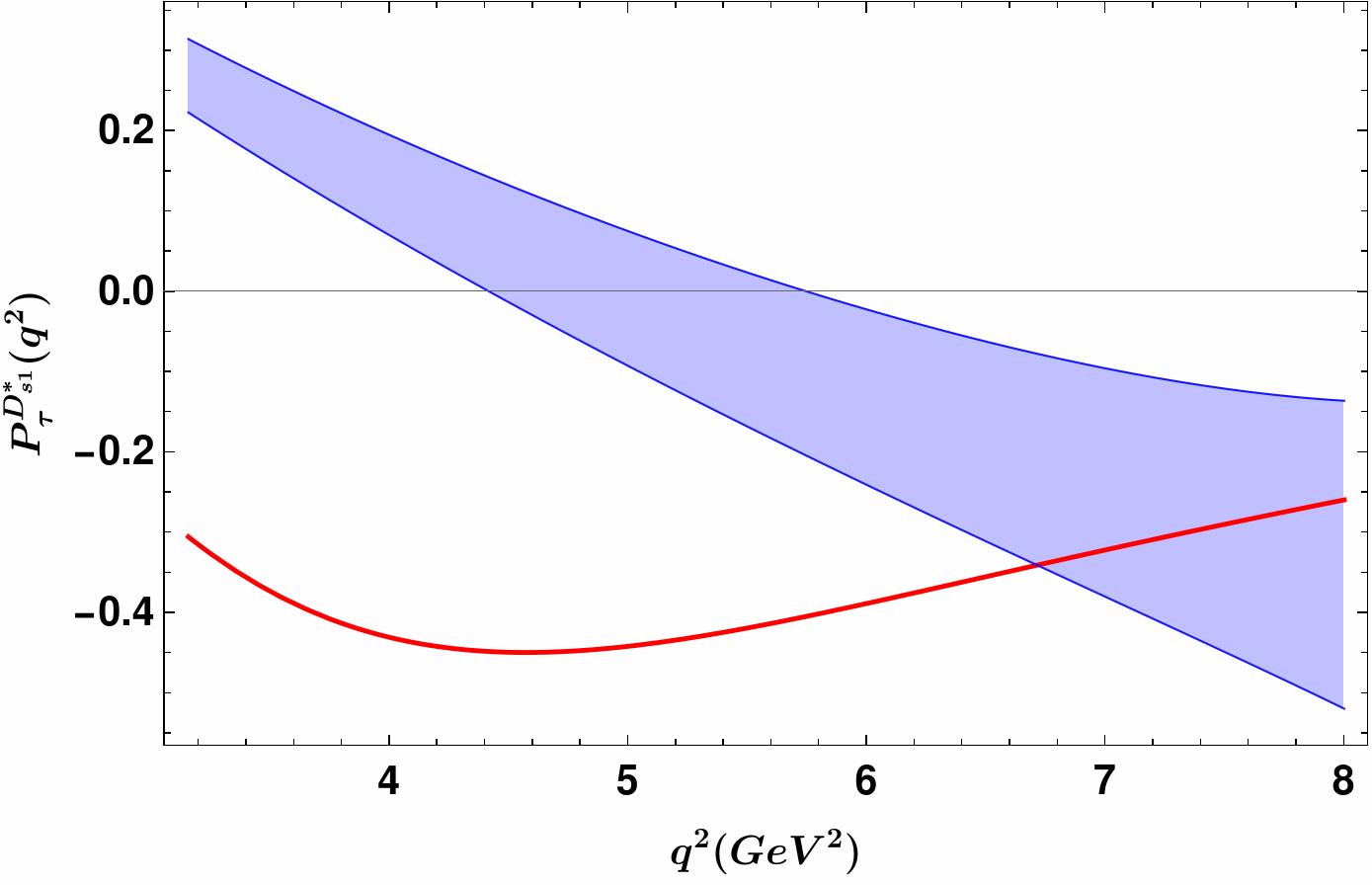}
	\hspace{0.01\textwidth}
	\includegraphics[width=0.31\textwidth]{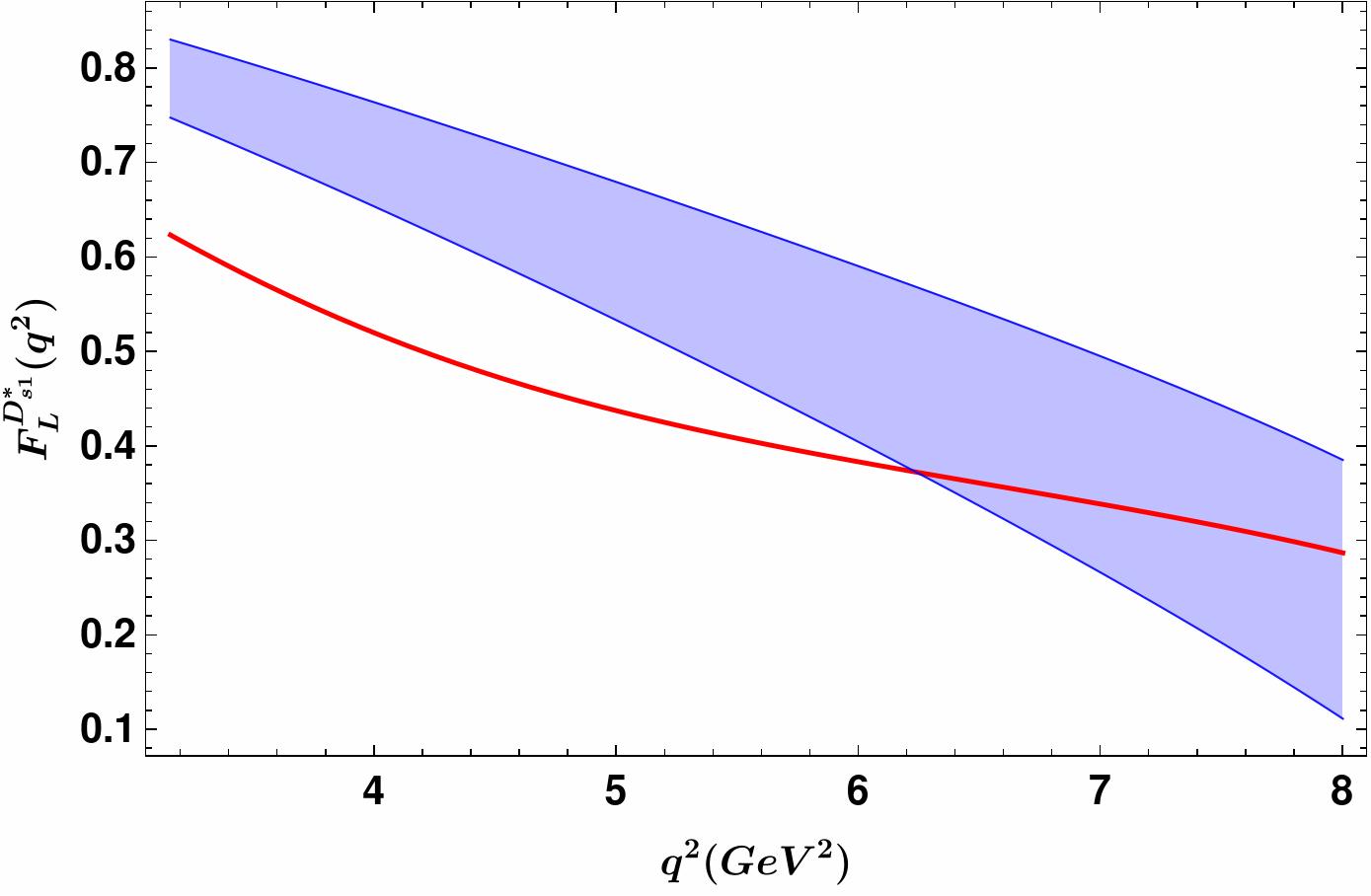}
	
	\vspace{0.3cm}
	
	\includegraphics[scale=0.34]{DsstNPscn2HDMleg.jpg}
\end{center}
\caption{$q^2$-distribution of $B_s \to D_{s1}^*\tau\nu_\tau$ observables in 2HDM.}\label{2HDM_Ds1s}
\end{figure}

\begin{figure}
\begin{center}
	\includegraphics[width=0.31\textwidth]{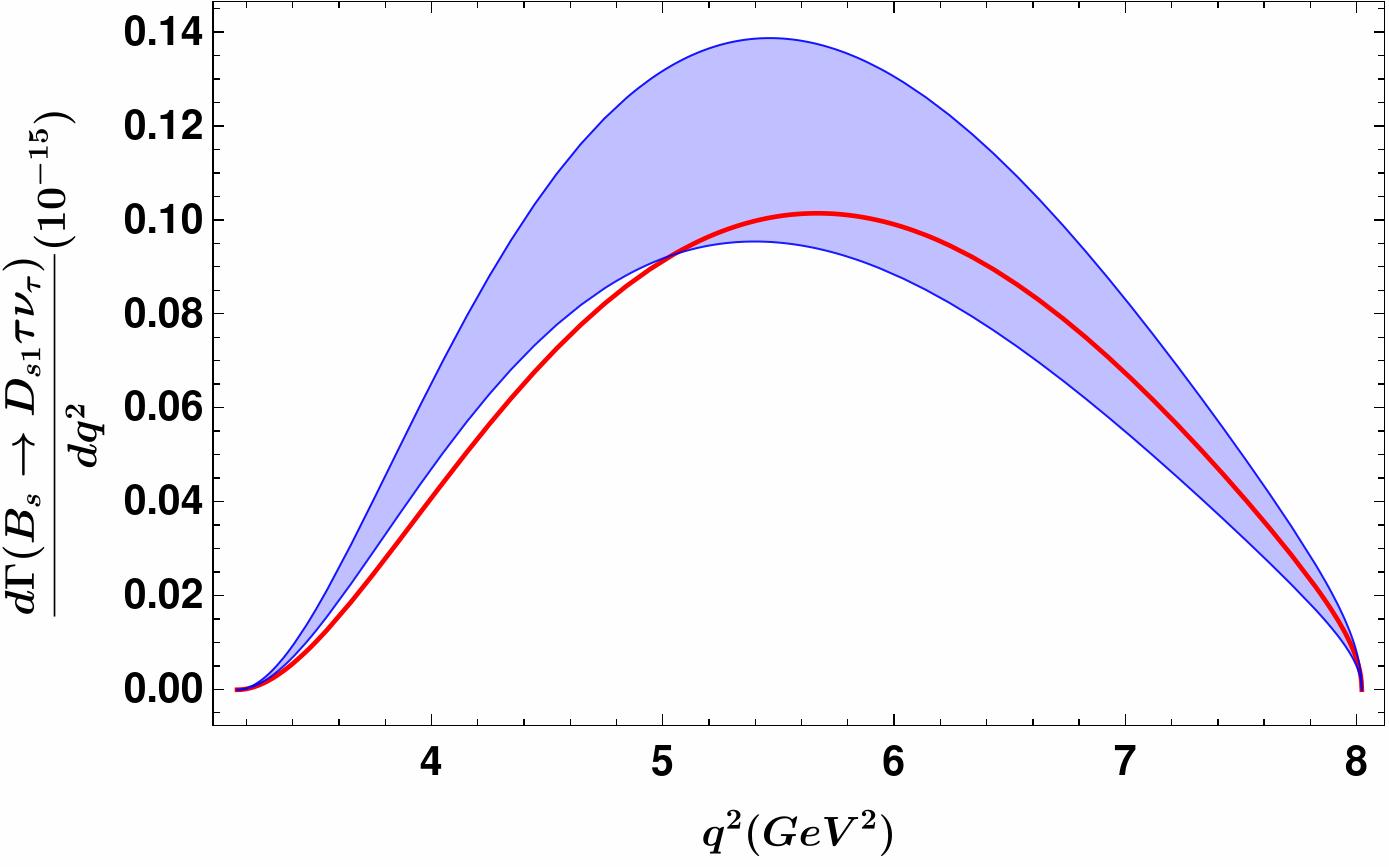}
	\hspace{0.01\textwidth}
	\includegraphics[width=0.31\textwidth]{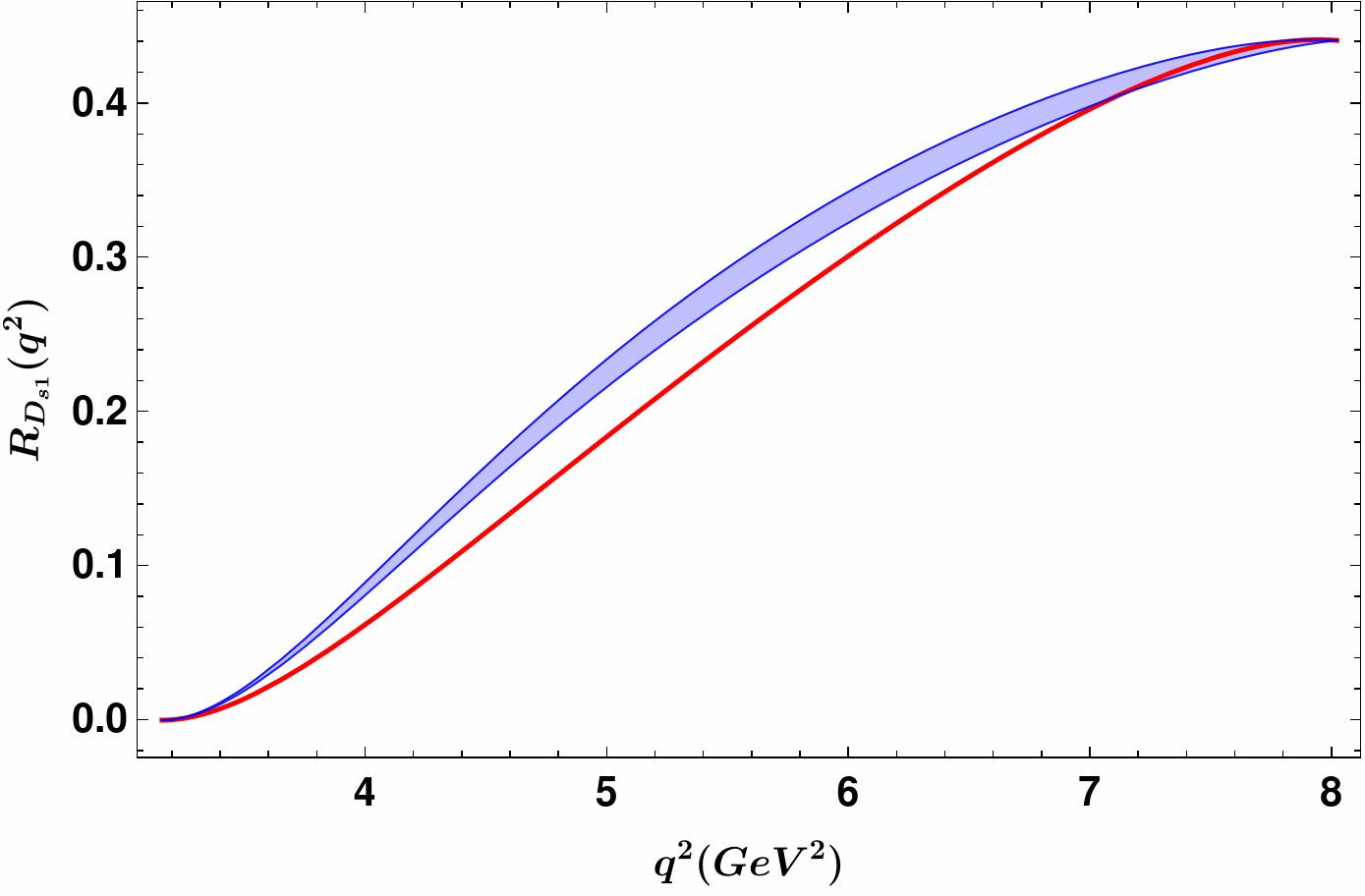}
	\hspace{0.01\textwidth}
	\includegraphics[width=0.31\textwidth]{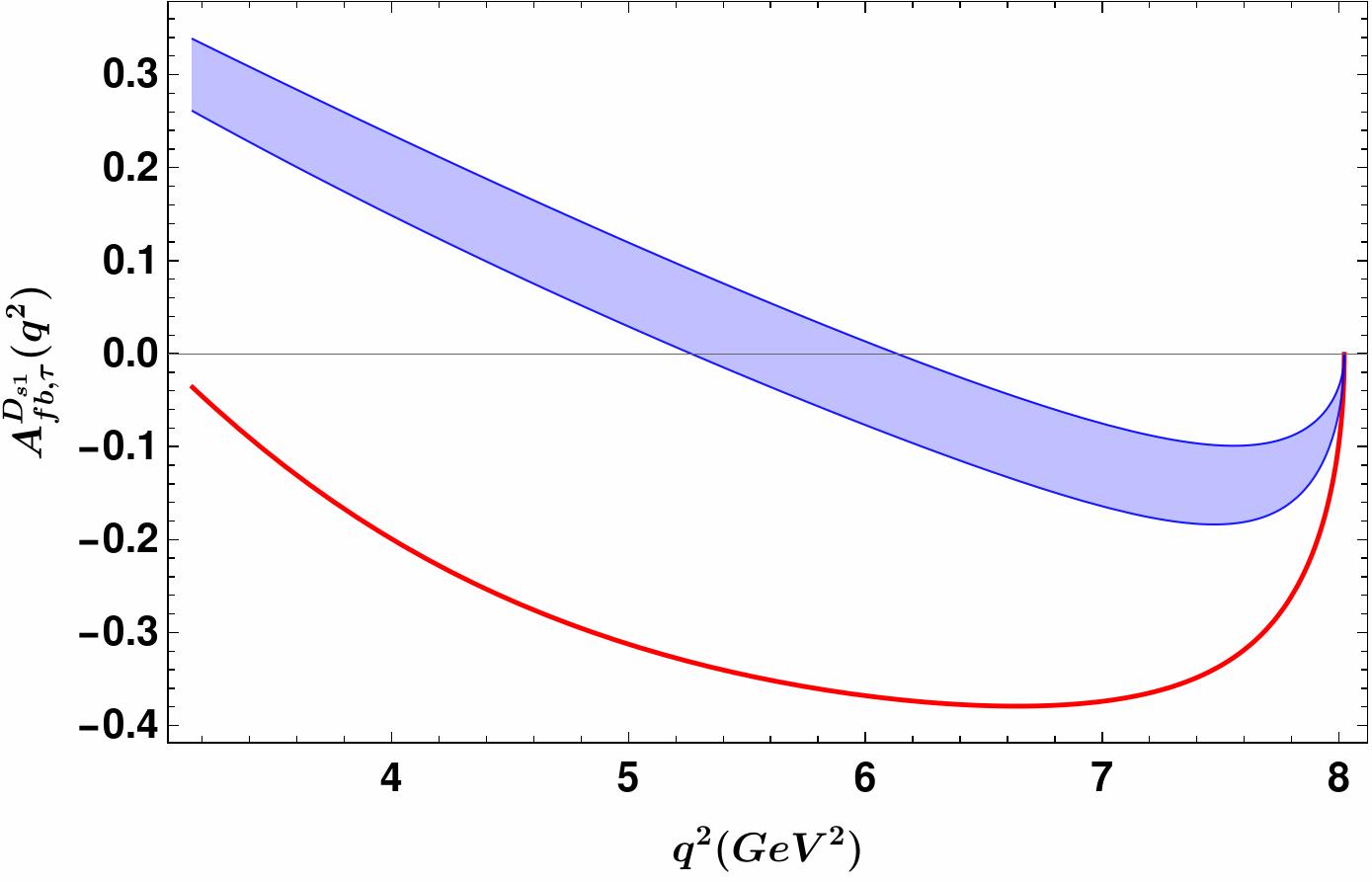}
	
	\vspace{0.3cm}
	
	\includegraphics[width=0.31\textwidth]{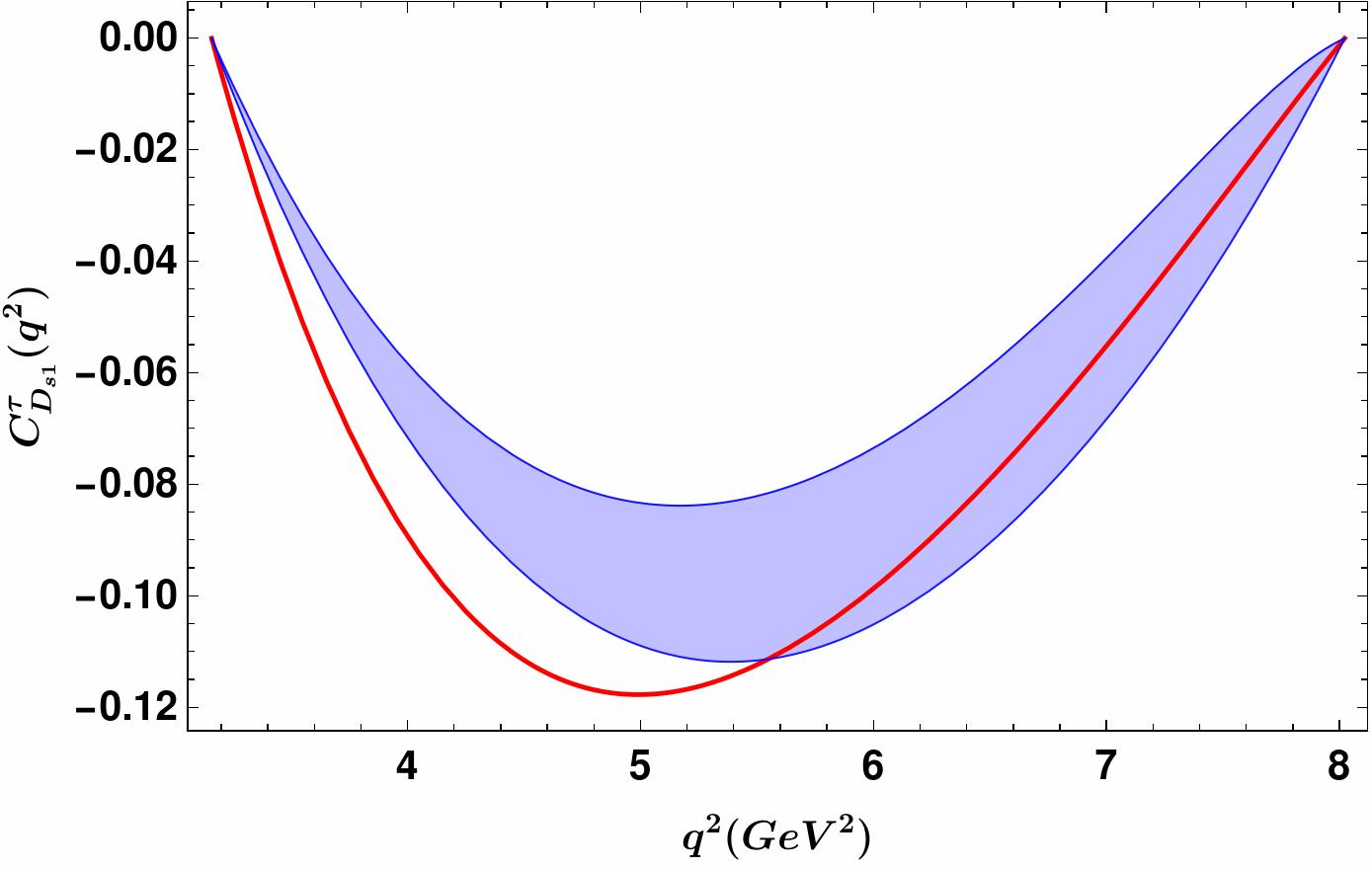}
	\hspace{0.01\textwidth}
	\includegraphics[width=0.31\textwidth]{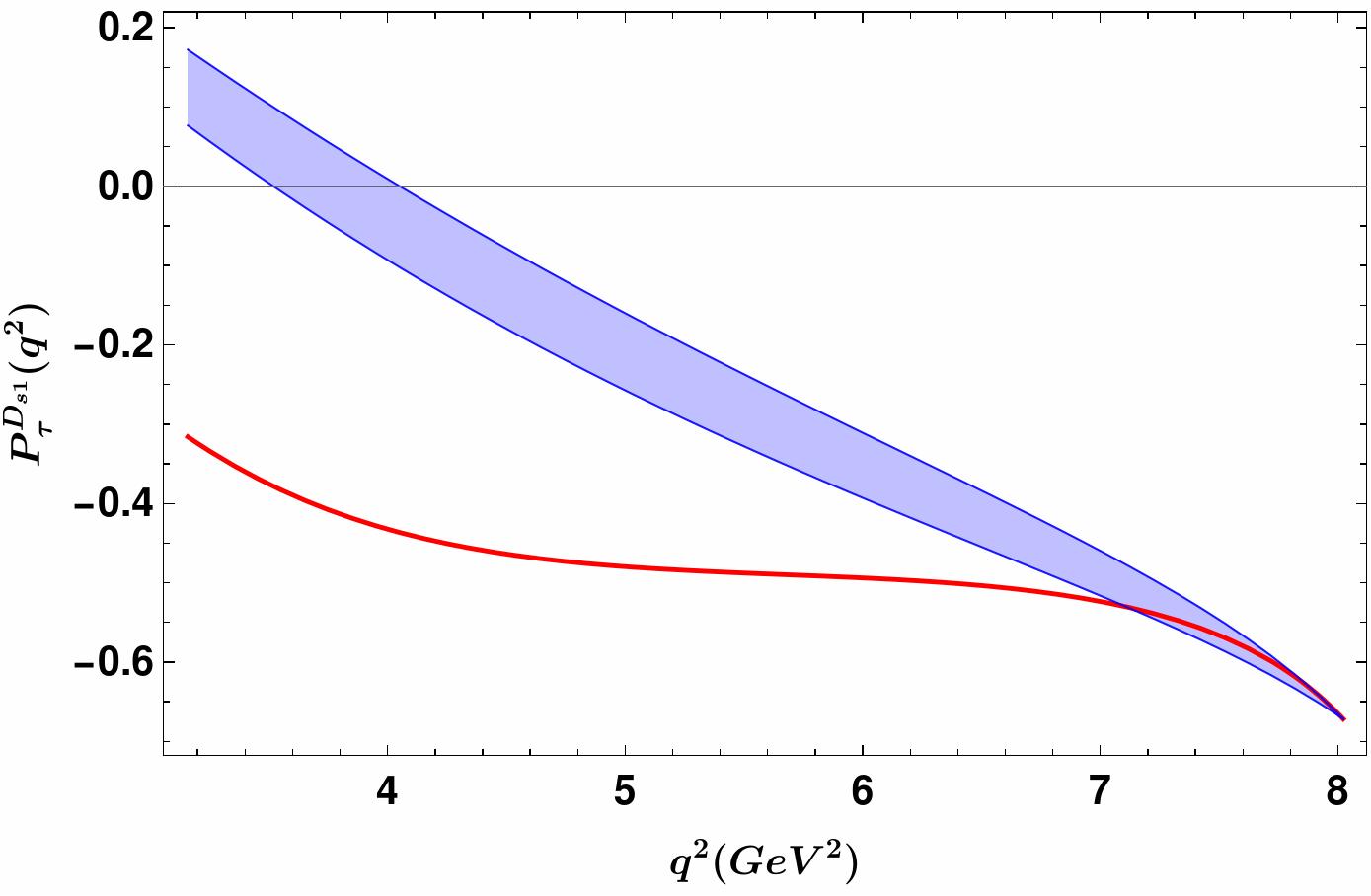}
	\hspace{0.01\textwidth}
	\includegraphics[width=0.31\textwidth]{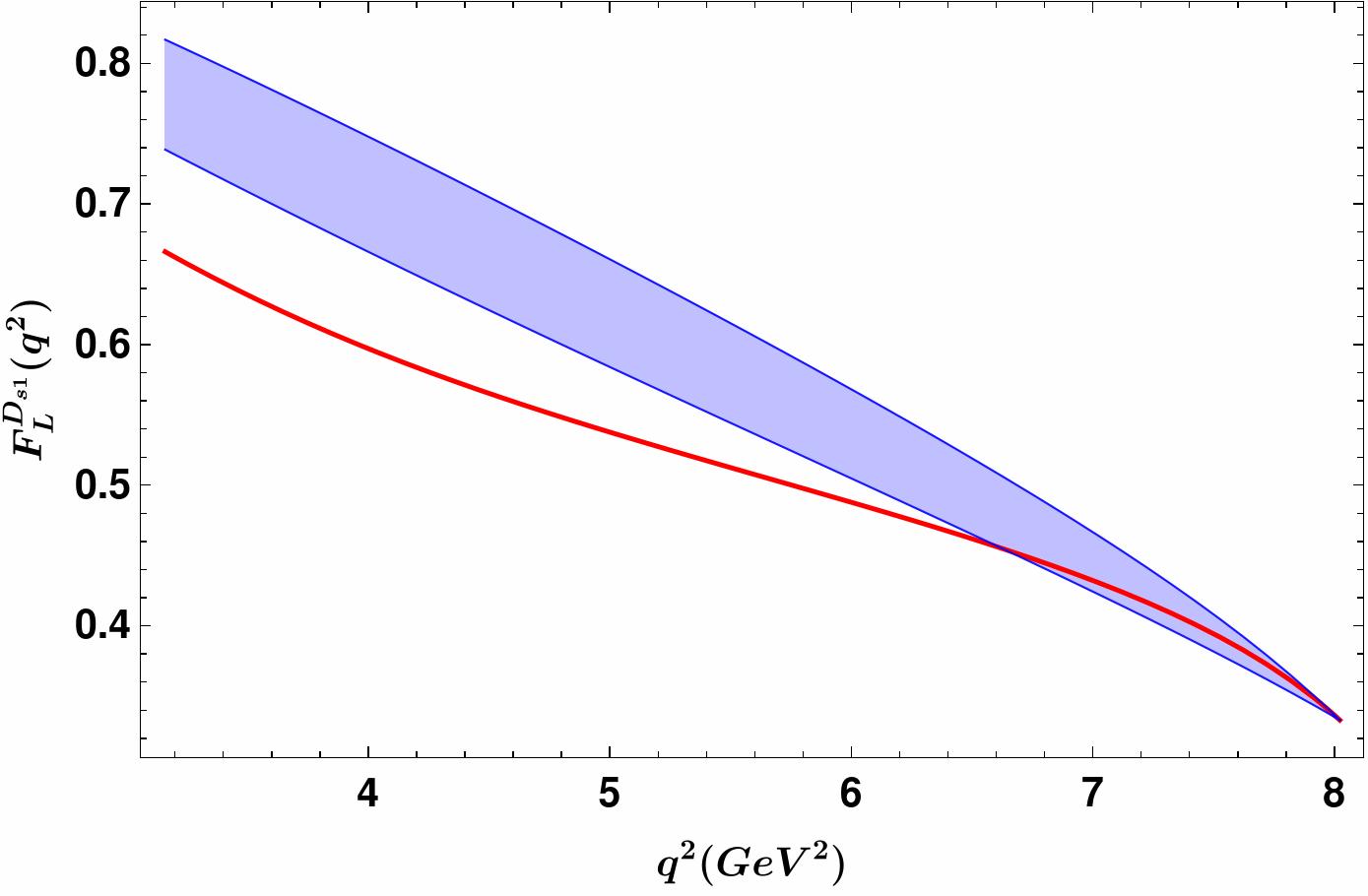}
	
	\vspace{0.3cm}
	
	\includegraphics[scale=0.34]{DsstNPscn2HDMleg.jpg}
\end{center}
\caption{$q^2$-distribution of $B_s \to D_{s1}\tau\nu_\tau$ observables in 2HDM.}\label{2HDM_Ds1}
\end{figure}

\begin{figure}
\begin{center}
	\includegraphics[width=0.31\textwidth]{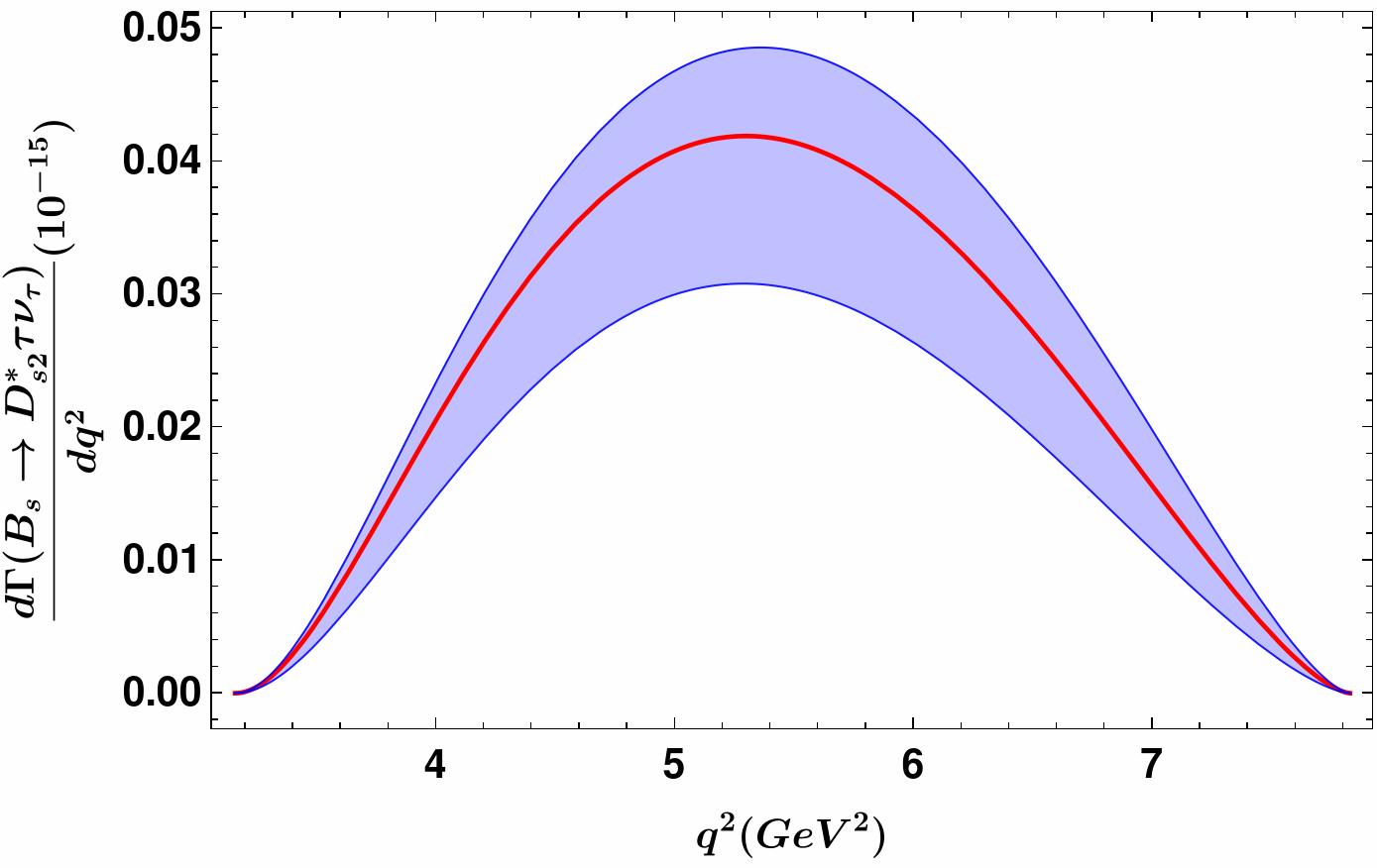}
	\hspace{0.01\textwidth}
	\includegraphics[width=0.31\textwidth]{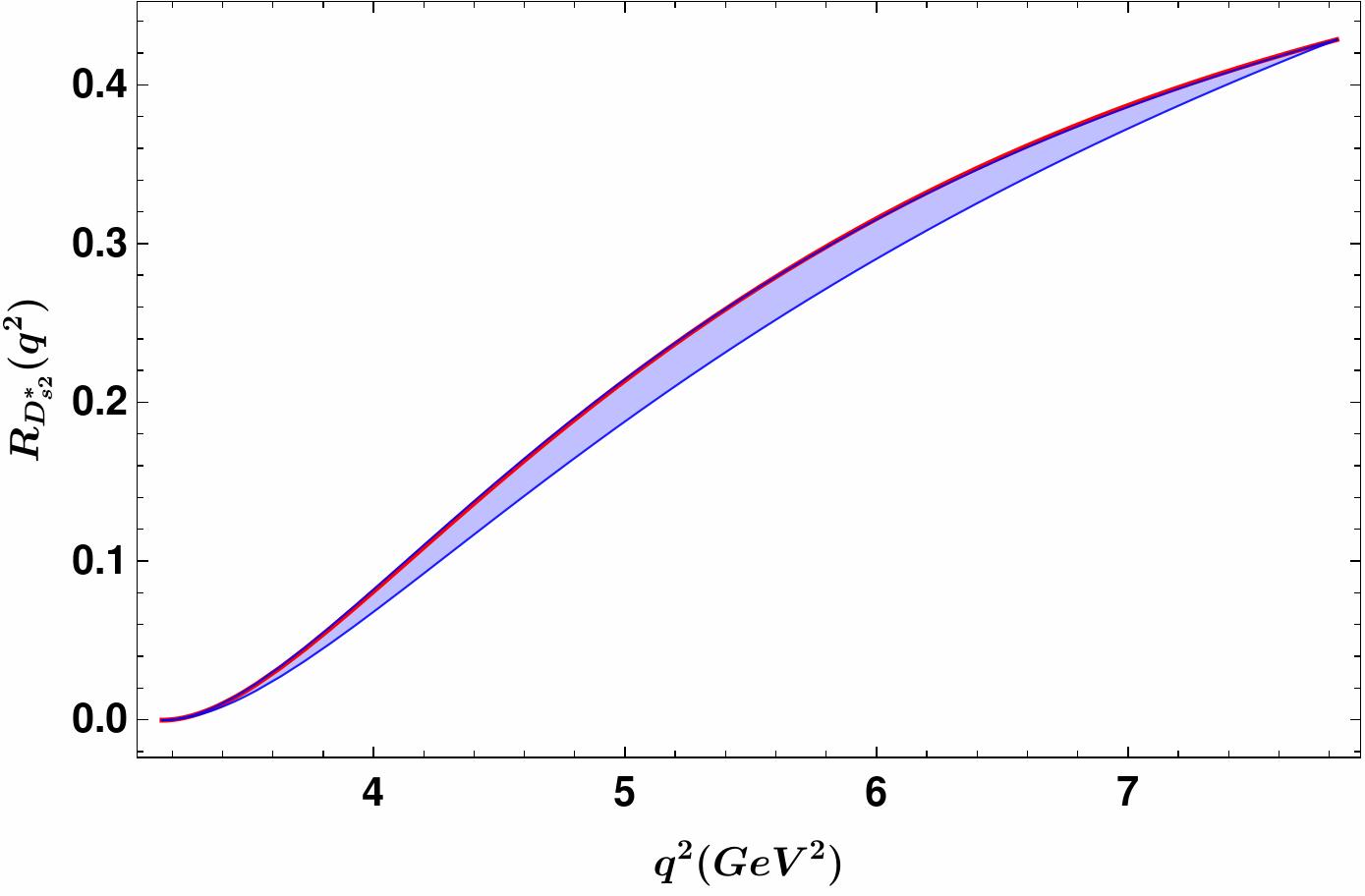}
	\hspace{0.01\textwidth}
	\includegraphics[width=0.31\textwidth]{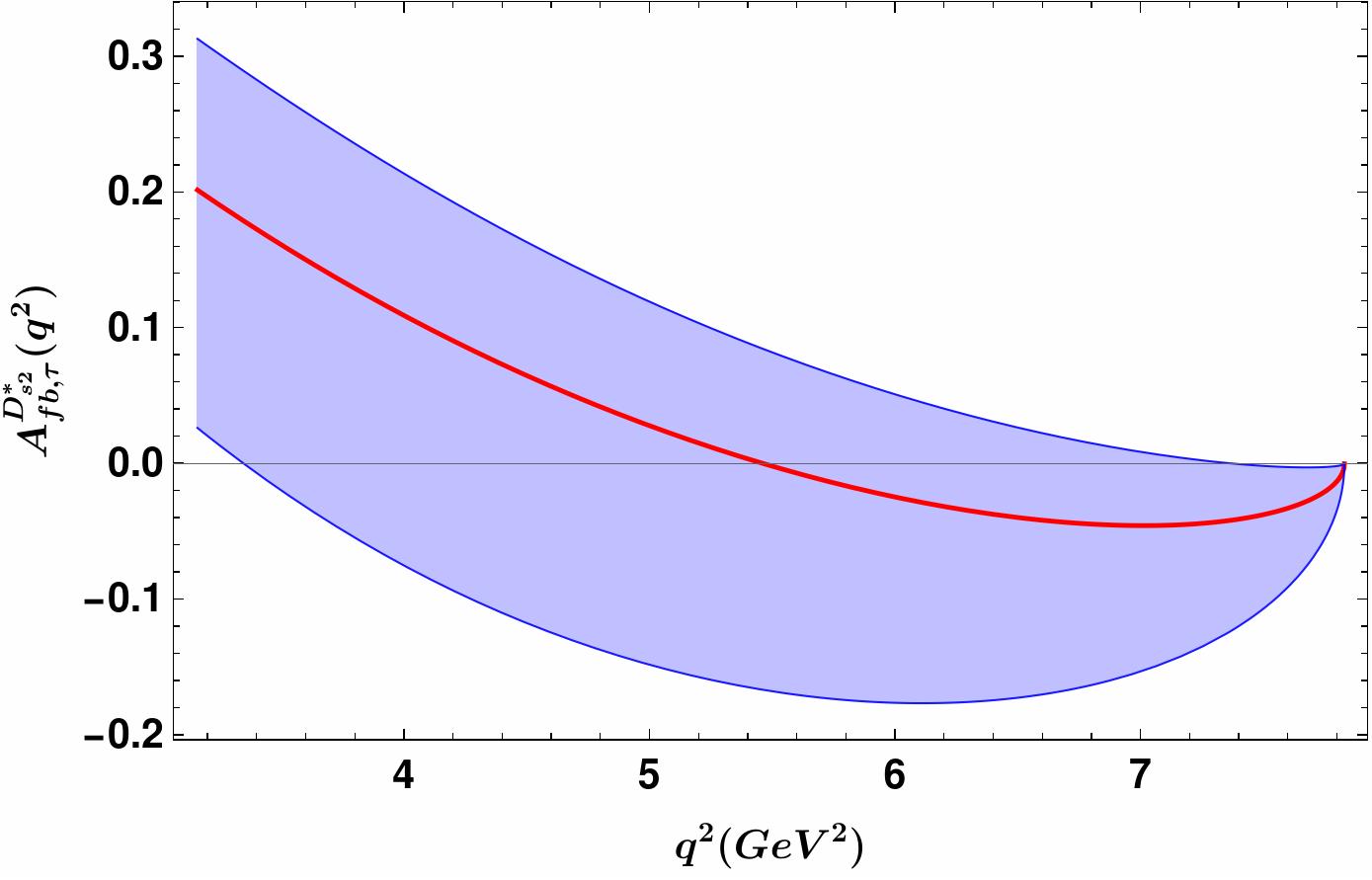}
	
	\vspace{0.3cm}
	
	\includegraphics[width=0.31\textwidth]{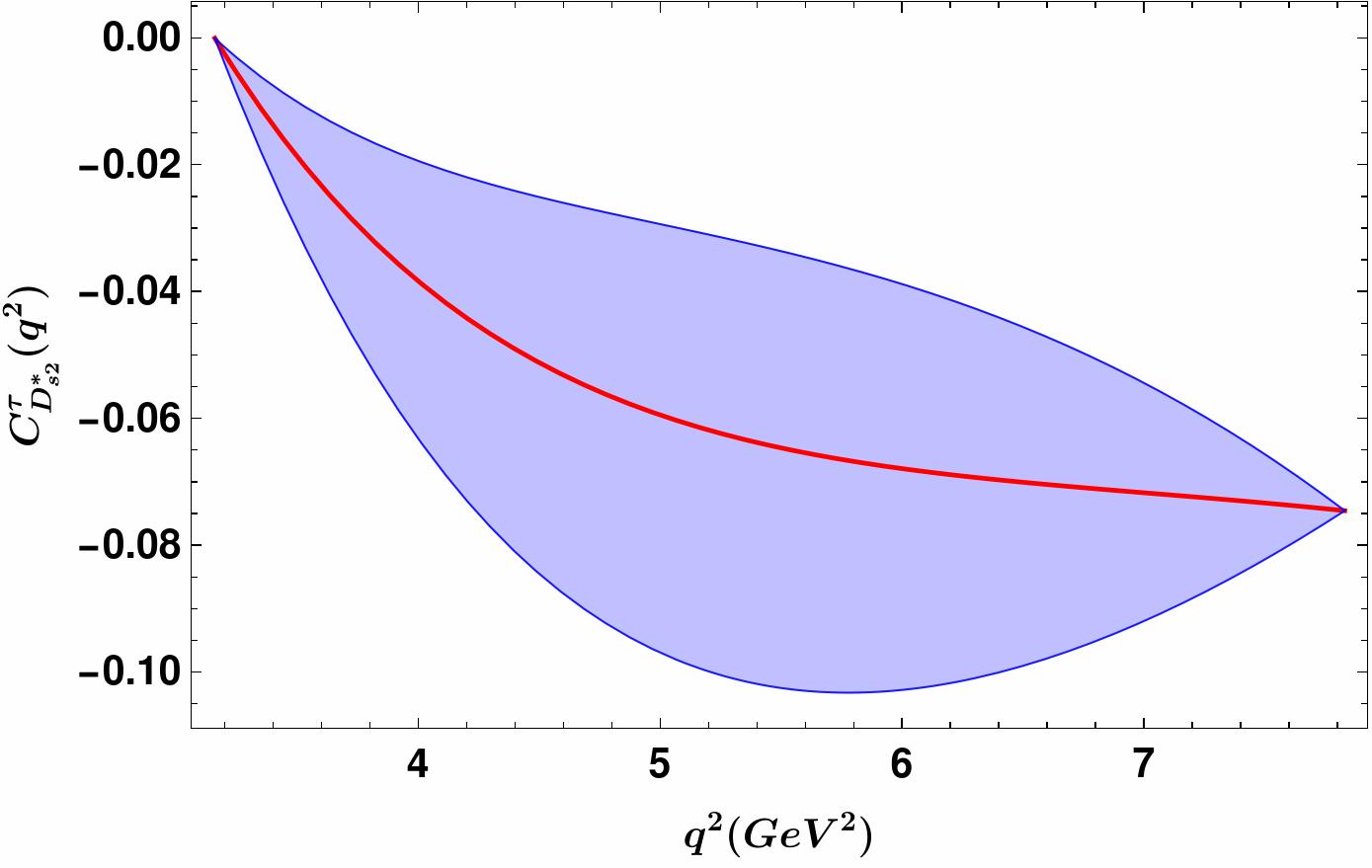}
	\hspace{0.01\textwidth}
	\includegraphics[width=0.31\textwidth]{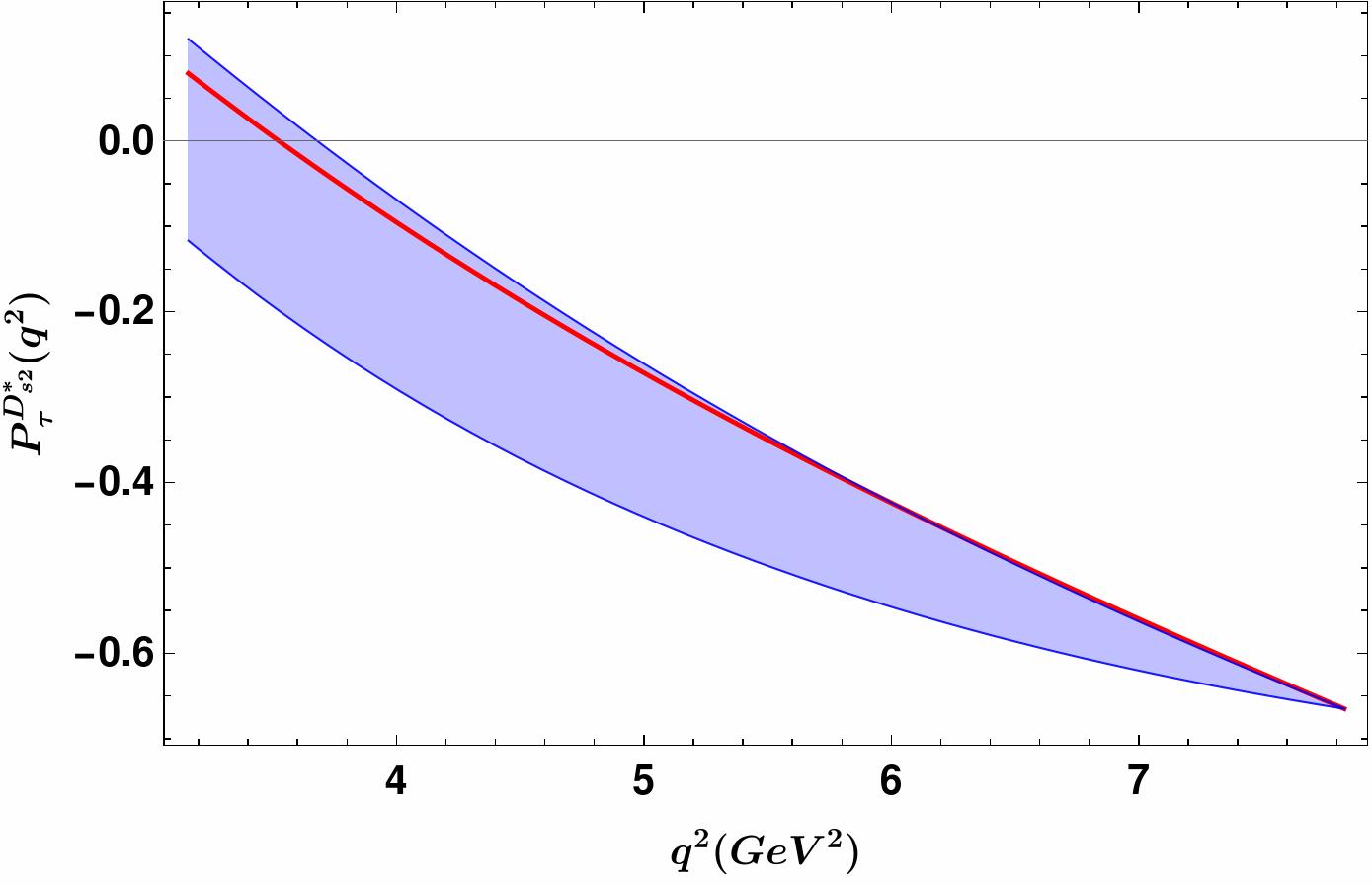}
	\hspace{0.01\textwidth}
	\includegraphics[width=0.31\textwidth]{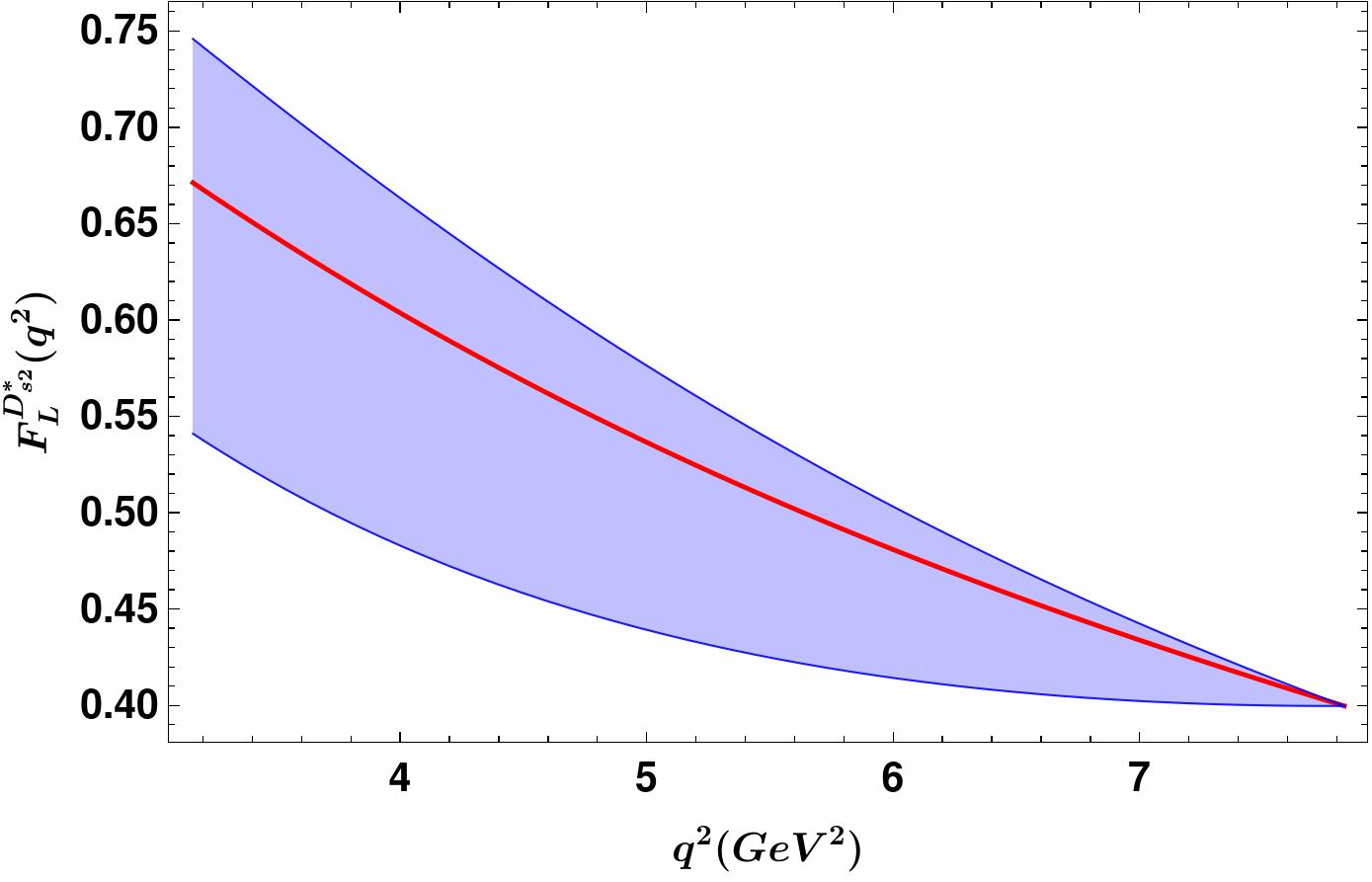}
	
	\vspace{0.3cm}
	
	\includegraphics[scale=0.34]{DsstNPscn2HDMleg.jpg}
\end{center}
\caption{$q^2$-distribution of $B_s \to D_{s2}^*\tau\nu_\tau$ observables in 2HDM.}\label{2HDM_Ds2s}
\end{figure}

To test NP sensitivity of the $B_s \to D_s^{**}\tau \nu_\tau$ modes, we fit the product couplings $\rho^{23}_d\rho^{33*}_\ell$ and $\rho^{32*}_u\rho^{33*}_\ell$, which is equivalent to a two-operator scenario with $C_{S_1}$ and $C_{S_2}$. The fit results are presented in table~\ref{tab:2HDM} and the $q^2$-distribution is shown in figures~\ref{2HDM_Ds0s}-\ref{2HDM_Ds2s}. As before, we compute the tension between the SM and 2HDM predictions using eq.~\ref{eq:tension_formula}. The results are presented in table~\ref{tab:tens_2HDM_zeta} for the $B_s \to D_{s0}^*\tau\nu_\tau$ and $B_s \to D_{s1}^*\tau\nu_\tau$ modes, and in table~\ref{tab:tens_2HDM_tau} for the $B_s \to D_{s1}\tau\nu_\tau$ and $B_s \to D_{s2}^*\tau\nu_\tau$ modes. Tensions exceeding $3\sigma$ are highlighted in bold in the tables. Figures~\ref{2HDM_Ds0s}-\ref{2HDM_Ds1} illustrate that most observables for the $B_s \to D_{s0}^*\tau\nu_\tau$, $B_s \to D_{s1}^*\tau\nu_\tau$ and $B_s \to D_{s1}\tau\nu_\tau$ modes are sensitive to the 2HDM couplings. However, table~\ref{tab:tens_2HDM_zeta} reveals the deviations from the SM are within $2\sigma$ for $B_s \to D_{s0}^*\tau\nu_\tau$. For $B_s \to D_{s1}^*\tau\nu_\tau$, the observable $C^\tau_{D_{s1}^*}$ is largely SM-like. From table~\ref{tab:tens_2HDM_zeta}, it is evident that $A_{fb,\tau}^{D_{s1}^*}$ deviates from the SM by more than $3\sigma$, and $P_\tau^{D_{s1}^*}$ by more than $2\sigma$. Also, it is particularly noticeable that these two observables have zero-crossings in the SM, but these crossings are absent in the 2HDM. For $B_s \to D_{s1}\tau\nu_\tau$, high NP sensitivity is observed in $A_{fb,\tau}^{D_{s1}}$ and $P_\tau^{D_{s1}}$, with deviations exceeding $3\sigma$ from the SM predictions, as shown in table~\ref{tab:tens_2HDM_tau}. The LFU ratio $R_{D_{s1}}$ also deviates by more than $2\sigma$ from the SM. Additionally, zero-crossings occur for $A_{fb,\tau}^{D_{s1}}$ and $P_\tau^{D_{s1}}$ in the SM, but are absent in the 2HDM. Figure~\ref{2HDM_Ds2s} shows that all observables of the $B_s \to D_{s2}^*\tau\nu_\tau$ mode are consistent with the SM predictions. The main reason for these observations is the combined influence of the $C_{S_1}$ and $C_{S_2}$ couplings on the modes: in pseudoscalar to axial-vector transitions, they appear as $C_{S_1} + C_{S_2}$, whereas in pseudoscalar to scalar (tensor) transitions, as $C_{S_1} - C_{S_2}$.  Table~\ref{tab:2HDM} and eq.~\ref{eqn:2HDM_match} show that both $C_{S_1}$ and $C_{S_2}$ share the same sign, making the effects of 2HDM couplings more significant in the $B_s\to D_{s1}^{(*)}\tau\nu_\tau$ modes.
	
\begin{table}[t]
	\begin{center}
		\renewcommand*{\arraystretch}{1.4}
		\resizebox{1.02 \textwidth}{!}{
			\begin{tabular}{|c|c|c|c|c|c|c|c|c|c|c|c|}
				\hline
				& \multicolumn{5}{c|}{$B_s \to D_{s0}^*\tau\nu_\tau$} & \multicolumn{6}{c|}{$B_s \to D_{s1}^*\tau\nu_\tau$} \\
				\hline
				& $\Gamma$ & $\langle R_{D_{s0}^*} \rangle$ & $\langle A^{D_{s0}^*}_{fb,\tau} \rangle$ & $\langle C^\tau_{D_{s0}^*} \rangle$ & $\langle P^{D_{s0}^*}_\tau \rangle$ & $\Gamma$ & $\langle R_{D_{s1}^*} \rangle$ & $\langle A^{D_{s1}^*}_{fb,\tau} \rangle$ & $\langle C^\tau_{D_{s1}^*} \rangle$ & $\langle P^{D_{s1}^*}_\tau \rangle$ & $\langle F_L^{D_{s1}^*} \rangle$ \\
				\hline
				2HDM & 0.56 & 1.06 & 0.46 & 1.25 & 1.31 & 1.05 & 1.51 & \textbf{4.17} & 0.03 & 2.47 & 1.22 \\
				\hline
			\end{tabular}
		}
	\end{center}
\caption{Computed tensions for observables of $B_s \to D_{s0}^*\tau\nu_\tau$ and $B_s \to D_{s1}^*\tau\nu_\tau$ within 2HDM (in units of $\sigma$).}\label{tab:tens_2HDM_zeta}
\end{table}

\begin{table}[t]
	\begin{center}
		\renewcommand*{\arraystretch}{1.4}
		\resizebox{1.02 \textwidth}{!}{
		\begin{tabular}{|c|c|c|c|c|c|c|c|c|c|c|c|c|}
			\hline
			& \multicolumn{6}{c|}{$B_s \to D_{s1}\tau \nu_\tau$} & \multicolumn{6}{c|}{$B_s \to D_{s2}^*\tau\nu_\tau$} \\
			\hline
			& $\Gamma$ & $\langle R_{D_{s1}} \rangle$ & $\langle A^{D_{s1}}_{fb,\tau} \rangle$ & $\langle C^\tau_{D_{s1}} \rangle$ & $\langle P^{D_{s1}}_\tau \rangle$ & $\langle F_L^{D_{s1}} \rangle$ & $\Gamma$ & $\langle R_{D_{s2}^*} \rangle$ & $\langle A^{D_{s2}^*}_{fb,\tau} \rangle$ & $\langle C^\tau_{D_{s2}^*} \rangle$ & $\langle P^{D_{s2}^*}_\tau \rangle$ & $\langle F_L^{D_{s2}^*} \rangle$ \\
			\hline
			2HDM & 0.51 & 2.07 & \textbf{5.42} & 0.67 & \textbf{3.02} & 1.39 & 0.16 & 0.27 & 0.23 & 0.08 & 0.55 & 0.30\\
			\hline
		\end{tabular}
	}
	\end{center}
\caption{Computed tensions for observables of $B_s \to D_{s1}\tau\nu_\tau$ and $B_s \to D_{s2}^*\tau\nu_\tau$ within 2HDM (in units of $\sigma$).}\label{tab:tens_2HDM_tau}
\end{table}

\section{Conclusion}\label{sec:concl}

This study examined the decays of the $B_s$ meson into various $D_s^{**}$ states, which are the lightest orbitally excited $c\bar{s}$ states. In the absence of lattice data, we employed a data-driven fit of the form factors for $ B_s \to D_s^{**}\ell\nu_\ell$ within the HQET framework. In HQET, all form factors for $B_s \to D_{s0}^*\ell\nu_\ell$ and $B_s \to D_{s1}^*\ell\nu_\ell$ are parametrized by a single Isgur-Wise function $\zeta(w)$, while those for $B_s \to D_{s1}\ell\nu_\ell$ and $B_s \to D_{s2}^*\ell\nu_\ell$ by the IW function $\tau(w)$. These IW functions were Taylor-expanded to order $(w-1)$, incorporating $\mathcal{O}(\alpha_s, \Lambda/m_{Q})$ corrections also. Initially, we performed a data-driven analysis to determine the HQET fit parameters. Using synthetic data based on this fit, we then applied the robust $z$-parametrization to the form factors in the BSZ basis. For $B_s \to D_{s0}^*\ell\nu_\ell$, we performed a combined fit with the available LCSR predictions at $q^2 = 0$. The approach used in this work enabled us to obtain precise predictions of the LFU ratios for these decay modes. Our Standard Model results can be further refined with improved experimental data on light lepton channels, and with calculations from LQCD and LCSR for the form factors. We also examined NP sensitivity of various observables predicted by our fits, assuming that NP effects manifest only in the $\tau$ channel. To assess NP sensitivity, we adopted benchmark values for the Wilson coefficients derived from a global analysis of available $b \to c\tau\nu_\tau$ data, considering frameworks like WET, SMEFT, and 2HDM. In the low-energy WET, all the decay modes were found to be sensitive to the complex $C_T$ coupling, while $ B_s \to D_{s0}^*\tau\nu_\tau$ was additionally sensitive to the complex $C_{S_2}$ coupling. Several observables such as $A_{fb,\tau}^{D_{s0}^*}$, $R_{D_{s1}}$, and $F_L^{D_{s1}}$ exhibited deviations exceeding $2\sigma$. The observable $P_\tau^{D_{s1}}$ features a zero-crossing in the SM, which is absent in the presence of complex $C_T$. These key results can be tested experimentally. In the SMEFT context, relevant at high energy scales, only $A_{fb,\tau}^{D_{s0}^*}$ impacted by $C_{lequ}^{(1)}$ showed a deviation over $2\sigma$ from the SM, while other observables aligned with the SM at this significance level. The zero-crossing of $P_\tau^{D_{s1}}$ in the SM is also absent with complex $C_{lequ}^{(3)}$, providing another testable probe of the underlying WCs. Within the 2HDM, we found significant deviations from the SM for observables in $B_s \to D_{s1}^{(*)}\tau\nu_\tau$, with $A_{fb,\tau}^{D_{s1}^*}$, $P_\tau^{D_{s1}^*}$, $R_{D_{s1}}$, $A_{fb,\tau}^{D_{s1}}$, and $P_\tau^{D_{s1}}$ differing from the SM predictions by over $2\sigma$. Notably, $A_{fb,\tau}^{D_{s1}}$ showed deviations as large as $5\sigma$ in the 2HDM, making it a promising NP signature test. The absence of zero-crossings for $A_{fb,\tau}^{D_{s1}^*}$, $P_\tau^{D_{s1}^*}$, $A_{fb,\tau}^{D_{s1}}$, and $P_\tau^{D_{s1}}$ in the 2HDM should also be experimentally testable. Overall, our findings demonstrate that the $ B_s \to D_s^{**}\ell\nu_\ell$ decay modes are highly sensitive to scalar and tensor NP couplings.

\acknowledgments
BM and SS acknowledge the support of the Anusandhan National
Research Foundation (ANRF), Government of India, under research grant no. ANRF/IRG/2024/000256/PS. TK is supported by the Council of Scientific and Industrial Research (CSIR),
Government of India, through a research fellowship (grant no. 09/0414(15918)/2022-EMR-I).
 
\appendix

\section{Angular Coefficients and Helicity Amplitudes}\label{app:ang_coeff}
We present only the helicity amplitudes and angular coefficients relevant to the SM calculations. The expressions in the presence of NP can be found in refs.~\cite{Sakaki:2013bfa,Mandal:2019vwq,Watanabe:2017mip,Becirevic:2020rzi,Becirevic:2019tpx}.

For $B_s \to D_{s0}^*\ell\nu_\ell$,
\begin{eqnarray}\label{eq:abc_Ds0s}
	a &=& \frac{G_F^2 V_{cb}^2}{256 \pi^3 m_{B_s}^3} q^2 \sqrt{\lambda(m_{B_s}^2,m_{D_{s0}^*}^2,q^2)} \left(1 - \frac{m_\ell^2}{q^2}\right)^2 \left[|H_0|^2 + \frac{m_\ell^2}{q^2}|H_t|^2\right]\, , \nonumber \\
	b &=& \frac{G_F^2 V_{cb}^2}{128 \pi^3 m_{B_s}^3} q^2 \sqrt{\lambda(m_{B_s}^2,m_{D_{s0}^*}^2,q^2)} \left(1 - \frac{m_\ell^2}{q^2}\right)^2  \frac{m_\ell^2}{q^2} H_0H_t \, , \nonumber \\
	c &=& -\frac{G_F^2 V_{cb}^2}{256 \pi^3 m_{B_s}^3} q^2 \sqrt{\lambda(m_{B_s}^2,m_{D_{s0}^*}^2,q^2)} \left(1 - \frac{m_\ell^2}{q^2}\right)^3 |H_0|^2 \, .
\end{eqnarray}
For $B_s \to D_{s1}^{(*)}\ell\nu_\ell$,
\begin{eqnarray}\label{eq:abc_Ds1s}
	a &=& \frac{G_F^2 V_{cb}^2}{64 (2\pi)^3 m_{B_s}^3} q^2 \sqrt{\lambda(m_{B_s}^2,m_{D_{s1}^{(*)}}^2,q^2)} \left(1 - \frac{m_\ell^2}{q^2}\right)^2 \bigg[(|H_+|^2 + |H_-|^2 + 2|H_0|^2) \nonumber\\ && + \frac{m_\ell^2}{q^2} (|H_+|^2 + |H_-|^2 + 2|H_t|^2) \bigg]\, , \nonumber\\
	b &=& \frac{G_F^2 V_{cb}^2}{32 (2\pi)^3 m_{B_s}^3} q^2 \sqrt{\lambda(m_{B_s}^2,m_{D_{s1}^{(*)}}^2,q^2)} \left(1 - \frac{m_\ell^2}{q^2}\right)^2 \bigg[ |H_+|^2 - ||H_-|^2 + 2 \frac{m_\ell^2}{q^2} H_0 H_t \bigg]\, , \nonumber\\
	c &=& \frac{G_F^2 V_{cb}^2}{64 (2\pi)^3 m_{B_s}^3} q^2 \sqrt{\lambda(m_{B_s}^2,m_{D_{s1}^{(*)}}^2,q^2)} \left(1 - \frac{m_\ell^2}{q^2}\right)^2 \bigg[(|H_+|^2 + |H_-|^2 - 2|H_0|^2) \nonumber\\ && - \frac{m_\ell^2}{q^2} (|H_+|^2 + |H_-|^2 - 2|H_0|^2) \bigg]\, .
\end{eqnarray}
For $B_s \to D_{s2}^{*}\ell\nu_\ell$,
\begin{eqnarray}\label{eq:abc_Ds2s}
	a &=& \frac{G_F^2 V_{cb}^2}{2^9 \pi^3 m_{B_s}^3} q^2 \sqrt{\lambda(m_{B_s}^2,m_{D_{s2}^*}^2,q^2)} \left(1 - \frac{m_\ell^2}{q^2}\right)^2 \bigg[(|H_+|^2 + |H_-|^2 + 2|H_0|^2) \nonumber\\ && + \frac{m_\ell^2}{q^2} (|H_+|^2 + |H_-|^2 + 2|H_t|^2) \bigg]\, ,\nonumber\\
	b &=& \frac{G_F^2 V_{cb}^2}{2^8 \pi^3 m_{B_s}^3} q^2 \sqrt{\lambda(m_{B_s}^2,m_{D_{s2}^{*}}^2,q^2)} \left(1 - \frac{m_\ell^2}{q^2}\right)^2 \bigg[ |H_+|^2 - ||H_-|^2 + 2 \frac{m_\ell^2}{q^2} H_0 H_t \bigg] \, , \nonumber\\
	c &=& \frac{G_F^2 V_{cb}^2}{2^9 \pi^3 m_{B_s}^3} q^2 \sqrt{\lambda(m_{B_s}^2,m_{D_{s2}^{*}}^2,q^2)} \left(1 - \frac{m_\ell^2}{q^2}\right)^2 \bigg[(|H_+|^2 + |H_-|^2 - 2|H_0|^2) \nonumber\\ && - \frac{m_\ell^2}{q^2} (|H_+|^2 + |H_-|^2 - 2|H_0|^2) \bigg]\, .
\end{eqnarray}

For $B_s \to D_{s0}^*\ell\nu_\ell$ decays,
\begin{eqnarray}
	H_0 &=& \sqrt{\frac{\lambda(m_{B_s}^2,m_{D_{s0}^{*}}^2,q^2)}{q^2}} f_+(q^2) \, , \nonumber\\
	H_t &=& \frac{m_{B_s}^2 - m_{D^*_{s0}}^2}{\sqrt{q^2}} f_0(q^2) \, .
\end{eqnarray}
For $B_s \to D_{s1}^{(*)}\ell\nu_\ell$,
\begin{eqnarray}
	H_+ &=& (m_{B_s} + m_{D_{s1}^{(*)}}) V_1(q^2) - \frac{\sqrt{\lambda(m_{B_s}^2,m_{D_{s1}^{(*)}}^2,q^2)}}{M_{B_s} + M_{D_{s1}^{(*)}}} A(q^2)\, , \nonumber\\
	H_- &=& (m_{B_s} + m_{D_{s1}^{(*)}}) V_1(q^2) + \frac{\sqrt{\lambda(m_{B_s}^2,m_{D_{s1}^{(*)}}^2,q^2)}}{M_{B_s} + M_{D_{s1}^{(*)}}} A(q^2)\, , \nonumber\\
	H_0 &=& \frac{m_{B_s} + m_{D_{s1}^{(*)}}}{2 m_{D_{s1}^{(*)}} \sqrt{q^2}} \left[-(m_{B_s}^2 - m_{D_{s1}^{(*)}}^2 - q^2) V_1(q^2) + \frac{\lambda(m_{B_s}^2,m_{D_{s1}^{(*)}}^2,q^2)}{m_{B_s} + m_{D_{s1}^{(*)}}} V_2(q^2) \right]\, , \nonumber\\
	H_t &=& -\sqrt{\frac{\lambda(m_{B_s}^2,m_{D_{s1}^{(*)}}^2,q^2)}{q^2}} V_0(q^2)\, .
\end{eqnarray}
For $B_s \to D_{s2}^*\ell\nu_\ell$,
\begin{eqnarray}
	H_+ &=& \frac{\sqrt{\lambda(m_{B_s}^2,m_{D_{s2}^{*}}^2,q^2)}}{2 \sqrt{2} m_{B_s} m_{D_{s2}^*}} \left( (m_{B_s} + m_{D^*_{s2}}) A_1(q^2) - \frac{\sqrt{\lambda(m_{B_s}^2,m_{D_{s2}^{*}}^2,q^2)}}{m_{B_s} + m_{D^*_{s2}}} V(q^2) \right)\, , \nonumber \\
	H_- &=& \frac{\sqrt{\lambda(m_{B_s}^2,m_{D_{s2}^{*}}^2,q^2)}}{2 \sqrt{2} m_{B_s} m_{D_{s2}^*}} \left( (m_{B_s} + m_{D^*_{s2}}) A_1(q^2) + \frac{\sqrt{\lambda(m_{B_s}^2,m_{D_{s2}^{*}}^2,q^2)}}{m_{B_s} + m_{D^*_{s2}}} V(q^2) \right)\, , \nonumber \\
	H_0 &=& \frac{\sqrt{\lambda(m_{B_s}^2,m_{D_{s2}^{*}}^2,q^2)}}{\sqrt{6} m_{B_s} m_{D_{s2}^*}}\frac{(m_{B_s} + m_{D_{s2}^*})}{2 m_{D_{s2}^*}\sqrt{q^2}} \bigg( (m_{B_s}^2 - m_{D_{s2}^*}^2 - q^2) A_1(q^2) \nonumber \\ && - \frac{\lambda(q^2)}{(m_{B_s} + m_{D_{s2}^*})^2} A_2(q^2) \bigg)\, , \nonumber\\
	H_t &=& \frac{\sqrt{\lambda(m_{B_s}^2,m_{D_{s2}^{*}}^2,q^2)}}{\sqrt{6} m_{B_s} m_{D_{s2}^*}} \sqrt{\frac{\lambda(m_{B_s}^2,m_{D_{s2}^{*}}^2,q^2)}{q^2}} A_0(q^2)\, .
\end{eqnarray}

\section{Details of HQET Form Factors}\label{App:ffhqet}
The form factor expressions presented below are taken from refs.~\cite{Bernlochner:2016bci,Bernlochner:2017jka,Bernlochner:2017jxt}.

For $B_s \to D_{s0}^*$,
\begin{eqnarray}\label{eq:ff_Ds0s}
	g_+ &=& \hat{\alpha}_s(\omega - 1) \frac{C_{A_2} + C_{A_3}}{2} \zeta - \epsilon_c \left[3\frac{\omega\bar{\Lambda}_s^*-\bar{\Lambda}_s}{\omega + 1} \zeta - 2(\omega - 1)\zeta_1\right] - \epsilon_b G_b \, ,\nonumber\\
	g_- &=& \zeta + \hat{\alpha}_s \left[C_{A_1} + (\omega-1) \frac{C_{A_2} - C_{A_3}}{2}\right] \zeta + \epsilon_c[6 \chi_1 - 2(w+1) \chi_2] \, .
\end{eqnarray}

For $B_s \to D_{s1}^*$,
\begin{eqnarray}\label{eq:ff_Ds1s}
	g_{V_1} &=& (\omega - 1) (1 + \hat{\alpha}_s C_{V_1}) \zeta + \epsilon_c \left[(\omega \bar{\Lambda}_s^* - \bar{\Lambda}_s) \zeta - 2(w - 1)\chi_1\right] - \epsilon_b(\omega + 1)G_b\, ,\nonumber\\
	g_{V_2} &=& -\hat{\alpha}_s C_{V_2} \zeta + \epsilon_c (2\zeta_1 - 2\chi_2)\, ,\nonumber\\		
	g_{V_3} &=& -\zeta - \hat{\alpha}_s (C_{V_1} + C_{V_3}) \zeta - \epsilon_c \left( \frac{\omega\bar{\Lambda}_s^* - \bar{\Lambda}_s}{\omega + 1} \zeta + 2\zeta_1 - 2\chi_1 + 2\chi_2\right) + \epsilon_b G_b\, , \nonumber\\
	g_A &=& \zeta + \hat{\alpha}_s C_{A_1} \zeta + \epsilon_c \left( \frac{\omega\bar{\Lambda}_s^* - \bar{\Lambda}_s}{\omega + 1} \zeta - 2\chi_1\right) - \epsilon_b G_b\, .
\end{eqnarray}

For $B_s \to D_{s1}$,
\begin{eqnarray}\label{eq:ff_Ds1}
	f_{V_1} &=& \frac{1}{\sqrt{6}} \bigg\{(1 - \omega^2) (1 + \hat{\alpha_s} C_{V_1}) \tau - \epsilon_b (\omega^2 - 1)F_b - \epsilon_c[4(\omega + 1)(\omega\bar{\Lambda}_s^\prime - \bar{\Lambda}_s) \tau \nonumber\\ &&  - (w^2-1) (3\tau_1 - 3\tau_2 + 2\eta_1 + 3\eta_3)] \bigg\} \, ,\nonumber\\
	f_{V_2} &=& \frac{1}{\sqrt{6}} \bigg\{-3\tau -\hat{\alpha}_s [3 C_{V_1} + 2(1 + \omega) C_{V_2}]\tau - \epsilon_b 3 F_b -\epsilon_c[(4w-1)\tau_1 + 5\tau_2 \nonumber\\ && + 10\eta_1 + 4(w-1)\eta_2 - 5\eta_3] \bigg\} \, ,\nonumber\\
	f_{V_3} &=& \frac{1}{\sqrt{6}} \bigg\{ (\omega - 2)\tau - \hat{\alpha}_s [(2 - \omega) C_{V_1} + 2(1 + \omega) C_{V_3}]\tau + \epsilon_b (2 + \omega) F_b  + \epsilon_c[4(\omega\bar{\Lambda}_s^\prime - \bar{\Lambda}_s)\tau \nonumber\\ && + (2+w)\tau_1  + (2+3w)\tau_2 - 2(6+w)\eta_1  -4(w-1)\eta_2 - (3w-2)\eta_3]\bigg\} \, ,\nonumber\\
	f_A &=& \frac{1}{\sqrt{6}} \bigg\{-(\omega + 1) (1 + \hat{\alpha}_s C_{A_1})\tau - \epsilon_b (\omega - 1) F_b - \epsilon_c[4(\omega\bar{\Lambda}_s^\prime - \bar{\Lambda}_s)\tau \nonumber\\ && -3(w-1)(\tau_1  - \tau_2)  - (w+1)(2\eta_1 + 3\eta_3)] \bigg\}\, .
\end{eqnarray}
For $B_s \to D_{s2}^*$,
\begin{eqnarray}\label{eq:ff_Ds2s}
	k_{V} &=& -\tau - \hat{\alpha_s} C_{V_1}\tau - \epsilon_b F_b - \epsilon_c[\tau_1 - \tau_2 - 2\eta_1 + \eta_3] \, ,\nonumber\\
	k_{A_1} &=& -(\omega + 1) (1 + \hat{\alpha_s} C_{A_1})\tau - \epsilon_b (\omega - 1) F_b - \epsilon_c[(w-1)(\tau_1 - \tau_2) - (w+1)(2\eta_1 - \eta_3)]\, , \nonumber\\
	k_{A_2} &=& \hat{\alpha}_s C_{A_2}\tau - \epsilon_c 2 (\tau_1 + \eta_2)\, , \nonumber\\
	k_{A_3} &=& \tau + \hat{\alpha}_s (C_{A_1} + C_{A_3})\tau + \epsilon_b F_b - \epsilon_c[\tau_1 + \tau_2 - 2\eta_1 - 2\eta_2 - \eta_3]\, ,
\end{eqnarray}
where
\begin{eqnarray*}
	G_b &=& \frac{(1 + 2\omega)\bar{\Lambda}^*_s - (2 + \omega)\bar{\Lambda}_s}{\omega + 1}\zeta - 2 (\omega - 1) \zeta_1 \, ,~~~
	F_b = (\bar{\Lambda}_s + \bar{\Lambda}^\prime_s)\tau - (2\omega+1)\tau_1 - \tau_2\, , \\ \hat{\alpha}_s &=& \alpha_s / \pi \, ,~~~ \epsilon_b = \frac{1}{2m_b} \, ,~~~ \epsilon_c = \frac{1}{2m_c}\, .
\end{eqnarray*}
In the above equations, the $C_i$s denote the Wilson coefficients that include the $\mathcal{O}(\alpha_s)$ QCD corrections arising from matching QCD onto HQET. Their expressions can be found in ref.~\cite{Bernlochner:2017jka}.

\section{Form Factor Relations Between Helicity and HQET Bases}\label{App:ffrel}
For $B_s \to D_{s0}^*\ell\nu_\ell$,
\begin{eqnarray}\label{eq:appDs0s}
	f_+(q^2) &=& \frac{-i}{2 \sqrt{m_{B_s} m_{D_{s0}^*}}} \left((m_{B_s} + m_{D_{s0}^*}) g_+(w) - (m_{B_s} - m_{D_{s0}^*}) g_-(w) \right)\, , \nonumber\\
	f_0(q^2) &=& \frac{i}{2\sqrt{m_{B_s} m_{D_{s0}^*}}} \left(\frac{Q_- \, g_-(w)}{m_{B_s} - m_{D_{s0}^*}} - \frac{Q_+ \, g_+(w)}{m_{B_s} + m_{D_{s0}^*}}\right) \, .
\end{eqnarray}
For $B_s \to D_{s1}\ell\nu_\ell$,
\begin{eqnarray}\label{eq:appDs1s}
	A(q^2) &=& \frac{i(m_{B_s} + m_{D_{s1}})}{2\sqrt{m_{B_s}m_{D_{s1}}}} f_A (w)\, , \nonumber \\
	V_0(q^2) &=& \frac{i}{2\sqrt{m_{B_s}m_{D_{s1}}}} \left( m_{B_s} f_{V_1} (w) + \frac{m_{B_s}^2 - m^2_{D_{s1}} + q^2}{2 m_{B_s}} f_{V_2} (w) + \frac{m_{B_s}^2 - m^2_{D_{s1}} - q^2}{2 m_{D_{s1}}} f_{V_3} (w) \right)\, , \nonumber \\
	V_1(q^2) &=& \frac{i\sqrt{m_{B_s} m_{D_{s1}}}}{m_{B_s} + m_{D_{s1}}} f_{V_1} (w)\, , \nonumber \\
	V_2(q^2) &=& -\frac{i(m_{B_s} + m_{D_{s1}})}{2\sqrt{m_{B_s}m_{D_{s1}}}} \left( f_{V_3} (w) + \frac{m_{D_{s1}}}{m_{B_s}} f_{V_2} (w) \right)\, .
\end{eqnarray}
For $B_s \to D_{s1}^*\ell\nu_\ell$, the above equations can be used with the replacements $f_A \to g_A$, $f_{V_1} \to g_{V_1}$, $f_{V_2} \to g_{V_2}$, and $f_{V_3} \to g_{V_3}$. \\

For $B_s \to D^*_{s2}\ell\nu_\ell$,
\begin{eqnarray}\label{eq:appDs2s}
	V(q^2) &=& \frac{i(m_{B_s} + m_{D_{s1}})}{2\sqrt{m_{B_s}m_{D_{s1}}}} f_V (w)\, , \nonumber \\
	A_0(q^2) &=& \frac{i}{2\sqrt{m_{B_s}m_{D_{s1}}}} \left( m_{B_s} f_{A_1} (w) + \frac{m_{B_s}^2 - m_{D_{s1}} + q^2}{2 m_{B_s}} f_{A_2} (w) + \frac{m_{B_s}^2 - m_{D_{s1}} - q^2}{2 m_{D_{s1}}} f_{A_3} (w) \right)\, , \nonumber\\
	A_1(q^2) &=& \frac{i\sqrt{m_{B_s} m_{D_{s1}}}}{m_{B_s} + m_{D_{s1}}} f_{A_1} (w)\, , \nonumber \\
	A_2(q^2) &=& -\frac{i(m_{B_s} + m_{D_{s1}})}{2\sqrt{m_{B_s}m_{D_{s1}}}} \left( f_{A_3} (w) + \frac{m_{D_{s1}}}{m_{B_s}} f_{A_2} (w) \right)\, .
\end{eqnarray}	

\bibliographystyle{JHEP.bst}
\bibliography{refs}

\end{document}